\documentclass{article}

\usepackage{ILD}
\usepackage{graphicx}
\usepackage{caption}
\usepackage{subcaption}
\usepackage{amsmath}
\usepackage{amssymb}
\usepackage{epstopdf}
\usepackage{array}
\usepackage{slashed} 
\usepackage{hyperref}
\usepackage{color}
\usepackage{braket}
\usepackage[english]{babel}
\usepackage{siunitx}
\usepackage{cellspace}
\usepackage[parfill]{parskip} 
\usepackage{float}
\usepackage{multirow}
\usepackage{booktabs}
\usepackage{lineno}
\graphicspath{
    {figs/}
}

\def\to{\rightarrow}

\def\bi{\begin{itemize}}
\def\ei{\end{itemize}}

\def\c1p{C1^\prime}
\def\msq3{\overline{m}_{\tilde{q}}(3)}

\def\tb{\tilde b}

\def\tst{\tilde t}

\def\tg{\tilde g}

\def\be{\begin{equation}}  
\def\ee{\end{equation}}  
\def\bea{\begin{eqnarray}}  
\def\eea{\end{eqnarray}}  
\def\tw{\tilde\chi}

\def\tz{\tilde\chi^0}
\newcommand{\alt}{\mbox{$\;\raisebox{-1mm}{$\stackrel{\scriptstyle<}{\scriptstyle\sim}$}\;$}}
\newcommand{\agt}{\mbox{$\;\raisebox{-1mm}{$\stackrel{\scriptstyle>}{\scriptstyle\sim}$}\;$}}

\newcommand{\LSP}{\ensuremath{\widetilde{\chi}_1^0}}
\newcommand{\neutralinotwo}{\ensuremath{\widetilde{\chi}_2^0}}
\newcommand{\neutralinothree}{\ensuremath{\widetilde{\chi}_3^0}}
\newcommand{\neutralinofour}{\ensuremath{\widetilde{\chi}_4^0}}
\newcommand{\neutralino}{\ensuremath{\widetilde{\chi}_i^0}}
\newcommand{\charginoone}{\ensuremath{\widetilde{\chi}_1^{\pm}}}
\newcommand{\charginoonep}{\ensuremath{\widetilde{\chi}_1^{+}}}
\newcommand{\charginoonem}{\ensuremath{\widetilde{\chi}_1^{-}}}
\newcommand{\charginotwo}{\ensuremath{\widetilde{\chi}_2^{\pm}}}

\newcommand{\stopone}{\widetilde{t}_{1}}
\newcommand{\stoptwo}{\widetilde{t}_{2}}

\newcommand{\sbottomone}{\widetilde{b}_{1}}
\newcommand{\sbottomtwo}{\widetilde{b}_{2}}
\newcommand{\supL}{\widetilde{u}_{L}}
\newcommand{\supR}{\widetilde{u}_{R}}
\newcommand{\gluino}{\widetilde{g}}

\newcommand{\GeV}{\text{ GeV}}

\def\beq{\begin{equation}}
\def\eeq#1{\label{#1}\end{equation}}
\def\eeqn{\end{equation}}

\newenvironment{Eqnarray}%
   {\arraycolsep 0.14em\begin{eqnarray}}{\end{eqnarray}}
\def\beqa{\begin{Eqnarray}}
\def\eeqa#1{\label{#1}\end{Eqnarray}}
\def\eeqan{\end{Eqnarray}}

\title{The ILC as a natural SUSY discovery machine \\
and precision microscope:\\
from light higgsinos to tests of unification}

\date{\today}

\addauthor{Howard Baer}{\institute{1}}
\addauthor{Mikael Berggren}{\institute{2}}
\addauthor{Keisuke Fujii}{\institute{3}}
\addauthor{Jenny List}{\institute{2}}
\addauthor{Suvi-Leena Lehtinen}{\institute{4}\institute{2}}
\addauthor{Tomohiko Tanabe}{\institute{5}}
\addauthor{Jacqueline Yan}{\institute{3}}

\addinstitute{1}{University of Oklahoma, Norman, OK 73019, USA}
\addinstitute{2}{DESY, Notkestrasse 85, 22607 Hamburg, Germany}
\addinstitute{3}{High Energy Accelerator Research Organization (KEK), 1-1 Oho, Tsukuba, Ibaraki, 305-0801, Japan}
\addinstitute{4}{Institut f\"ur Experimental Physik, University of Hamburg, Luruper Chaussee 149,
22761 Hamburg, Germany}
\addinstitute{5}{ICEPP, The University of Tokyo, 7-3-1 Hongo, Bunkyo-ku, Tokyo, 113-0033, Japan}

\abstract{
The requirement of electroweak naturalness in simple supersymmetric models 
implies the existence of a cluster of four light higgsinos with 
mass $\sim 100-300$\,GeV, the lighter the better. 
While such light compressed spectra may be challenging to observe at LHC, 
  the International Linear $e^+e^-$ Collider (ILC) with $\sqrt{s}>2m_{\rm higgsino}$ 
would serve as both a SUSY discovery machine and a precision microscope.
We study higgsino pair production signatures at the ILC based on full, \texttt{Geant4-}based 
simulation of the ILD detector concept.
We examine several benchmark scenarios that may be challenging for discovery
at HL-LHC due to mass differences between the higgsino states between $20$ and $4$\,GeV.
Assuming $\sqrt{s}= 500$\,GeV and 1000\,fb$^{-1}$ of integrated luminosity, 
the individual higgsino masses can be measured to $1-2\%$ precision in case of the larger mass differences,
and at the level of $5\%$ for the smallest mass difference case.
The higgsino mass splittings are sensitive to the electroweak gaugino masses 
and allow extraction of gaugino masses to 
$\sim 3-20\%$ (depending on the model). 
Extrapolation of gaugino masses via renormalization group running
can test the hypothesis of gaugino mass unification. 
We also examine a case with natural generalized mirage mediation where the 
unification of gaugino masses at an intermediate scale
apparently gives rise to a natural SUSY spectrum somewhat beyond the reach of 
HL-LHC.
}

\titlecomment{This work was carried out in the framework of the ILD detector
concept group}

\begin{document}



\titlepage

\section{Introduction}

The Standard Model (SM) of particle physics has been spectacularly confirmed 
across a broad array of measurements and often to very high precision at the LHC.
The crowning achievement was to establish the existence of a physical scalar (Higgs) boson with 
mass $m_h=125.09\pm 0.24$\,GeV~\cite{lhc_h}. 
In spite of this impressive success, the narrative brings with it cause for concern:
quantum mechanical contributions to the Higgs mass rapidly exceed $m_h$ for energy fluctuations
of order $\Lambda\sim 1$\,TeV~\cite{Susskind:1978ms}. 
These {\em quadratic divergences} necessitate ever more incredulous
fine-tunings to maintain $m_h\simeq 125$ GeV as the excluded energy scale of new physics increases.
In addition, the SM is lacking the necessary ingredients to explain 
e.g.\  cosmic inflation, the existence of dark matter and dark energy in the universe, 
the origin of the matter-antimatter asymmetry and a suppression of $CP$-violation in the strong interactions.

A rather minimal extension of the SM -- moving from the Poincar\'e group to the more general 
super-Poincar\'e group (supersymmetry or SUSY) of space-time symmetries -- 
allows for solutions or improvements of all these problems. 
The added spacetime supersymmetry guarantees cancellation of the offending quadratic divergences 
to all orders in perturbation theory thus rendering the Higgs field natural. 
The allowance for a vast assortment of scalar fields in SUSY, as expected from string theory, 
allows for many possible inflaton candidate fields and for a non-zero minimum of the 
scalar potential, yielding a cosmological constant. 
The lightest SUSY particle and/or the inclusion of an axion 
(necessary for solving the strong $CP$ problem)
yields dark matter while scalar field flat direction (Affleck-Dine)  baryogenesis and 
various other thermal and non-thermal leptogenesis mechanisms seem automatic in SUSY.
In addition, SUSY receives indirect support from 
(1) the measured values of gauge couplings which unify under 
Minimal Supersymmetric Standard Model (MSSM) renormalization group evolution, 
(2) the measured value of the top mass, which is just right
to produce a radiative breakdown of electroweak gauge symmetry, and 
(3) the measured value of $m_h\simeq 125$\,GeV which lies 
squarely within the prediction of $m_h\alt 135$\,GeV required by the MSSM.  

In spite of these theoretical successes, many physicists have developed a large degree of skepticism 
regarding the eventual emergence of SUSY at experimental facilities. 
This arises due to 
(1) the lack of evidence for superpartners at LHC and 
(2) the rather large value of $m_h$ that has been found. 
The first of these is exemplified by the latest gluino mass limits: that $m_{\tg}\agt 2$\,TeV 
in many simplified models, which may be compared against early projections by Barbieri-Giudice (BG)~\cite{Ellis:1986yg,Barbieri:1987fn} 
where electroweak naturalness requires $m_{\tg}\alt 400$ GeV for fine-tuning parameter $\Delta_{\rm BG}\alt 30$. 
Secondly, a value of $m_h\simeq 125$\,GeV requires~\cite{Carena:2002es} 
within the MSSM the existence of highly mixed (large trilinear soft parameters $A_t$) 
TeV-scale top squarks $\tst_{1,2}$. 
This may be contrasted with Dimopoulos-Giudice naturalness~\cite{Dimopoulos:1995mi} that $m_{\tst_1}\alt 350$ GeV 
for $\Delta_{\rm BG}<30$ or that
$m_{\tst_{1,2},\tb_1}\alt 500$\,GeV from requiring 
$\delta m_{H_u}^2\alt m_h^2$~\cite{Kitano:2006gv,Papucci:2011wy}.
Thus, in the LHC era, the question of electroweak naturalness has been elevated to one of 
prime importance which can serve as a guide for construction of future 
experimental facilities.\footnote{
In Ref.~\cite{Arkani-Hamed:2015vfh}, it is declared that 
``Settling the ultimate fate of naturalness is perhaps the most profound theoretical question of our time $\cdots$
and will largely dictate the future of fundamental physics in this century.''}

The most direct connection between the weak scale, as exemplified by the weak gauge and 
Higgs boson masses $m_{W,Z,h}$ and the SUSY Lagrangian parameters,
arises from the scalar potential minimization condition~\cite{Baer:2006rs}
\begin{equation}
 \frac{1}{2}m_Z^2=\frac{(m_{H_d}^2+\Sigma_d^d)-(m_{H_u}^2+\Sigma_u^u)\tan^2\beta}{(\tan^2\beta
  -1)} - \mu^2 \simeq -m_{H_u}^2-\Sigma_u^u-\mu^2
\label{eq:mzs}
\end{equation}
where the latter partial equality arises for moderate to large values of the ratio of Higgs 
vevs $\tan\beta\equiv v_u/v_d$.
The $\mu$ term arises as a mass term in the MSSM superpotential; 
thus, it is supersymmetry conserving and feeds mass both to
the SM particles $W,\ Z$ and $h$ and also the SUSY higgsinos. 
The {\em weak scale} soft SUSY breaking term
$m_{H_u}^2$ feeds mass just to $W$, $Z$ and $h$ (and other SUSY Higgs via suppressed mixing). 
The $\Sigma_u^u$ are radiative corrections (for a full listing, see Ref.~\cite{Baer:2012cf}),
the largest of which typically arise from the top-squark contributions. 
The MSSM may be considered as natural if there
are no large, unnatural cancellations (fine-tunings) on the right-hand-side of Eq.~\ref{eq:mzs}.
A naturalness measure $\Delta_{\rm EW}$ has been proposed which considers the ratio of the largest element on the 
right-hand-side (RHS) of Eq.~\ref{eq:mzs} to $m_Z^2/2$. 
Fine-tuning of $m_Z$ sets in for values of $\Delta_{\rm EW}\agt 20-30$ and is visually displayed 
in Fig. 1 of Ref.~\cite{Baer:2015rja}.

The validity of the early naturalness estimates using the BG measure has been challenged in that 
these calculations are performed using multiple-soft-parameter effective theories instead of more 
fundamental theories in which the soft parameters
are all related~\cite{Baer:2013gva}. 
Using correlated soft parameters, the BG measure reduces to the EW measure~\cite{Baer:2014ica}. 
The validity of naturalness estimates using $\delta m_{H_u}^2/m_h^2$ has been challenged in that, 
in an effort to simplify,  several contributions to $m_h^2$
and $\delta m_{H_u}^2$ have been set to zero. By including these pieces, then one allows for 
{\em radiatively-driven naturalness} (RNS)~\cite{Baer:2012up} wherein large, 
seemingly unnatural high scale values of $m_{H_u}^2$ are driven to natural values at the weak scale. 
The revised measure is thus brought into accord with $\Delta_{\rm EW}$~\cite{Baer:2013gva,Baer:2014ica}.

From Eq.~\ref{eq:mzs}, the requirements for electroweak naturalness are then
\begin{itemize}
\item The superpotential $\mu$ parameter, bounded from below by $\mu\agt 100$\,GeV due to chargino 
searches at LEP2, 
is not too far from $m_Z$: $\mu\sim 100-300$\,GeV, the lower the better. 
This immediately implies the existence of several higgsino-like
electroweakinos in SUSY with 
$m_{\tw_1^\pm}\sim m_{\tz_{1,2}}\sim |\mu |$.\footnote{It is possible that non-holonomic soft terms may arise allowing for higher mass higgsinos
without compromising naturalness\cite{Ross:2016pml}. 
Such ``semi-soft'' mass terms are expected to be of order 
$m_{weak}^2/m_P$\cite{Martin:1999hc} but in the case where the 
mediation scale is arranged 
to be far lower than the usual Planck scale 
(as expected for gravity-mediation), then these terms can become much larger.} 
\item The soft term $m_{H_u}^2$, which must be driven to negative values to initiate a breakdown of electroweak symmetry, 
is driven to small and not large negative values. 
\item The radiative corrections $\Sigma_u^u(\tst_{1,2})$ are actually minimized for 
TeV-scale highly mixed top squarks. 
These same conditions lift the Higgs mass to $m_h\simeq 125$\,GeV.
Detailed evaluations require $m_{\tst_1}\alt 3$ TeV for $\Delta_{\rm EW}<30$~\cite{Baer:2012cf}.
\item The gluino mass contributes at two-loop level to Eq.~\ref{eq:mzs} via the $\Sigma_u^u(\tst_{1,2})$. 
Detailed evaluations require $m_{\tg}\alt 6$\,TeV for $\Delta_{\rm EW}<30$.\footnote{In the case
of natural anomaly-mediated SUSY breaking~\cite{Baer:2018hwa}, 
the gluino mass bound increases to $m_{\tg}\alt 9$\,TeV.} 
This may be compared to the ultimate reach of HL-LHC which extends to about $m_{\tg}\sim 2.8$\,TeV 
(at the $5\sigma$ discovery level~\cite{Baer:2018hpb}).
\end{itemize}
Thus, HL-LHC will probe only a portion of natural SUSY parameter space via gluino 
and top squark pair production searches.

\begin{figure}[tbp]
\centering
\includegraphics[height=0.35\textheight]{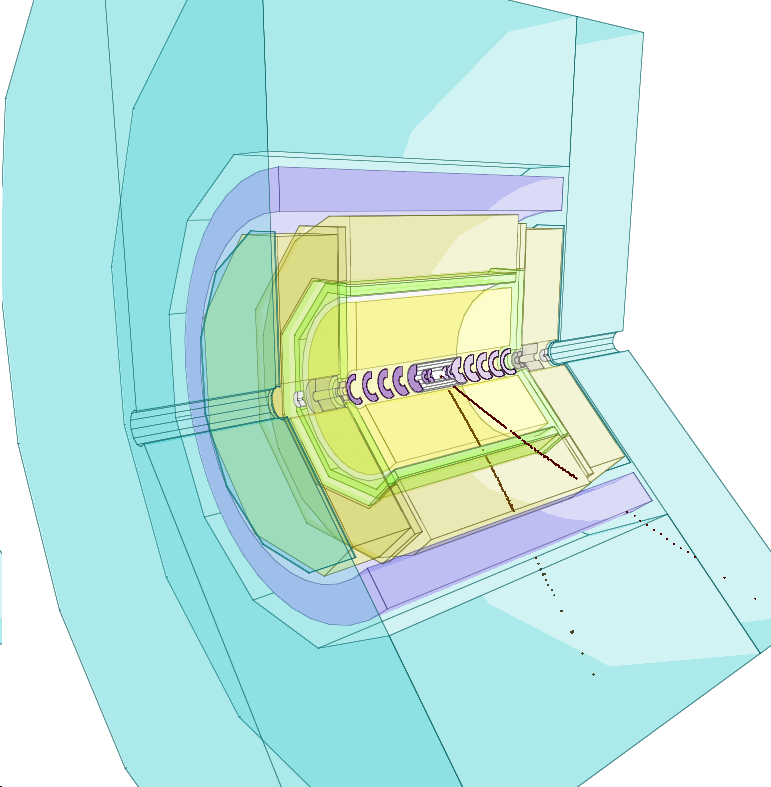}
\caption{ILD event display of a simulated $e^+e^-\rightarrow\tz_1\tz_2$ event
with $\tz_2\rightarrow\mu^+\mu^-\tz_1$.
\label{fig:display}}
\end{figure}

The naturalness-required light higgsinos may be produced at decent rates at LHC 
but their relatively compressed spectra imply low visible energy release from their decays. 
Thus, light higgsinos are very challenging to see at LHC~\cite{Baer:2014kya,Baer:2018hpb}. 
In contrast, the International Linear $e^+e^-$ Collider (ILC) with 
$\sqrt{s}>2m(higgsino)$  would be a {\em higgsino factory} in addition to being a
Higgs factory. 
The reactions $e^+e^-\to \tw_1^+\tw_1^-$ and $\tz_2\tz_1$ should occur at rates 
comparable to muon pair production and at rates exceeding $Zh$ production~\cite{Baer:2014yta}.
The expected mass gaps $m_{\tw_1}-m_{\tz_1}\sim m_{\tz_2}-m_{\tz_1}\sim 3-20$\,GeV lead to 
events which are easily identified at ILC: see Fig.~\ref{fig:display} for a
simulated $e^+e^-\rightarrow\tz_1\tz_2$ with $\tz_2\rightarrow\mu^+\mu^-\tz_1$
event display with light higgsinos in the ILD detector.

The cleanliness of ILC higgsino pair production events along with tunable beam energy 
and beam polarization should allow for a rich program of higgsino measurements.
While the higgsino masses should be comparable to the superpotential $\mu$ parameter, 
thus allowing for a determination of $\mu$, the higgsino {\em mass splittings} depend 
sensitively on the weak scale gaugino masses $M_1$ (bino) and $M_2$ (wino). 
Thus, precision measurements of $m_{\tw_1^\pm}$ and $m_{\tz_{1,2}}$ should allow for an 
extraction of $M_1$ and $M_2$ to good precision.
Once the soft breaking gaugino masses are known, then 
the physical masses of the heavier neutralinos and charginos 
can also be found. The fitted values of $M_1$ and $M_2$  
can be extrapolated to high energies to test the hypothesis of 
gaugino mass unification. 
If gluinos are discovered at LHC, then $M_3$ (gluino) may be 
extracted\cite{Baer:2016wkz} and unification of all three gaugino masses 
may be tested.

In this paper, we first present in Sec.~\ref{sec:bm} two natural SUSY benchmark 
models labeled ILC1 and ILC2 that arise from the non-universal Higgs model (NUHM2)\cite{nuhm2}. 
These models allow for $\mu$ as an input parameter so that
SUSY spectra with a low value of $\Delta_{\rm EW}$ can easily be generated. 
The NUHM2 model incorporates gaugino mass unification so that under renormalization
group (RG) evolution $M_1$ and $M_2$ should unify at the scale $m_{\rm GUT}\simeq 2\times 10^{16}$\,GeV.
We also propose a natural generalized {\em mirage mediation} (nGMM) 
benchmark model which instead has gaugino mass unification at the
mirage scale $\mu_{mir}=m_{\rm GUT}e^{-8\pi^2/\alpha} \sim 5\times 10^7$\,GeV where $\alpha =4$ 
parametrizes the relative amounts of modulus-mediation versus anomaly-mediation
in SUSY breaking. 
In this case, by determining the mirage unification scale $\mu_{mir}$, 
ILC can measure the strength $\alpha$ of moduli- vs. anomaly-mediation.
If the gaugino masses are extrapolated beyond the mirage scale to the GUT scale, 
then ILC can also indirectly measure the underlying {\em gravitino mass} 
$m_{3/2}$~\cite{Baer:2016hfa}!
Thus, in such cases ILC would allow for a window into the nature of the 
laws of physics at energy scales far beyond $\sqrt{s}\sim 0.5-1$\,TeV.

All three benchmarks have been studied in a detailed, \texttt{Geant4-}based simulation of the ILD detector 
concept~\cite{ILD}, which we introduce in Sec.~\ref{sec:sim}. 
In  Sec.~\ref{sec:event_selection}, we present a detailed portrait of various higgsino 
pair production measurements at ILC with $\sqrt{s}=500$\,GeV. 
Continuum measurements of energy and invariant mass distributions of higgsino decay 
products should allow for extraction of higgsino masses to percent level 
or better precision.

In Sec.~\ref{sec:fit}, we present results from our calculations using the \texttt{Fittino}~\cite{Bechtle:2004pc} program 
to extract fits of fundamental {\em weak scale} MSSM Lagrangian parameters, 
especially the gaugino masses $M_1$ and $M_2$. 
We also obtain predictions for the masses of many of the 
kinematically inaccessible superparticles. 
We then can extract the underlying GUT scale parameters if we assume a particular 
high scale SUSY model such as NUHM2, NUHM3 or nGMM.
If the gluino is discovered at LHC, the extracted gaugino masses may be augmented 
with the $SU(3)$ gaugino mass $M_3$.
We also show that the thermally-produced relic density of WIMPs may be extracted, 
thus testing the WIMP-only versus mixed axion-WIMP dark matter hypotheses.
In Sec.~\ref{sec:test}, we show results from running the gaugino masses to 
high energy scales, thus offering a test of the unification hypothesis and the
underlying SUSY breaking mechanism.
A summary and conclusions are presented  in Sec.~\ref{sec:conclude}.

\section{Benchmark models}
\label{sec:bm}  

In this Section, we present three natural SUSY benchmark points 
which have been used for the detailed ILD studies described in Sec.~\ref{sec:sim} and Sec.~\ref{sec:event_selection}.

\subsection{ILC1 benchmark model}
\label{ssec:ILC1}

\begin{figure}[tbp]
\centering
\includegraphics[height=0.35\textheight]{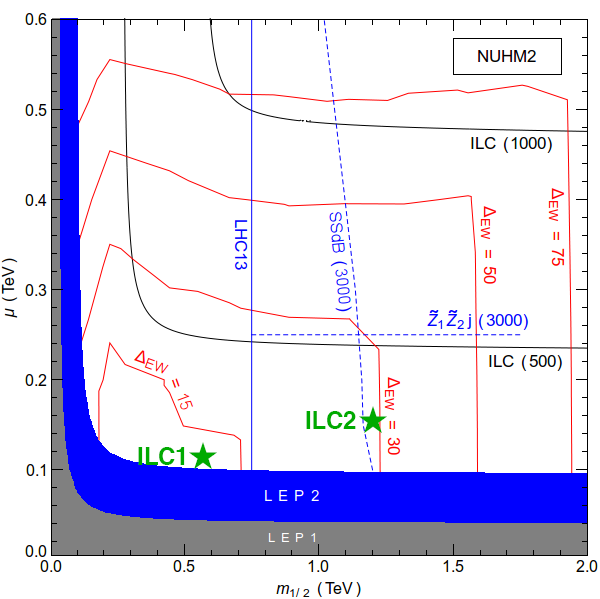}
\caption{The $M_{1/2}$ vs. $\mu$ plane in the NUHM2 model for $\tan\beta =15$, 
$M_0=5$\,TeV, $A_0=-8$\,TeV and $M_A=1$\,TeV. 
We show contours of $\Delta_{\rm EW}$ along with some limits from LHC13 
and the future reach of ILC with $\sqrt{s}=500$ and 1000 GeV and 
HL-LHC (via same-sign diboson production labeled 
SSdB (3000) and via neutralino associated production labeled $\widetilde{Z}_1\widetilde{Z}_2j$ (3000) in the nomenclature of Ref. \cite{Baer:2016usl} where
$\widetilde{Z}_i$ denotes neutralino eigenstate $i$).
Location of benchmark points is indicated in green. 
To aid the reader, we note that $m_{\tg}\sim 2.5\,M_{1/2}$.
The figure is adapted from Ref.~\cite{Baer:2016usl}.
\label{fig:pspace}}
\end{figure}

The ILC1 benchmark point, whose parameters are listed in Tab.~\ref{tab:bm}, 
has been presented previously~\cite{Baer:2012up} and has been used for 
some detailed ILC studies using a toy detector simulation~\cite{Baer:2014yta}. 
The ILC1 benchmark point was generated within the NUHM2 model 
with input parameters and output masses as listed.  
While the various matter scalars are essentially decoupled for LHC 
and ILC physics, the spectrum does contain light higgsinos
with mass $m_{\tw_1}=117.3$\,GeV and $m_{\tz_{1,2}}=102.7$ and $124$\,GeV, respectively,
so that higgsino pair production should already turn on for ILC with $\sqrt{s}>227$\,GeV.
The associated mass gaps, which play a central role in these analyses, are 
$m_{\tw_1^\pm}-m_{\tz_1}=14.6$\,GeV and $m_{\tz_2}-m_{\tz_1}=21.3$\,GeV.
The model contains a light Higgs scalar with $m_h=125.3$\,GeV due to highly mixed TeV-scale 
top squarks with $m_{\tst_1}=1893.3$\,GeV (well beyond the reach of HL-LHC~\cite{Baer:2012vr}
where the 95\% CL exclusion reach extends to $m_{\tst_1}\sim 1400$\,GeV~\cite{ATLAS:2013hta}).
The model is highly electroweak natural with $\Delta_{\rm EW}=14$ corresponding to just
7\% fine-tuning. 
This benchmark is now likely excluded by LHC: The gluino mass $m_{\tg}=1563.5$\,GeV is
excluded by LHC13 searches with 
$\sim 36$\,fb$^{-1}$. 
The searches for higgsinos with small mass-splittings performed
by CMS  ~\cite{Sirunyan:2018iwl} and ATLAS \cite{Aaboud:2017leg} are done for
a spectrum different from our benchmark - in particular the mass of $\tilde{\chi}^{\pm}_{1}$ is
assumed to be exactly halfway between those of $\tilde{\chi}^0_{1}$ and $\tilde{\chi}^0_{2}$  -  
but are nevertheless
likely to exclude it.
However, we retain the point for comparison with
previous work. 
If one adopts modest gaugino mass non-universality, then a small 
increase in $M_3(weak) > 2$\,TeV would bring the point in accord with LHC searches.
The relic density of thermally produced higgsino-like LSPs is a factor 13 below
the measured value. Requiring also naturalness in the QCD sector, then one must
bring the axion into the model, and axionic dark matter may constitute the bulk of
the dark matter~\cite{Bae:2013bva}. The location of the ILC1 benchmark point is denoted by one of the green stars in the
$\mu$ vs. $M_{1/2}$ parameter space plane of the NUHM2 model shown in 
Fig.~\ref{fig:pspace}. 
%

\begin{table}[htbp]
\centering
\caption{Input parameters and mass spectrum and rates for
benchmark points ILC1, ILC2 and nGMM1. 
All masses and dimensionful parameters are in GeV units. 
All values have been obtained with {\tt ISASUGRA}.
}
\label{tab:bm}
\begin{tabular}{lcccc}
\hline\hline
\addlinespace[2pt]
                               & units & ILC1          & ILC2    & nGMM1  \\[2pt]
\hline
\addlinespace[2pt]
$M_{0}$                        & [GeV] & 7025.0        & 5000    & -- \\
$M_{1/2}$                      & [GeV] & 568.3         & 1200    & -- \\
$A_{0}$                        & [GeV] & -10427        & -8000   & -- \\
$m_{3/2}$                      & [GeV] &  --           & --      & 75000 \\
$M_1,M_2,M_3$                  & [GeV] &  --           & --      & 3382.5, 2124.4, 1225.8 \\
$\tan\beta$                    & --    &  10           & 15      & 10 \\
$a_3$                          & --    &  --           & --      & 3 \\
$c_m$                          & --    &  --           & --      & 6.9 \\
$\alpha$                       & --    &  --           & --      & 4 \\[2pt]
\hline
\addlinespace[2pt]
$m_h$                          & [GeV] &  125.3        & 125.4   & 124.9  \\
$m_{A}$                        & [GeV] & 1000.0        & 1000    & 2000   \\
$m_{H}$                        & [GeV] & 1006.8        & 1006.7  & 2013.3 \\
$m_{H^{\pm}}$                  & [GeV] & 1003.2        & 1003.2  & 2001.6 \\
$\mu$                          & [GeV] &  115.0        & 150     & 150    \\[2pt]
\hline
\addlinespace[2pt]
$m_{\tilde{g}}$                & [GeV] & 1563.5        & 2832.6        & 2856.5         \\
$m_{\tilde{\chi}^{\pm}_{1}}$, $m_{\tilde{\chi}^{\pm}_{2}}$
                               & [GeV] & 117.3, 513.0  & 158.3, 1017.5 & 158.7, 1791.6  \\
$m_{\tilde{\chi}^0_{1}}$, $m_{\tilde{\chi}^0_{2}}$     
                               & [GeV] & 102.7, 124.0  & 148.1, 157.8  & 151.4, 155.8   \\
$m_{\tilde{\chi}^0_{3}}$, $m_{\tilde{\chi}^0_{4}}$
                               & [GeV] & 267.0, 524.2  & 538.7, 1031.1 & 1526.9, 1799.4 \\[2pt]
\hline 
\addlinespace[2pt]
$m_{ \tilde{u}_{L}}$, $m_{ \tilde{u}_{R}}$
                               & [GeV] & 7021, 7254    & 5440, 5566    & 5267, 5399 \\
$m_{\tilde{t}_{1}}$, $m_{\tilde{t}_{2}}$          
                               & [GeV] & 1893, 4919    & 1774, 3878    & 1433, 3732 \\[2pt]
\hline 
\addlinespace[2pt]
$m_{ \tilde{d}_{L}}$, $m_{ \tilde{d}_{R}}$         
                               & [GeV] & 7022, 6999    & 5441, 5384    & 5267, 5229 \\
$m_{\tilde{b}_{1}}$, $m_{\tilde{b}_{2}}$          
                               & [GeV] & 4959, 6893    & 3903, 5204    & 3770, 5124 \\[2pt]
\hline
\addlinespace[2pt]
$m_{ \tilde{e}_{L}}$, $m_{ \tilde{e}_{R}}$         
                               & [GeV] & 7152, 6759    & 5149, 4817    & 5128, 4825 \\
$m_{\tilde{\tau}_{1}}$, $m_{\tilde{\tau}_{2}}$        
                               & [GeV] & 6657, 7103    & 4652, 5072    & 4749, 5094 \\[2pt]
\hline
\addlinespace[2pt]
$\Omega_{\tz}^{\rm TP}h^2$     & --    & 0.009            & 0.007     & 0.005 \\
$\langle\sigma v\rangle(v\rightarrow 0)$ & [cm$^3$\,s$^{-1}$] & 2.2$\times 10^{-25}$ & 2.9$\times 10^{-25}$ & 3.1$\times 10^{-25}$   \\
$\sigma^{\rm SI}(\tz p)\times 10^{9}$ & [pb]  & 6.8 & 1.5 & 0.3 \\[2pt]
\hline
\addlinespace[2pt]
$a_\mu^{\rm SUSY} \times 10^{10}$              & -- & 0.03 & 0.13 & 0.06 \\
$BF(b\rightarrow s\gamma )\times 10^4$         & -- & 3.3  & 3.3  & 3.1  \\
$BF(B_S\rightarrow \mu^+\mu^- )\times 10^9$    & -- & 3.8  & 3.8  & 3.8  \\
$BF(B_u\rightarrow \tau\nu_\tau )\times 10^4$  & -- & 1.3  & 1.3  & 1.3  \\[2pt]
\hline
\addlinespace[2pt]
$\Delta_{\rm EW}$                & -- & 14             & 28      & 15 \\[2pt]
\hline\hline
\end{tabular}
\end{table}

\begin{table}[htbp]
\centering
\caption{Higgsino and gaugino fractions of the lightest neutralino $\LSP$.
  The fractions are expressed so that they satisfy the relation
  $R_{\widetilde{H}}^2 + R_{\widetilde{W}}^2 + R_{\widetilde{B}}^2 = 1$.
}
\label{tab:ewkino-composition}
\begin{tabular}{cSSS}
\hline\hline
\addlinespace[2pt]
  & \multicolumn{1}{c}{ILC1}
  & \multicolumn{1}{c}{ILC2}
  & \multicolumn{1}{c}{nGMM1} \\[2pt]
\hline
\addlinespace[2pt]
  $R_{\widetilde{H}}$ &  0.97 & 0.99 & 0.999 \\
  $R_{\widetilde{W}}$ & -0.14 & 0.07 & 0.04 \\
  $R_{\widetilde{B}}$ &  0.19 & 0.08 & 0.02 \\[2pt]
\hline\hline
\end{tabular}
\end{table}

\subsection{ILC2 benchmark}
\label{ssec:ILC2}

The ILC2 benchmark point is also generated within the NUHM2 SUSY model 
with parameter values as listed in Tab.~\ref{tab:bm}.
The location of ILC2 is indicated by the other green star in
Fig.~\ref{fig:pspace} and is found to lie just beyond
the HL-LHC reach for the same-sign diboson signature arising from wino pair production~\cite{Baer:2013yha}.
The higgsino pair signature from $pp\to \tz_1\tz_2j$ production followed by 
$\tz_2\rightarrow \tz_1\ell^+\ell^-$ decay 
should be viable since the cluster of higgsinos lies in the vicinity of $\mu= 150$ GeV~\cite{Baer:2014kya}. 
The value of $m_{\tg}=2832$\,GeV appears just beyond the HL-LHC 
reach for gluino pair production 
(where the $5\sigma$ reach extends to $m_{\tg}\sim 2800$\,GeV\cite{Baer:2016wkz}).
The higher value of gaugino masses in ILC2 -- as compared to benchmark point ILC1 -- 
is reflected in the reduced 
inter-higgsino mass gaps where we find $m_{\tw_1}-m_{\tz_1}=10.2$\,GeV and $m_{\tz_2}-m_{\tz_1}=9.7$\,GeV. 
The naturalness measure $\Delta_{\rm EW}=28$ leads to $\sim 3\%$ electroweak fine-tuning.
The thermally-produced abundance of dark matter 
$\Omega_{\tz_1}^{\rm TP}h^2\sim 0.007$, well below the measured value of 0.12. So, as for ILC1, 
the axions needed to solve the strong CP problem can be expected to make up the remainder.

\subsection{Natural mirage mediation (NMM) benchmark}
\label{ssec:nGMM1}

Mirage mediated (MM) SUSY breaking models are motivated by string model 
compactifications with moduli fields stabilized by fluxes and where an 
uplifted scalar potential leads to a de Sitter vacuum 
(as required by cosmology) with a small breaking of 
supersymmetry\cite{Choi:2004sx,Choi:2005ge,Choi:2007ka}. 
In such cases, it is expected that the SUSY breaking soft terms arise with
comparable moduli-mediated and anomaly-mediated contributions. 
In the gaugino sector (and in the scalar sector for particular modular weight choices) the  
GUT scale soft masses are offset from each other by contributions containing their 
gauge group beta functions. 
As a consequence, the running of the gaugino masses exactly compensates the high scale mass splitting 
leading to an apparent unification at the intermediate (mirage) scale 
$\mu_{mir}=m_{\rm GUT}e^{-8\pi^2/\alpha }$ where
the parameter $\alpha$ is introduced to parametrize the relative amounts
of anomaly- versus moduli-mediation. A value $\alpha =0$ corresponds to pure
anomaly-mediation (with tachyonic sleptons) while, as $\alpha$ gets large, 
the soft terms become increasingly universal.

Initially, simple MM models predicted scalar masses involving discrete 
values of scalar field modular weights which depend on the compactification geometry 
and upon which branes harbored the various visible sector fields. 
This class of models, over a wide range of choices for modular weights, was 
found to be unnatural when $m_h\simeq 125$ GeV was required\cite{Baer:2014ica}.
However, in more general compactification and stabilization schemes, then the
previously discrete parameter choices $c_m$, $c_{H_u}$, $c_{H_d}$ and $a_3$ 
become continuous, allowing for
the construction of models with low values of $\Delta_{\rm EW}$\cite{Baer:2016hfa}. 

In Tab.~\ref{tab:bm}, we show one such example point from natural
generalized mirage mediation or nGMM in column 4. 
The gaugino masses unify at the mirage scale $\mu_{mir}\sim 5\times 10^7$\,GeV. 
The rather large value of $m_{\tg}=2856.5$\,GeV means the winos and bino
are also rather massive so that both gluino pair production and wino
pair production appear out of reach of HL-LHC. 
The higgsinos are clustered with masses around $\mu\sim 150$\,GeV 
but with even smaller mass splittings than ILC2: 
$m_{\tw_1}-m_{\tz_1}= 7.3$\,GeV and $m_{\tz_2}-m_{\tz_1}=4.4$\,GeV. 
Such small neutralino mass splittings may also push the soft dilepton plus jet
signature from $\tz_1\tz_2j$ production out of reach of HL-LHC~\cite{Baer:2018hpb}.
Nonetheless, the model is highly natural with $\Delta_{\rm EW}=15$ or
6.7\% fine-tuning. The LSP is more purely higgsino-like than the
ILC1 or ILC2 benchmarks leading to a reduced thermally-produced dark matter relic density 
and reduced direct dark matter detection rates. Thus, direct detection of WIMP dark matter 
from the nGMM1 benchmark may  require multi-ton noble liquid detectors for discovery.


\section{Software Tools and Observables}
\label{sec:sim}

In this section we describe the main features of the ILD detector as used for the simulation study and introduce the characteristics of the higgsino signal on which the strategies for event selection and reconstruction will be based.

\subsection{Event generation}
\label{ssec:production}

The physical masses for the three benchmark points have been calculated by  {\tt ISASUGRA}.
The SUSY and SM events have been generated using \texttt{WHIZARD} 1.95~\cite{Whizard},
which considers both resonant and non-resonant production, as well as their interference.
\texttt{WHIZARD} also generates the amount and spectrum of ISR appropriate for each considered channel,
and takes beam-polarisation fully into account.
The dedicated setup of the generator provided by the ILC Generator Group was used,
and all types SM $e^+e^-$ interactions yielding up to six fermions in the final state were considered.
In addition,
all $e\gamma$ interactions yielding three or five fermions and all $\gamma\gamma$ interactions yielding up to four fermions
were also included.
The initial photons in the latter cases might be virtual (in which case the beam remnants come in addition to
the final fermions),
or real from the photon component of the beams (in which case there is no beam remnants).
The electron and positron beams have an initial energy-spread,
which is further smeared by the effects of beamstrahlung.
The resulting spectra as well as the flux and energy spectra of the beam photons are simulated according to the
parameters in the ILC Technical Design Report (TDR)~\cite{TDR}, using {\tt GuineaPig}\cite{Schulte:1999tx}.

Pure left-handed or right-handed beam polarisations are used for the event generation.
These samples are then weighted according to the nominal ILC beam polarisations for our simulation study.
We introduce the following notation for beam polarizations,
$\mathcal{P}\equiv( \mathcal{P}_{e^-}, \mathcal{P}_{e^+} )$,
and define the pure beam polarisations as
$\mathcal{P}_{LR}\equiv(-1, +1)$ and $\mathcal{P}_{RL}\equiv(+1, -1)$.
The nominal beam polarisations for the ILC are defined as
$\mathcal{P}_{-+}\equiv(-0.8, +0.3)$ and $\mathcal{P}_{+-}\equiv(+0.8, -0.3)$.

Table~\ref{tab:xsec} shows the production cross sections for chargino and neutralino pairs for the three benchmark models
introduced in Sec.~\ref{ssec:ILC1}-\ref{ssec:nGMM1} for 100\% polarised beams at several center-of-mass energies.
Table~\ref{tab:BR} shows the decay branching ratios in the same three benchmarks.

The results of the simulation study assume
a center-of-mass energy of $\sqrt{s}=500$\,GeV
and an integrated luminosity of $\mathcal{L}=500$~fb$^{-1}$ for each beam polarisation;
these results are then scaled according to the operation scenarios listed in
Tab.~\ref{tab:operating-scenarios} for the parameter fit.
In case of ILC1, where all three higgsinos would already be kinematically accessible at $\sqrt{s}=250$\,GeV,
the assumed integrated luminosities correspond to the H20 operating scenario,
while for the other two benchmarks the I20 scenario was assumed,
since in these cases the higgsinos are only accessible at $\sqrt{s}=350$\,GeV.

\begin{table}[htbp]
\centering
\caption{
  Chargino and neutralino production cross sections
  for the three benchmark points calculated
  using {\tt WHIZARD} at various center-of-mass energies.
  The ILC beam energy spectrum and ISR effects are included.
  Pure beam polarizations are assumed.
}
\label{tab:xsec}
\begin{tabular}{ccccccc}
\hline\hline
\addlinespace[2pt]
           &         &           & \multicolumn{3}{c}{Cross Section [fb]} \\[2pt]
  $\sqrt{s}$ & Process & $\mathcal{P}$ & ILC1 & ILC2 & nGMM1 \\[2pt]
\hline     
\addlinespace[2pt]
\multirow{4}{*}{250 GeV}
& \multirow{2}{*}{$e^- e^+ \rightarrow\widetilde{\chi}^+_1\widetilde\chi^-_1$}
  & $\mathcal{P}_{LR}$  & 2618   &  --  & -- \\
& & $\mathcal{P}_{RL}$  &  397.1 &  --  & -- \\[2pt]
\cline{2-6}
\addlinespace[2pt]
& \multirow{2}{*}{$e^- e^+ \rightarrow\widetilde{\chi}^0_1\widetilde\chi^0_2$}
  & $\mathcal{P}_{LR}$  & 1044   &  --  & -- \\
& & $\mathcal{P}_{RL}$  &  804.8 &  --  & -- \\[2pt]
\hline
\addlinespace[2pt]
\multirow{4}{*}{350 GeV}
& \multirow{2}{*}{$e^- e^+ \rightarrow\widetilde{\chi}^+_1\widetilde\chi^-_1$}
  & $\mathcal{P}_{LR}$  & 3094   & 1602   & 1571   \\
& & $\mathcal{P}_{RL}$  &  538.8 &  302.8 &  301.4 \\
\cline{2-6}
\addlinespace[2pt]
& \multirow{2}{*}{$e^- e^+ \rightarrow\widetilde{\chi}^0_1\widetilde\chi^0_2$}
  & $\mathcal{P}_{LR}$  &  897.0 &  578.5 &  576.0 \\
& & $\mathcal{P}_{RL}$  &  691.5 &  446.0 &  444.1 \\[2pt]
\hline
\addlinespace[2pt]
\multirow{4}{*}{500 GeV}
& \multirow{2}{*}{$e^- e^+ \rightarrow\widetilde{\chi}^+_1\widetilde\chi^-_1$}
  & $\mathcal{P}_{LR}$  & 1801   & 1531   & 1520   \\
& & $\mathcal{P}_{RL}$  &  334.8 &  307.2 &  309.2 \\
\cline{2-6}
\addlinespace[2pt]
& \multirow{2}{*}{$e^- e^+ \rightarrow\widetilde{\chi}^0_1\widetilde\chi^0_2$}
  & $\mathcal{P}_{LR}$  &  491.4 &  458.9 &  463.3 \\
& & $\mathcal{P}_{RL}$  &  379.8 &  353.8 &  357.1 \\[2pt]
\hline\hline
\end{tabular}
\end{table}
\begin{table}[htbp]
\centering
\caption{Chargino and neutralino decay branching ratios for the three benchmark points calculated using {\tt ISASUGRA}.
For the final-state leptons, only the electrons and muons are included ($\ell=e,\mu$).}
\label{tab:BR}
\begin{tabular}{lccc}
\hline\hline  
& ILC1 & ILC2 & nGMM1 \\
\hline
$BR(\widetilde{\chi}^+_1\rightarrow\widetilde{\chi}^0_1 qq')$               & 67$\%$       & 67$\%$    & 66$\%$ \\
$BR(\widetilde{\chi}^+_1\rightarrow\widetilde{\chi}^0_1 \ell^+\nu_\ell)$    & 22$\%$       & 22$\%$    & 22$\%$ \\
\hline
$BR(\widetilde{\chi}^0_2\rightarrow\widetilde{\chi}^0_1 qq')$               & 58$\%$       & 63$\%$    & 51$\%$ \\
$BR(\widetilde{\chi}^0_2\rightarrow\widetilde{\chi}^0_1 \ell^+\ell^-)$      &7.4$\%$       &8.0$\%$    &7.5$\%$ \\
\hline\hline
\end{tabular}
\end{table}
\begin{table}[htbp]
\centering
\caption{Total integrated luminosities for various operation scenarios for the ILC~\cite{Barklow:2015tja}.
  H20 is assumed for ILC1, while I20 is assumed for ILC2 and nGMM1.
  See Ref.~\cite{Barklow:2015tja} for the assumed timelines and machine upgrades.
}
\label{tab:operating-scenarios}
\begin{tabular}{ccccc}
\hline\hline
\addlinespace[2pt]
  Scenario & $\sqrt{s}$ [GeV]& $\mathcal{P}$ & $\mathcal{L}$ [fb$^{-1}$] \\
\hline
  \multirow{6}{*}{H20} & \multirow{2}{*}{250} & $\mathcal{P}_{-+}$ &  900 \\
		       &                      & $\mathcal{P}_{+-}$ &  900 \\
\cline{2-4}
		       & \multirow{2}{*}{350} & $\mathcal{P}_{-+}$ &   90 \\
		       &                      & $\mathcal{P}_{+-}$ &   90 \\
\cline{2-4}
		       & \multirow{2}{*}{500} & $\mathcal{P}_{-+}$ & 1600 \\
		       &                      & $\mathcal{P}_{+-}$ & 1600 \\
\hline
  \multirow{6}{*}{I20} & \multirow{2}{*}{250} & $\mathcal{P}_{-+}$ &  225 \\
		       &                      & $\mathcal{P}_{+-}$ &  225 \\
\cline{2-4}
		       & \multirow{2}{*}{350} & $\mathcal{P}_{-+}$ &  765 \\
		       &                      & $\mathcal{P}_{+-}$ &  765 \\
\cline{2-4}
		       & \multirow{2}{*}{500} & $\mathcal{P}_{-+}$ & 1600 \\
		       &                      & $\mathcal{P}_{+-}$ & 1600 \\
\hline\hline
\addlinespace[2pt]
\end{tabular}
\end{table}

\subsection{The ILD detector model}
\label{ssec:detector}
The ILD concept is one of the two detectors being designed for the ILC.
ILD employs a hybrid tracking system comprised of a time projection chamber and silicon strip sensors for tracking, and silicon pixel sensors as vertex detectors.
Outside of the tracking system sits a highly granular calorimeter system optimized for particle flow reconstruction.
A superconducting solenoid with a magnetic field of $3.5$\,T encases the calorimeters.
An iron yoke outside the solenoid coil returns the magnetic flux, and is instrumented with scintillator-based muon detectors.
In the low angle region, charged particles will be efficiently tracked down to 7 degrees.
Dedicated calorimeters are placed in the forward region for detecting particles at even lower angles
to the beam~\cite{Abramowicz:2010bg}.
The most forward component of this system - the BeamCal -
has holes for the beam-pipe, which constitutes the only region outside detector acceptance,
and corresponds to $5.6$\,mrad.

The simulation and reconstruction tools used in this study are part of the \texttt{iLCSoft} framework (v01-16-02) \cite{ILCSOFT}.
The beam crossing angle of $14$\,mrad and the response of the ILD detector in its version ILD\_o1\_v05 as used for the ILC TDR~\cite{ILD} are simulated using \texttt{Mokka}~\cite{Mokka} based on \texttt{GEANT4}.
The event reconstruction is performed using the Marlin \cite{Marlin} framework, including the particle flow algorithm \texttt{PandoraPFA}~\cite{Pandora} for calorimeter clustering and the analysis of track and calorimeter information.

\subsection{Signal processes and key observables}
\label{observables}
We study the pair production of the two light charginos ($\widetilde\chi^+_1$, $\widetilde\chi^-_1$)
and two light neutralinos ($\widetilde\chi^0_1$, $\widetilde\chi^0_2$).
In our benchmark models, the higgsino component is strongly dominant for these four light states.
Their masses are shown in Tab.~\ref{tab:bm}.
The charginos and neutralinos are both produced dominantly via the $s$-channel exchange since the sleptons are assumed to be heavy.
The chargino pair production proceeds as $e^+e^-\rightarrow\widetilde{\chi}^+_1\widetilde\chi^-_1$ through $\gamma/Z$ exchange,
while the neutralino associated production $e^+e^-\rightarrow\widetilde{\chi}^0_1\widetilde\chi^0_2$ undergoes via the $Z$ boson exchange.
While in a real analysis at ILC more decay modes of the \neutralinotwo\ and the \charginoone\ can be utilized, we focus here on the semi-leptonic channel for the charginos, i.e.\ $\charginoonep \charginoonem \to q \bar{q}' \LSP l \nu \LSP$, and on the leptonic channel for the neutralinos, i.e.\ $\neutralinotwo \LSP \to l^+ l^- \LSP$. We restrict $l = e, \mu$ in this study.

The key target observables are the three masses ($M_{\LSP}$, $M_{\neutralinotwo}$, and $M_{\charginoone}$)
and, in this study, four polarized cross sections: chargino and neutralino production for the two opposite-sign beam polarization configurations. In a real ILC analysis, the like-sign combinations would be included as well, at least to serve as background-enriched control samples.  

\begin{figure}[htb]
\centering
\begin{subfigure}{0.32\linewidth}
\includegraphics[width=\linewidth]{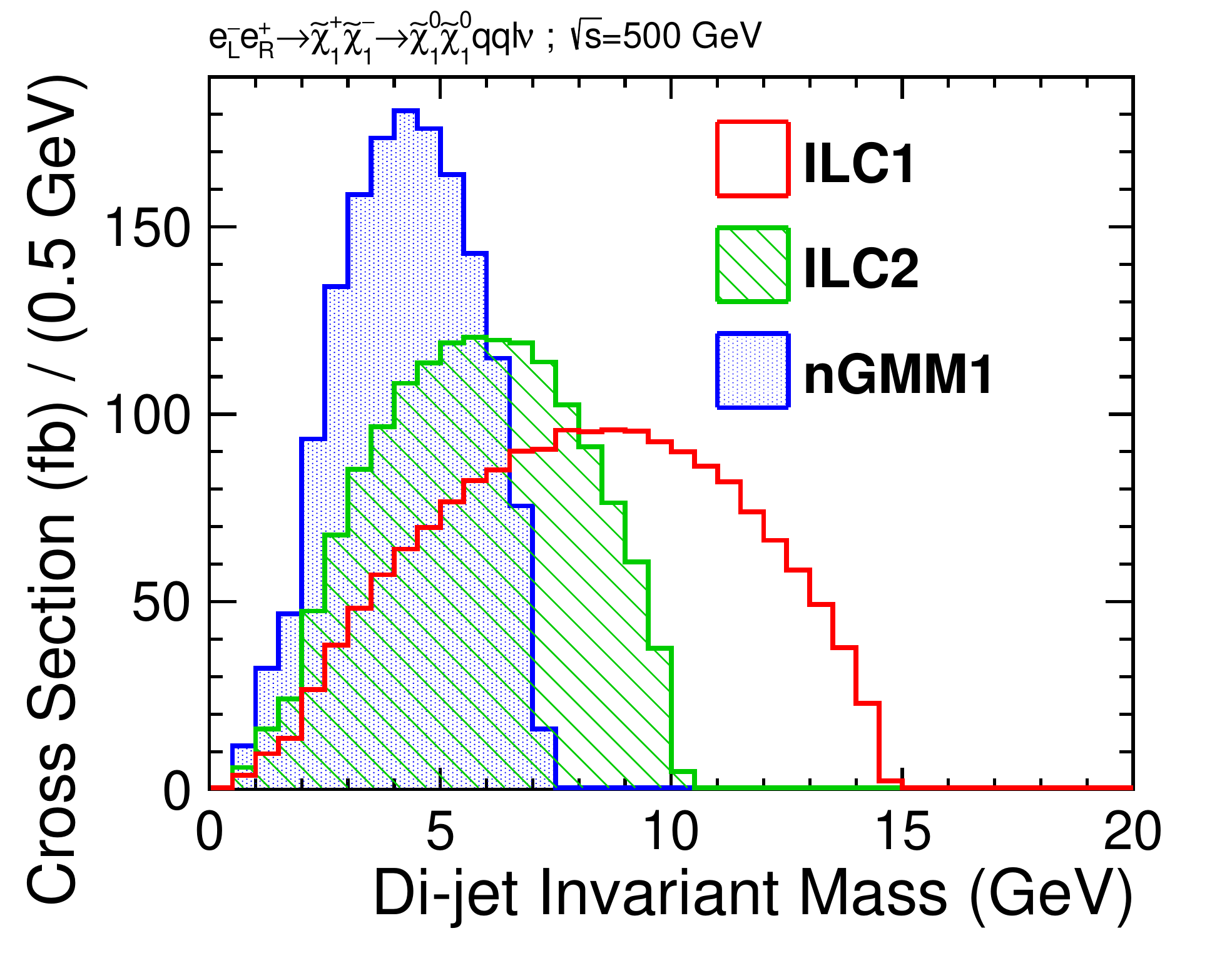}
\caption{chargino channel \label{fig:generator_level_distributions:a}}
\end{subfigure}
\begin{subfigure}{0.32\linewidth}
\includegraphics[width=\linewidth]{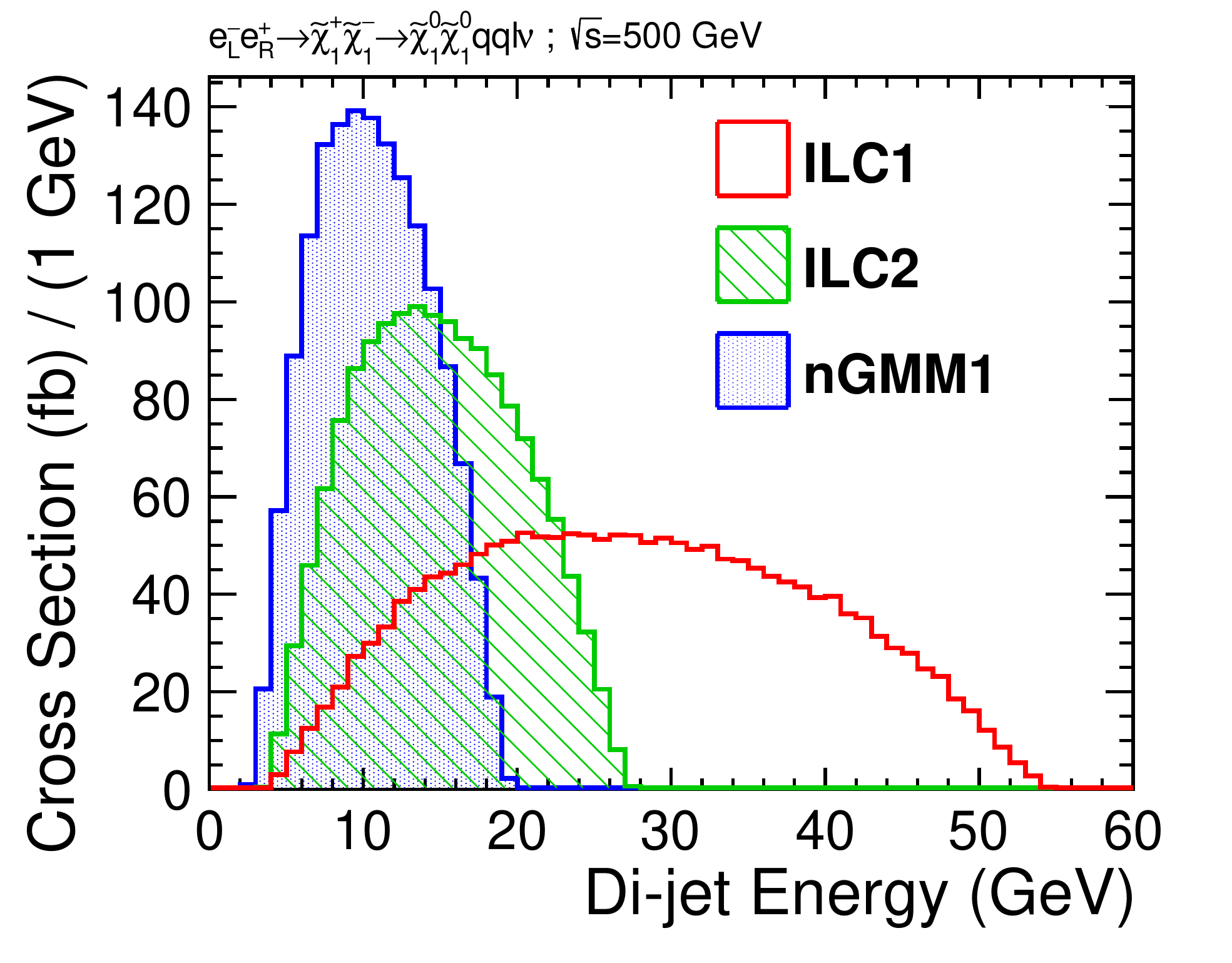}
\caption{chargino channel \label{fig:generator_level_distributions:b}}
\end{subfigure}
\begin{subfigure}{0.32\linewidth}
\includegraphics[width=\linewidth]{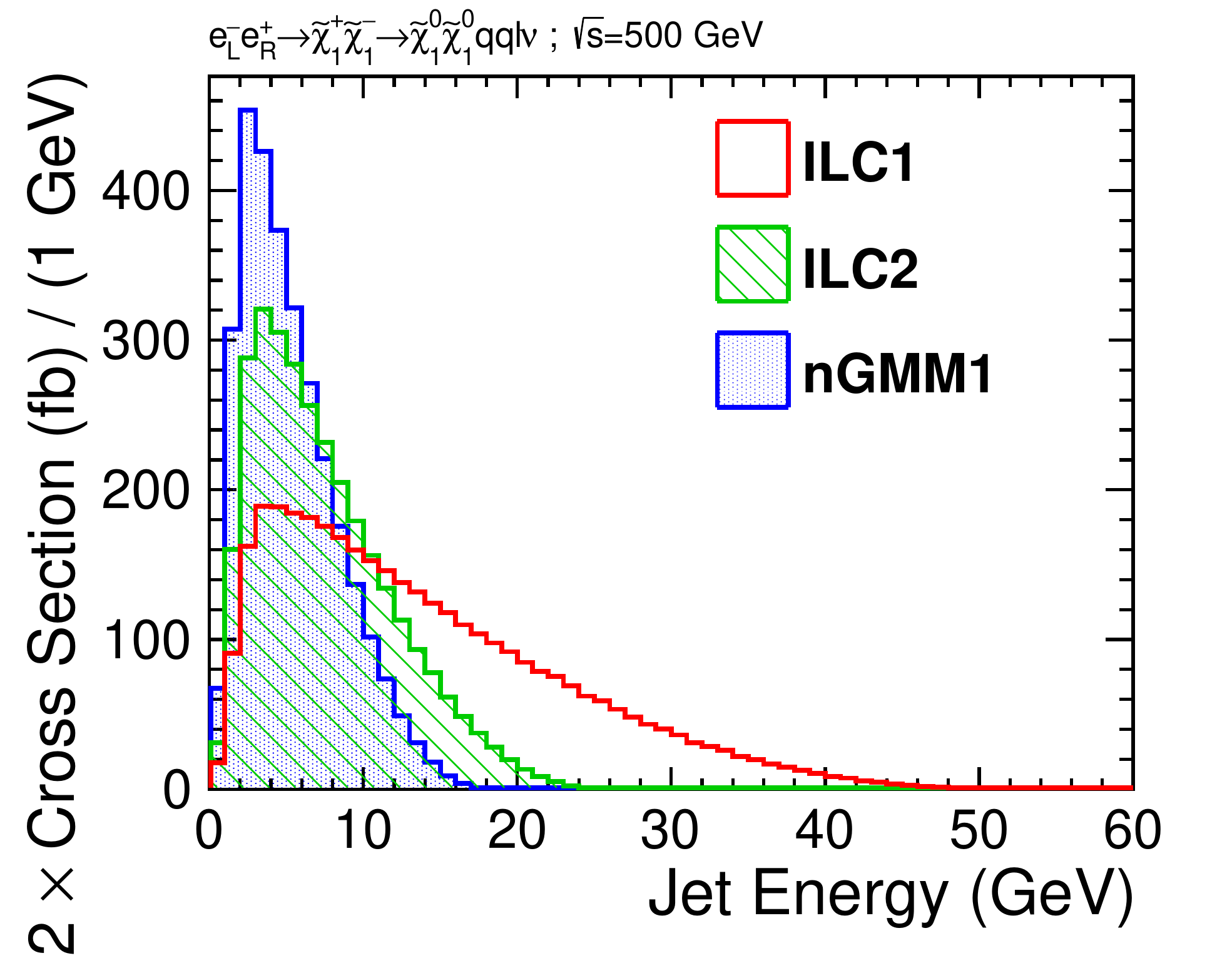}
\caption{chargino channel \label{fig:generator_level_distributions:c}}
\end{subfigure}
\begin{subfigure}{0.32\linewidth}
\includegraphics[width=\linewidth]{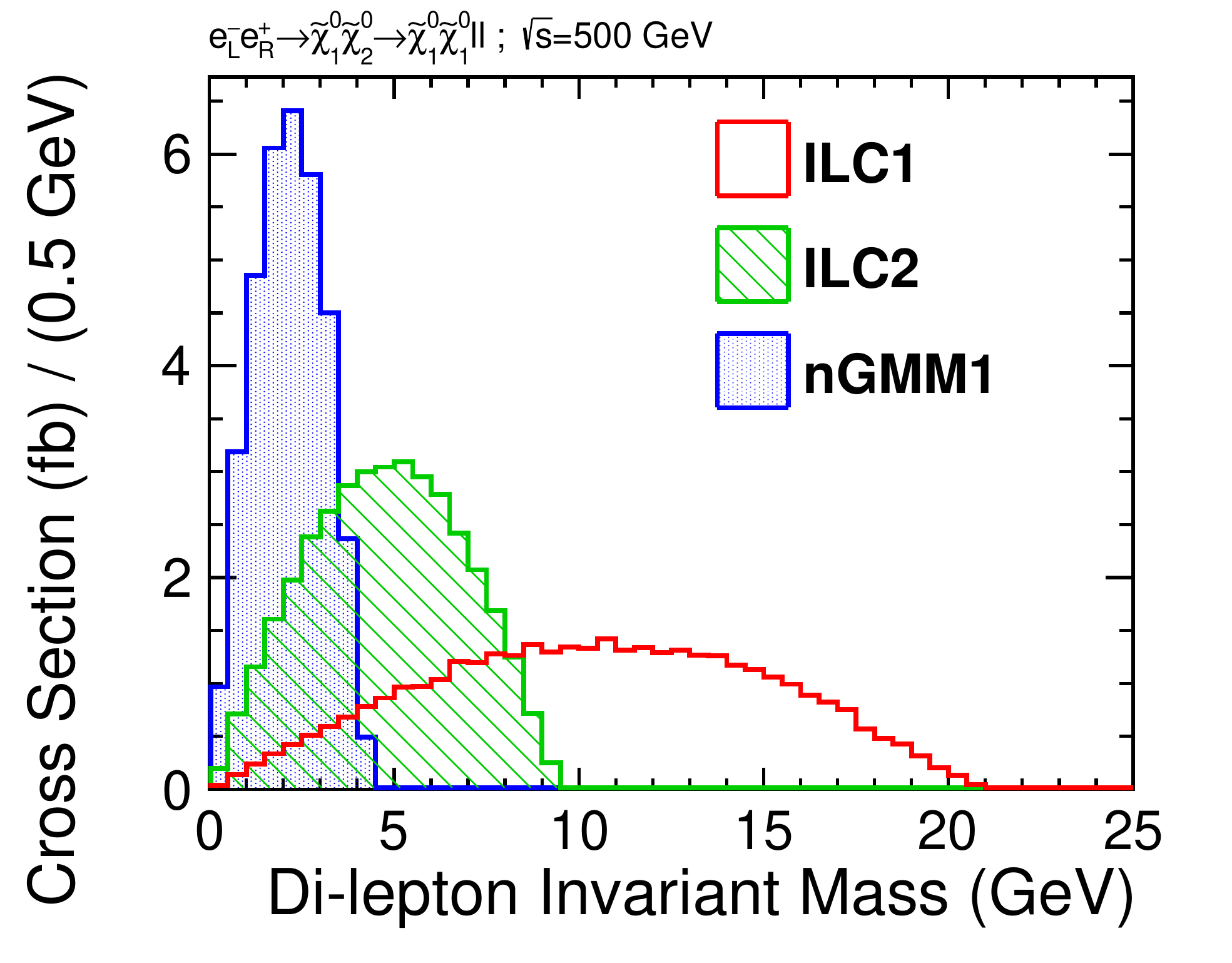}
\caption{neutralino channel \label{fig:generator_level_distributions:d}}
\end{subfigure}
\begin{subfigure}{0.32\linewidth}
\includegraphics[width=\linewidth]{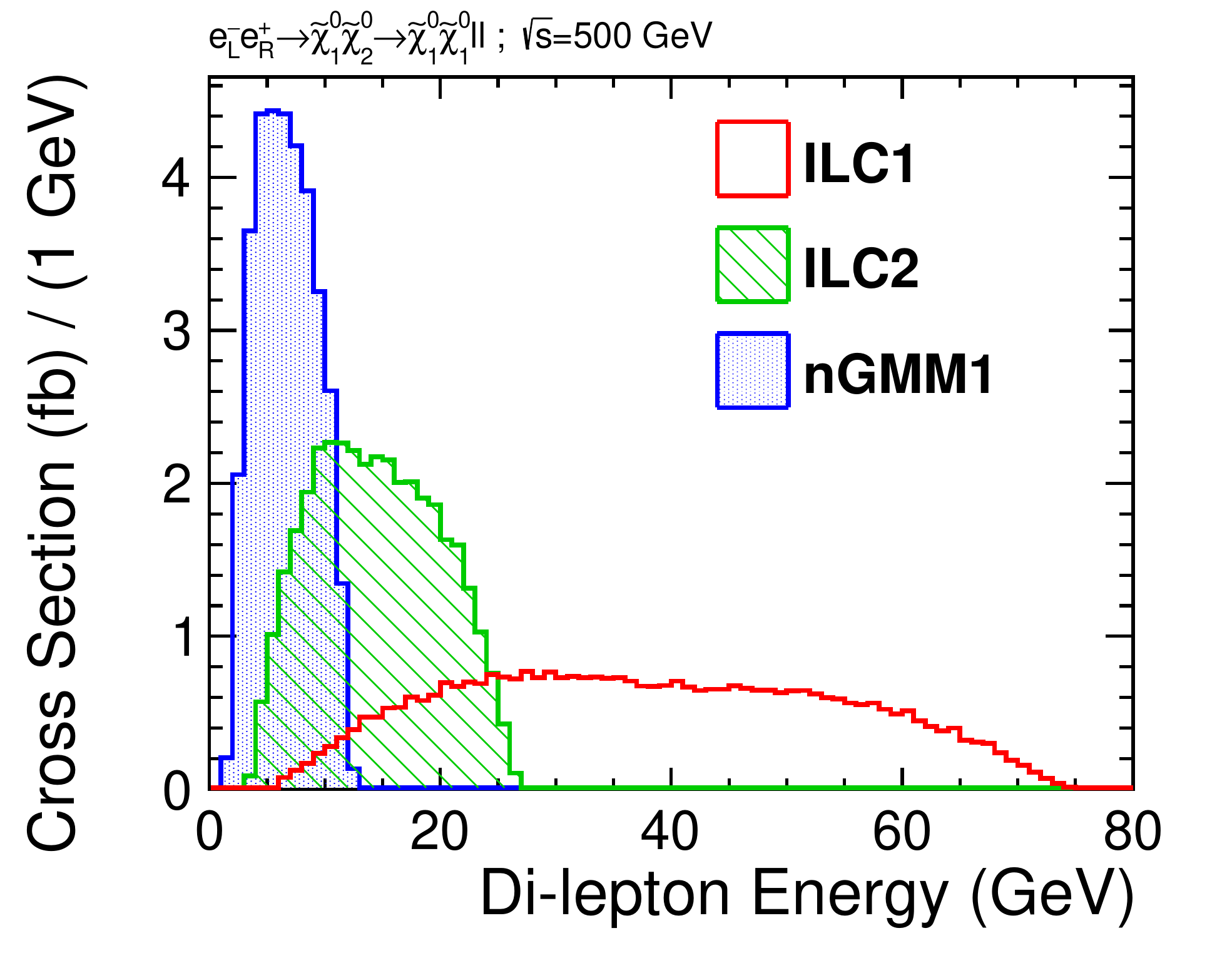}
\caption{neutralino channel \label{fig:generator_level_distributions:e}}
\end{subfigure}
\begin{subfigure}{0.32\linewidth}
\includegraphics[width=\linewidth]{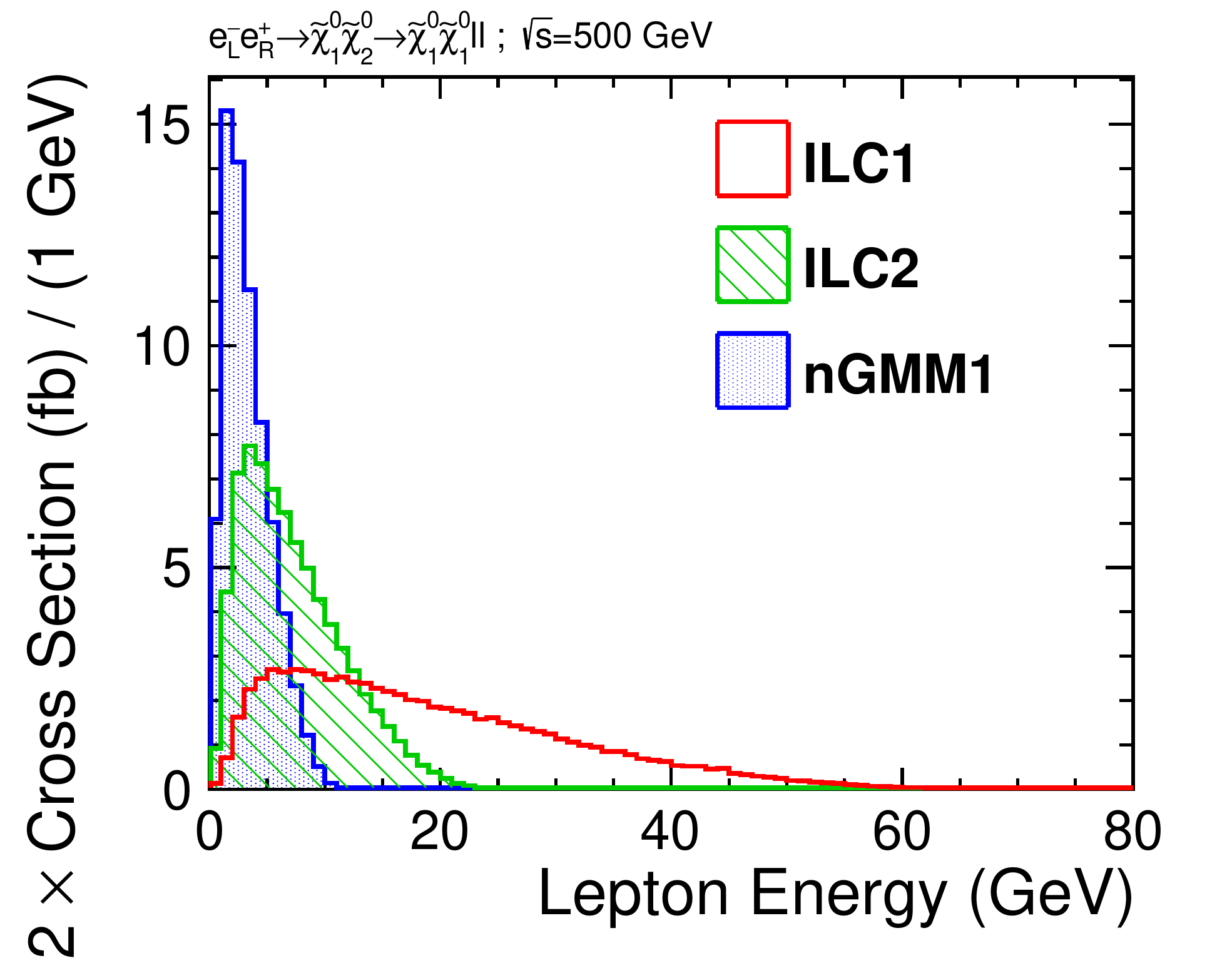}
\caption{neutralino channel \label{fig:generator_level_distributions:f}}
\end{subfigure}
\caption{Generator-level distributions, given for the beam polarization with $\mathcal{P}_{LR}$.
  The distributions for the other beam polarization $\mathcal{P}_{RL}$ are similar,
up to the normalization due to the cross section.
}
\label{fig:generator_level_distributions}
\end{figure}

The three masses can be extracted from the maximum endpoints of the kinematic distributions shown at generator-level in Fig.~\ref{fig:generator_level_distributions}(a),(b),(d) and (e).
Specifically, we will rely on the maximum invariant mass and energy of the visible decay products of the $\neutralinotwo \to l^+l^- \LSP$ and $\charginoone \to q \bar{q}' \LSP$. We find that the minimum endpoints, typically used in other SUSY studies, are too small to be useful in this study, since the resulting detector response is challenging to model in the soft spectrum and, in the case of the neutralino channel, it has large overlap with irreducible backgrounds.

The maximum energy $E_{\mathrm{max}}$ of the di-jet (or di-lepton) system seen in the laboratory frame is given by\footnote{
The value for $E_{\mathrm{max}}$ given in formula Eq.~\ref{eq:Emax} is attained when the dilepton or 
dijet invariant mass is zero. The complete formulae relating the maximum and minimum energy to the invariant mass may be found on p. 439 of Ref. \cite{Baer:2006rs}. The sensitivity of the analysis could even be improved by evaluating the maximum energy on an event-by-event
basis, taking into account the dijet / dilepton invariant mass measured in each event individually.}
\begin{equation}
E_{\mathrm{max}} = \frac{\gamma(1+\beta)}{2}\left(1+\frac{M}{M'}\right)\Delta M,
\label{eq:Emax}
\end{equation}
where 
$M$ is the LSP ($\LSP$) mass and $M'$ is the mass of $\charginoone$ ($\neutralinotwo$) for the chargino (neutralino) channel.
The mass difference is given by $\Delta M = M'-M$.
As the decays studied are three body decays, it follows that the Lorentz-invariant
mass of any pair of final-state particles has a maximum equal 
to the mass-difference
between the decaying particle and the mass of the third 
decay product, see e.g. sect. 47.4.4.1 of \cite{Tanabashi:2018oca}.
In other words, the maximum of the the di-jet (or di-lepton) mass is
a direct measure of  $\Delta M$.
The boost factors $\beta$ and $\gamma$ are given according to $\gamma = (1-\beta^2)^{-\frac{1}{2}}$
and $\beta = \frac{p}{\sqrt{M'^2+p^2}}$,
where the maximum momentum $p$ in the laboratory frame is given by:
\begin{equation}
p = \frac{\sqrt{s}}{2} \sqrt{1-2\left[\left(\frac{M}{\sqrt{s}}\right)^2+\left(\frac{M'}{\sqrt{s}}\right)^2\right]+\left[\left(\frac{M'}{\sqrt{s}}\right)^2-\left(\frac{M}{\sqrt{s}}\right)^2\right]^2}.
\end{equation}
For any given channel, the measurements of $\Delta M$ and $E$ yield the masses $M$ and $M'$ by numerically solving the relations above.
In our study, we obtain several measurements of $\Delta M$ and $E$,
specifically for the two different lepton final states ($\ell=e$ and $\mu$),
and for the two different beam polarizations $\mathcal{P}_{-+}$ and $\mathcal{P}_{+-}$.
These measurements can be readily combined individually for the chargino and neutralino channels.
Because the chargino and neutralino measurements both include the LSP mass $M_{\LSP}$ in the observables,
a final combination is performed using a fit to extract the uncertainty of the three masses.
%

Figure~\ref{fig:generator_level_distributions} shows the generator-level distributions
of the mass and energy distributions for the di-jet (di-lepton) system
for the chargino (neutralino) channel.
The three benchmark points, ILC1, ILC2, and nGMM1, give progressively softer distributions due to the smaller mass gaps.
The visible part of the chargino/neutralino decay will be very soft;
for example, most of the jets and leptons have energies less than 20 GeV in the case of the ILC2 benchmark, and less than 10-15 GeV in the case of the nGMM1 benchmark.


\subsection{Parameter Fitting}
\label{subsec:parfit_tools}
In the final step of the study we investigate the possibility to extract SUSY parameters from the projected measurement precisions obtained from the detector simulation. This will be addressed using a Markov chain technique as implemented in the program \texttt{Fittino}. Unless stated otherwise, the length of the Markov chains is $10^6$ for each fitted configuration. While the MC samples used in the full detector simulation were based on \texttt{Isajet}, \texttt{Fittino} employs \texttt{SPheno} as a spectrum calculator during the fit. More details about the fitting procedure can be found in~\cite{Lehtinen:PhDThesis}.

\begin{figure}[htb]
\begin{subfigure}{0.33\linewidth}
\includegraphics[trim={0cm 1cm 1cm 0cm},clip,width=1.00\textwidth]{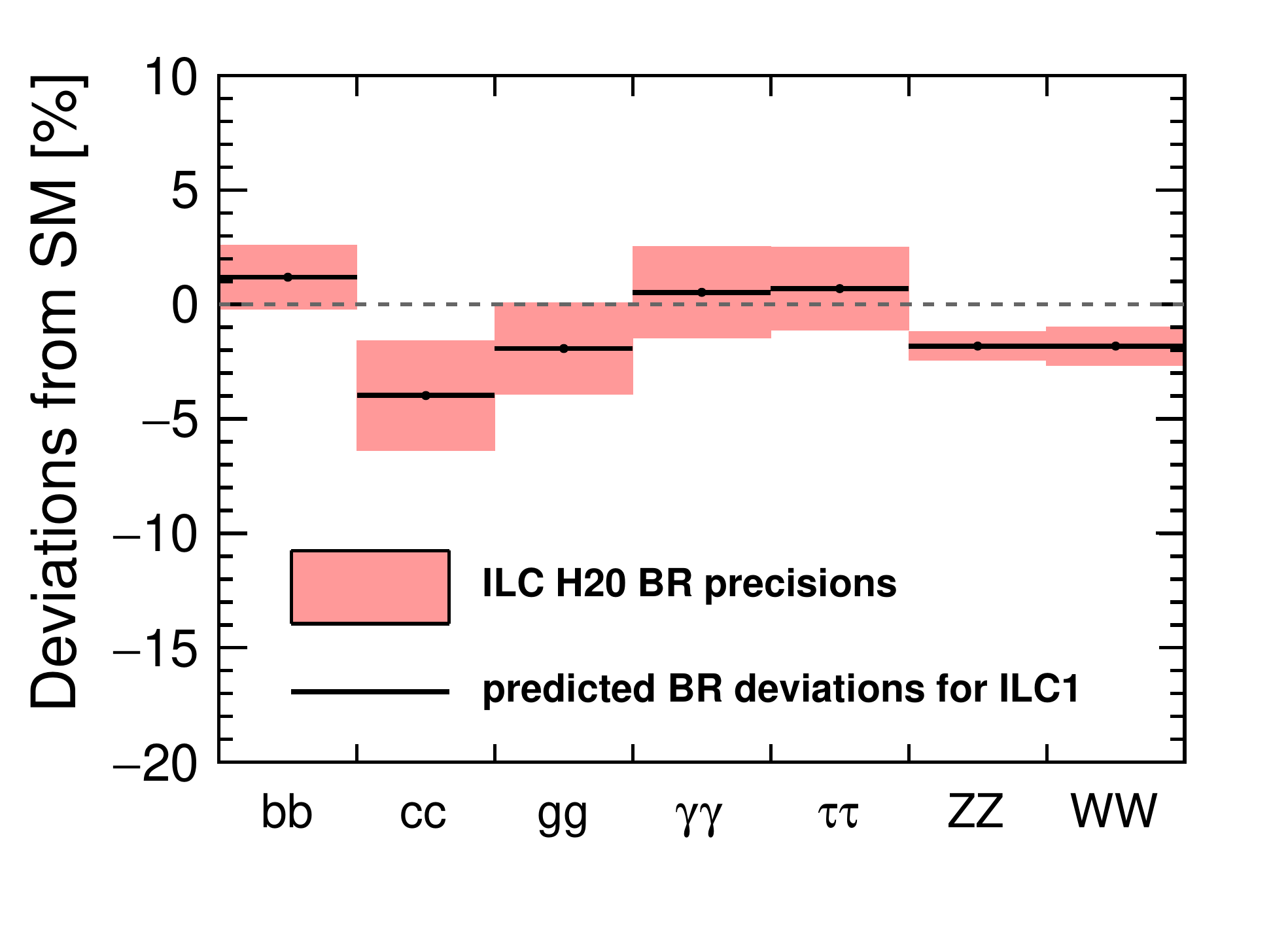}
\caption{ILC1\label{fig:higgsBRdeviations:ILC1}}
\end{subfigure}
\begin{subfigure}{0.33\linewidth}
\includegraphics[trim={0cm 1cm 1cm 0cm},clip,width=1.00\textwidth]{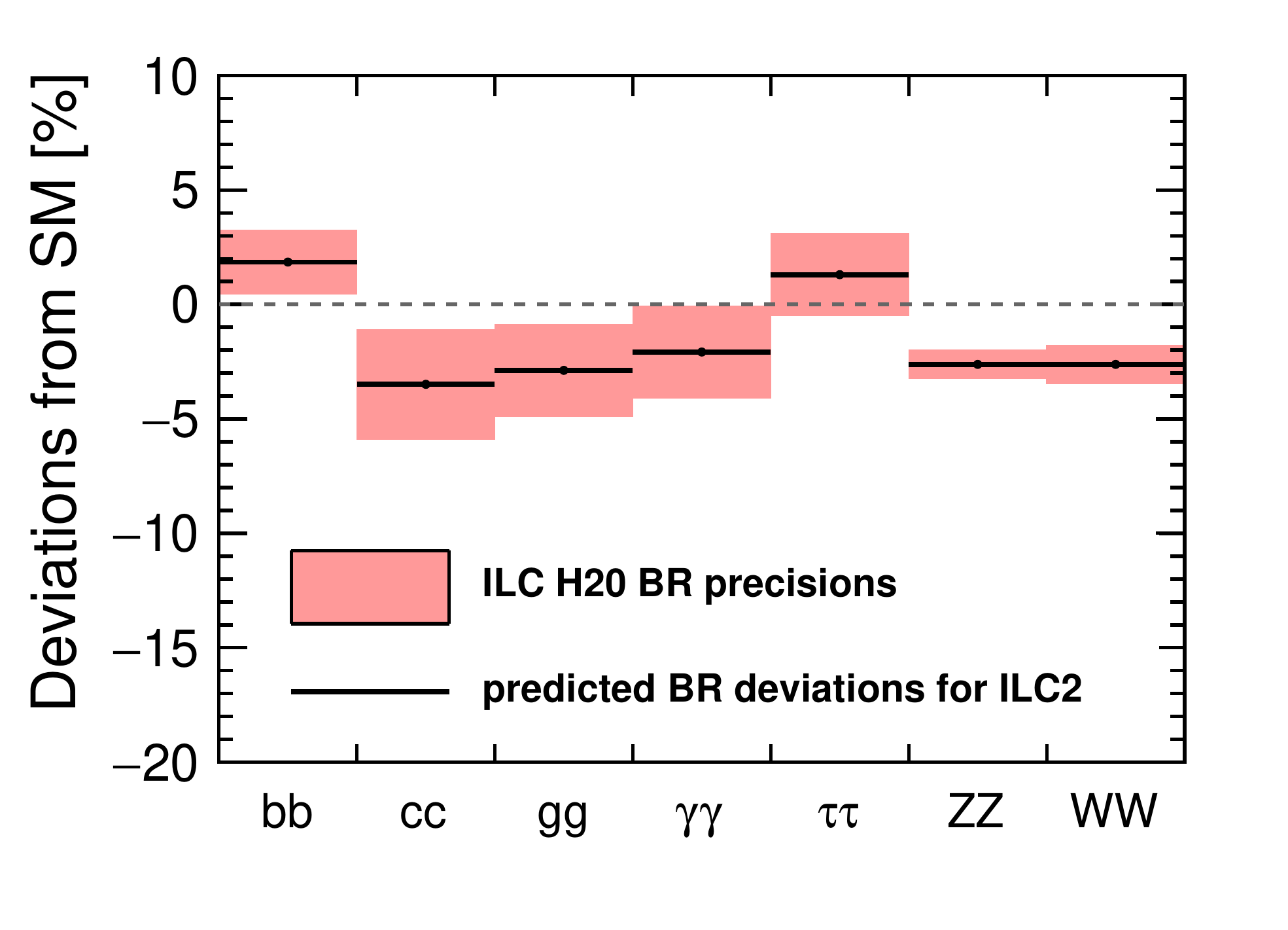}
\caption{ILC2\label{fig:higgsBRdeviations:ILC2}}
\end{subfigure}
\begin{subfigure}{0.33\linewidth}
\includegraphics[trim={0cm 1cm 1cm 0cm},clip,width=1.00\textwidth]{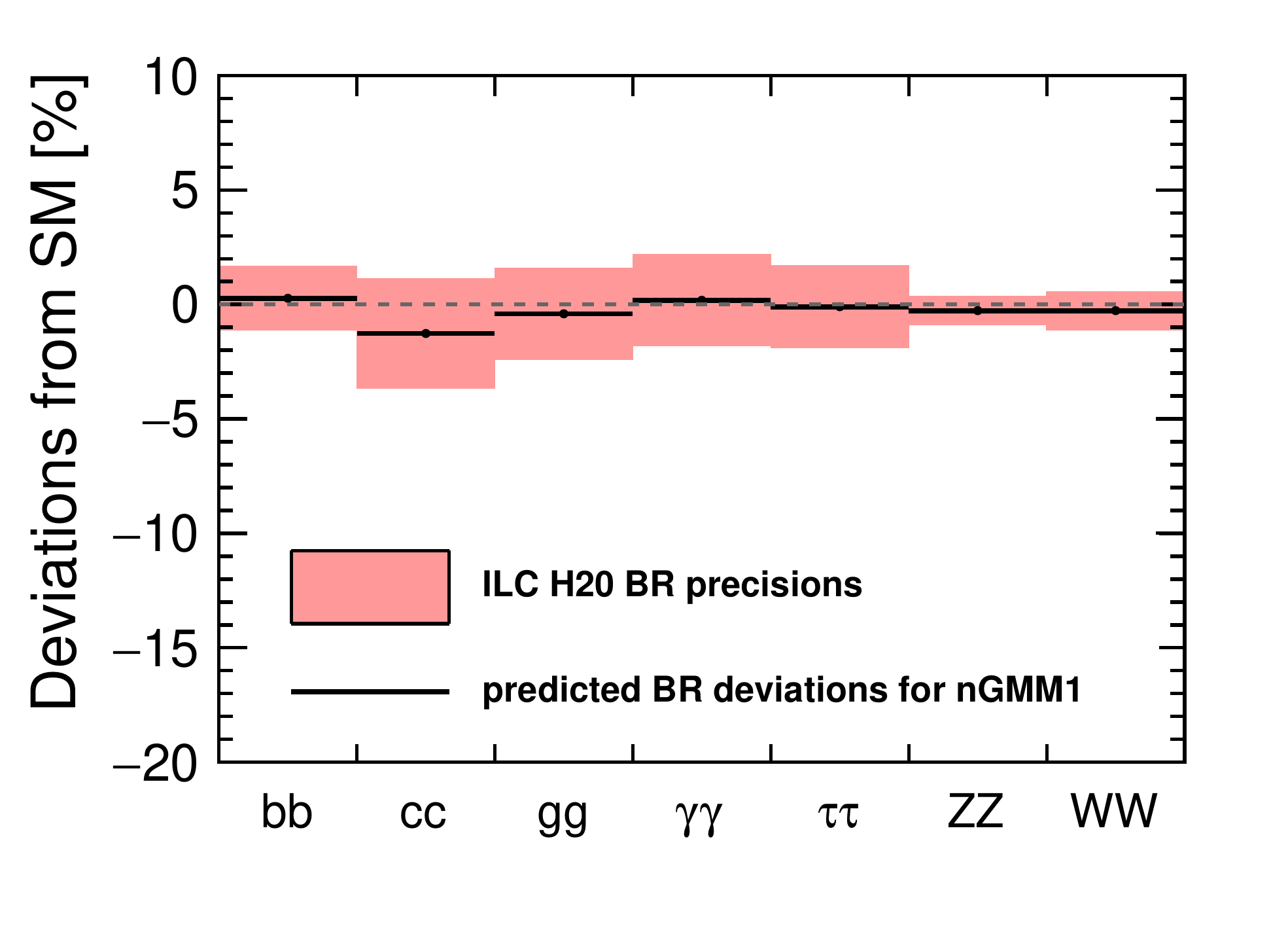}
\caption{nGMM1\label{fig:higgsBRdeviations:nGMM1}}
\end{subfigure}
\caption{Deviations of the branching fractions of the SUSY light Higgs from the Standard Model expectations, as obtained with \texttt{FeynHiggs2.10.4}. The uncertainty bands illustrate the expected measurement precisions after the full 250 and 500\,GeV ILC program, assuming the H20 scenario~\cite{Fujii:2015jha}. Note that these are somewhat more conservative than the most recent estimates from~\cite{Bambade:2019fyw}.}
\label{fig:higgsBRdeviations}
\end{figure}

In addition to the mass and cross-section projections from this study, 
which will be described in detail in Sec.~\ref{sec:event_selection}, 
a standard set of projected Higgs precision observables from the ILC was used to constrain the fit: 
These projections assume the H20 running scenario for the ILC~\cite{Barklow:2015tja},
and include the Higgs mass (with a precision of 15\,MeV as obtained in a ILD full simulation study~\cite{Yan:2016xyx}), 
and a set of Higgs branching ratio precisions obtained from the model-independent 
coupling fit results in~\cite{Fujii:2015jha}, 
based on the so-called $\kappa$-framework.\footnote{Note that this includes the cross section measurements which in effect dominate the coupling precisions for the weak gauge bosons. For technical reasons the coupling precisions could not be used directly in the fit.} 
The resulting precisions are shown in Fig.~\ref{fig:higgsBRdeviations} in comparison to the expected deviations from the SM branching ratios in our three benchmarks. The Higgs mass and branching-ratio values are taken from \texttt{FeynHiggs2.10.4} \cite{bib:feynhiggs} for each of the SUSY models 
in question. 
For the NUHM2-based benchmarks, the most significant deviations would be observed in the Higgs couplings to the $W$ and $Z$ bosons, although they would be hardly convincing as a discovery on their own. 
In case of nGMM1, all Higgs precision measurements agree perfectly with the SM.

\section{Full Detector Simulation Study}
\label{sec:event_selection}

For each benchmark point, we select separately the chargino and neutralino channels. These are considered as background to each other, including {\em all} decay modes.
A common event selection is performed and provides sufficient sensitivity in all three benchmark cases. In a real experiment, once a signal has been discovered, the selection could be optimized based on initial mass estimates, see e.g.~\cite{Caiazza:2018suv} for a discussion of how to boot-strap a selection of an a priori unknown signal. In the following sections, we describe the event selection for the chargino and neutralino channels.

\subsection{Chargino channel}
\label{ssec:chargino_measurement}

For the chargino pair production, we study the semileptonic final state
$e^-e^+\rightarrow \widetilde\chi^+_1\widetilde\chi^-_1\rightarrow \widetilde\chi^0_1\widetilde\chi^0_1qq'\ell\nu_\ell$
where $\ell=e$ or $\mu$.
The strategy here is to look for a single isolated lepton accompanied by two jets and large missing energy.
We reconstruct the invariant mass and energy distributions of the di-jet system and extract their endpoints.

First, an isolated lepton candidate is identified according to the following criteria.
Electron identification requires that the total energy measured in the calorimetric system $E_{\rm tot}$ is consistent with the momentum measured in the tracker $p_{\rm trk}$, such that they satisfy $0.5 <E_{\rm tot}/p_{\rm trk}<1.3$.
In addition, the ECAL energy deposit $E_{\rm ECAL}$ must be dominant over the HCAL energy deposit $E_{\rm HCAL}$, so that we have $E_{\rm ECAL}/(E_{\rm ECAL}+E_{\rm HCAL})>0.9$.
For muon identification, we require that a charged track is associated with signals in the muon detector.
In addition, lepton candidates with large impact parameter significance ($>$5$\sigma$) are rejected in order to suppress backgrounds due to $\tau$ or heavy quark decays.
For the isolation requirement, we define an isolation cone around the lepton candidate with half-angle $\alpha$ such that $\cos\alpha=0.95$. We require that the total energy of charged particles within the cone (not including the candidate itself) is less than 0.2~GeV.
The isolated lepton candidate with the highest transverse momentum is selected as the isolated lepton in the event.

Next, we deal with high cross section $\gamma\gamma$ processes that produce soft hadrons 
that overlap with our signal.
Jet finders are used in two steps, following the procedures in~\cite{ILD}.
We apply the $k_{t}$ jet finder algorithm with the jet radius parameter $R=1.4$, forcing all reconstructed particles of the event apart from the isolated lepton into two jets, plus two additional beam jets;
particles that are clustered into the beam jets are removed in the remainder of the event reconstruction~\cite{Catani:1993hr,Ellis:1993tq}.
The value $R=1.4$ was chosen to yield dijet mass distributions which are optimal for the extraction of the kinematic endpoint.
The remaining particles are used to reconstruct the chargino that decayed hadronically.
They are forced into two jets using the Durham jet finding algorithm~\cite{Catani:1991hj}.

\begin{figure}[htbp]
\centering
\begin{subfigure}{0.34\linewidth}
\includegraphics[width=\linewidth]{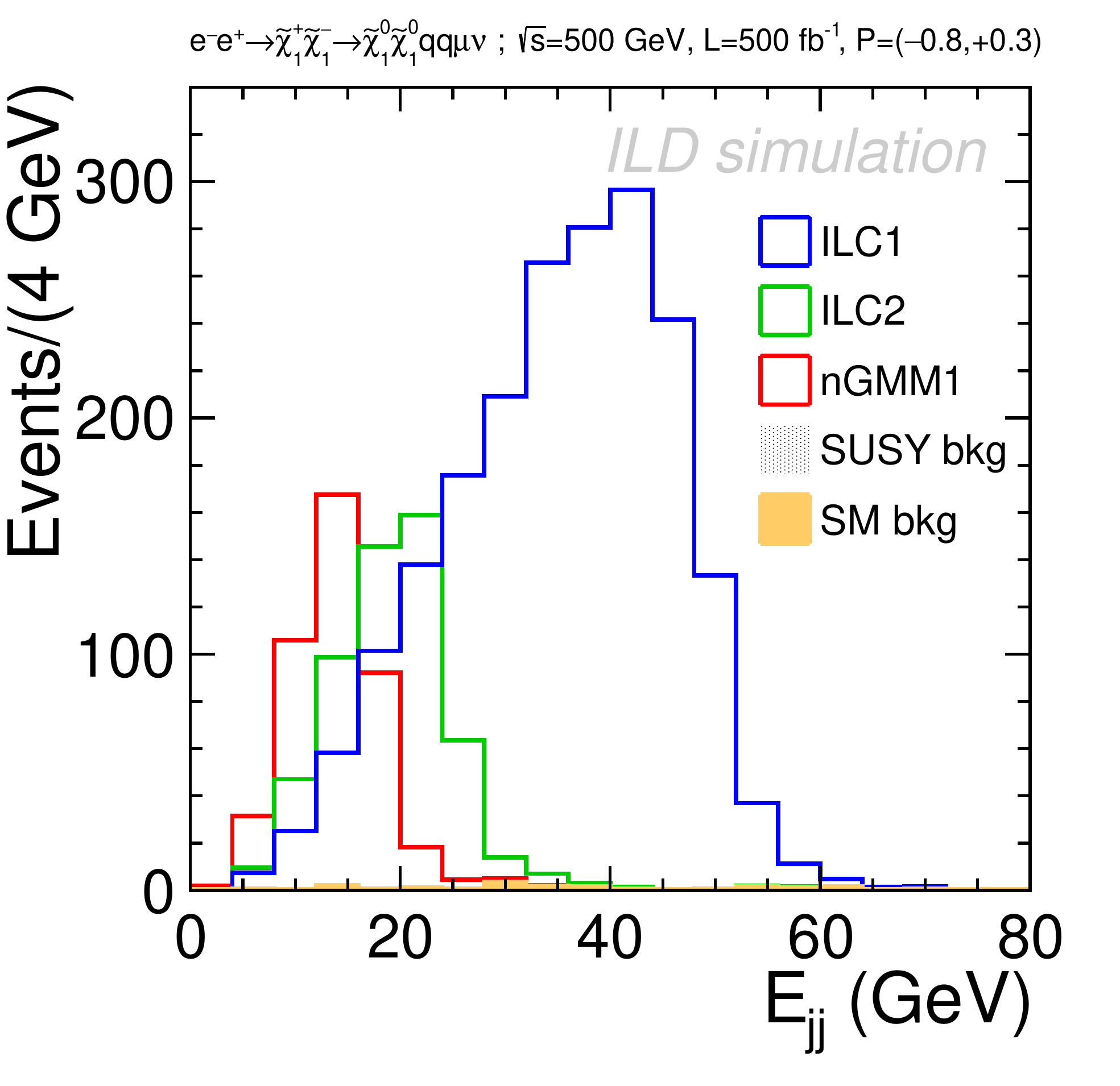}
\caption{di-jet energy \label{fig:chargino_measurement:a}}
\end{subfigure}
\hspace{1cm}
\begin{subfigure}{0.34\linewidth}
\includegraphics[width=\linewidth]{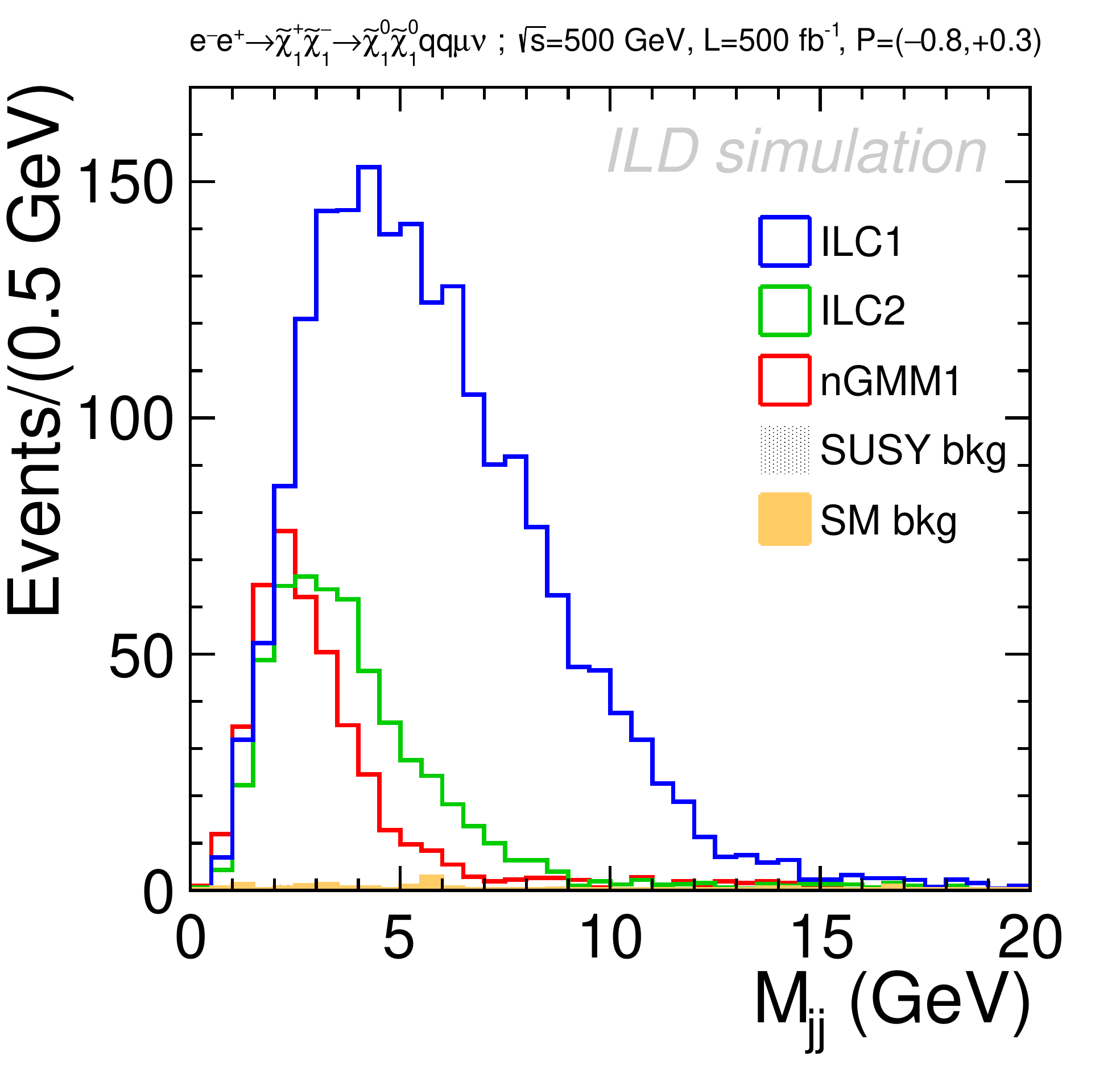}
\caption{di-jet mass \label{fig:chargino_measurement:b}}
\end{subfigure}
\caption{Example of reconstructed distributions in the chargino channel
$e^+e^-\rightarrow \tilde{\chi}_1^+\tilde{\chi}_1^-\rightarrow 
\tilde{\chi}_1^0\tilde{\chi}_1^0q\bar{q}^\prime\mu\nu_{\mu}$
  with beam polarization $\mathcal{P}_{-+}$.
The contributions from SUSY and SM backgrounds are very small. The signal histograms are stacked on top of the backgrounds.}
\label{fig:chargino_measurement}
\end{figure}

The event selection proceeds as follows.
We select events with exactly one isolated lepton candidate, and its lepton type is identified.
We reject events containing particles that are reconstructed in the BeamCal~\cite{Abramowicz:2010bg}.
The transverse momentum of the lepton is required to be 5\,GeV or greater.
The number of reconstructed charged particles in each jet must be 2 or greater. It was tested whether a tighter cut on the track multiplicity would help to reject background from e.g.\ 3-prong $\tau$ decays, but due to low jet energies, especially in the ILC2 and nGMM1 benchmarks, c.f.\  Fig.~\ref{fig:generator_level_distributions:c},  the resulting loss in signal was too severe.
Both of the reconstructed jets should not be very forward, so that the polar angle of each jet $\theta_j$ is such that $|\cos\theta_j|<0.98$.
We require the coplanarity of the two jets as defined by the difference of the azimuthal angle to be $\Delta\phi=|\phi_2-\phi_1|<1.0$.
The angle between the two jets $\theta_{jj}$ is required to satisfy $|\cos\theta_{jj}|<0.2$.
The visible energy in the event is required to be less than 80\,GeV.
The missing energy in the event is required to be greater than 400\,GeV.
The polar angle of the missing momentum $\theta_{\rm miss}$ is required to satisfy $|\cos\theta_{\rm miss}|<0.99$.
The expected number of signal and background events after the event selection is shown in Tab.~\ref{tab:chargino_cutflow}. Very few background events survive after the event selection.
An example of the resulting distributions is shown for the
$qq\mu\nu$ channel with beam polarization $\mathcal{P}_{-+}$ in Fig.~\ref{fig:chargino_measurement}.
The number of events at various steps of the event selection
and the distributions for all studied channels can be found in Appendix~\ref{sec:appendix_event_selection}.

\begin{table}[htbp]
\caption{Expected number of events after the event selction for the chargino signal and major backgrounds, normalized to an integrated luminosity of 500~fb$^{-1}$.
For each benchmark model the SUSY background is given in the column ``Bkg.''. 
}
\label{tab:chargino_cutflow}
\centering
\resizebox{0.99\textwidth}{!}{
\begin{tabular}{ll|rr|rr|rr|rrrrr}
\hline\hline
\multicolumn{2}{c|}{$e^+e^-\rightarrow\widetilde\chi^+_1\widetilde\chi^-_1$} &
\multicolumn{2}{c|}{ILC1} & \multicolumn{2}{c|}{ILC2} & \multicolumn{2}{c|}{nGMM1} & \multicolumn{5}{c}{SM bkg.} \\
Process & $\mathcal{P}$ & Sig. & Bkg. & Sig. & Bkg. & Sig. & Bkg. & $e^+e^-\rightarrow$ 2f & $e^+e^-\rightarrow$ 4f & $\gamma\gamma \rightarrow $ 2f & $e\gamma\rightarrow$ 3f & $\gamma\gamma\rightarrow$ 4f \\
\hline
\multirow{2}{*}{$qqe\nu$} & $\mathcal{P}_{-+}$
&    1463 &      85	&     392 &      23	&     283 &      15	&     5.9 &      64 &     0.0 &      22 &     2.0\\
& $\mathcal{P}_{+-}$
&     404 &      23	&      96 &     4.6	&      73 &     5.1	&     7.4 &      16 &     0.0 &     8.0 &     2.0\\
\hline
\multirow{2}{*}{$qq\mu\nu$} & $\mathcal{P}_{-+}$
&    1862 &     108	&     509 &      28	&     389 &      29	&      33 &      37 &     0.0 &     0.0 &     7.0\\
& $\mathcal{P}_{+-}$
&     524 &      34	&     127 &     8.5	&     101 &     8.2	&     8.2 &     7.2 &     0.0 &     0.0 &     7.0\\
\hline\hline
\end{tabular}
}
\end{table}



\begin{figure}[htbp]
\centering
\begin{subfigure}{0.32\linewidth}
\includegraphics[width=\linewidth]{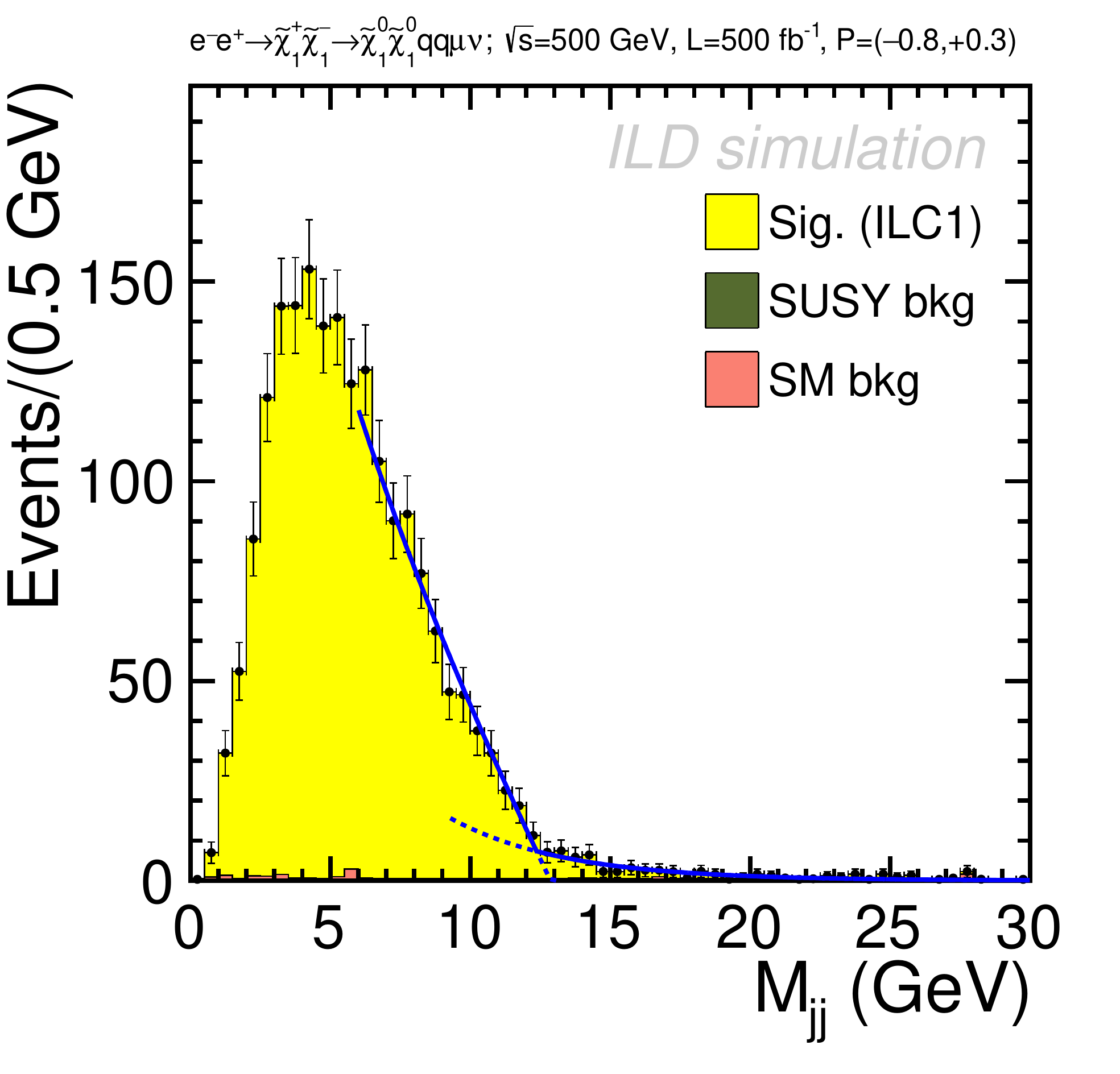}
\caption{ILC1\label{fig:chargino_fit_mass:a}}
\end{subfigure}
\begin{subfigure}{0.32\linewidth}
\includegraphics[width=\linewidth]{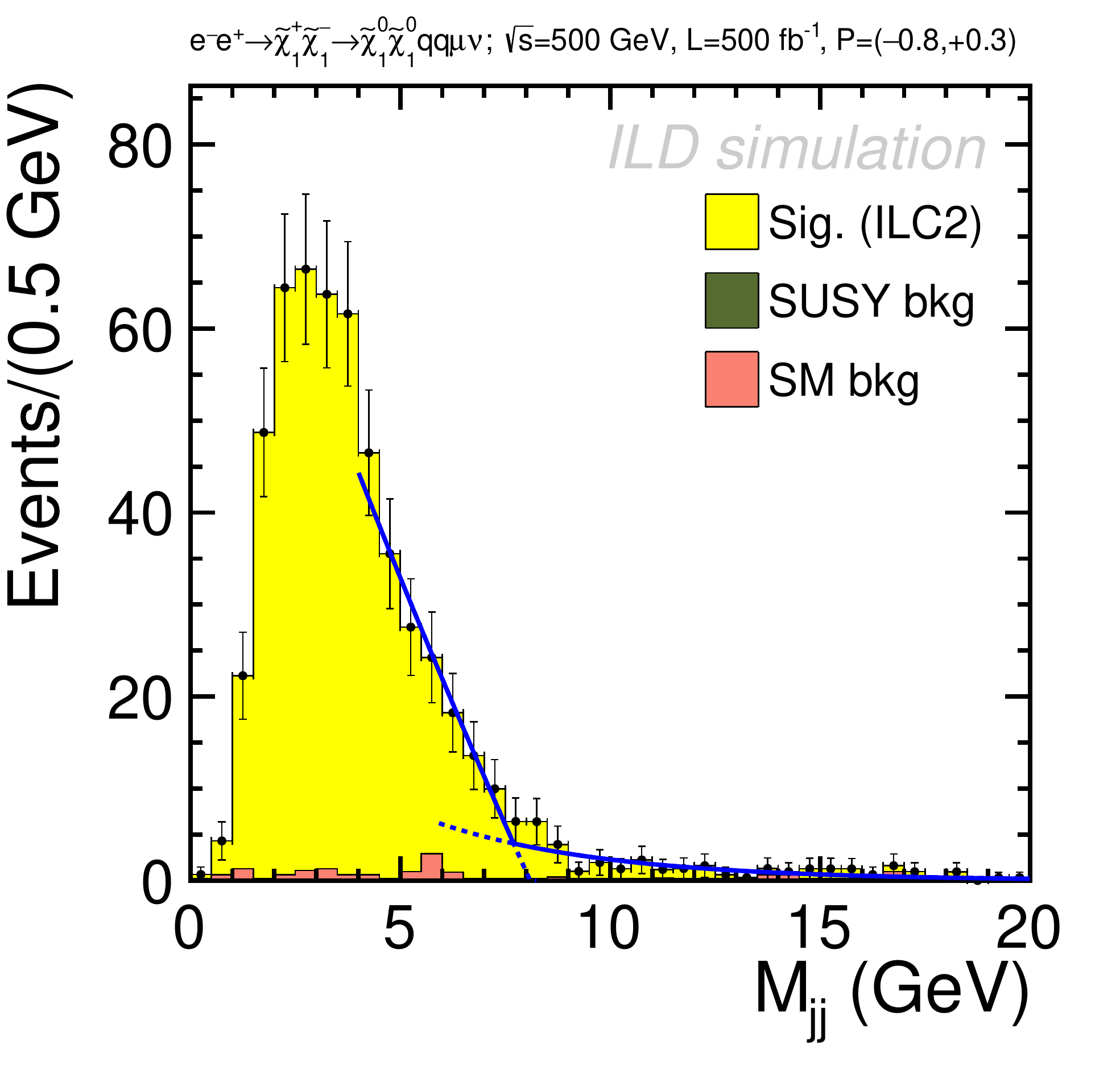}
\caption{ILC2\label{fig:chargino_fit_mass:b}}
\end{subfigure}
\begin{subfigure}{0.32\linewidth}
\includegraphics[width=\linewidth]{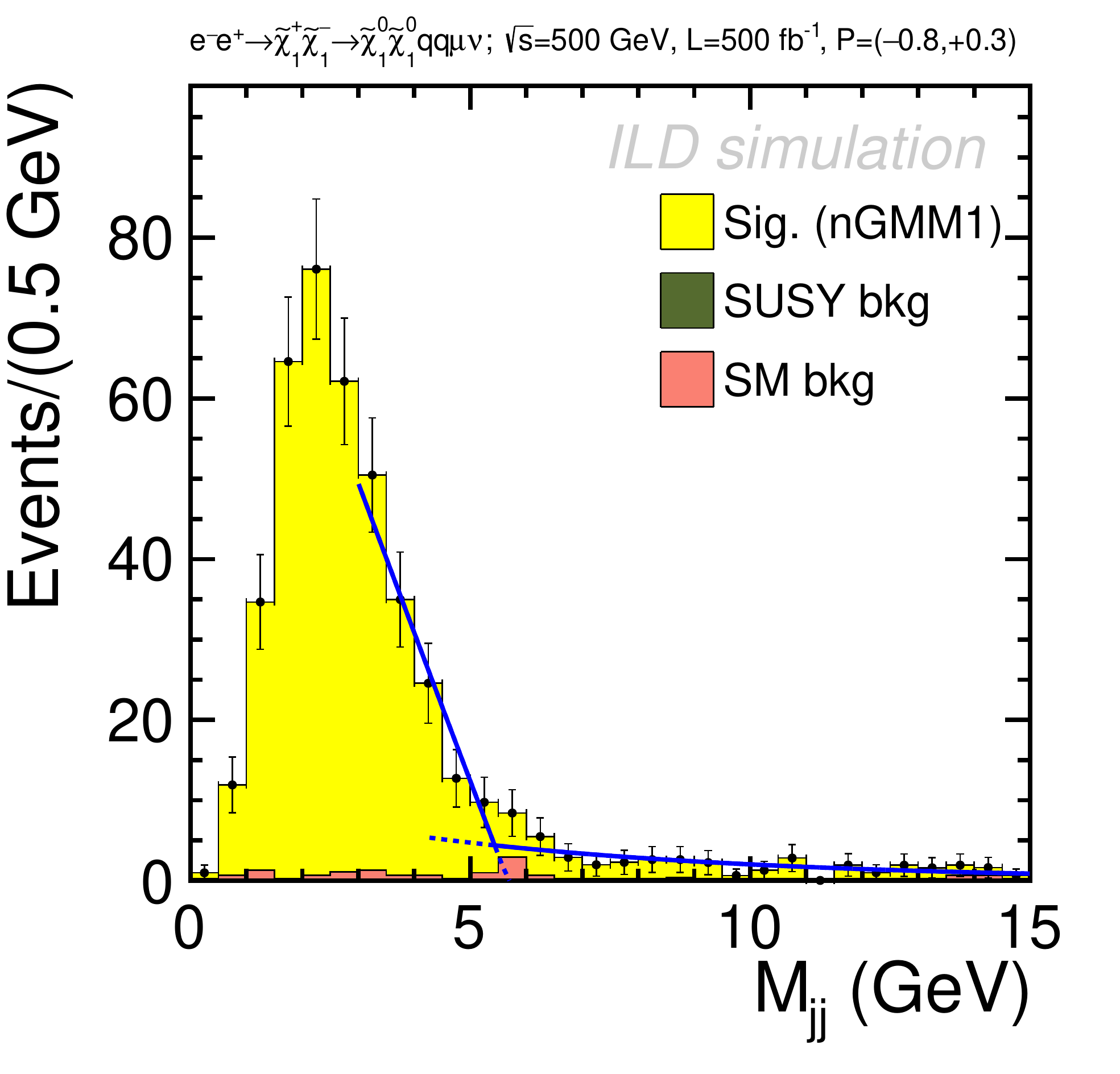}
\caption{nGMM1\label{fig:chargino_fit_mass:c}}
\end{subfigure}
\caption{Example of the endpoint extraction of the di-jet invariant mass system
in the chargino channel
$e^-e^+\rightarrow \tilde{\chi}_1^+\tilde{\chi}_1^-\rightarrow 
\tilde{\chi}_1^0\tilde{\chi}_1^0q\bar{q}^\prime\mu\nu_{\mu}$
with beam polarization $\mathcal{P}_{-+}$
for the three benchmarks.}
\label{fig:chargino_fit_mass}
\end{figure}

\begin{figure}[htbp]
\centering
\begin{subfigure}{0.32\linewidth}
\includegraphics[width=\linewidth]{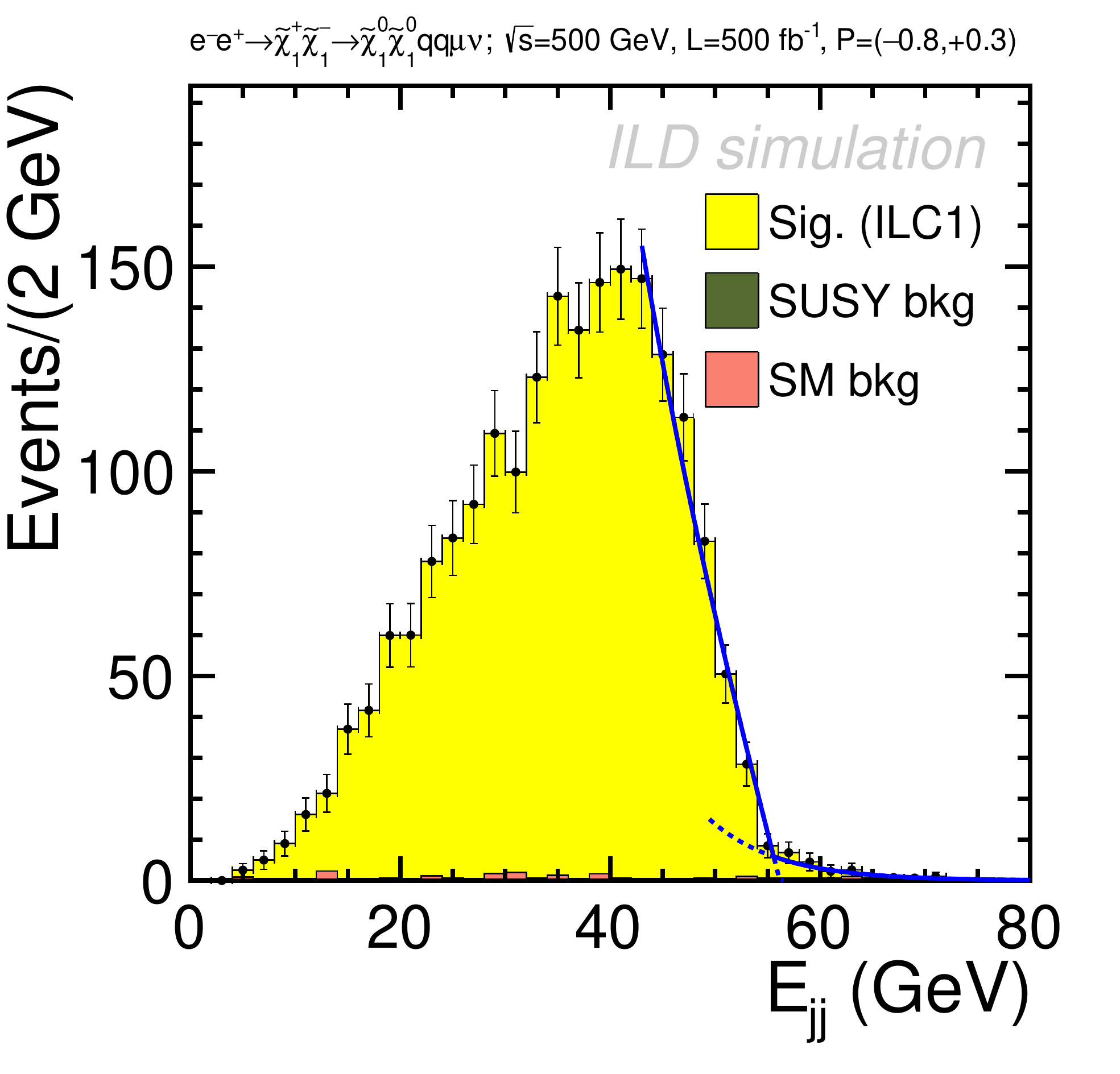}
\caption{ILC1\label{fig:chargino_fit_energy:a}}
\end{subfigure}
\begin{subfigure}{0.32\linewidth}
\includegraphics[width=\linewidth]{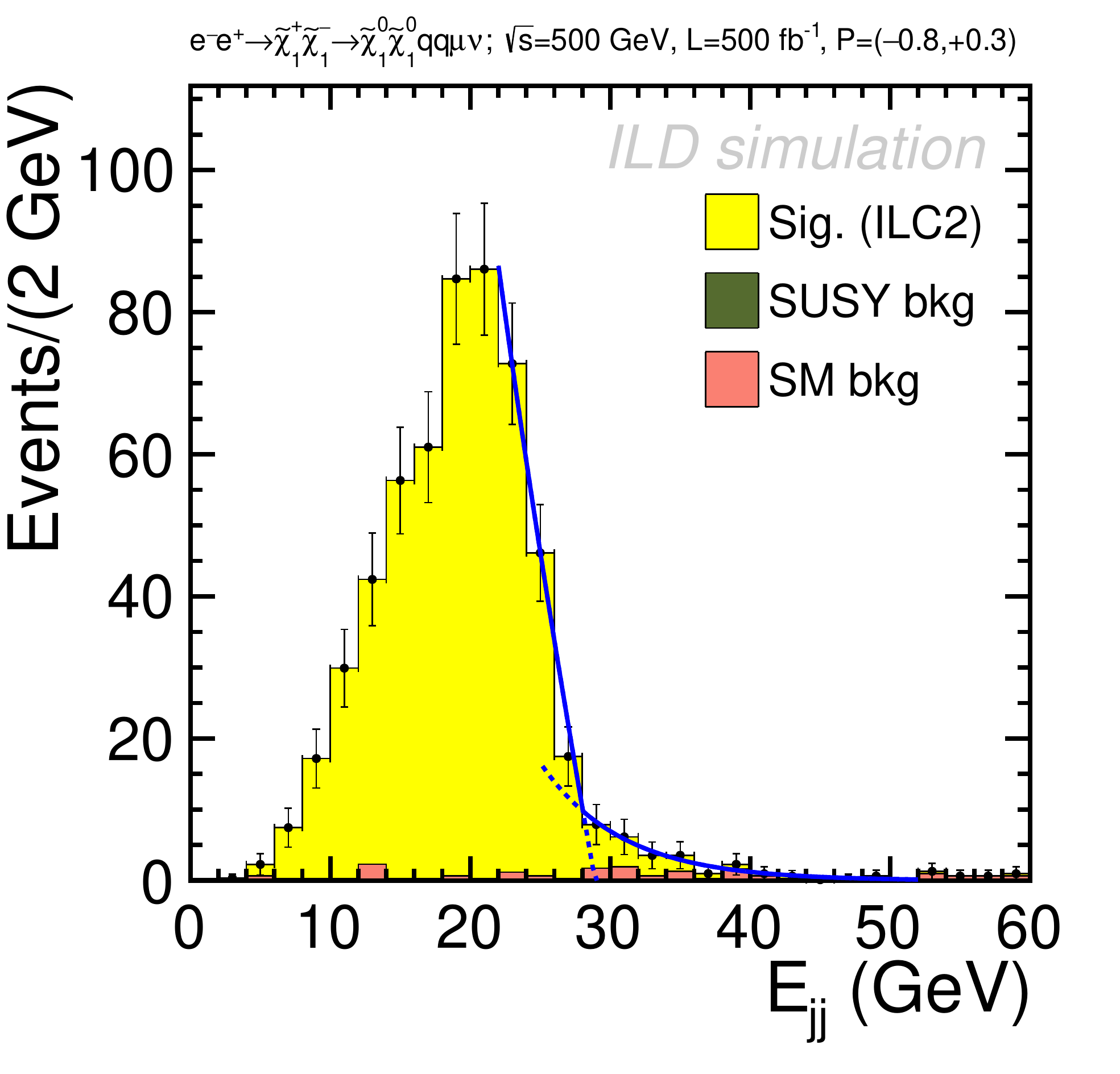}
\caption{ILC2\label{fig:chargino_fit_energy:b}}
\end{subfigure}
\begin{subfigure}{0.32\linewidth}
\includegraphics[width=\linewidth]{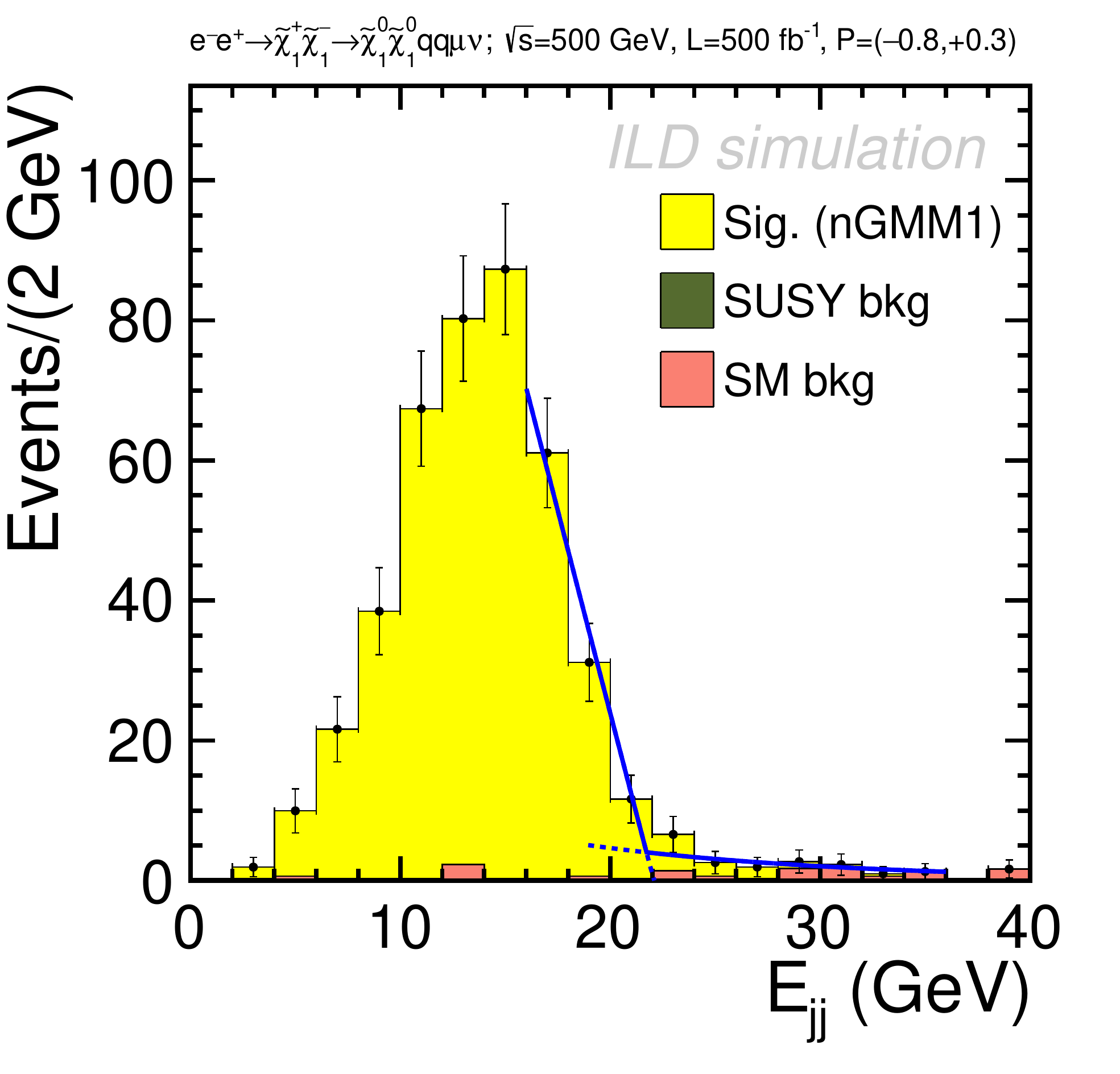}
\caption{nGMM1\label{fig:chargino_fit_energy:c}}
\end{subfigure}
\caption{Example of the endpoint extraction of the di-jet energy system
in the chargino channel
$e^-e^+\rightarrow \tilde{\chi}_1^+\tilde{\chi}_1^-\rightarrow 
\tilde{\chi}_1^0\tilde{\chi}_1^0q\bar{q}^\prime\mu\nu_{\mu}$
with beam polarization $\mathcal{P}_{-+}$
for the three benchmarks.}
\label{fig:chargino_fit_energy}
\end{figure}

The maximum endpoints of the dijet energy ($E_{jj}$) and mass ($M_{jj}$) distributions are extracted using a fit.
Figures~\ref{fig:chargino_fit_energy}-\ref{fig:chargino_fit_mass} show some examples of such a fit.
Although the samples are almost free of backgrounds, the signal distribution has a tail which is caused by the failure to properly reconstruct the energy of soft neutral particles.
We use an exponential curve to model such an effect, combined with a linear function to model the steep drop leading to the kinematic endpoint.
The point where the two functions meet after the fit was used to estimate the kinematic endpoint.
The statistical uncertainty of this value is estimated by performing toy Monte-Carlo experiments that repeatedly fit statistically fluctuated versions of these parent distributions.
The extracted maximum endpoint of the dijet mass distributions is seen to have a systematic shift from the actual mass difference,
which requires a correction at the level of 10-20\% before they are used input to the final fit for the masses.
It is assumed here that such a calibration procedure does not add any significant systematic uncertainties to our results, as described in \cite{Berggren:2013vfa}.
The endpoints from the dijet energy distributions are used without corrections in the mass fit.

\subsection{Neutralino channel}
\label{ssec:neutralino_measurement}

For the neutralino mixed production, we choose the clean leptonic decay of $\neutralinotwo$ as the final state:
$e^-e^+\rightarrow\widetilde\chi^0_1\widetilde\chi^0_2\rightarrow \widetilde\chi^0_1\widetilde\chi^0_1\ell^+\ell^-$
where $\ell=e$ or $\mu$.
The strategy is to look for a pair of isolated leptons with large missing energy.
The invariant mass and the energy of the di-lepton system provides information about the neutralino masses.
The isolated leptons are selected in the same way as in the chargino channel.
This time, we require two oppositely-charged leptons of the same flavor, instead of one.

The expected number of signal and major background events are summarized in Tab.~\ref{tab:neutralino_cutflow};
the full tables of event selection can be found in Appendix~\ref{sec:appendix_event_selection}.
At the preselection stage, we require that two oppositely charged leptons are found, each having a transverse momentum of at least 2\,GeV.
Then, the lepton flavor is required to be either an electron or a muon pair, and the total number of reconstructed charged particles (including the leptons) in the event is required to be exactly two.
We reject events containing particles that are reconstructed in the BeamCal.
The requirement on the transverse momentum of both leptons are further tightened to 2.3~GeV or greater.
The polar angle of each lepton's momentum $\theta_\ell$ is required to satisfy $|\cos\theta_\ell|<0.95$.
The coplanarity of the two leptons $\Delta\phi = |\phi_1-\phi_2|$ is required to satisfy $\Delta\phi<0.8$.
The visible energy of the event is required to be less than 25\,GeV.
The missing energy in the event must be greater than 300\,GeV.
The polar angle of the missing momentum angle $\theta_{\rm miss}$ must satisfy $|\cos\theta_{\rm miss}|<0.98$.

An example of the $M_{\ell\ell}$ and $E_{\ell\ell}$ distributions after this selection is shown in Fig.~\ref{fig:neutralino_measurement};
the full distributions can be found in Appendix~\ref{sec:appendix_event_selection}.
In contrast to the chargino channel,
the neutralino channel has sizable SM backgrounds after the event selection,
since due to the much smaller number of signal events the cuts cannot be as tight as in the chargino case.
The dominant backgrounds are the
$e^+e^-\rightarrow \ell^+\ell^-\nu_{\ell'}\overline{\nu}_{\ell'}$ processes,
where $\ell$ is the same lepton flavor as the final state leptons of the signal.
The SUSY backgrounds remain negligible.

The maximum endpoints of the energy ($E_{\ell\ell}$) and mass ($M_{\ell\ell}$) distributions of the di-lepton system are extracted using a fit.
We show some examples of the fit in Figs.~\ref{fig:neutralino_fit_mass}-\ref{fig:neutralino_fit_energy}.
We use an exponential curve to model the background near the endpoint.
A linear function is used to model the signal part.
The intersection of the two functions is used to extract the kinematic endpoint.
Again, the uncertainty of this value is estimated using toy Monte-Carlo experiments.
Fitting the invariant mass distribution in the nGMM1 benchmark point requires special care due to the $J/\psi$ resonance from the neutralino decay,
which sits on the falling end of the distribution.
The fit is done in two steps.
First, a Gaussian distribution with a narrow width is used to fit the narrow peak in the small window of the $J/\psi$ resonance.
The fitted yield and width of the resonance are fixed in the second, overall fit, which extracts the maximum endpoint.
As was the case for the chargino channel, the extracted maximum endpoint of the dilepton mass distributions
requires a correction at the level of 10-20\% before they are used input to the final fit for the masses,
while the endpoints from the dilepton energy distributions are used without corrections.

\begin{table}[htbp]
\caption{
Expected number of events after the event selction for the neutralino signal and major backgrounds, normalized to an integrated luminosity of 500~fb$^{-1}$.
For each benchmark model the SUSY background is given in the column ``Bkg.''. }
\label{tab:neutralino_cutflow}
\centering
\resizebox{0.99\textwidth}{!}{
\begin{tabular}{ll|rr|rr|rr|rrrrr}
\hline\hline
\multicolumn{2}{c|}{$e^+e^-\rightarrow\widetilde\chi^0_1\widetilde\chi^0_2$} &
\multicolumn{2}{c|}{ILC1} & \multicolumn{2}{c|}{ILC2} & \multicolumn{2}{c|}{nGMM1} & \multicolumn{5}{c}{SM bkg.} \\
Process & $\mathcal{P}$ & Sig. & Bkg. & Sig. & Bkg. & Sig. & Bkg. & $e^+e^-\rightarrow$ 2f & $e^+e^-\rightarrow$ 4f & $\gamma\gamma \rightarrow $ 2f & $e\gamma\rightarrow$ 3f & $\gamma\gamma\rightarrow$ 4f \\
\hline
\multirow{2}{*}{$ee$} & $\mathcal{P}_{-+}$
&    1621 &     185	&    1250 &     226	&     490 &     207	&      14 &    3875 &      14 &     371 &      19\\
& $\mathcal{P}_{+-}$
&    1284 &      69	&    1017 &     111	&     409 &     119	&      13 &     508 &      14 &      83 &      19\\
\hline
\multirow{2}{*}{$\mu\mu$} & $\mathcal{P}_{-+}$
&    1939 &     176	&    1496 &     197	&     640 &      91	&     0.0 &    5506 &      77 &     100 &     9.6\\
& $\mathcal{P}_{+-}$
&    1521 &      49	&    1222 &      67	&     516 &      40	&     0.0 &     672 &      77 &     100 &     9.6\\
\hline\hline
\end{tabular}
}
\end{table}

\begin{figure}[htbp]
\centering
\begin{subfigure}{0.34\linewidth}
\includegraphics[width=\linewidth]{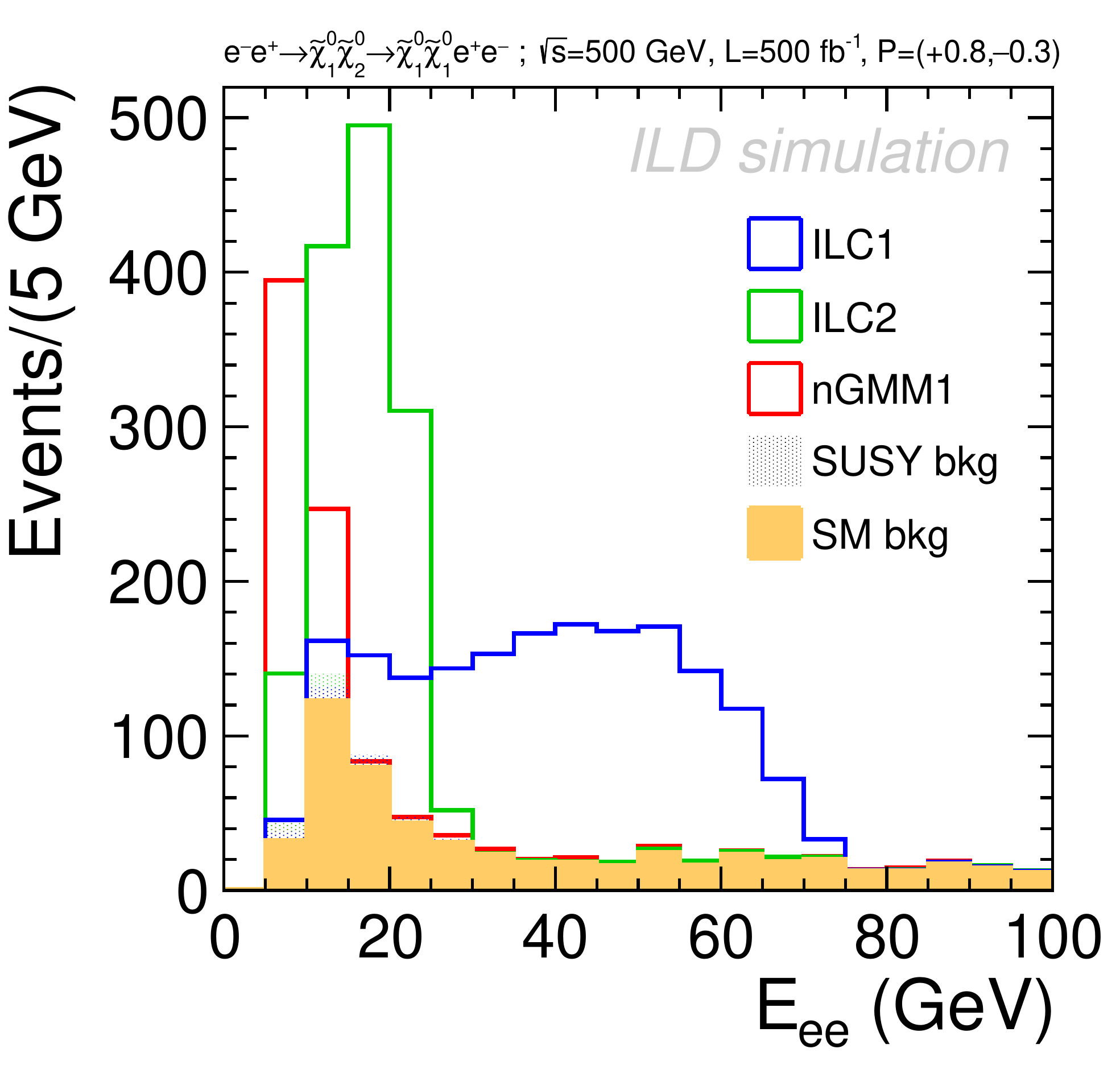}
\caption{di-jet energy \label{fig:neutralino_measurement:a}}
\end{subfigure}
\hspace{1cm}
\begin{subfigure}{0.34\linewidth}
\includegraphics[width=\linewidth]{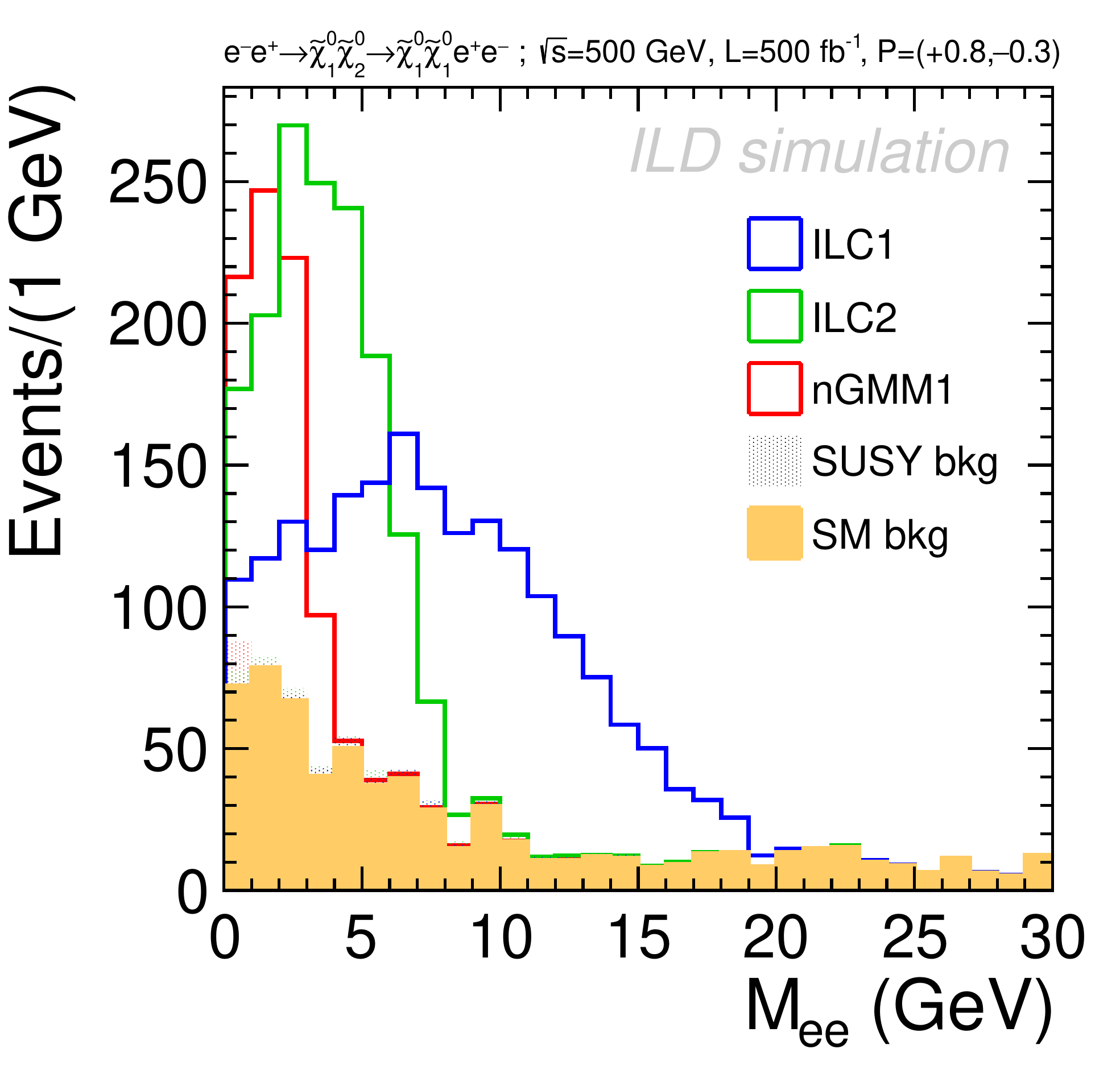}
\caption{di-jet mass \label{fig:neutralino_measurement:b}}
\end{subfigure}
\caption{Example of reconstructed distributions in the neutralino channel
$e^-e^+\rightarrow \tilde{\chi}_1^0\tilde{\chi}_2^0\rightarrow 
\tilde{\chi}_1^0\tilde{\chi}_1^0 e^+e^-$
  with beam polarization $\mathcal{P}_{+-}$.
The signal histograms are stacked on top of the backgrounds.
The remaining background is fully dominated by the SM contribution.
}
\label{fig:neutralino_measurement}
\end{figure}

\begin{figure}[htbp]
\centering
\begin{subfigure}{0.32\linewidth}
\includegraphics[width=\linewidth]{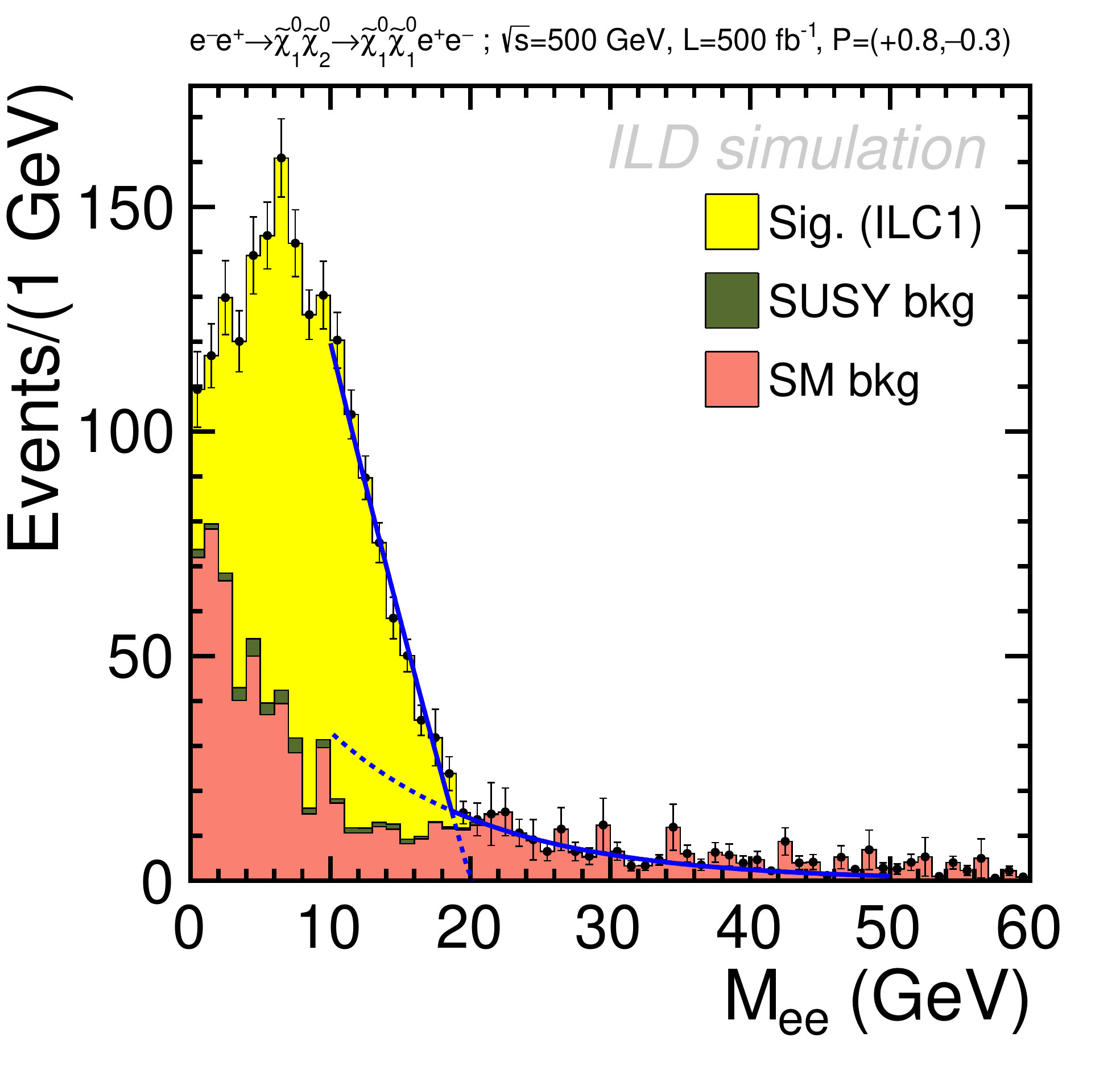}
\caption{ILC1\label{fig:neutralino_fit_mass:a}}
\end{subfigure}
\begin{subfigure}{0.32\linewidth}
\includegraphics[width=\linewidth]{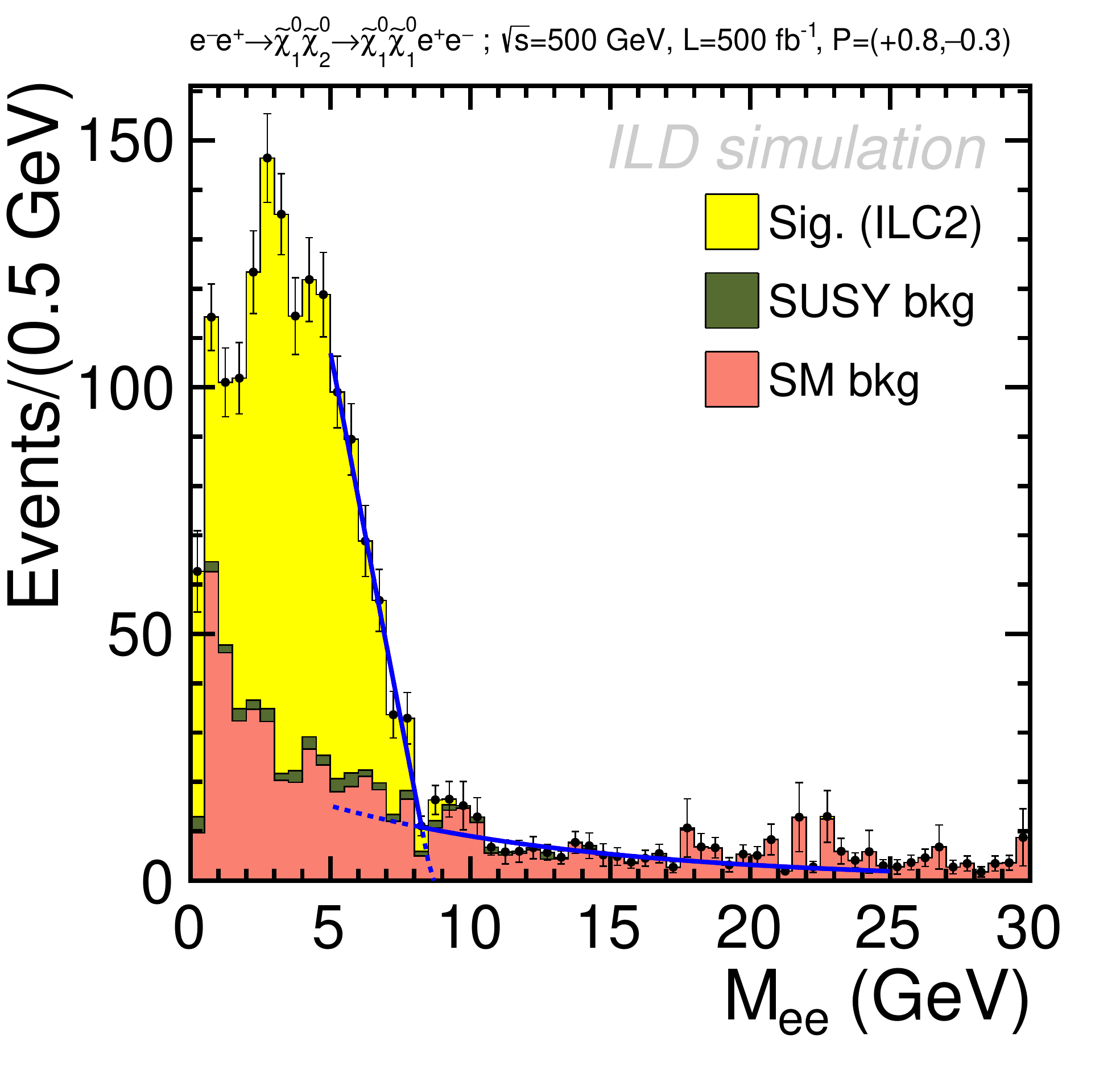}
\caption{ILC2\label{fig:neutralino_fit_mass:b}}
\end{subfigure}
\begin{subfigure}{0.32\linewidth}
\includegraphics[width=\linewidth]{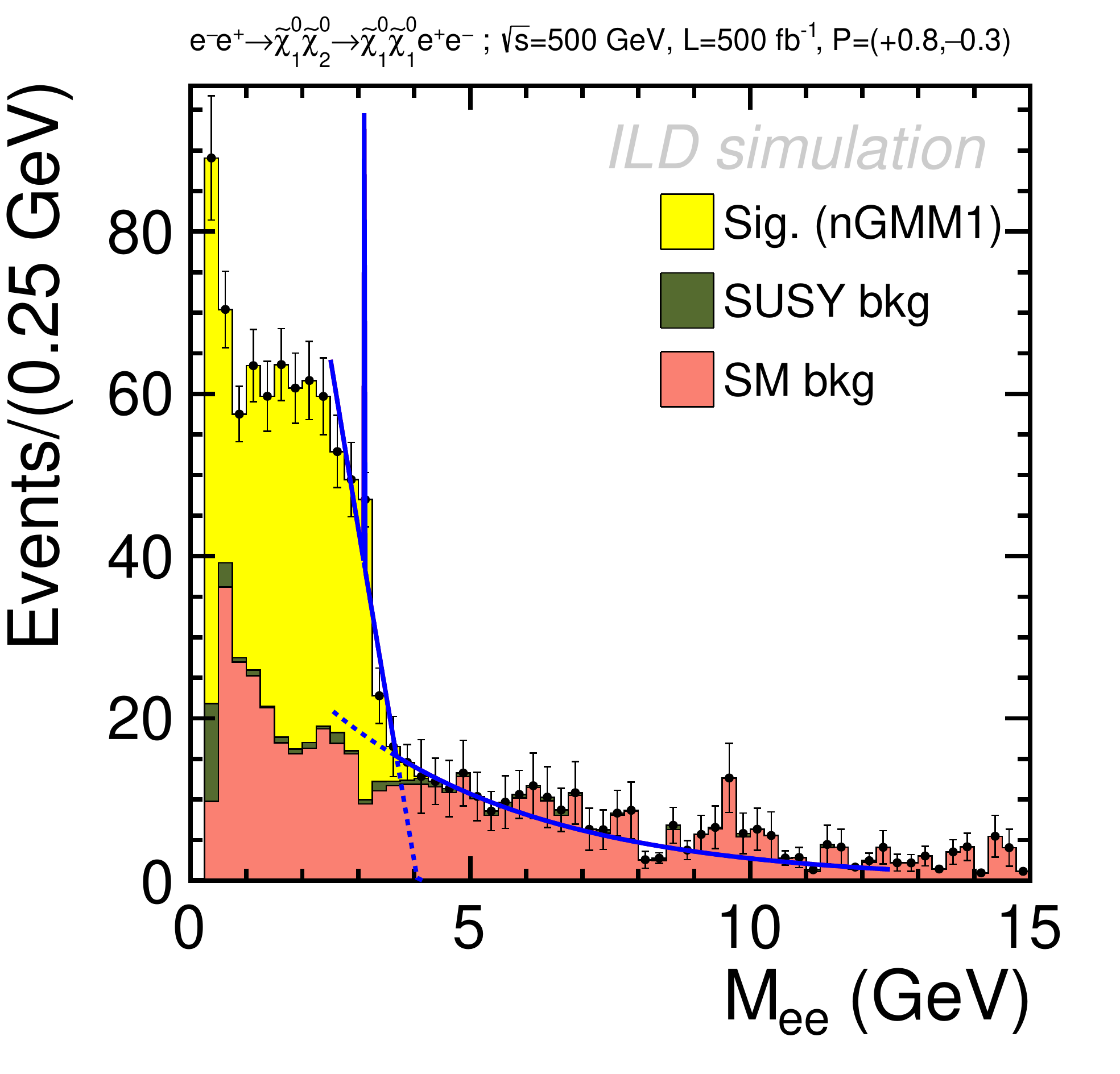}
\caption{nGMM1\label{fig:neutralino_fit_mass:c}}
\end{subfigure}
\caption{Example of the endpoint extraction of the di-lepton invariant mass system
in the neutralino channel
$e^-e^+\rightarrow \tilde{\chi}_1^0\tilde{\chi}_2^0\rightarrow 
\tilde{\chi}_1^0\tilde{\chi}_1^0 e^+e^-$
with beam polarization $\mathcal{P}_{+-}$
for the three benchmarks.
For the nGMM1 benchmark, the $J/\psi$ peak is included in the fit.
}
\label{fig:neutralino_fit_mass}
\end{figure}

\begin{figure}[htbp]
\centering
\begin{subfigure}{0.32\linewidth}
\includegraphics[width=\linewidth]{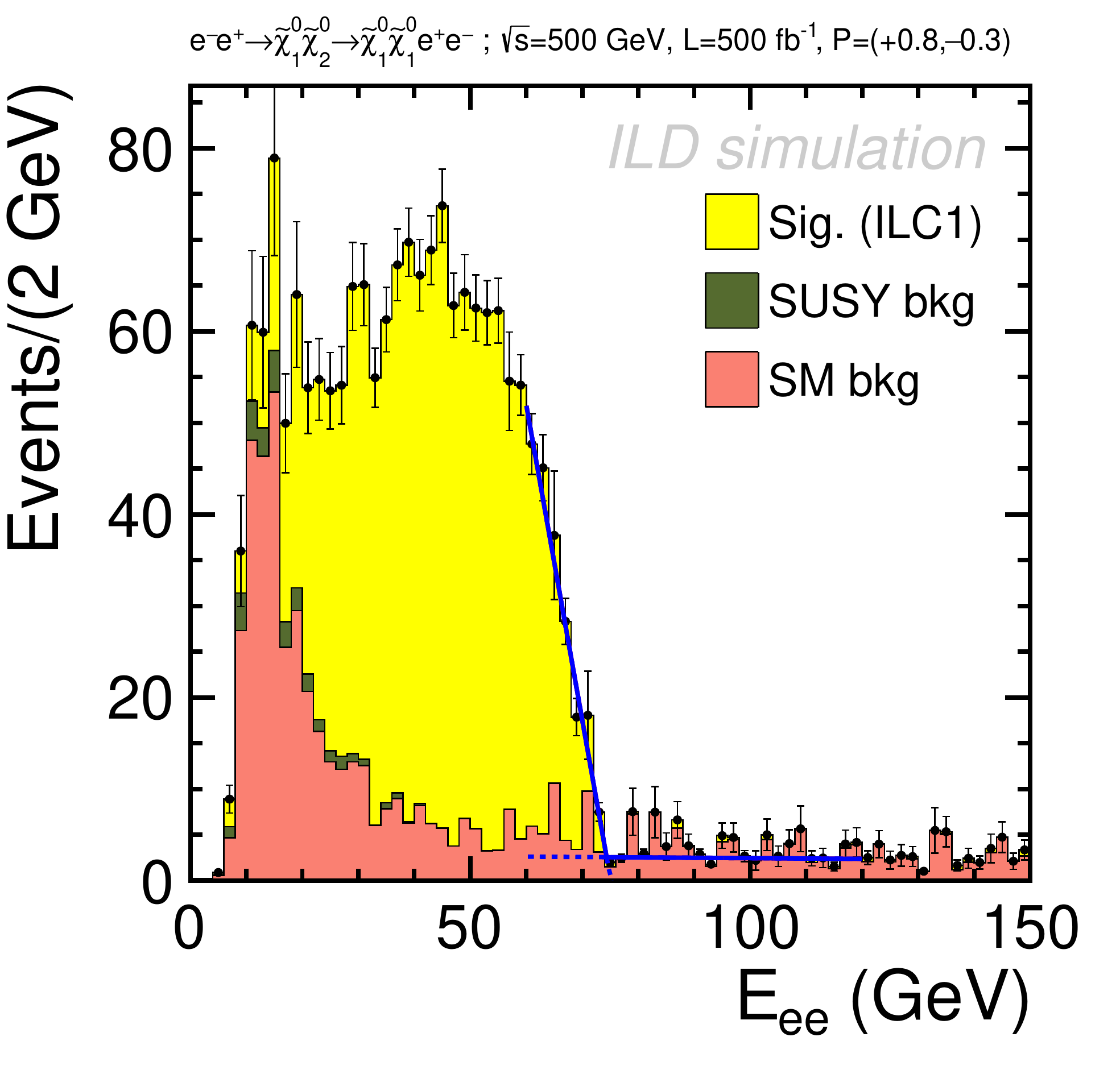}
\caption{ILC1\label{fig:neutralino_fit_energy:a}}
\end{subfigure}
\begin{subfigure}{0.32\linewidth}
\includegraphics[width=\linewidth]{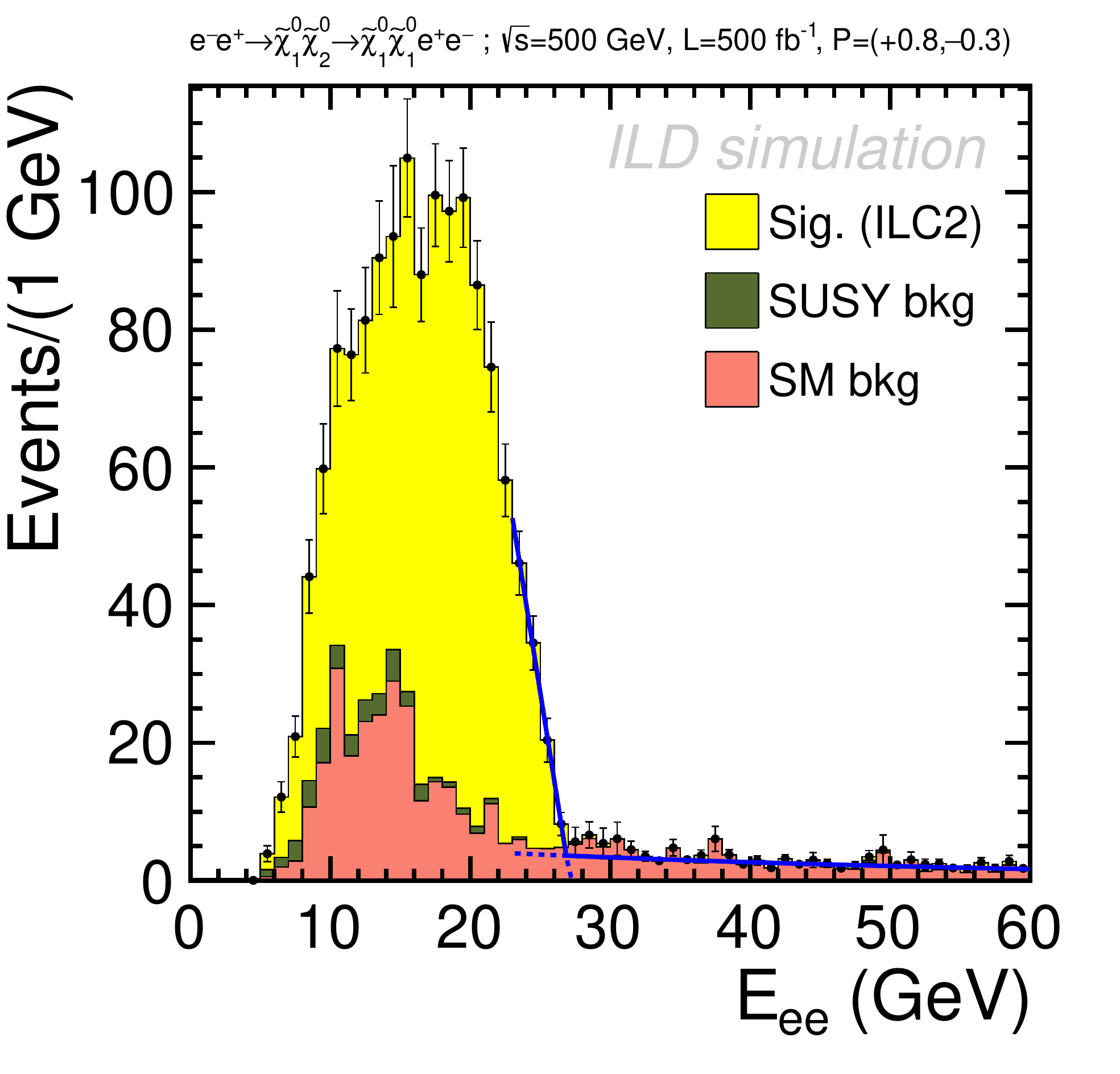}
\caption{ILC2\label{fig:neutralino_fit_energy:b}}
\end{subfigure}
\begin{subfigure}{0.32\linewidth}
\includegraphics[width=\linewidth]{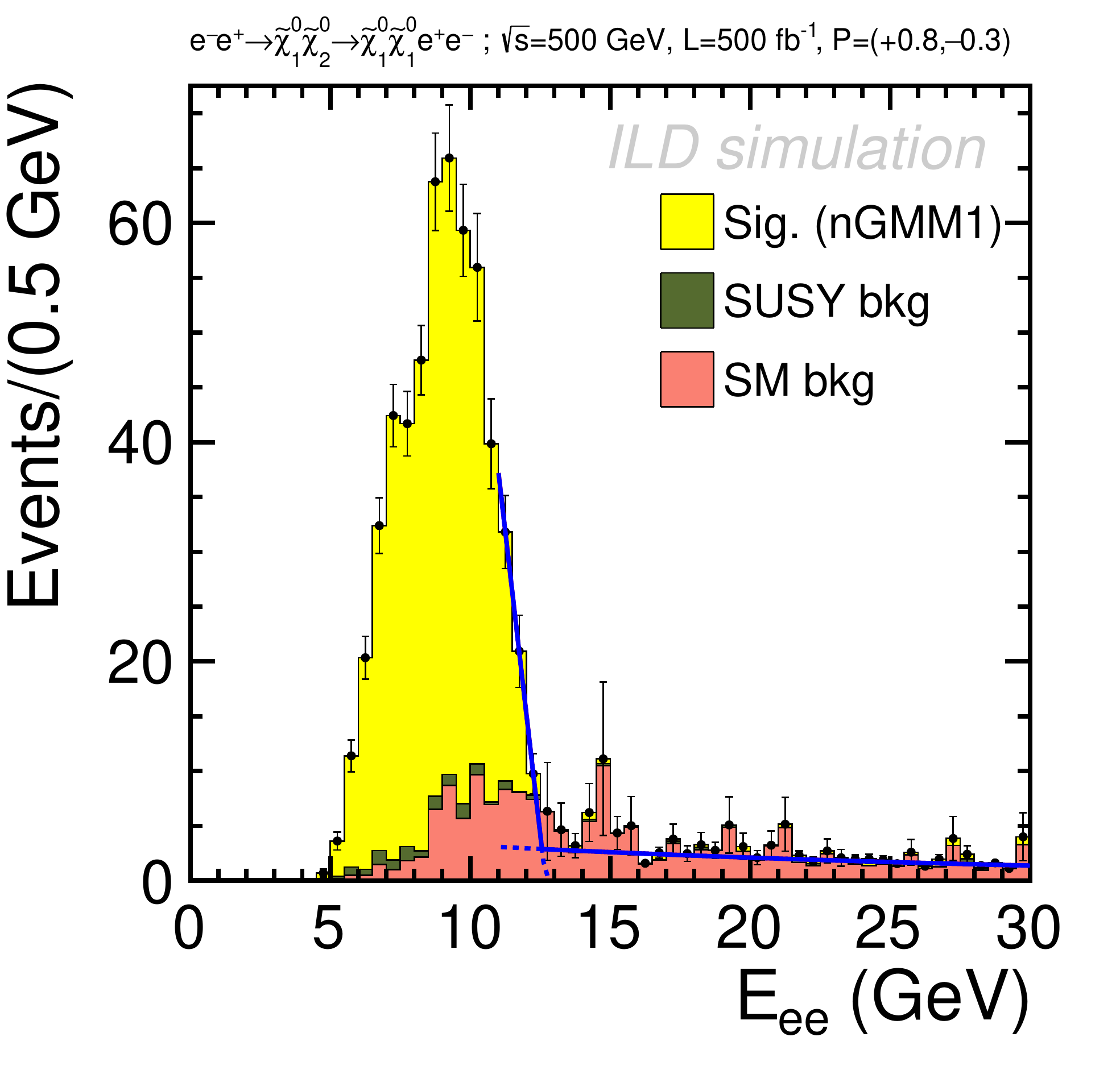}
\caption{nGMM1\label{fig:neutralino_fit_energy:c}}
\end{subfigure}
\caption{Example of the endpoint extraction of the di-lepton energy system
in the neutralino channel
$e^-e^+\rightarrow \tilde{\chi}_1^0\tilde{\chi}_2^0\rightarrow 
\tilde{\chi}_1^0\tilde{\chi}_1^0 e^+e^-$
with beam polarization $\mathcal{P}_{+-}$
for the three benchmarks.}
\label{fig:neutralino_fit_energy}
\end{figure}


\subsection{Results from the full detector simulation study}
\label{ssec:fitinputs}

We present the combined result of the mass measurements in Tab.~\ref{tab:resultmasses}.
Assuming an integrated luminosity of 500~fb$^{-1}$ at each of the two beam polarizations P($e^{+}$, $e^{-}$) = ($\pm$30$\%$ ,$\mp$80$\%$), it is shown that the chargino and neutralino masses can be measured to about 0.5-0.7$\%$ for benchmarks with mass gaps of 10 GeV or larger, and better than 1$\%$ for benchmarks with mass gaps of a few GeV.

\begin{table}[htpb]{}
\centering
\caption{ 
ILC1, ILC2 and nGMM1 MSSM model masses from \texttt{Isajet} (see also Tab.~\ref{tab:bm}).
Experimental mass precision from $\sqrt{s}=500$\,GeV and $\mathcal{L}=500$\,fb$^{-1}$ combining both beam polarizations.
It is assumed that the same precision is valid for these masses as the simulation shows for the \texttt{Isajet} masses.
The scaled precision for 1600 fb$^{-1}$ for each of the two opposite-sign polarisation configurations at $\sqrt{s}=500 \GeV$,
ignoring the data sets with other centre-of-mass energies in the H20 and I20 operating scenarios.
}
\label{tab:resultmasses}
\begin{tabular}{lcccc}
\hline \hline 
\addlinespace[2pt]
  $\sqrt{s}=500$\,GeV only
  && ILC1 & ILC2 & nGMM1 \\[2pt]
\hline
\addlinespace[2pt]
  \multirow{3}{*}{Model mass [GeV]}
  & $ m_{\LSP}          $ & 102.7 & 148.1 & 151.4 \\
  & $ m_{\neutralinotwo}$ & 124.0 & 157.8 & 155.8 \\
  & $ m_{\charginoone}  $ & 117.3 & 158.3 & 158.7 \\[2pt]
\hline
\addlinespace[2pt]
  Precision                                                   & $\delta m_{\LSP}/m_{\LSP}$                       & 0.5  \% & 0.7  \% & 1.0  \% \\
  ($\mathcal{P}_{-+}$, $\mathcal{L}=500$~fb$^{-1}$)           & $\delta m_{\neutralinotwo}/m_{\neutralinotwo}$   & 0.5  \% & 0.7  \% & 1.0  \% \\
  $\oplus$ ($\mathcal{P}_{+-}$, $\mathcal{L}=500$~fb$^{-1}$)  & $\delta m_{\charginoone}/m_{\charginoone}$       & 0.5  \% & 0.7  \% & 1.0  \% \\[2pt]
\hline
\addlinespace[2pt]
  Scaled precision                                            & $\delta m_{\LSP}/m_{\LSP}$                       & 0.3  \% & 0.4  \% & 0.5\%\\
  ($\mathcal{P}_{-+}$, $\mathcal{L}=1600$~fb$^{-1}$)          & $\delta m_{\neutralinotwo}/m_{\neutralinotwo}$   & 0.3  \% & 0.4  \% & 0.5\%\\
  $\oplus$ ($\mathcal{P}_{+-}$, $\mathcal{L}=1600$~fb$^{-1}$) & $\delta m_{\charginoone}/m_{\charginoone}$       & 0.3  \% & 0.4  \% & 0.5\%\\[2pt]
\hline \hline
\end{tabular}
\end{table}

In the last column, the relative precisions on the masses have been scaled to the full luminosity foreseen to be collected at $\sqrt{s}=500$\,GeV according to the H20 and I20 running scenarios of the ILC~\cite{Barklow:2015tja}. These values are considered to be conservative as they neglect further improvements from data sets at lower center-of-mass energies or from dedicated threshold scans.
The relative precisions on the masses range from 0.3\% in case of ILC1 with the largest mass differences to about 0.6\% for nGMM1 as the case with the smallest mass differences.

The precision expected for the cross section times branching ratio measurements
at $\sqrt{s}=500$\,GeV is estimated from the statistical significance
computed from the number of signal and background events
using the di-lepton and di-jet energy distributions.
In the case of the neutralino channel, an additional cut on the di-lepton mass distribution is applied
in order to remove the SM backgrounds in the high mass region;
namely, events with $M_\mathrm{M_{\ell\ell}}$ less than 25, 15, and 5\,GeV are selected for the ILC1, ILC2, and nGMM1 benchmarks, respectively.
The acepted region of the energy distribution is optimized to yield the best statistical significance for each channel.
These were extrapolated to lower center-of-mass energies based on cross section and luminosity scaling.
In most cases, the cross sections can be measured with a precision of a few $\%$, as summarized in Tab.~\ref{tab:higgsinoxsmeasurements}. Notorical exception are the chargino cross sections for the $\mathcal{P}_{+-}$ case, which are typically a factor two worse than the other precisions.

\begin{table}[htbp]{}
\centering
\caption{Estimated experimental precisions for the three benchmark points,
  for the four different final states and the two beam polarizations.
  The full simulation results, performed for $\sqrt{s}=500$~GeV,
  are given for $\mathcal{L}=500$~fb$^{-1}$.
  The scaled precisions for the various center-of-mass energies are shown assuming
  the H20 scenario for ILC1 and the I20 scenario for ILC2 and nGMM1.}
\label{tab:higgsinoxsmeasurements}
\begin{tabular}{c|cc|cc|cc|cc}
\hline \hline 
\addlinespace[2pt]
\multirow{2}{*}{$\Delta(\sigma \times BR)$ [\%]}
  & \multicolumn{2}{c|}{$\sqrt{s}=500$\GeV}
  & \multicolumn{2}{c|}{$\sqrt{s}=500$\GeV}
  & \multicolumn{2}{c|}{$\sqrt{s}=250$\GeV}
  & \multicolumn{2}{c }{$\sqrt{s}=350$\GeV}
  \\[2pt]
  & $\mathcal{P}_{-+}$ & $\mathcal{P}_{+-}$
  & $\mathcal{P}_{-+}$ & $\mathcal{P}_{+-}$
  & $\mathcal{P}_{-+}$ & $\mathcal{P}_{+-}$
  & $\mathcal{P}_{-+}$ & $\mathcal{P}_{+-}$
  \\[2pt]
\hline
\addlinespace[2pt]
  {\textbf{ILC1}}
  & \multicolumn{2}{c|}{$\mathcal{L}=500$ fb$^{-1}$}
  & \multicolumn{6}{c}{Scaled (H20)}
  \\[2pt]
\hline
\addlinespace[2pt]
  $\LSP \neutralinotwo \to \LSP \LSP ee$                                      & 3.98 & 3.13 & 2.22 & 1.75 & 2.04 & 1.60 & 6.94 & 5.47 \\
  $\LSP \neutralinotwo \to \LSP \LSP \mu\mu$                                  & 3.81 & 2.97 & 2.13 & 1.66 & 1.95 & 1.52 & 6.66 & 5.18 \\
  $\widetilde{\chi}^+_1\widetilde{\chi}^-_1  \to \LSP \LSP qq e \nu_e$        & 2.59 & 4.94 & 1.45 & 2.76 & 1.61 & 3.22 & 4.66 & 9.04 \\
  $\widetilde{\chi}^+_1\widetilde{\chi}^-_1  \to \LSP \LSP qq \mu \nu_\mu$    & 2.27 & 4.30 & 1.27 & 2.40 & 1.41 & 2.80 & 4.09 & 7.87
  \\[2pt]
\hline
\addlinespace[2pt]
  {\textbf{ILC2}}
  & \multicolumn{2}{c|}{$\mathcal{L}=500$ fb$^{-1}$}
  & \multicolumn{6}{c}{Scaled (I20)}
  \\[2pt]
\hline
\addlinespace[2pt]
  $\LSP \neutralinotwo \to \LSP \LSP ee$                                       & 3.92 &  3.50 & 2.19 & 1.96 & - & - & 2.82 & 2.52 \\
  $\LSP \neutralinotwo \to \LSP \LSP \mu\mu$                                   & 3.90 &  3.33 & 2.18 & 1.86 & - & - & 2.81 & 2.40 \\
  $\widetilde{\chi}^+_1\widetilde{\chi}^-_1  \to \LSP \LSP qq e \nu_e$         & 5.17 & 10.30 & 2.89 & 5.76 & - & - & 4.09 & 8.28 \\
  $\widetilde{\chi}^+_1\widetilde{\chi}^-_1  \to \LSP \LSP qq \mu \nu_\mu$     & 4.39 &  8.84 & 2.45 & 4.94 & - & - & 3.47 & 7.10
  \\[2pt]
\hline
\addlinespace[2pt]
  {\textbf{nGMM1}}
  & \multicolumn{2}{c|}{$\mathcal{L}=500$ fb$^{-1}$}
  & \multicolumn{6}{c}{Scaled (I20)}
  \\[2pt]
\hline
\addlinespace[2pt]
  $\LSP \neutralinotwo \to \LSP \LSP ee$                                       &  5.30 &  4.98 & 2.96 & 2.78 & - & - & 3.84 & 3.61 \\
  $\LSP \neutralinotwo \to \LSP \LSP \mu\mu$                                   &  5.05 &  4.64 & 2.82 & 2.59 & - & - & 3.66 & 3.36 \\
  $\widetilde{\chi}^+_1\widetilde{\chi}^-_1  \to \LSP \LSP qq e \nu_e$         &  6.20 & 11.73 & 3.47 & 6.56 & - & - & 4.94 & 9.48 \\
  $\widetilde{\chi}^+_1\widetilde{\chi}^-_1  \to \LSP \LSP qq \mu \nu_\mu$     &  4.99 &  9.90 & 2.79 & 5.53 & - & - & 3.98 & 8.00
  \\[2pt]
\hline \hline
\end{tabular}
\end{table}

\section{Fitting fundamental parameters}
\label{sec:fit}

In this section we will pursue the question of whether the projected precisions 
on the physics observables will be sufficient to discriminate between different 
SUSY models, and to determine the parameters of the correct model. 
To this purpose, assumed measurements of Higgsino masses and polarized cross sections are presented to \texttt{Fittino} along with their projected uncertainties. It should be noted that the asummed measurements have not been varied randomly around their true values.
Thus, in all cases where the correct underlying model is fitted, the expected $\chi^2$ is zero, apart from effects of finite numerical precision and the finite length of the Markov chains. As discussed in Sec.~\ref{subsec:parfit_tools}, \texttt{SPheno} had to be used instead of \texttt{Isajet}
as a spectrum calculator in the fitting step.\footnote{Here we 
switch to the spectrum generator \texttt{SPheno} since there exists 
a direct interface between \texttt{SPeno} and \texttt{Fittino} 
while no such interface exists for \texttt{Fittino} and \texttt{Isajet}. 
The mass spectra generated from \texttt{Isajet} and
\texttt{SPheno} are slightly different due to different algorithms used by 
the code authors.}  

The fits include the following inputs based on ILC simulation studies:
\begin{itemize}
\item The higgsino masses obtained with \texttt{SPheno3.3.9beta}
as listed in Tab.~\ref{tab:fitinputmasses}, together with their estimated precisions based on a preliminary version of 
the full simulation study at $\sqrt{s}=500$\,GeV, listed in Tab.~7.11 of Ref.~\cite{Lehtinen:PhDThesis}. Note that these 
are between 30\% and 100\% more conservative than the results given in Tab.~\ref{tab:resultmasses} of this paper. 
We will study and discuss the relevance of these differences in Sec.~\ref{ssec:test:nGMM1}.
\item The polarised total cross sections for chargino and neutralino production at all relevant center-of-mass 
energies with precisions as given in Tab.~7.12-7.14 of Ref.~\cite{Lehtinen:PhDThesis}. As for the masses, 
these are between 20\% and 100\% more conservative than the full simulation results listed in 
Tab.~\ref{tab:higgsinoxsmeasurements}, and we will study and discuss the relevance of these differences in 
Sec.~\ref{ssec:test:nGMM1}. Note that in ILC1, higgsino production is kinematically accessible at center-of-mass 
energies as low as $250$\,GeV, while for ILC2 and nGMM1 $\sqrt{s}=350$\,GeV is the lowest ILC energy stage which 
allows higgsino production. Therefore, we consider the alternative running scenario I20~\cite{Barklow:2015tja} 
for these benchmarks.
\item The mass of the lightest $CP$-even Higgs boson, with the ILC precision according to~\cite{Yan:2016xyx}, as 
discussed in Sec.~\ref{subsec:parfit_tools}.
\item The branching ratios $BR(h \to b\bar{b})$, $BR(h \to c\bar{c})$, $BR(h \to \tau^+\tau^-)$, $BR(h \to gg)$, 
$BR(h \to W^+W^-)$, $BR(h \to ZZ)$ and $BR(h \to \gamma \gamma)$, with the ILC precisions according 
to~\cite{Barklow:2015tja}, as discussed in Sec.~\ref{subsec:parfit_tools}.
\end{itemize}
In total, these amount to 25 observables. In the following, we will start by discussing fits of different 
GUT-scale models in Sec.~\ref{ssec:gutscale_fits},
and before proceeding to the determination of weak-scale parameters in Sec.~\ref{ssec:weakscale_fits}.
Finally, we will address the predictions of the LSP's relic density in Sec.~\ref{ssec:omegaDM}.

\begin{table}[htpb]{}
\centering
\caption{
ILC1, ILC2 and nGMM1 MSSM model masses from \texttt{SPheno3.3.9beta} together with the input precisions assumed in the fit.
The assumed input precisions are given for $\sqrt{s}=500$\,GeV and $\mathcal{L}=1600$~fb$^{-1}$.
Data sets with other center-of-mass energies in the H20 and I20 operating scenarios are neglected.}
\label{tab:fitinputmasses}
\begin{tabular}{lcccc}
\hline \hline 
\addlinespace[2pt]
  $\sqrt{s}=500$\,GeV only
  && ILC1 & ILC2 & nGMM1 \\[2pt]
\hline
\addlinespace[2pt]
  \multirow{3}{*}{Model mass [GeV]}
  & $ m_{\LSP}          $ & 104.8 & 151.3 & 154.9 \\
  & $ m_{\neutralinotwo}$ & 127.5 & 162.4 & 160.2 \\
  & $ m_{\charginoone}  $ & 116.0 & 157.0 & 157.4 \\[2pt]
\hline
\addlinespace[2pt]
  Assumed precision                                           & $\delta m_{\LSP}/m_{\LSP}$                      & 0.5\% & 0.7\% & 1.0\% \\
  ($\mathcal{P}_{-+}$, $\mathcal{L}=1600$~fb$^{-1}$)          & $\delta m_{\neutralinotwo}/m_{\neutralinotwo}$  & 0.4\% & 0.7\% & 1.0\% \\
  $\oplus$ ($\mathcal{P}_{+-}$, $\mathcal{L}=1600$~fb$^{-1}$) & $\delta m_{\charginoone}/m_{\charginoone}$      & 0.5\% & 0.7\% & 1.0\% \\[2pt]
\hline \hline
\end{tabular}
\end{table}

\subsection{Fitting GUT-scale parameters}
\label{ssec:gutscale_fits}
The fitting of GUT-scale parameters requires strong assumptions on the underlying SUSY breaking scheme.
Since our benchmarks cover two very different approaches to unification, it is interesting to study whether these
can be distinguished by directly fitting different GUT-scale models.

\subsubsection{Results of fitting NUHM2} 

In case of NUHM2, we fit the parameters $M_{1/2}$, $\mu$, $\tan \beta$, $M_0$, $A_0$ and $m_A$ to the observables described in Sec.~\ref{ssec:fitinputs}. Table~\ref{tab:ILC1nuhm2fittedparam} shows the best fit point and its 1 and 2$\sigma$ confidence intervals obtained in the case of the ILC1 benchmark, in comparison to the input model parameters. Tables~\ref{tab:ILC2nuhm2fittedparam} and~\ref{tab:nGMM1nuhm2fittedparam} give the analogous information for the ILC2 and nGMM1 benchmarks, respectively. In case of the ILC1 and ILC2 benchmarks, where NUHM2 is the correct underlying model, 
the $\chi^2$ of the best fit point is very small, and all fitted parameters agree well with their true
input values. The 1$\sigma$ uncertainties for $M_{1/2}$, $\mu$, $\tan \beta$ are typically 10\% or better, while $M_0$, $A_0$ and $m_A$, which enter only at loop-level into the considered observables, are still determined within about 20\%.

\begin{table}[htbp]{}
\centering
$
\renewcommand{\arraystretch}{1.4}
\begin{array}{  c c c c c c c }
\hline  \hline
\text{parameter} & \text{ILC1 NUHM2 true} &  \text{best fit point } & 1\sigma \text{ CL} & 2\sigma \text{ CL} \\ \hline  
M_{1/2} & 568.3 & 556.7 & \ ^{+24.3}_{-20.3} &  \ ^{+37.7}_{-43.1}   \\
\mu &  115.0  & 105.3  & \ ^{+12.8}_{-8.2} &  \ ^{+14.0}_{-14.5}  \\
\tan \beta &  10.0 & 11.4  & \ ^{+5.6}_{-1.6} &  \ ^{+11.4}_{-1.6}  \\
m_{A} & 1000 & 968  & \ ^{+167}_{-65} &  \ ^{+288}_{-130}   \\
M_{0} & 7025 & 7685  & \ ^{+1243}_{-1917} &  \ ^{+2311}_{-2095}  \\
A_{0} & -10427  & -11064  & \ ^{+2695}_{-1422} &  \ ^{+2927}_{-2698}  \\ \hline
\chi^2 & 0.0013 & 0.0011 & & \\ 
\hline \hline
\end{array}
$
\caption{Fitted parameters in the fit of NUHM2 parameters to ILC1 observables in the H20 scenario. All values in GeV apart from $\tan{\beta}$.  Note that the $\chi^2$ value of the model point is increased from 0 by the rounding errors of the observables in the inputs.}
\label{tab:ILC1nuhm2fittedparam}
\end{table}

\begin{table}[htbp]{}
\centering
$
\renewcommand{\arraystretch}{1.4}
\begin{array}{  c c c c c c c }
\hline \hline
\text{parameter} & \text{ILC2 NUHM2 true} &  \text{best fit point } & 1\sigma \text{ CL} & 2\sigma \text{ CL} \\ \hline  
M_{1/2} & 1200 & 1194 & \ ^{+107}_{-68} &  \ ^{+164}_{-129}   \\
\mu &   150.0 & 150.7 & \ ^{+4.3}_{-4.5} &  \ ^{+7.2}_{-5.2}     \\
\tan \beta &  15.0 & 16.0 &  \ ^{+26.2}_{-6.6} &  \ ^{+28.8}_{-6.6}   \\
m_{A} & 1000 & 1008 &  \ ^{+141}_{-118} &  \ ^{+256}_{-196}    \\
M_{0} & 5000 & 4788 &  \ ^{+2546}_{-3137} &  \ ^{+3566}_{-3283}   \\
A_{0} & -8000  & -7663 &  \ ^{+3817}_{-3926} &  \ ^{+3817}_{-5342}    \\ \hline
\chi^2 & 0.0007 & 0.02848 & & \\
\hline \hline
\end{array}
$
\caption{Best fit point and confidence intervals of the NUHM2 parameters fitted to ILC2 SUSY and Higgs measurements in the I20 operating scenario. Note that the $\chi^2$ value of the model point is increased from 0 by the rounding errors of the observables in the inputs.}
\label{tab:ILC2nuhm2fittedparam}
\end{table}

In case of the nGMM1 benchmark, the $\chi^2$ of the best fit point is somewhat larger than for the other two benchmarks, but still so small that an NUHM2 interpretation of this benchmark cannot be rejected. This is not surprising as it has been constructed to have the physical observables very similar to ILC2. However, the best fit point is found for $M_{1/2}$ about a factor two bigger, and $M_A$ about 60\% larger, than in ILC2. 
This implies that a direct observation of the heavy Higgs bosons 
and the electroweakinos could distinguish the two models. 
However this also 
raises the question of whether the weak-scale fits based on input from the higgsino properties alone will be able to identify the nGMM1 benchmark unambiguously as a non-NUHM2 model, with a completely different underlying SUSY breaking mechanism. We will investigate this in the next section.

\begin{table}[htbp]{}
\centering
$
\renewcommand{\arraystretch}{1.4}
\begin{array}{  c c  c c c c }
\hline \hline
\text{parameter} &   \text{best fit point } & 1\sigma \text{ CL} & 2\sigma \text{ CL} \\ \hline  
M_{1/2} &  2407 & \ ^{+150}_{-135} &  \ ^{+356}_{-215}   \\
\mu &    155.6 & \ ^{+1.5}_{-1.9} &  \ ^{+4.0}_{-2.9}   \\
\tan \beta & 10.0     & \ ^{+2.1}_{-0.5} &  \ ^{+2.4}_{-0.7}   \\
m_{A} &  1603 & \ ^{+528}_{-279} &  \ ^{+1026}_{-469}   \\
M_{0} &  3422 & \ ^{+3309}_{-820} &  \ ^{+4435}_{-1196}   \\
A_{0} &  -7409 & \ ^{+666}_{-3756} &  \ ^{+887}_{-5304}   \\ \hline
\chi^2 & 0.233 & & \\
\hline \hline
\end{array}
$
\caption{Best fit point and confidence intervals of the NUHM2 parameters fitted to nGMM1 SUSY and Higgs measurements in the I20 operating scenario. }
\label{tab:nGMM1nuhm2fittedparam}
\end{table}

\begin{figure}[htbp]
\begin{subfigure}{0.33\linewidth} \includegraphics[width=\textwidth]{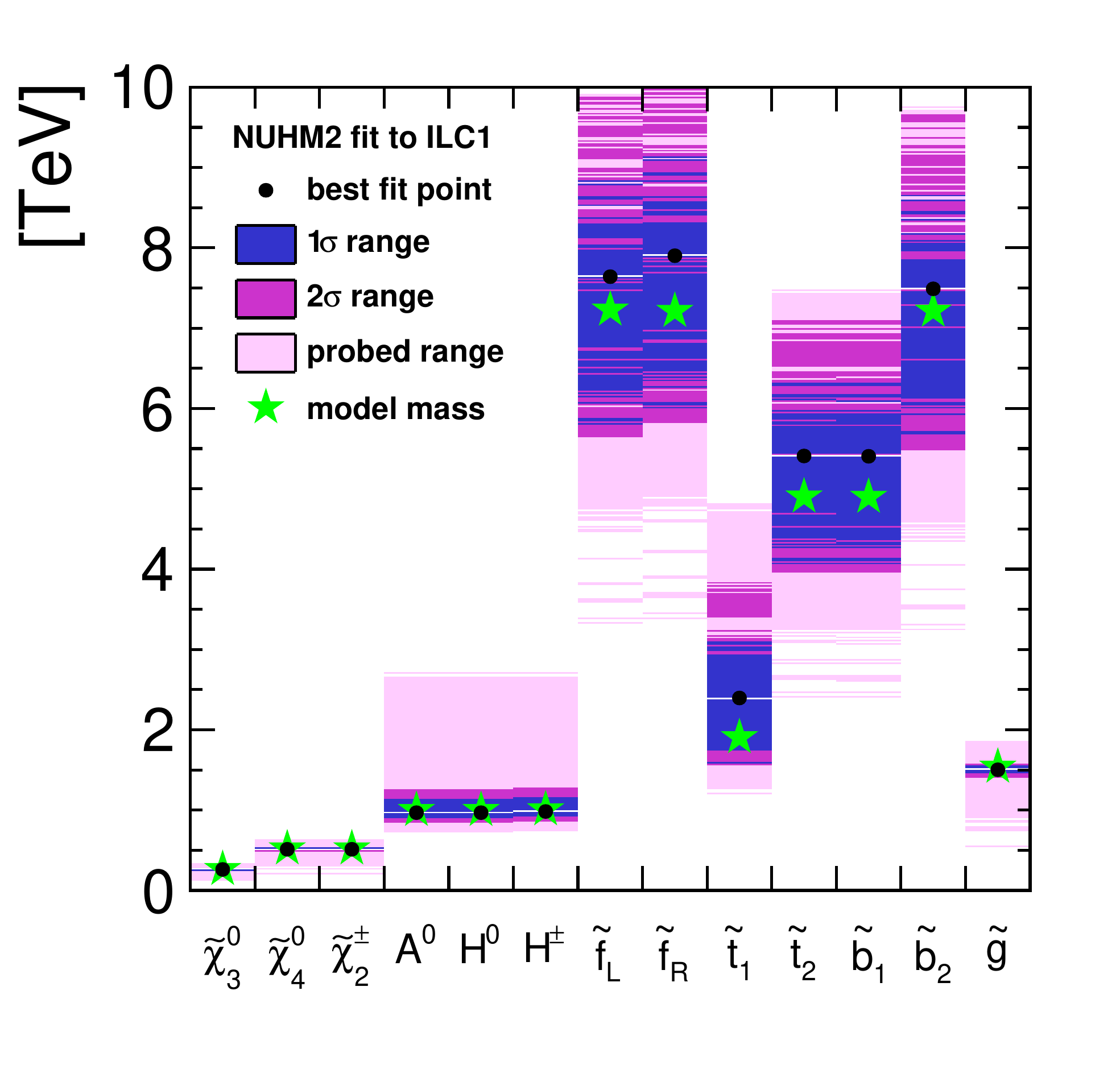}
 \caption{ILC1\label{sfig:nuhm2fittedmassesILC1}}
 \end{subfigure}
\begin{subfigure}{0.33\linewidth} \includegraphics[width=\textwidth]{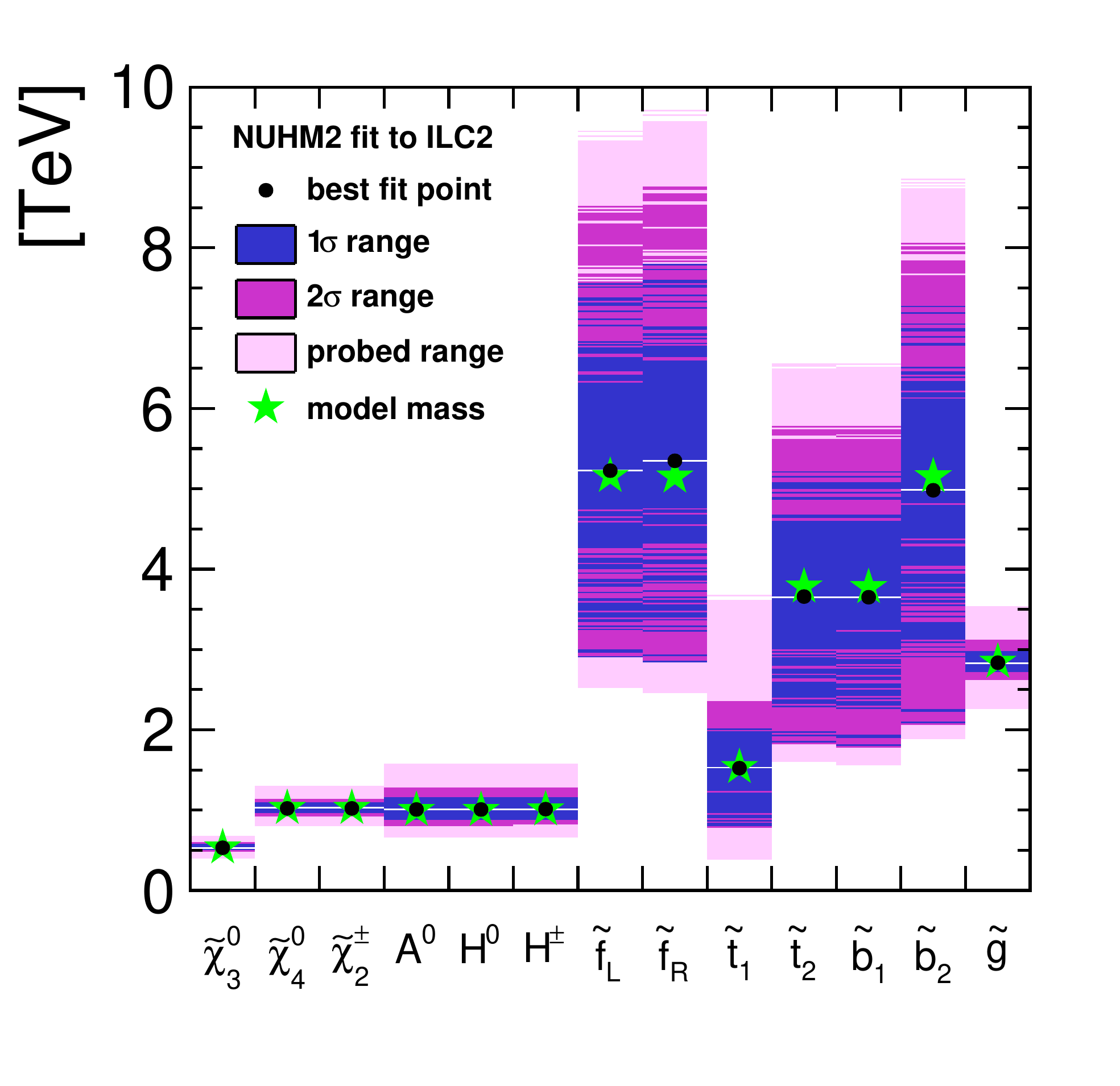}
 \caption{ILC2 \label{sfig:nuhm2fittedmassesILC2}}
 \end{subfigure}
\begin{subfigure}{0.33\linewidth} \includegraphics[width=\textwidth]{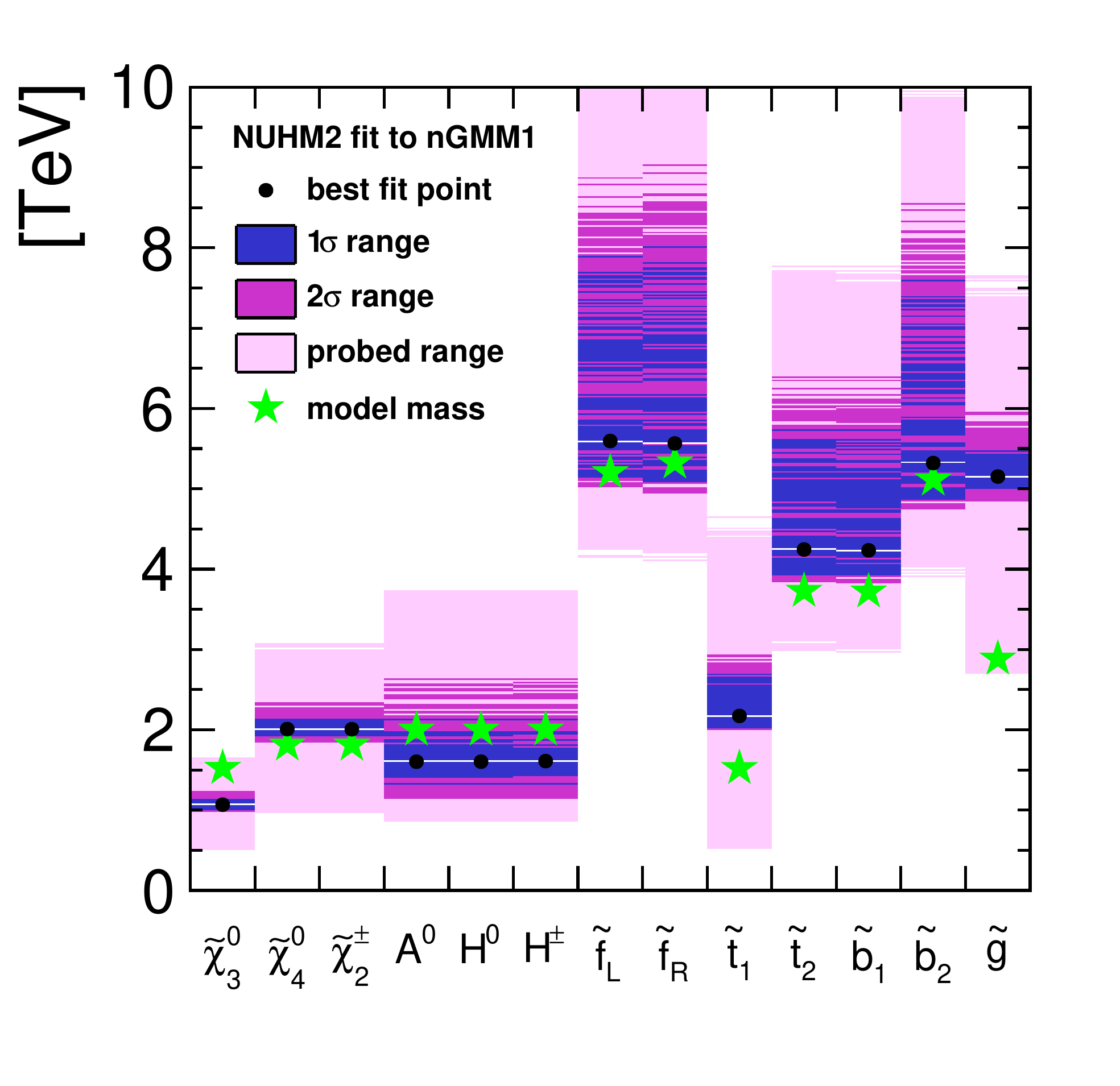}
 \caption{nGMM1 \label{sfig:nuhm2fittedmassesnGMM1}}
 \end{subfigure}
\caption{Predicted mass ranges for all the unobserved sparticles from the NUHM2 fit the observables of the 
three benchmark models. The green star indicates the true model mass, while the black dot shows the best 
fit point. 
}
\label{fig:nuhm2fittedmasses}
\end{figure}

Based on the fitted NUHM2 parameters and their uncertainties, the mass spectrum of the unobserved sparticles can be predicted for all three benchmark cases. This is illustrated in Fig.~\ref{fig:nuhm2fittedmasses}. In the cases of ILC1 and ILC2, clear predictions for the masses of the electroweakinos and the heavy Higgs bosons are obtained in excellent agreement with the true model masses, providing motivation and an energy scale for further upgrades of the ILC. Due to the modelling of all gauginos by a single $M_{1/2}$ parameter, the gluino mass can also be firmly predicted, which can give important inputs to LHC analyses. If the predicted gluino were not observed, or observed at a very different mass, this would give strong support to the idea that nature does not describe all gauginos by one mass parameter at the GUT scale. The other sfermions are less well constrained than the 
Higgs bosons and gauginos, but upper limits on their masses can still be obtained. Such information would provide important motivation --- and a target energy scale --- for a future hadron collider.

Even in case of the nGMM1 benchmark, the NUHM2 fit predicts the general pattern of the mass spectrum correctly, albeit with less precision and significantly worse agreement between true model masses and the best fit point. 
The worse agreement is not surprising as we're fitting a wrong model hypothesis in this case. Nonetheless, upper limits on all sparticle masses are obtained, which shows that even in the case that the wrong model is assumed such important information for the planning of future colliders or upgrades can be obtained.

\subsubsection{Results of fitting NUHM1 and CMSSM} 

Before turning to the weak-scale fits, we investigate whether the three benchmarks could also 
be described by other widely used constrained models, in particular NUHM1~\cite{Baer:2004fu} 
and the CMSSM, which have one or even two fewer parameters to model the Higgs and higgsino 
sectors: in NUHM1, instead of $M_A$ and $\mu$ (or $M^0_{H_u}$ and $M^0_{H_d}$) only one parameter, 
$M^0_{H}$, is used to describe the Higgs and higgsino sectors, while in the CMSSM this reduces 
further to only the sign of $\mu$ being a free choice, while its absolute value is derived 
from the other model parameters. Table~\ref{tab:constrainedbestfitpoints} gives the best fit 
point obtained when fitting NUHM1 and CMSSM to the ILC1 and ILC2 benchmarks, which give very 
large values of $\chi^2/\textrm{dof}$. These interpretations could be ruled out at the $95\%$\,C.L.\ 
already with about $0.1\%$ the total integrated luminosity.

\begin{table}[htbp]{}
\centering
$
\begin{array}{  c c c c c  }
\hline \hline
 \text{best fit points} \\
\text{parameter} & \text{ILC1 CMSSM}  & \text{ILC2 CMSSM} &  \text{ILC1 NUHM1} & \text{ILC2 NUHM1}  \\ \hline 
M_{1/2} [\GeV] & 128.8  & 155.9 & 136.9 &  159.9  \\
M_{0} [\GeV]  & 3585 & 5631 & 1796 & 4264     \\
A_{0} [\GeV]  & -6873 & -10873 & -4396 & -10086    \\
\tan \beta  & 13.7 & 16.2 & 16.2 & 14.8    \\
M^0_{H_0} [\GeV] & - & - & 47659 &  56553   \\ \hline
\chi^2_{\textrm{min}} & 238046 & 93155 & 235014 & 85174 \\
\chi^2/\textrm{dof} & 11336  & 4436 & 11751 & 4259  \\
\hline \hline
\end{array}
$
\caption{Best fit points in CMSSM and NUHM1 fits of ILC1 and ILC2 observables, including SUSY and Higgs measurements in the H20 operation for ILC1 and I20 operation for ILC2. }
\label{tab:constrainedbestfitpoints}
\end{table}


%
%
%
%
%
%

\subsection{Weak scale fit results}
\label{ssec:weakscale_fits}

In the following, the results of various weak scale fits 
to the ILC1, ILC2 and nGMM1 observables are discussed.
The most general model considered is pMSSM-10, the MSSM with 10 weak scale input parameters: 
$M_{1}$, $M_{2}$, $M_{3}$, $\mu$, $\tan\beta$, $m_A$, $A_t=A_b=A_{\tau}$ and 
$M_{Q_3}$, $M_{U_3}=M_{D_3}$, $M_{L}=M_{L_{1,2,3}}=M_{E_{1,2,3}}=M_{Q_{1,2}}=M_{U_{1,2}}=M_{D_{1,2}}$. We use this model to test whether it is possible to constrain a comprehensive set of parameters from the observables of the higgsino sector alone, and to study the influence of the parameters in which the higgsino sector enters only at loop level.
If the pMSSM-10 fit is successful and reproduces the input parameters at a satisfactory level, we proceed to investigate the precision achievable when fitting only tree-level higgsino parameters. For this we use the phenomenological MSSM with a reduced number of 4 free parameters, the pMSSM-4, which fixes the 
squark, slepton, heavy Higgs boson and gluino parameters to their true values and so only depends on the four weak scale parameters $M_{1,2}$, $\mu$ and $\tan\beta$. In a real analysis, the ``true'' values are of course unknown, but instead the best fit point of pMSSM-10 fit could be used, which agrees with the model point to usually much better than 1$\sigma$. The possible bias from fixing the 6 non-higgsino parameters was tested explicitly in case of ILC1 by fixing them to some point in the 2$\sigma$ region.  The best fit gaugino masses are nearly the same in the two pMSSM-4 fits irrespective of the fixed parameters. Their difference is only 2.5\,GeV for $\neutralinothree$ and 0.8\,GeV for $\neutralinofour$ and $\charginotwo$. The two sets of best fit masses agree within the 1$\sigma$ uncertainties~\cite{Lehtinen:PhDThesis}. In the following, the results for the individual benchmarks will be presented.

\subsubsection{ILC1 Benchmark}
\label{ssec:weakscale_fits:ILC1}

\begin{figure}[htpb]
\begin{center}
\begin{subfigure}{0.4\linewidth}
\includegraphics[width=\textwidth]{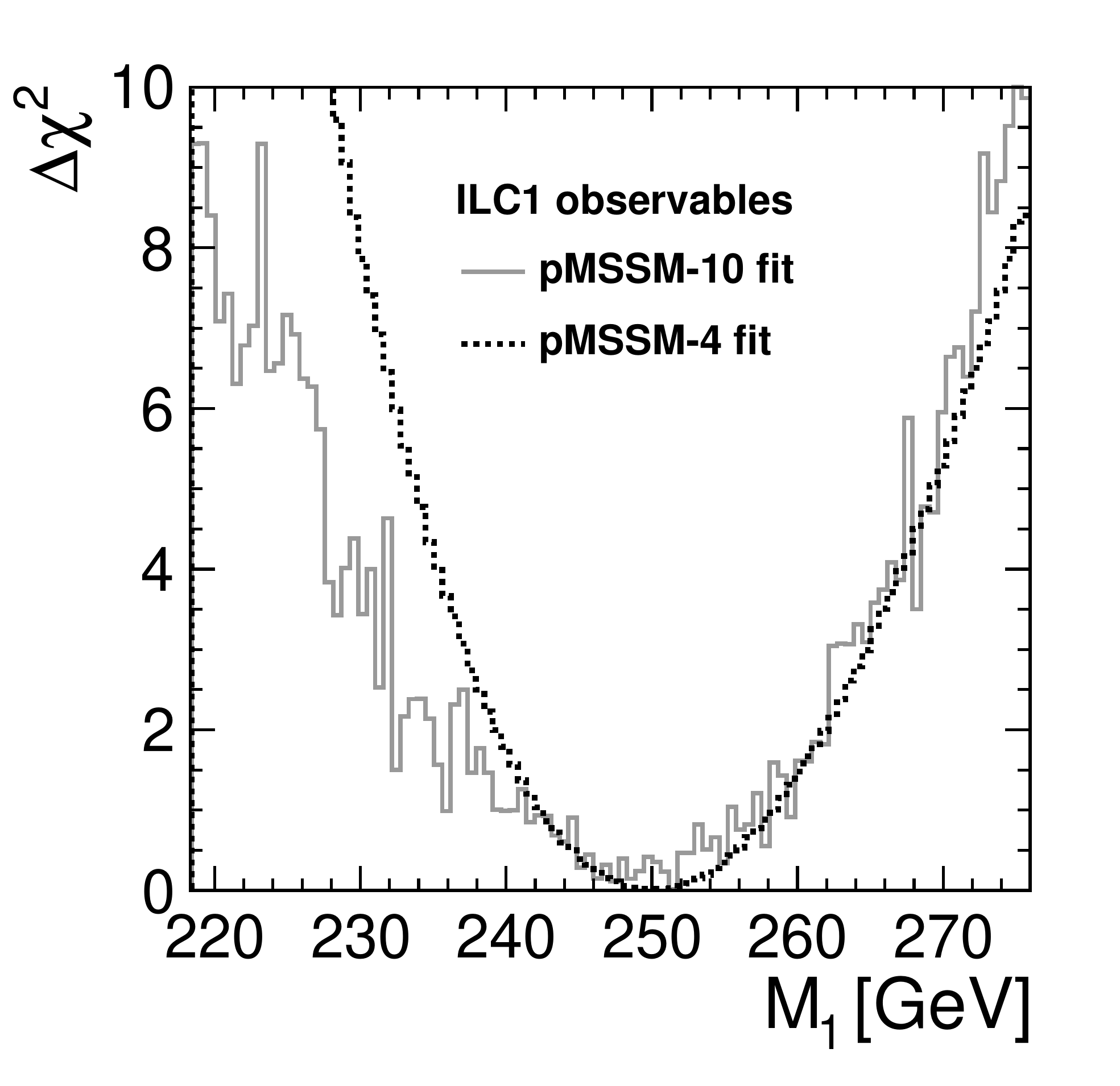}
\caption{}
\label{fig:ILC1H20parabolae:M1}
\end{subfigure}
\begin{subfigure}{0.4\linewidth}
\includegraphics[width=\textwidth]{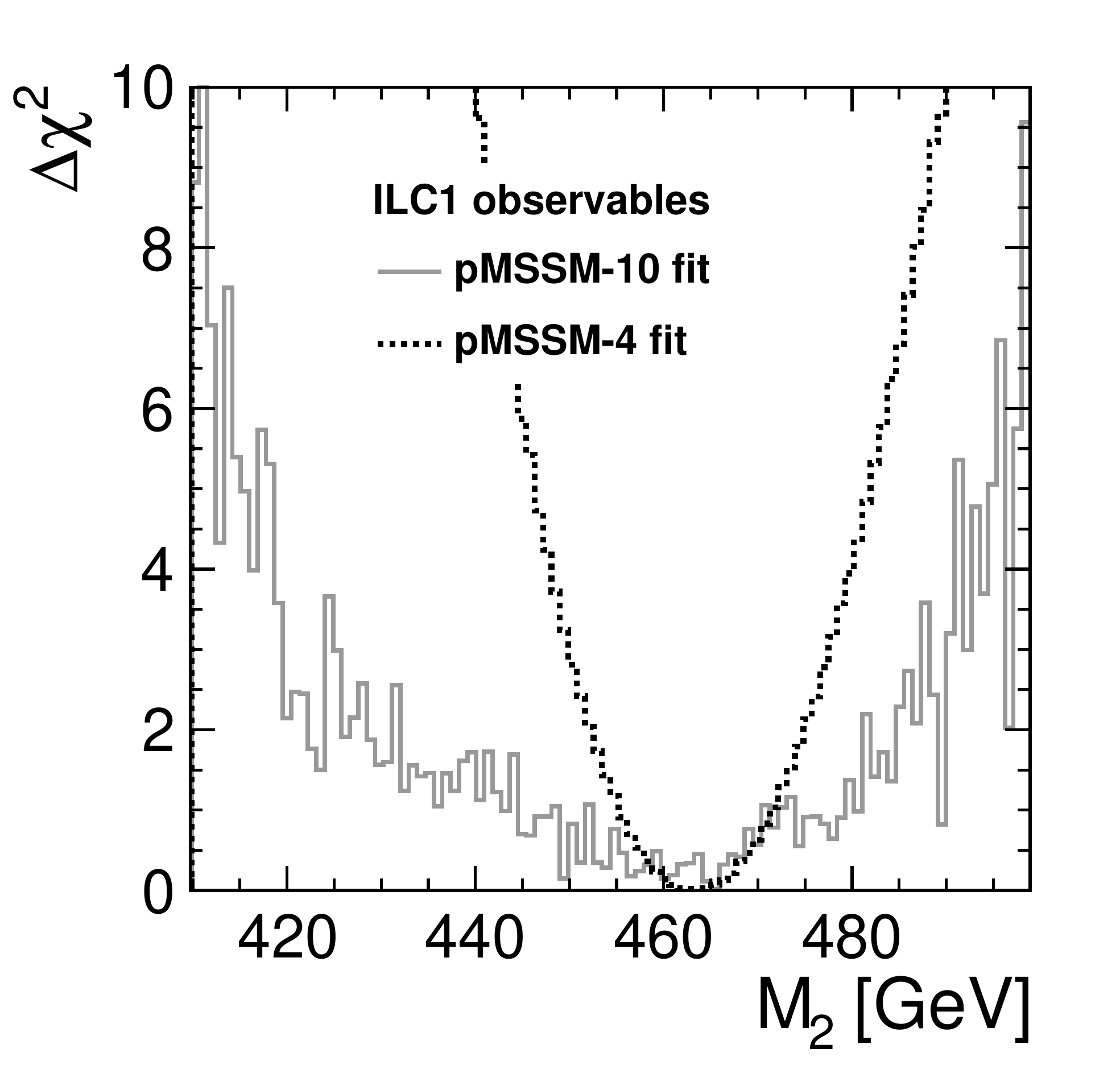}
\caption{}
\label{fig:ILC1H20parabolae:M2}
\end{subfigure}
\begin{subfigure}{0.4\linewidth}
\includegraphics[width=\textwidth]{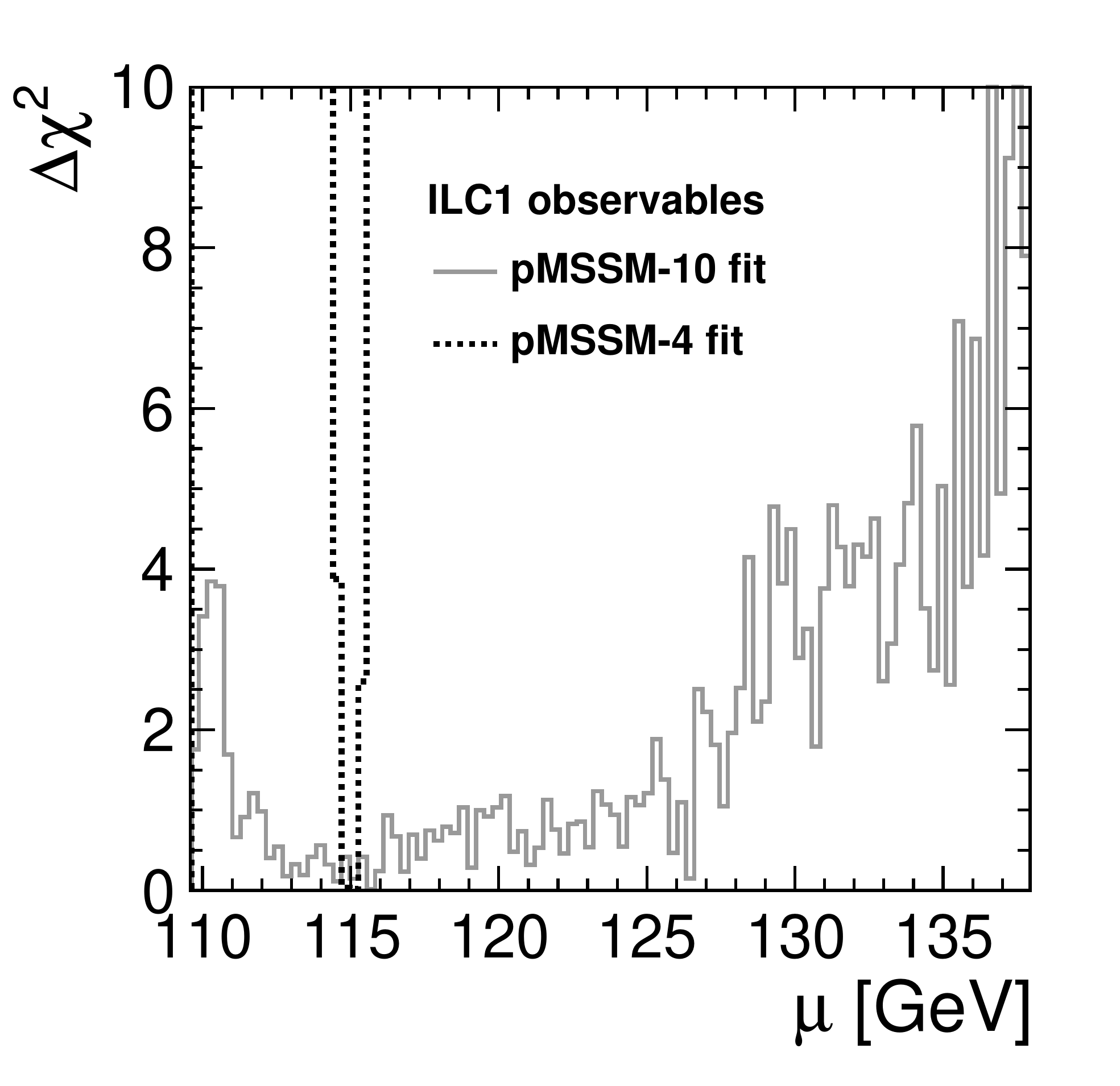}
\caption{}
\label{fig:ILC1H20parabolae:mu}
\end{subfigure}
\begin{subfigure}{0.4\linewidth}
\includegraphics[width=\textwidth]{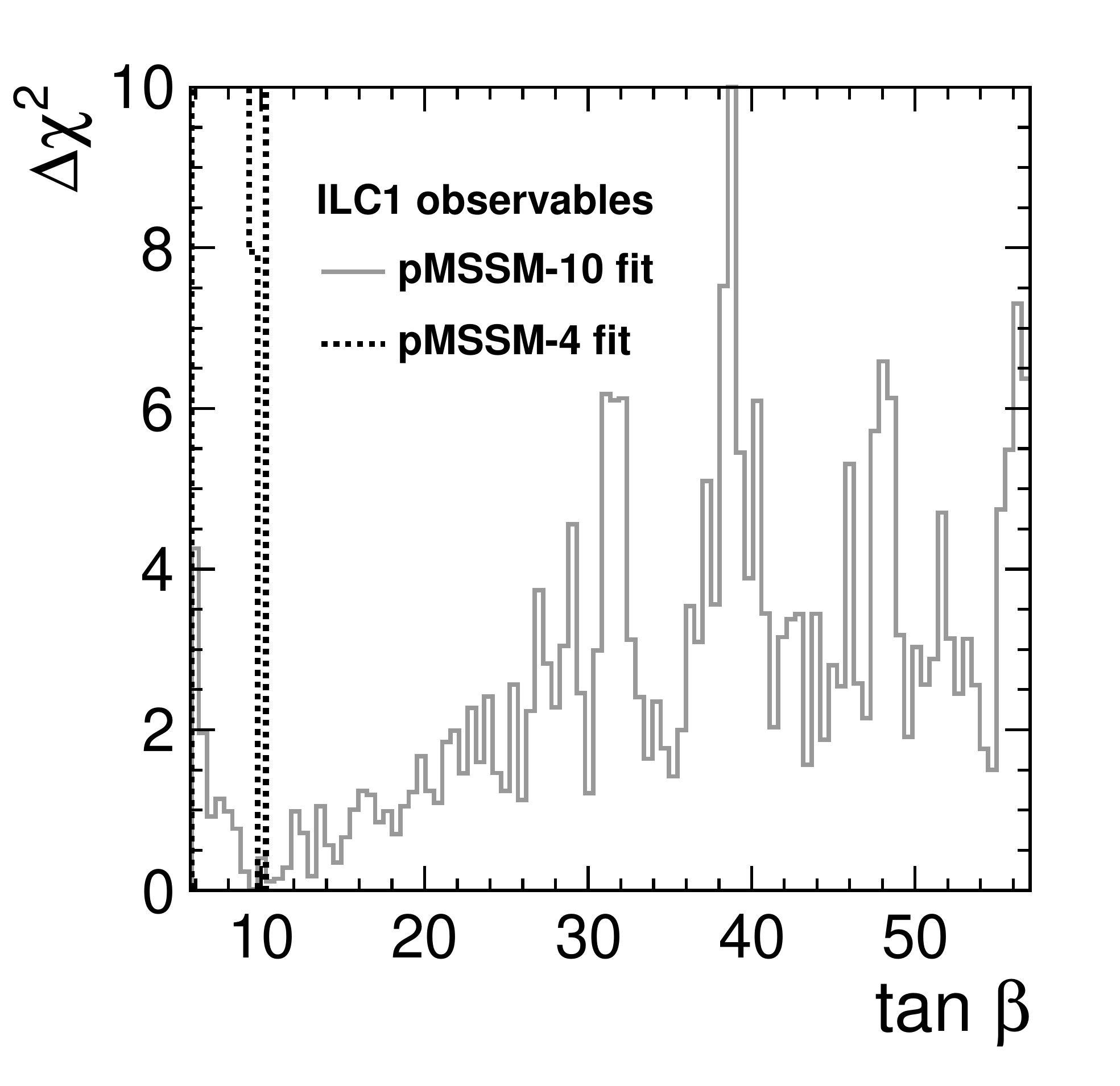}
\caption{}
\label{fig:ILC1H20parabolae:tb}
\end{subfigure}
\end{center}
\caption{ILC1: Minimum $\chi^2$ as a function of $M_1$, $M_2$ and $\mu$ and $\tan \beta$ in pMSSM-4 fit (dashed black line) and pMSSM-10 fit (solid grey line). For each bin, the minimum $\chi^2$ of all Markov chain points which have the $x$ axis quantity in that bin is plotted.}
\label{fig:ILC1H20parabolae}
\end{figure}

Figure~\ref{fig:ILC1H20parabolae} shows the minimum $\chi^2$ as a function of $M_1$, $M_2$ and $\mu$ and $\tan \beta$ in the pMSSM-10 and pMSSM-4 fits. 
Due to the much smaller parameter space to be sampled in case of the 4-dimensional fit, 
the resulting curve is much smoother than in the 10-dimensional case. 
The precision on $M_1$ is nearly identical in both cases. 
$M_2$ is somewhat better constrained in the pMSSM-4 fit, while the determination of $\mu$ and $\tan{\beta}$ improves drastically.


The resulting best fit values for the pMSSM parameters and their 1 and 2$\sigma$ intervals are compared to the input values in Tab.~\ref{tab:ILC1fittedparameters}, quantifying the effect which could already be seen qualitatively in Fig.~\ref{fig:ILC1H20parabolae}. 
In the case of the pMSSM-10 fit, it should be noted that also for the parameters of the coloured sector some constraints, and especially upper bounds, can be obtained. This even applies for the squark mass parameters, which might seem surprising at the first glance, but is due to the 2-loop RGEs included in \texttt{SPheno}. 
If a hypothetical gluino mass measurement with $11\%$ uncertainty from the LHC\cite{Baer:2016wkz}
is included in the fit, the constraint on $M_3$ improves accordingly to about $10\%$. 
All other parameters, including the squark mass parameters, show only minor improvements.

\begin{table}[htbp]{}
\centering
\resizebox{0.99\textwidth}{!}{
$\begin{array}{  c c | c c c  | c c c  }
\hline 
& & \text{pMSSM-4}  & & & \text{pMSSM-10}  & & \\
  \text{parameter}   & \text{ILC1 pMSSM true}  & \text{best fit point} & 1\sigma \text { CL} & 2\sigma \text { CL} & \text{best fit point} & 1\sigma \text { CL}&  2\sigma \text { CL}  \\ \hline \hline
M_{1}  & 250 & 250.2& \ ^{+8.2}_{-7.7}	 & \ ^{+17.1}_{-15.1} & 251.3 & \ ^{+8.6}_{-15.7} & \ ^{+17.2}_{-23.7} \\
M_{2} & 463  & 463.3 &  \ ^{+8.0}_{-8.1}	 & \ ^{+16.2}_{-14.9}  &  465.8 &  \ ^{+24.2}_{-23.0}	 & \ ^{+31.4}_{-49.8}  \\
\mu &  115.0   & 115.0 &  \ ^{+0.2}_{-0.2}	 & \ ^{+0.3}_{-0.3}  & 115.7 &  \ ^{+10.9}_{-4.7}	 & \ ^{+20.3}_{-6.1}  \\
\tan \beta &  10.0 & 10.0 & \ ^{+0.1}_{-0.1}	 & \ ^{+0.2}_{-0.2} & 9.7 &  \ ^{+8.8}_{-3.0}	 & \ ^{+45.3}_{-3.5}  \\
m_A & 1000 & & & &  1050 &  \ ^{+310}_{-180}	 & \ ^{+607}_{-296}  \\
M_3 &  1270  & & & & 1412 &  \ ^{+1791}_{-1104}	 & \ ^{+1411}_{-2843}  \\
M_{L} & 7150 & & & & 7063 &  \ ^{+2029}_{-4311}	 & \ ^{+2645}_{-5632}  \\
M_{U(3)} &  1670 & & & & 1751 &  \ ^{+2414}_{-628} & \ ^{+4498}_{-740}  \\
M_{Q(3)} & 4820 & & & & 4951 &  \ ^{+2324}_{-3226}	 & \ ^{+3858}_{-3226}  \\
A_{t=b=\tau} & -4400 & & & &  -4591 &  \ ^{+1371}_{-973}	 & \ ^{+1647}_{-2949} \\ \hline
\chi^2    &    & 0.0011 &&  &  0.1360 & &  \\
\hline \hline
\end{array}$
}
\caption{Fitted parameters in ILC1 pMSSM-4 and pMSSM-10. All units in GeV except for $\tan \beta$ and $\chi^2$.}
\label{tab:ILC1fittedparameters}
\end{table}

\begin{figure}[htbp]
\begin{subfigure}{0.49\linewidth}
\includegraphics[width=\textwidth]{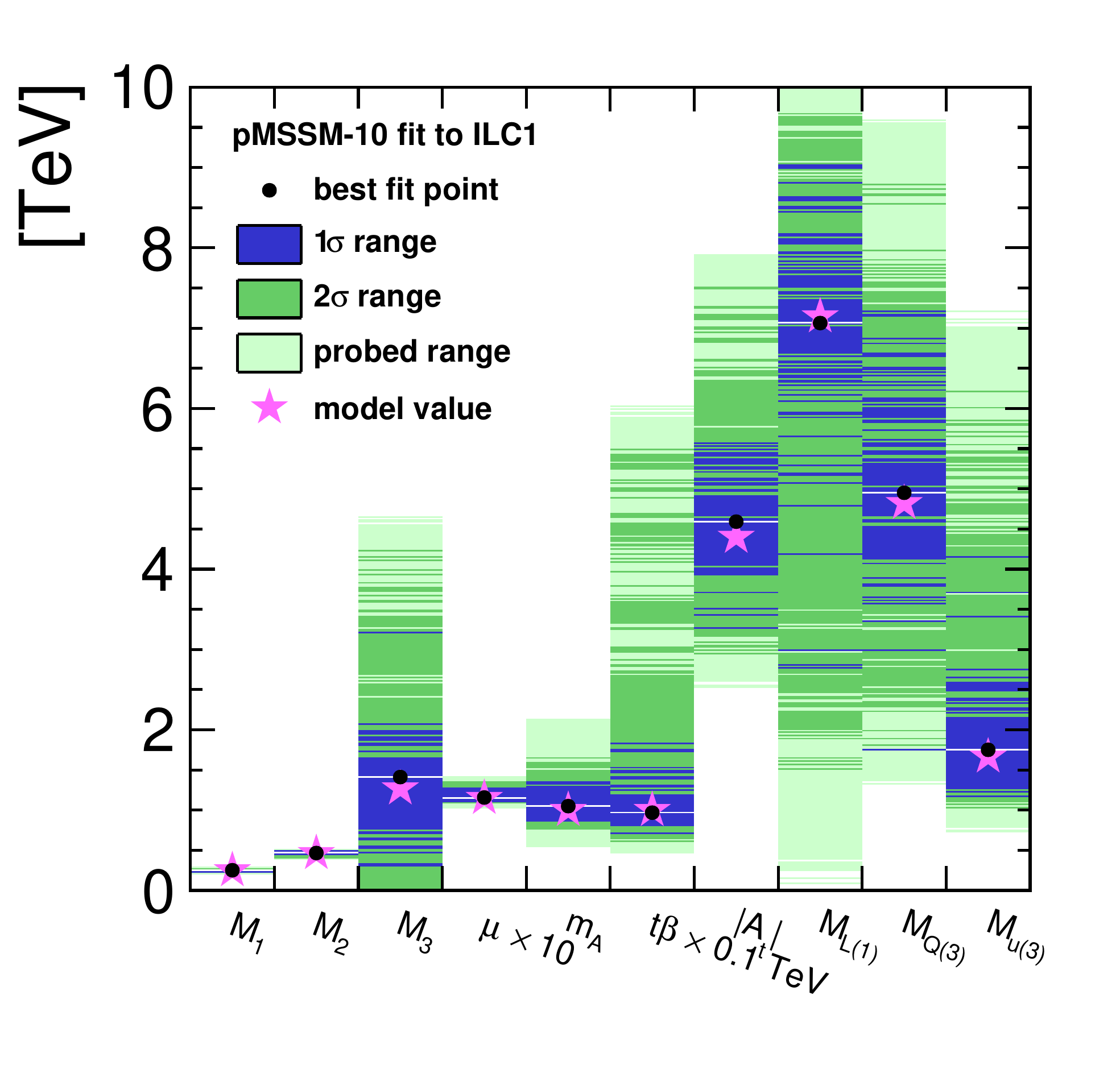}
\caption{without gluino observation}
\label{fig:ILC1fittedparameters10p}
\end{subfigure}
\begin{subfigure}{0.49\linewidth}
\includegraphics[width=\textwidth]{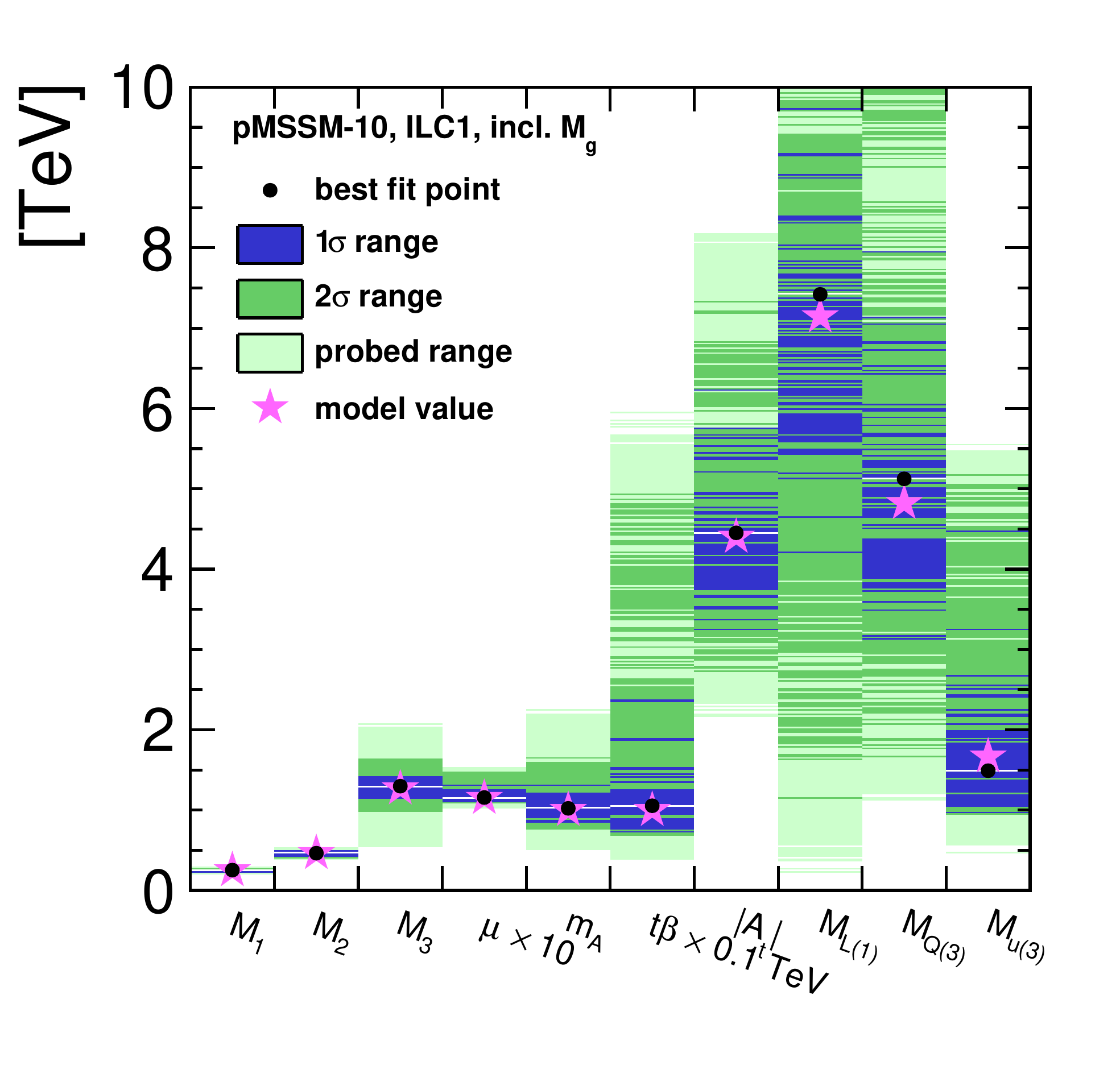}
\caption{with gluino observation}
\label{fig:ILC1fittedparameters10pglu}
\end{subfigure}
\caption{Predicted SUSY parameter ranges from the pMSSM-10 fit to ILC1. 
The magenta star indicates the true model values, while the black dot shows the best fit point. 
}
\label{fig:ILC1fitted10p}
\end{figure}

Figures~\ref{fig:ILC1fittedparameters10p} and~\ref{fig:ILC1fittedparameters10pglu} illustrate the
precisions obtained on the pMSSM-10 parameters, without and with assuming a gluino mass measurement from the LHC, respectively. Thereby, $\tan{\beta}$ is displayed as if it were in GeV. 
It can clearly be seen that the precision on $M_3$ is improved considerably by the gluino mass measurement, while the precision on all the other parameters don't change significantly.

\begin{figure}[htbp]
\begin{subfigure}{0.49\linewidth}
\includegraphics[width=\textwidth]{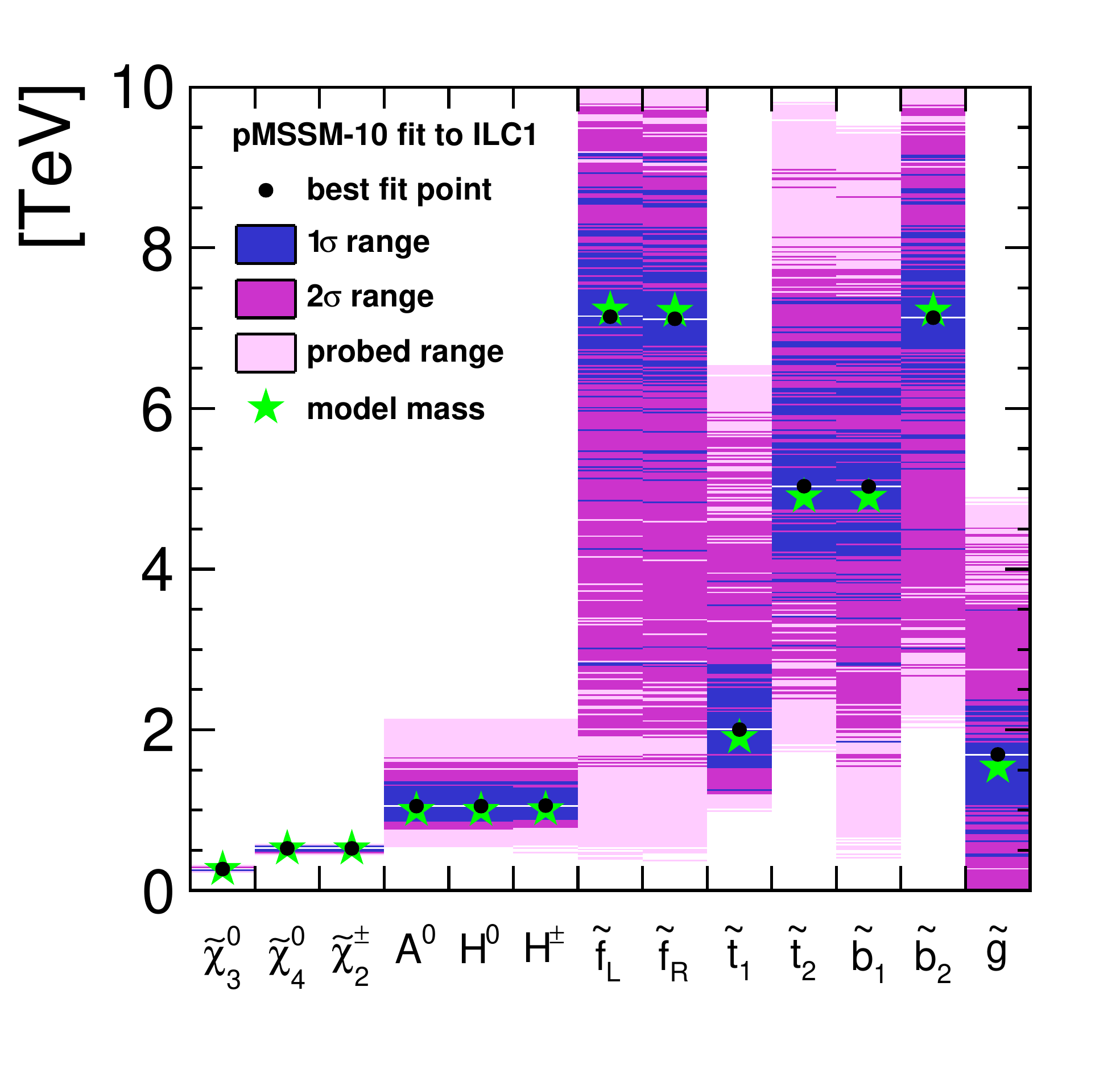}
\caption{without gluino observation}
\label{fig:ILC1fittedmasses10p}
\end{subfigure}
\begin{subfigure}{0.49\linewidth}
\includegraphics[width=\textwidth]{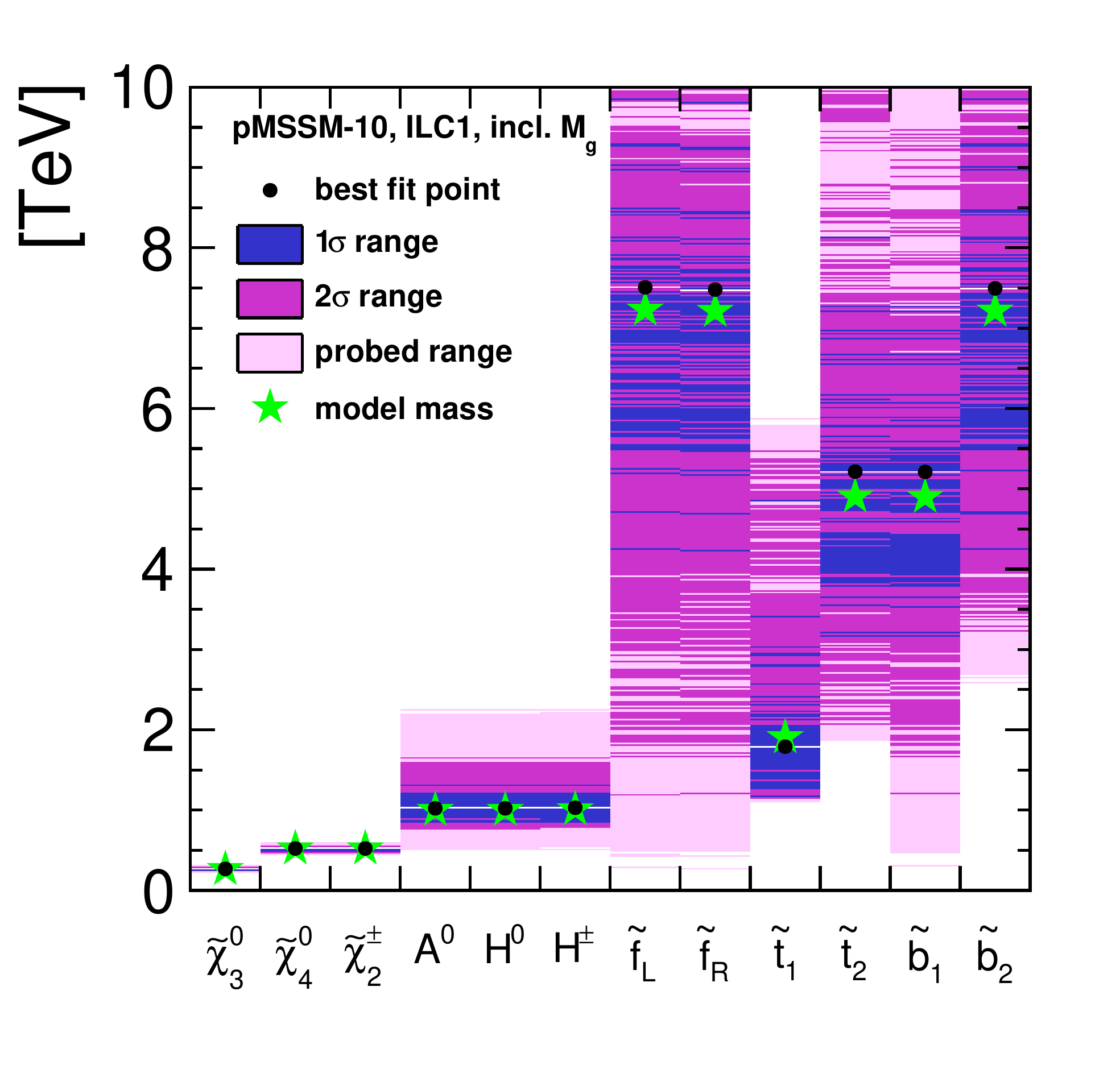}
\caption{with gluino observation}
\label{fig:ILC1fittedmasses10pglu}
\end{subfigure}
\caption{Predicted mass ranges from the pMSSM-10 fit to ILC1. The green star indicates the true model values, 
while the black dot shows the best fit point. 
}
\label{fig:ILC1fitted10m}

\end{figure}

The determined parameters can be used to predict the masses of the yet unobserved sparticles, as shown for the pMSSM-10 fit in Tab.~\ref{tab:ILC1mssm10pfit-fittedmasses} and Fig.~\ref{fig:ILC1fitted10m}, again without and with assuming a gluino mass measurement from the LHC. As expected from Fig.~\ref{fig:ILC1fitted10p}, the effect of the gluino measurement on the other predicted masses is small.

\begin{table}[htbp]{}
\centering
$\begin{array}{  c c c c c    }
  \hline \hline
    \text{prediction} & \multicolumn{4}{c}{\text{ILC1}} \\ 
    & \text{model value} & \text{best fit} & 1\sigma \text{ CL} & 2 \sigma \text{ CL}  \\ \hline \hline
m_{\neutralinothree} & 265.4 & 267        &     \ _{-16}^{+8 } &        \ _{-26 }^{+16}    \\
m_{\neutralinofour}  & 521.4 & 524        &     \ _{-26 }^{+ 20} &      \ _{-55 }^{+26}   \\
m_{\charginotwo}     & 521.2 &  524  &  \ _{-26 }^{+19 } &      \ _{-55 }^{+25}    \\
m_{H_0}              & 1001 &  1050  & \ _{-190 }^{+310 } &    \ _{-290 }^{+610}     \\
m_{A_0}              & 1000 &  1050  & \ _{-190 }^{+310 } &    \ _{-290 }^{+610}    \\
m_{H^{\pm}}          & 1008 &  1056  & \ _{-176 }^{+304 } &    \ _{-276 }^{+604}    \\
m_{\supL}            & 7229 &  7143  & \ _{-4343 }^{+2037 } &  \ _{-5603 }^{+2657}      \\
m_{\supR}            & 7203 &  7117  & \ _{-4337 }^{+2023 } &  \ _{-5577 }^{+2643}    \\
m_{\stopone}         & 1906 &   2003 & \ _{-763 }^{+1857 } &   \ _{-803 }^{+3957}    \\
m_{\stoptwo}         & 4903 &   5033 & \ _{-1993 }^{+2347 } &  \ _{-2653 }^{+3947}    \\
m_{\sbottomone}      & 4899 &  5028  & \ _{-3188 }^{+2352 } &  \ _{-3488 }^{+3912}     \\
m_{\sbottomtwo}      & 7216 &  7130  & \ _{-4310 }^{+2030 } &  \ _{-4470 }^{+2650}     \\
m_{\gluino}          & 1539 &  1693  & \ _{-1273 }^{+1807 } &  \ _{-1693 }^{+2827}      \\
  \hline \hline
  \end{array}$
\caption{True and fitted masses as well as their uncertainties from a pMSSM-10 fit to ILC1 
observables. All values in GeV.}
\label{tab:ILC1mssm10pfit-fittedmasses}
\end{table}

\begin{figure}[htbp]
\begin{subfigure}{0.49\linewidth}
\includegraphics[width=\textwidth]{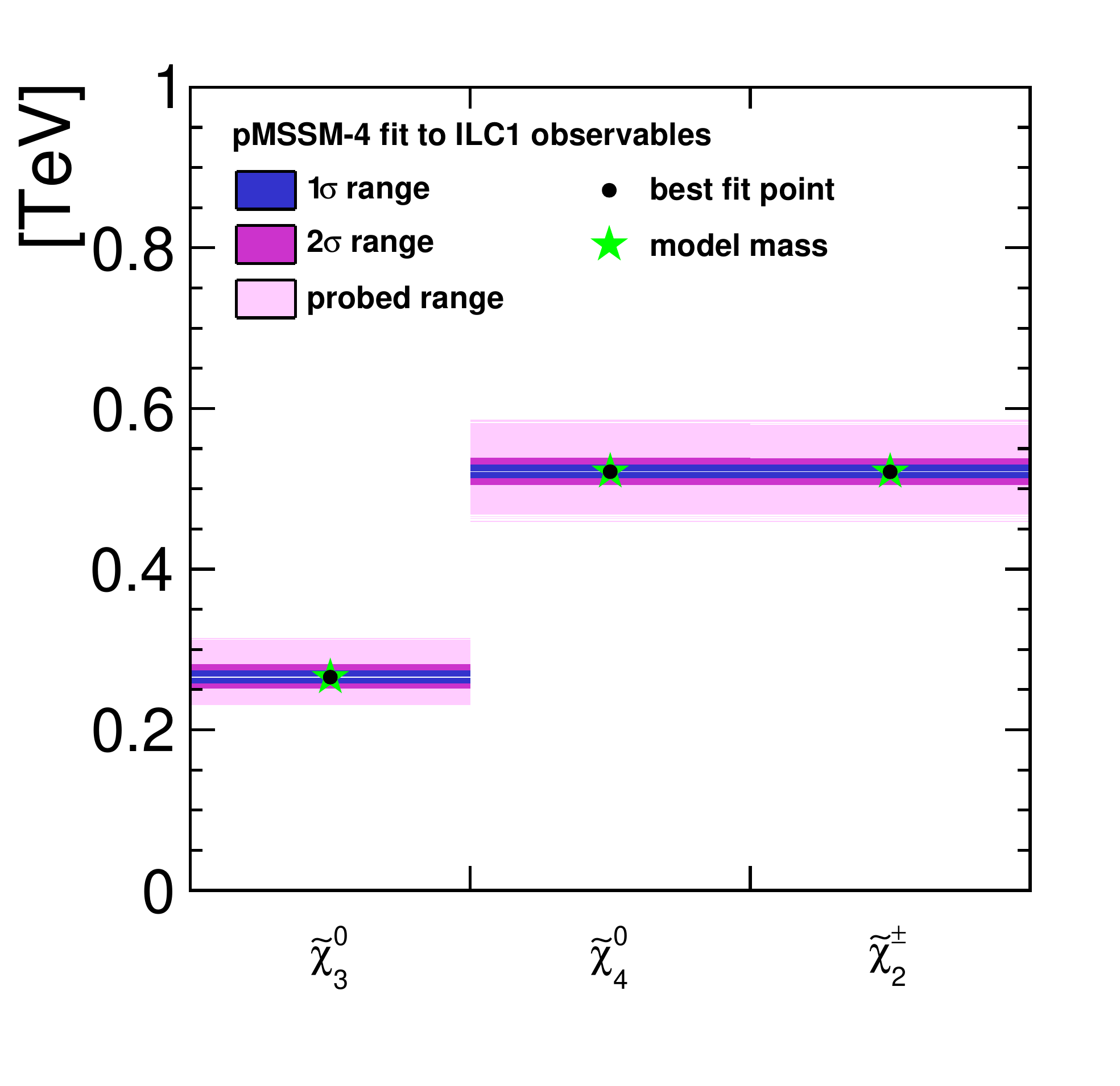}
\caption{Fitted masses}
\label{fig:ILC1fittedmasses4p}
\end{subfigure}
\begin{subfigure}{0.49\linewidth}
\includegraphics[width=\textwidth]{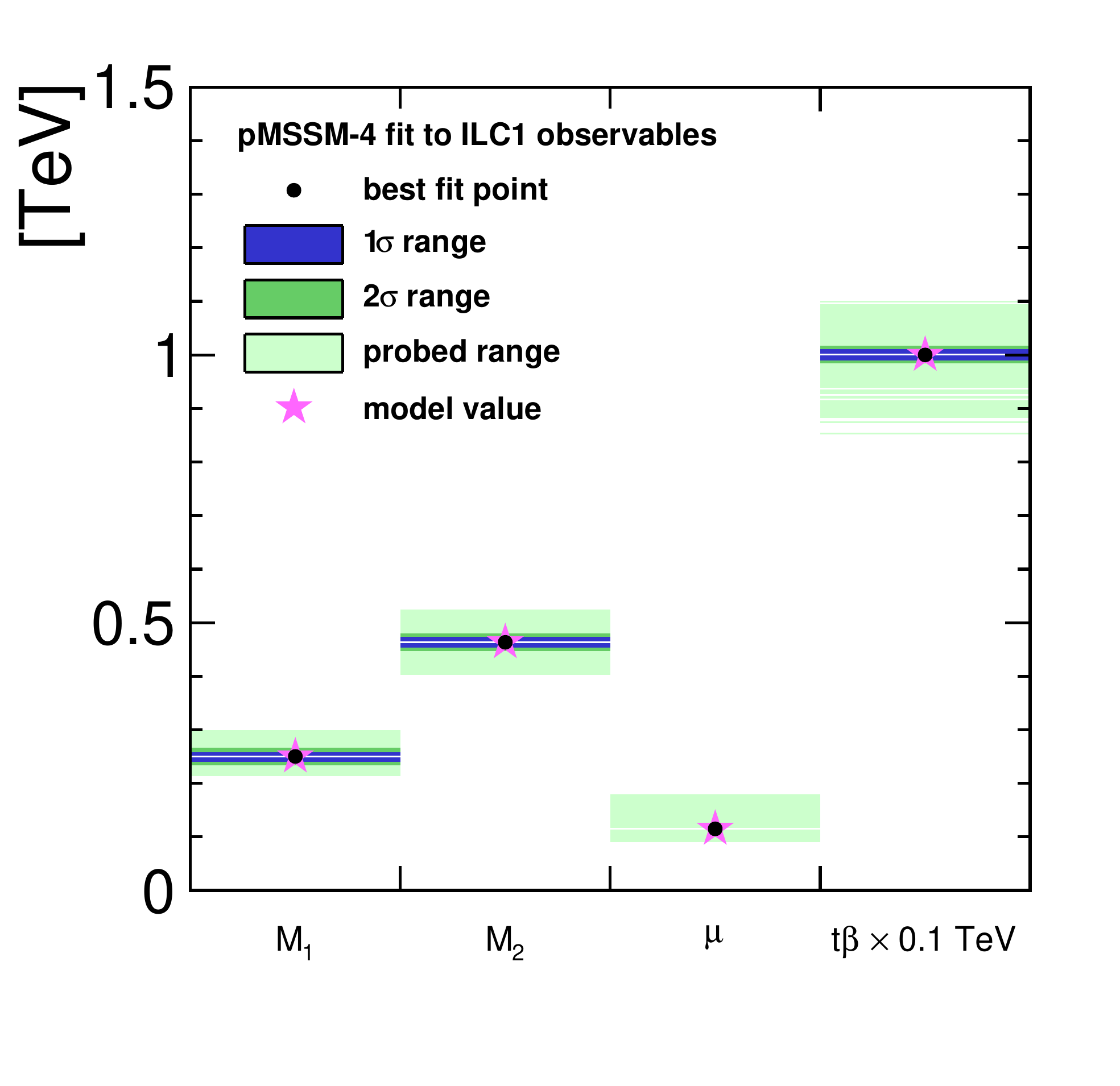}
\caption{Fitted parameters}
\label{fig:ILC1fittedparam4p}
\end{subfigure}
\caption{Predicted mass and SUSY parameter ranges from the pMSSM-4 fit to ILC1. The green/magenta star 
indicates the true model values, while the black dot shows the best fit point. 
}
\label{fig:ILC1fitted4p}
\end{figure}

Figure~\ref{fig:ILC1fitted4p} illustrates the result of the corresponding pMSSM-4 fit with $M_1$, $M_2$, $\mu$ and $\tan \beta$ only. All four parameters can be determined accurately as also shown in Tab.~\ref{tab:ILC1fittedparameters}. This results in predictions for the masses of the heavier electroweakinos with precisions between 1.6 and 3\%.

%
%
%
%

\subsubsection{ILC2 Benchmark}
\label{ssec:weakscale_fits:ILC2}
\begin{figure}[htpb]
\begin{center}
\begin{subfigure}{0.4\linewidth}
\includegraphics[width=\textwidth]{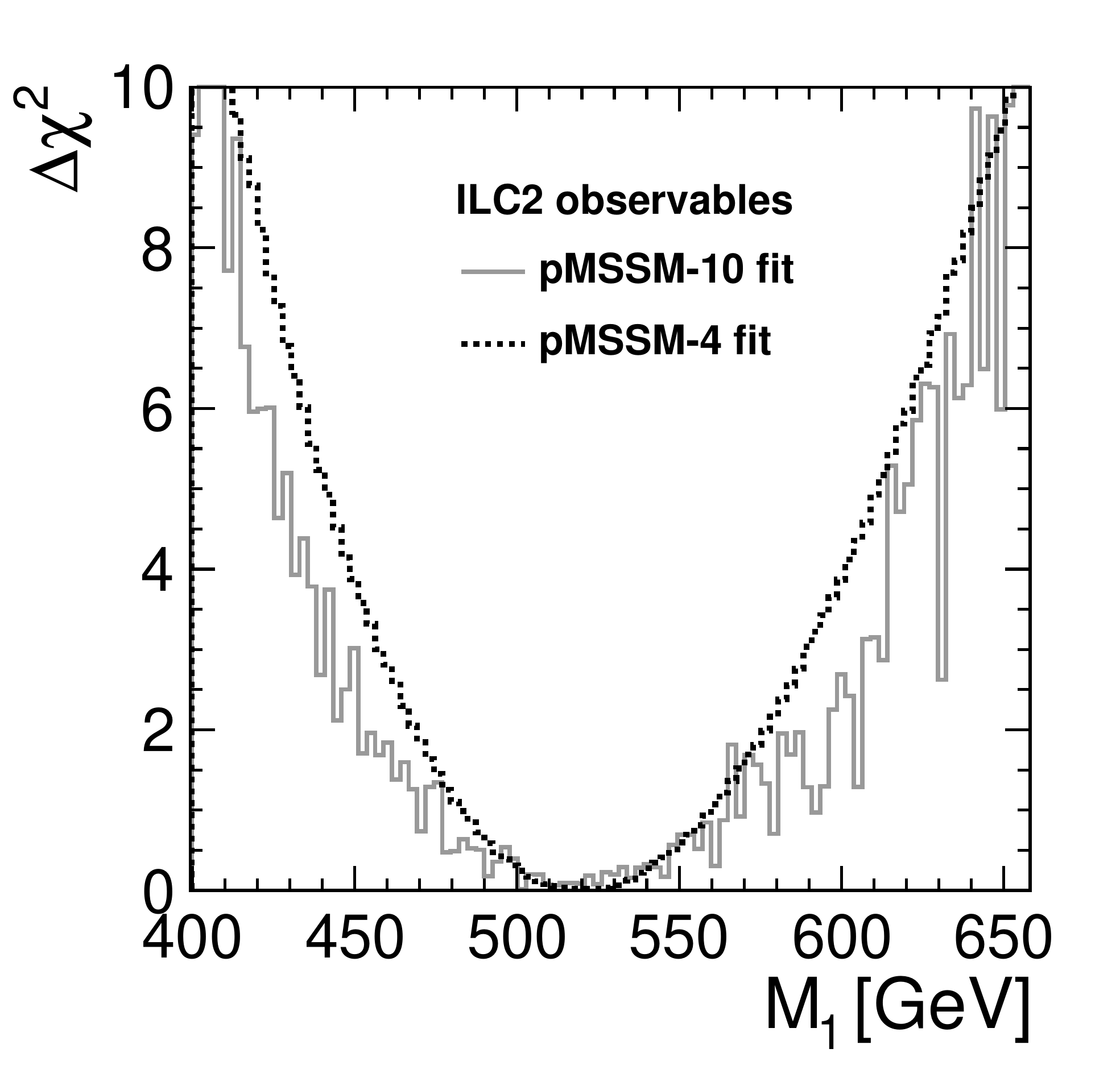}
\caption{}
\label{fig:ILC2I20parabolae:M1}
\end{subfigure}
\begin{subfigure}{0.4\linewidth}
\includegraphics[width=\textwidth]{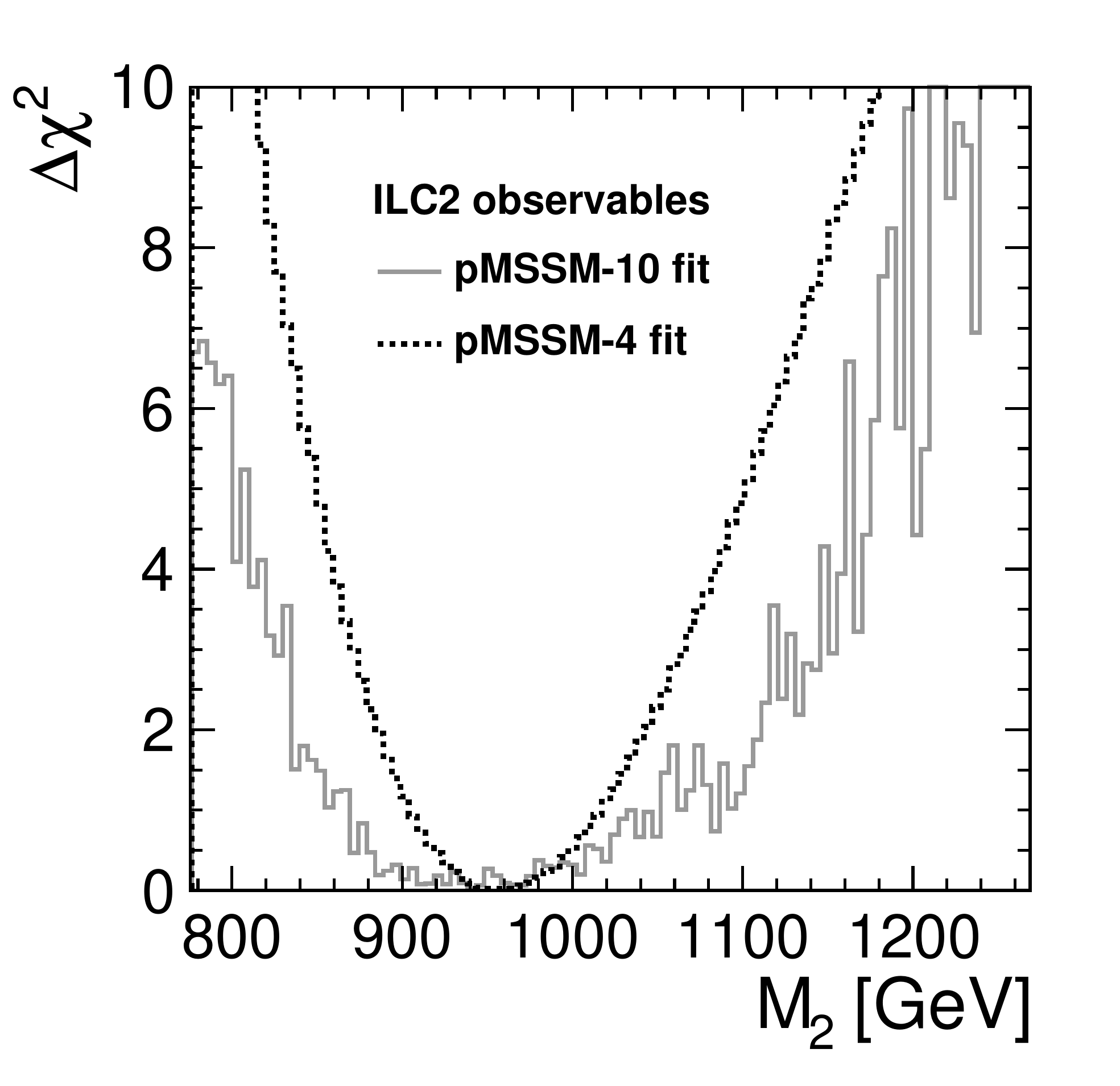}
\caption{}
\label{fig:ILC2I20parabolae:M2}
\end{subfigure}
\begin{subfigure}{0.4\linewidth}
\includegraphics[width=\textwidth]{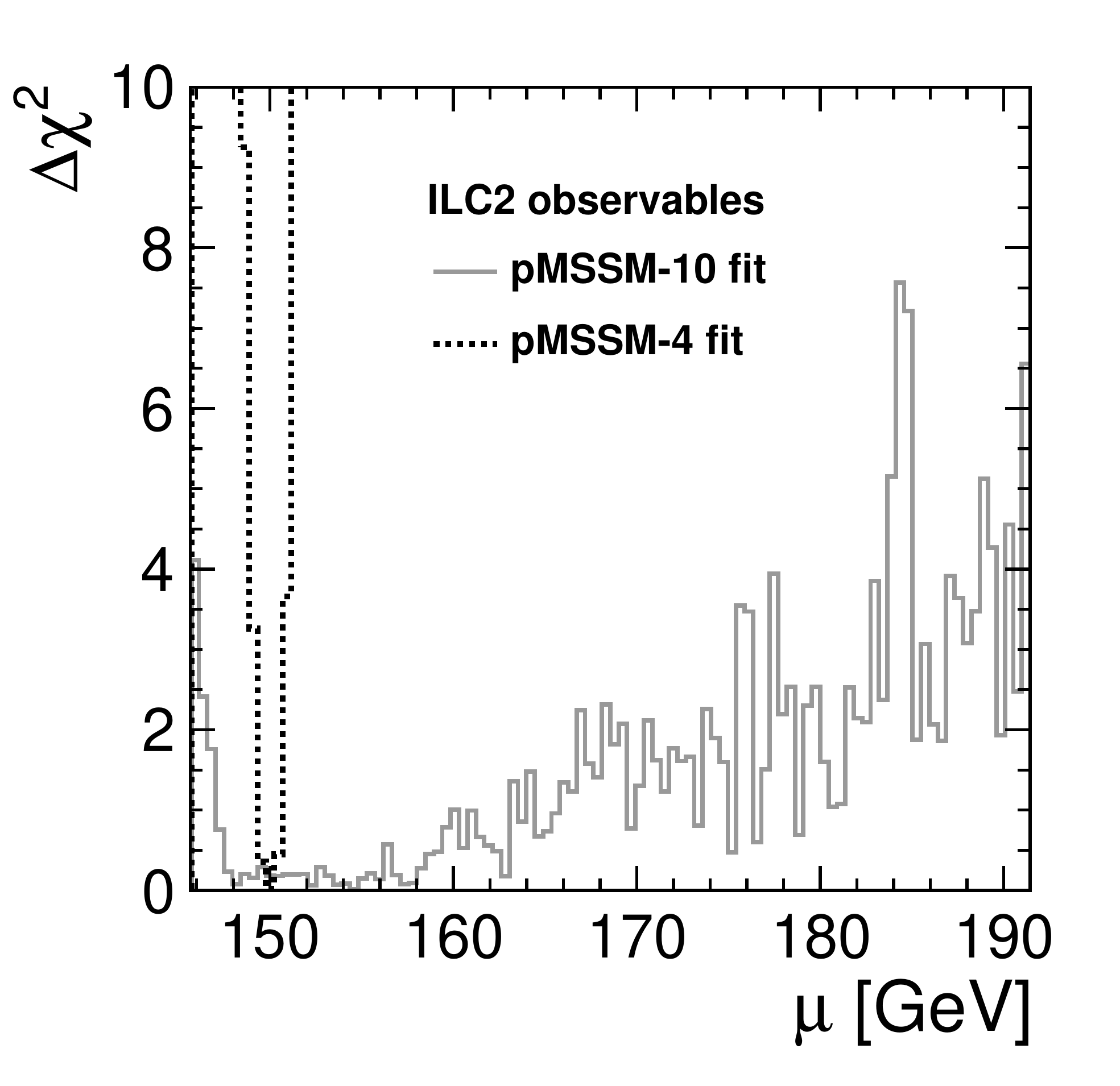}
\caption{}
\label{fig:ILC2I20parabolae:mu}
\end{subfigure}
\begin{subfigure}{0.4\linewidth}
\includegraphics[width=\textwidth]{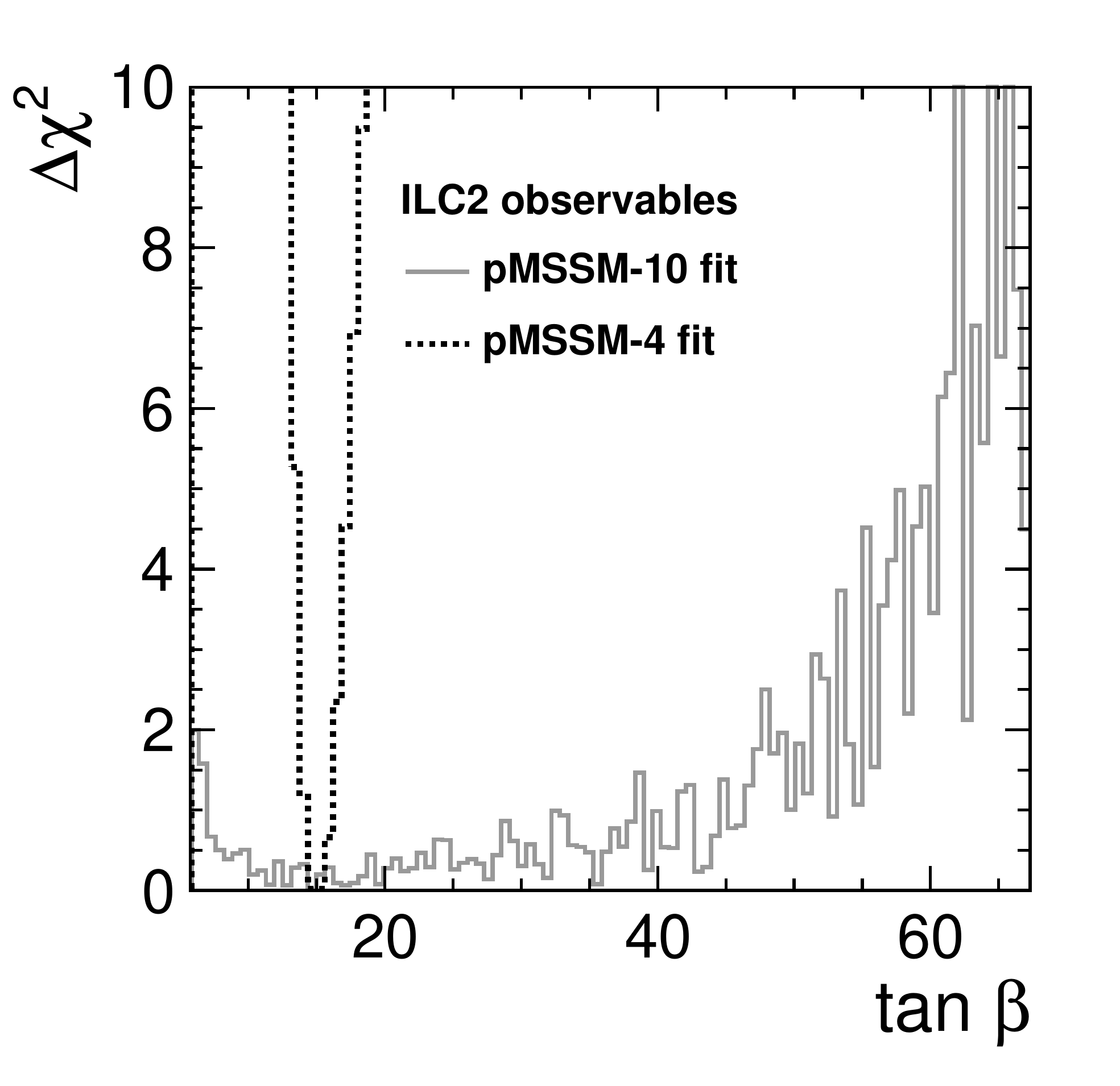}
\caption{}
\label{fig:ILC2I20parabolae:tb}
\end{subfigure}
\end{center}
\caption{ILC2: Minimum $\chi^2$ as a function of $M_1$, $M_2$ and $\mu$ and $\tan \beta$ in pMSSM-4 fit (dashed black line) and pMSSM-10 fit (solid grey line). For each bin, the minimum $\chi^2$ of all Markov chain points which have the $x$ axis quantity in that bin is plotted.}
\label{fig:ILC2I20parabolae}
\end{figure}

In the case of the ILC2 benchmark, the overall situation is similar 
to the case of ILC1. The minimum $\chi^2$ as a function of $M_1$, $M_2$ and $\mu$ and $\tan \beta$ is displayed in Fig.~\ref{fig:ILC2I20parabolae} for the pMSSM-10 and pMSSM-4. However, this time the I20 running scenario was assumed, c.f.\ Sec.~\ref{ssec:fitinputs}. 
Also here, the resulting curve for the 4-parameter fit is much smoother than for the 10-parameter version due to the much smaller parameter space to be sampled. 
Like for ILC1, the precision on $M_2$ improves somewhat in the pMSSM-4 fit, 
while $\mu$ and $\tan{\beta}$ are significantly better constrained.


\begin{figure}[htbp]
\begin{subfigure}{0.49\linewidth}
\includegraphics[width=\textwidth]{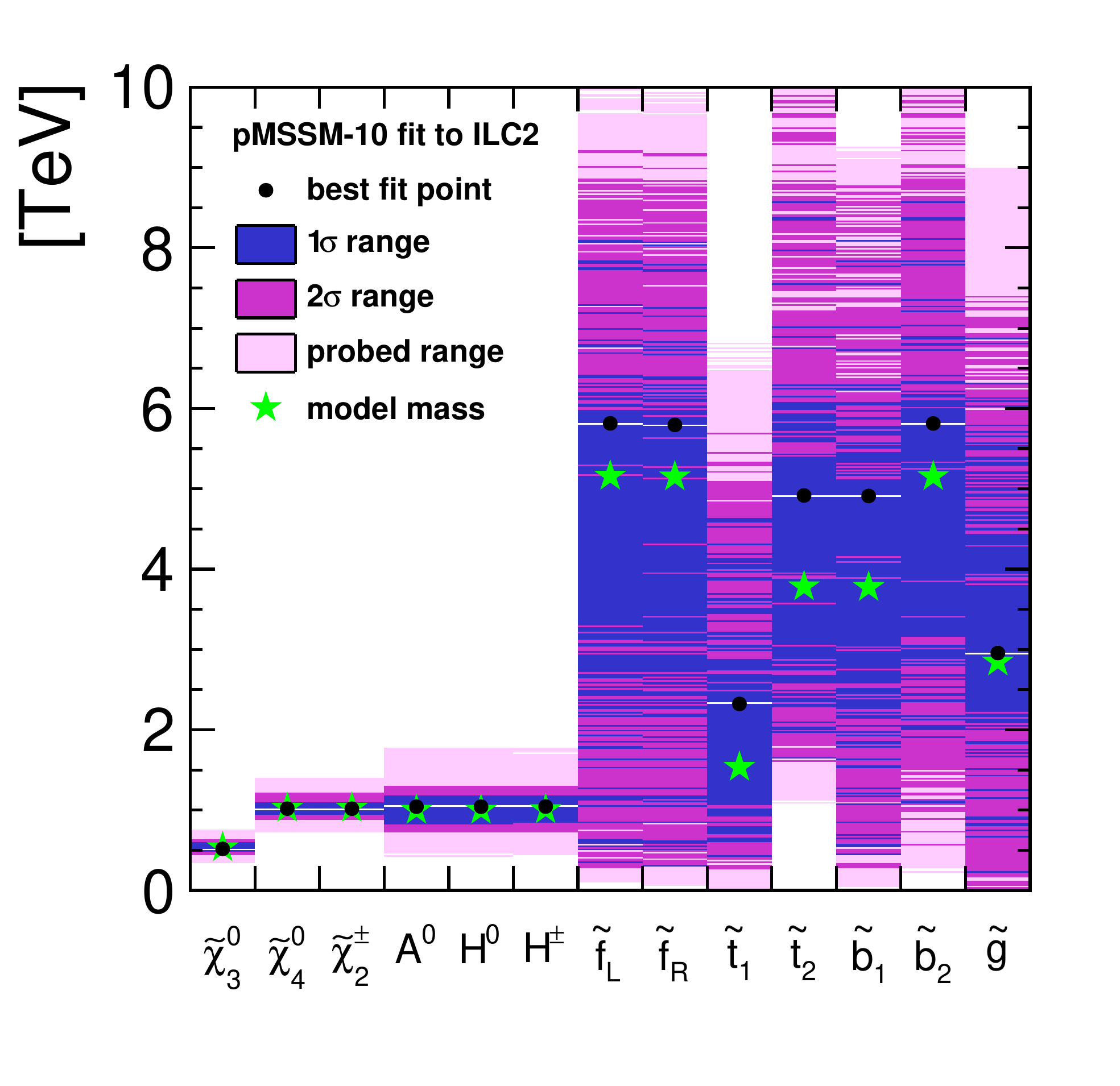}
\caption{Fitted masses}
\label{fig:ILC2fittedmasses10p}
\end{subfigure}
\begin{subfigure}{0.49\linewidth}
\includegraphics[width=\textwidth]{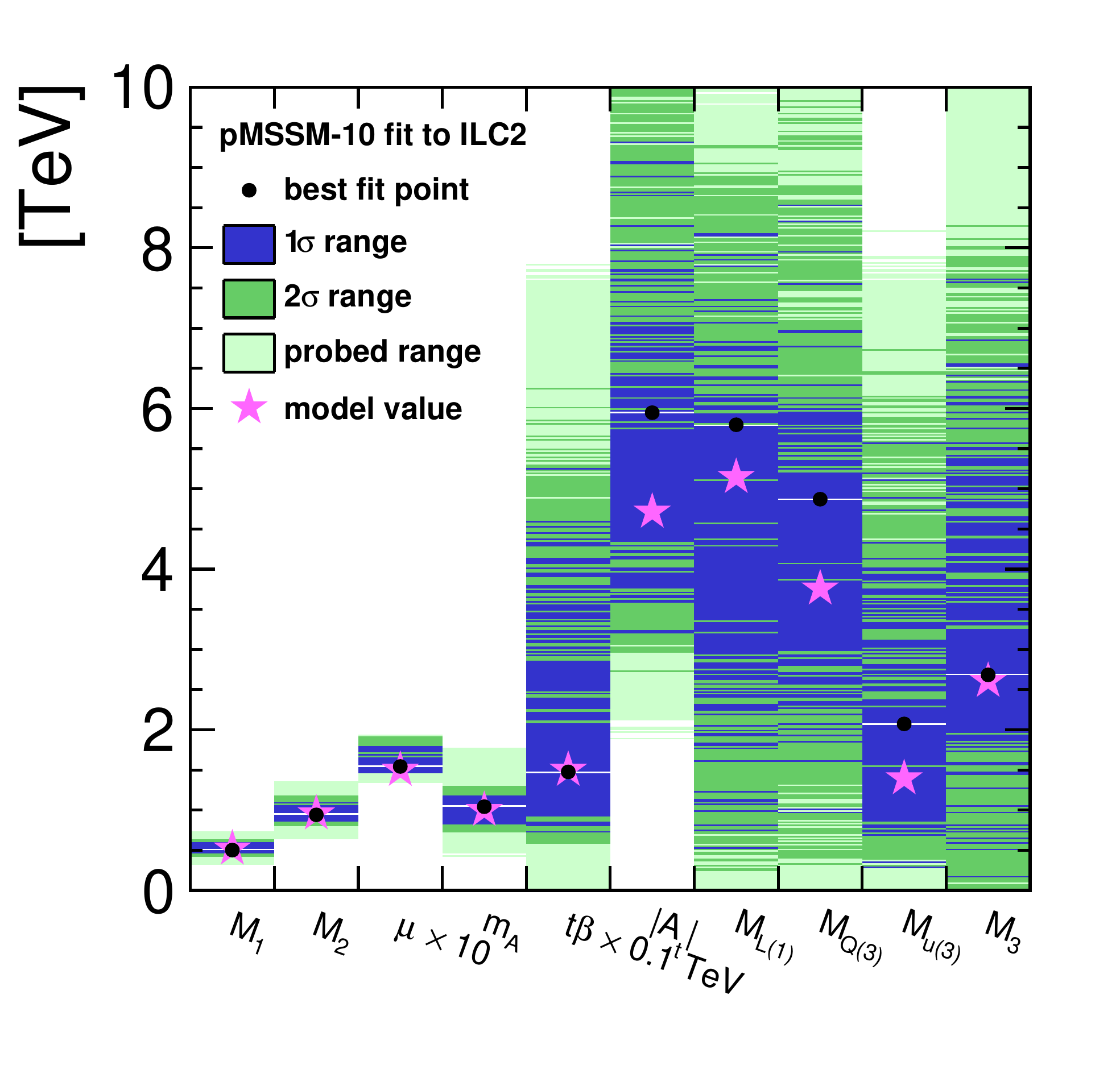}
\caption{Fitted parameters}
\label{fig:ILC2fittedparameters10p}
\end{subfigure}
\caption{Predicted mass and SUSY parameter ranges from the pMSSM-10 fit to ILC2. 
The green/magenta star indicates the true model values, while the black dot shows the best fit point. 
}
\label{fig:ILC2fitted10p}

\end{figure}

The resulting best fit values for the pMSSM parameters and their 1 and 2$\sigma$ intervals are compared 
to the input values in Tab.~\ref{tab:ILC2fittedparam10p}, quantifying the effect which could already 
be seen qualitatively in Fig.~\ref{fig:ILC2I20parabolae}. Again constraints on the sfermion sector can 
be derived due to their loop contributions. In contrast to the perfect agreement of the best fit point 
with the input parameter values in ILC1, the best fit point for ILC2 visibly overestimates the sfermion mass 
parameters. However the true values still remain within the 1$\sigma$ interval.
Figure~\ref{fig:ILC2fittedparameters10p} displays the
precisions obtained on the pMSSM-10 parameters. In case of ILC2, the gluino is most likely outside the reach of LHC, therefore $M_3$ is only constrained via its loop effects on the higgsino sector.

\begin{table}[htpb]{}
\centering
\resizebox{0.99\textwidth}{!}{
$\begin{array}{  c c | c c c  | c c c  }
\hline 
& & \text{pMSSM-4}  & & & \text{pMSSM-10}  & & \\
  \text{parameter}   & \text{ILC2 pMSSM true}  & \text{best fit point} & 1\sigma \text { CL} & 2\sigma \text { CL} & \text{best fit point} & 1\sigma \text { CL}&  2\sigma \text { CL}  \\ \hline \hline

 M_1 &   520.3  & 520.7 & \ ^{+38.6}_{-37.6} &  \ ^{+79.1}_{-71.0} & 502.1 & \ ^{+91.3}_{-32.9} &  \ ^{+130.1}_{-71.7}  \\
  M_2 &  957.2 & 959.42 &  \ ^{+55.4}_{-53.1} &  \ ^{+124.1}_{-100.1} & 941.0 &  \ ^{+145.4}_{-71.7} &  \ ^{+229.2}_{-130.9} \\
  \mu &  150.0  & 150.0 & \ ^{+0.4}_{-0.4} &  \ ^{+0.7}_{-0.8} & 154.4 & \ ^{+24.7}_{-7.3} &  \ ^{+36.6}_{-8.2} \\
  \tan \beta &  15.0 & 15.0 & \ ^{+0.7}_{-0.6} &  \ ^{+1.7}_{-1.2} & 14.8 &  \ ^{+38.4}_{-7.8} &  \ ^{+48.2}_{-9.0}  \\
   m_A &   1000  & & & &  1043 &  \ ^{+135}_{-203} &  \ ^{+240}_{-325}  \\
     M_3 &   2607 & & & &  2684 &   \ ^{+4990}_{-2585} &  \ ^{+5670}_{-2682}  \\
   M_{L} &  5146 & & & &  5797 &   \ ^{+2402}_{-5359} &  \ ^{+3511}_{-5544}  \\
   M_{U(3)} &   1395 & & & &  2073  &  \ ^{+3518}_{-1805} &  \ ^{+4716}_{-1805}  \\
   M_{Q(3)} &  3757 & & & &  4871 &  \ ^{+3680}_{-3933} &  \ ^{+5030}_{-4608}  \\
   A_t &  -4714 & & & &   -5948 & \ ^{+2734}_{-3387} &  \ ^{+3250}_{-4050}  \\ \hline
   \chi^2 &  & 0.0026 & & &   0.1627 & &  \\ \hline \hline
 \end{array}$
}
\caption{Fitted parameters in ILC2 pMSSM-4 and pMSSM-10. All units in GeV except for $\tan \beta$ and $\chi^2$.}
\label{tab:ILC2fittedparam10p}
\end{table}

\begin{table}[htbp]{}
\centering
$\begin{array}{  c c c c c  }
  \hline \hline
    \text{prediction} & \multicolumn{4}{c}{\text{ILC2}} \\ 
    & \text{model masses} & \text{best fit} & 1\sigma & 2 \sigma 
 \\ \hline \hline
m_{\neutralinothree} & 534.6 & 518        &     \ _{-34 }^{+72 } &      \ _{-74 }^{+110}    \\
m_{\neutralinofour}  & 1026 & 1018       &     \ _{-76 }^{+82 } &      \ _{-134 }^{+190}   \\
m_{\charginotwo}     & 1026 & 1018       &     \ _{-76 }^{+82 } &      \ _{-134 }^{+190}    \\
m_{H_0}              & 1000 & 1043       &     \ _{-223 }^{+137 } &    \ _{-323 }^{+257}     \\
m_{A_0}              & 1000 &  1043  & \ _{-223 }^{+137 } &    \ _{-323 }^{+257}    \\
m_{H^{\pm}}          & 1003 &  1045  & \ _{-205 }^{+135 } &    \ _{-325 }^{+255}    \\
m_{\supL}            & 5158 &  5814  & \ _{-5474 }^{+2286 } &  \ _{-5534 }^{+3406}      \\
m_{\supR}            & 5143 &  5795       &     \ _{-5495 }^{+2285 } &  \ _{-5495 }^{+3385}    \\
m_{\stopone}         & 1535 &   2322 & \ _{-1902 }^{+2318 } &  \ _{-2062 }^{+3378}    \\
m_{\stoptwo}         & 3782 &   4917  &        \ _{-3277 }^{+3663 } &  \ _{-3317 }^{+4983}    \\
m_{\sbottomone}      & 3774 &   4911       &     \ _{-4471 }^{+3189 } &  \ _{-4631 }^{+3869}     \\
m_{\sbottomtwo}      & 5154 &   5814       &     \ _{-4734 }^{+2766 } &  \ _{-5254 }^{+4086}     \\
m_{\gluino}          & 2846 &   2955  & \ _{-2735 }^{+3925 } &  \ _{-2935 }^{+4445}      \\
  \hline \hline
  \end{array}$
\caption{True and fitted masses as well as their uncertainties from a pMSSM-10 fit to ILC2 observables.
All values in GeV.}
\label{tab:ILC2mssm10pfit-fittedmasses}
\end{table}

\begin{figure}[htbp]
\begin{subfigure}{0.49\linewidth}
\includegraphics[width=\textwidth]{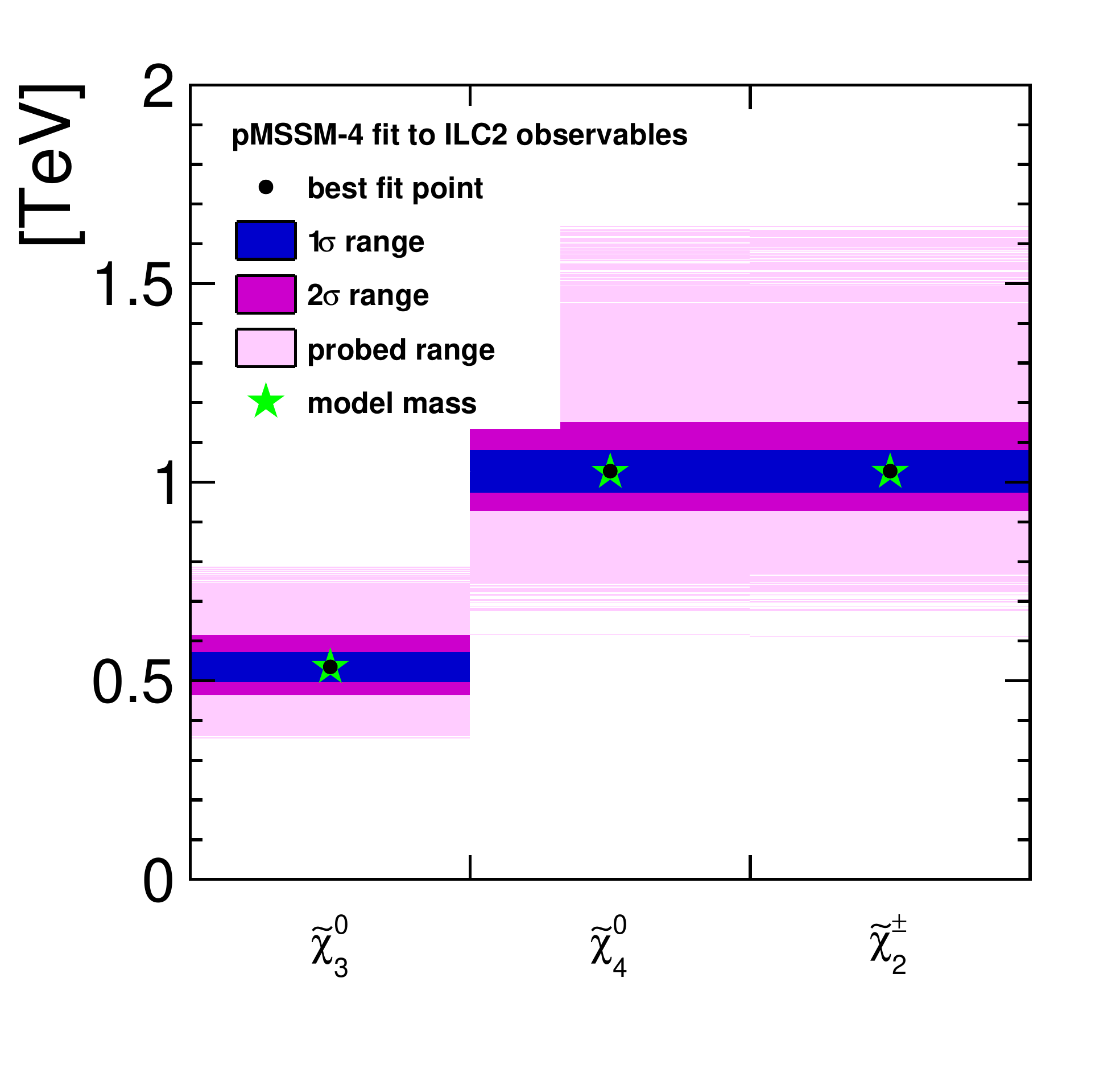}
\caption{Fitted masses}
\label{fig:ILC2fittedmasses4p}
\end{subfigure}
\begin{subfigure}{0.49\linewidth}
\includegraphics[width=\textwidth]{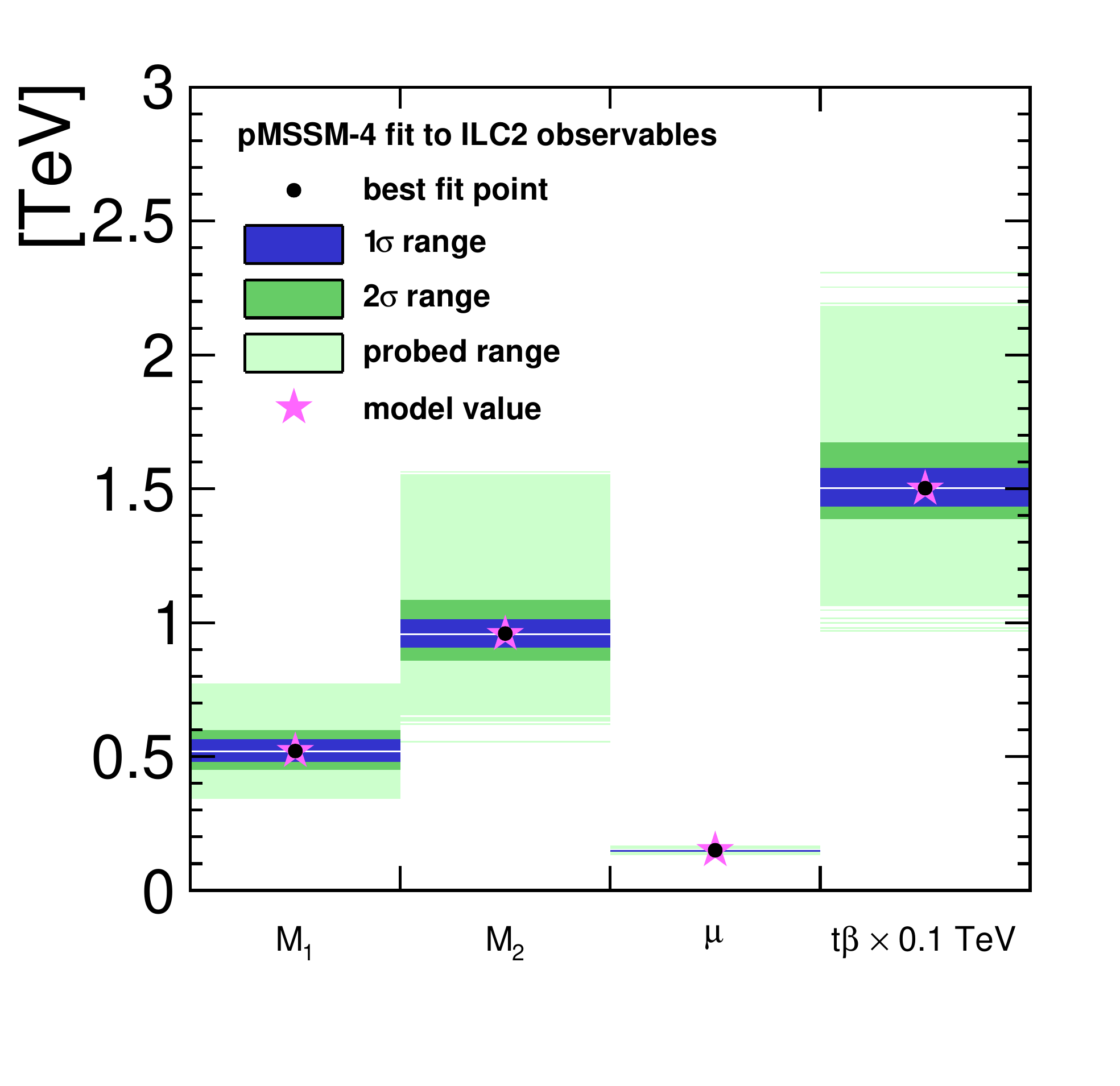}
\caption{Fitted parameters}
\label{fig:ILC2fittedparam4p}
\end{subfigure}
\caption{Predicted mass and SUSY parameter ranges from the pMSSM-4 fit to ILC2. The green/magenta 
star indicates the true model values, while the black dot shows the best fit point. 
}
\label{fig:ILC2fitted4p}

\end{figure}

As in the ILC1 case, the determined parameters can be used to predict the 
masses of the as-yet unobserved sparticles, 
as shown for the pMSSM-10 fit in Tab.~\ref{tab:ILC2mssm10pfit-fittedmasses} and Fig.~\ref{fig:ILC2fittedmasses10p}. Finally, Fig.~\ref{fig:ILC2fittedmasses4p} shows the result of the 4 parameter fit of
 $M_1$, $M_2$, $\mu$, $\tan \beta$. Again, the remaining parameters fixed to model values,
based on the assumption that the best fit point of the 10-parameter fit is sufficiently close to the
true point that the effect of fixing to the true values is negligible.

%
%
%
%

\subsubsection{nGMM1 Benchmark}
\label{ssec:weakscale_fits:nGMM1}

\begin{figure}[htpb]
\begin{center}
\begin{subfigure}{0.4\linewidth}
\includegraphics[width=\textwidth]{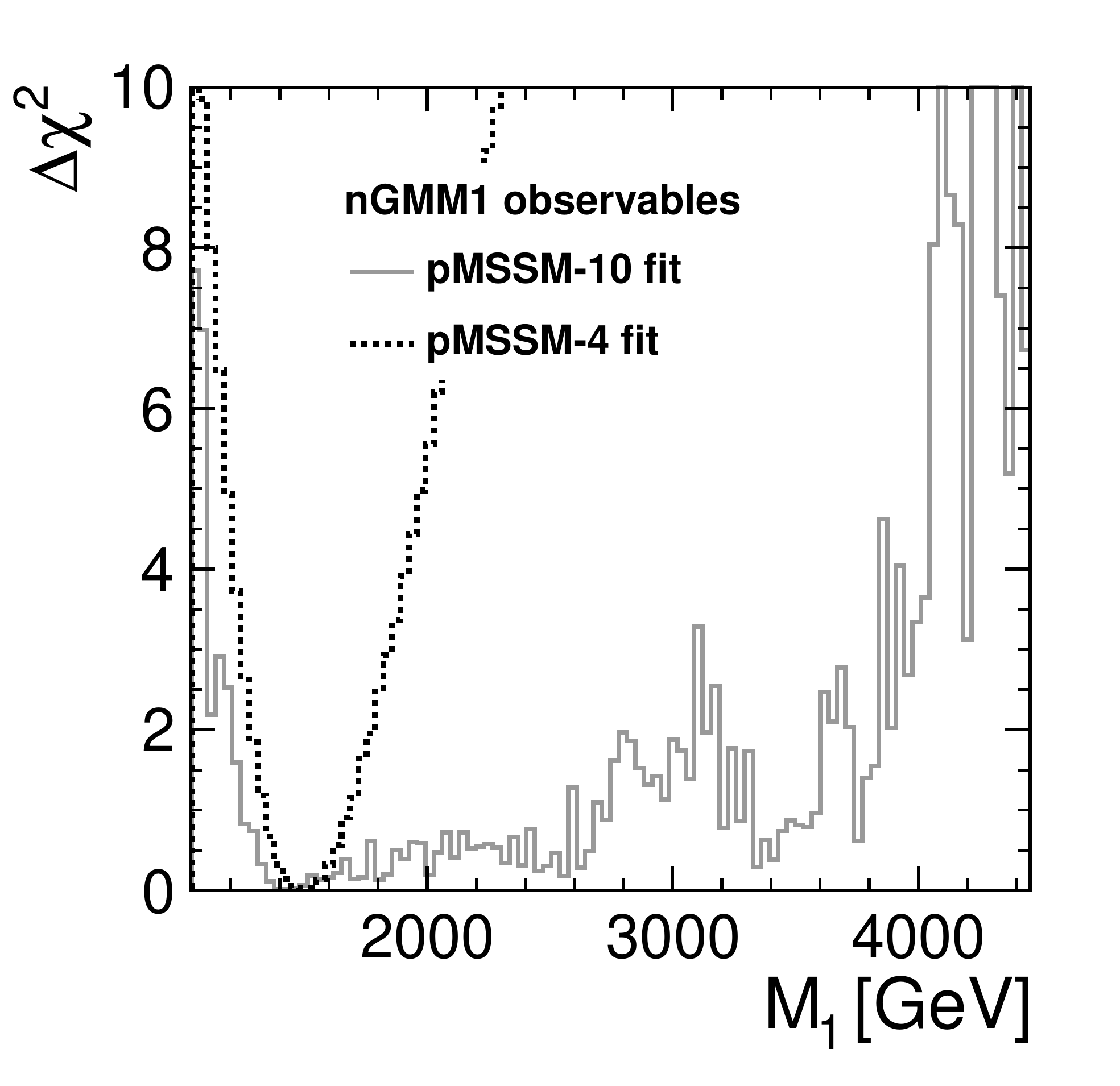}
\caption{}
\label{fig:nGMM1I20parabolae:M1}
\end{subfigure}
\begin{subfigure}{0.4\linewidth}
\includegraphics[width=\textwidth]{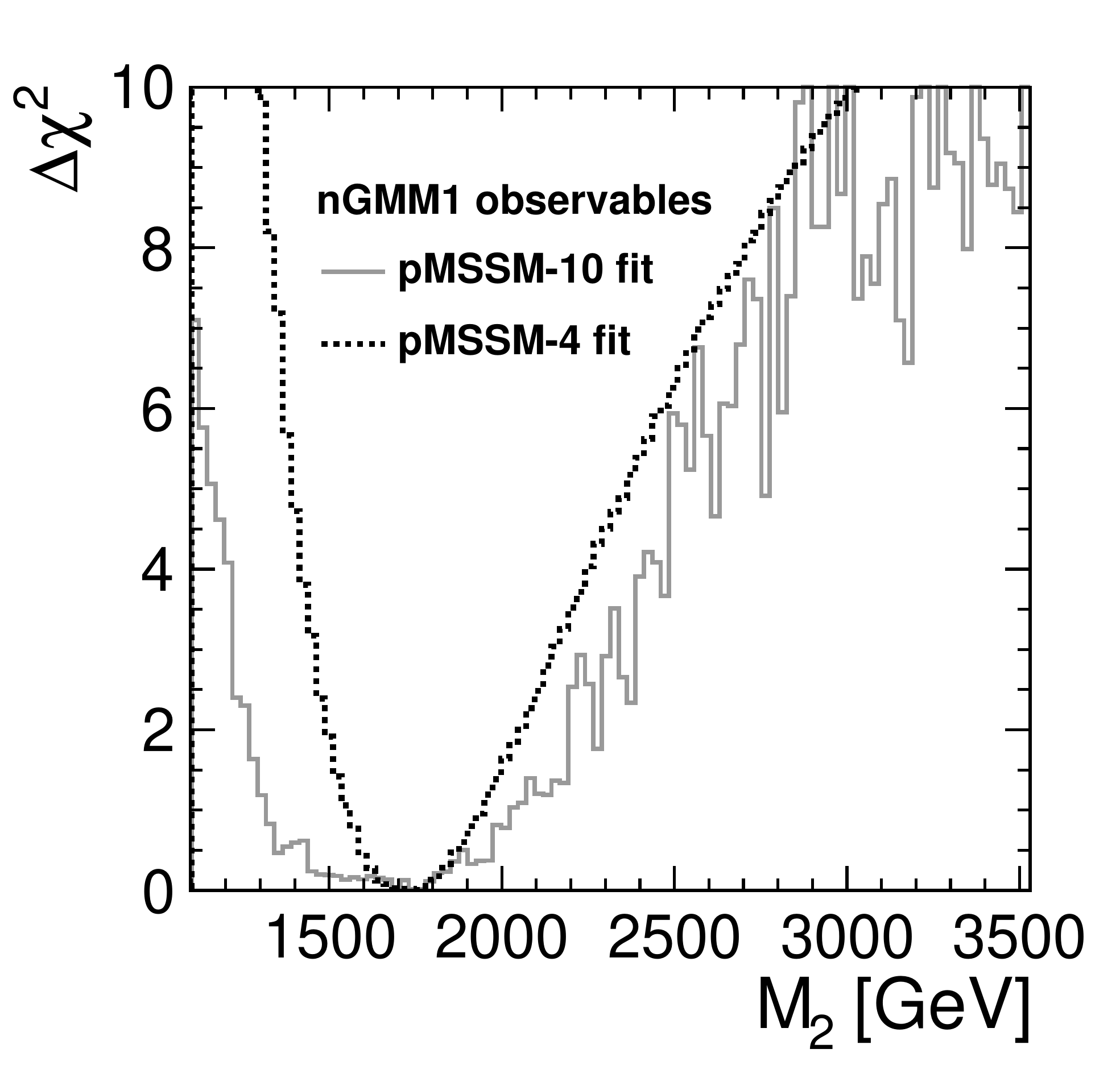}
\caption{}
\label{fig:nGMM1I20parabolae:M2}
\end{subfigure}
\begin{subfigure}{0.4\linewidth}
\includegraphics[width=\textwidth]{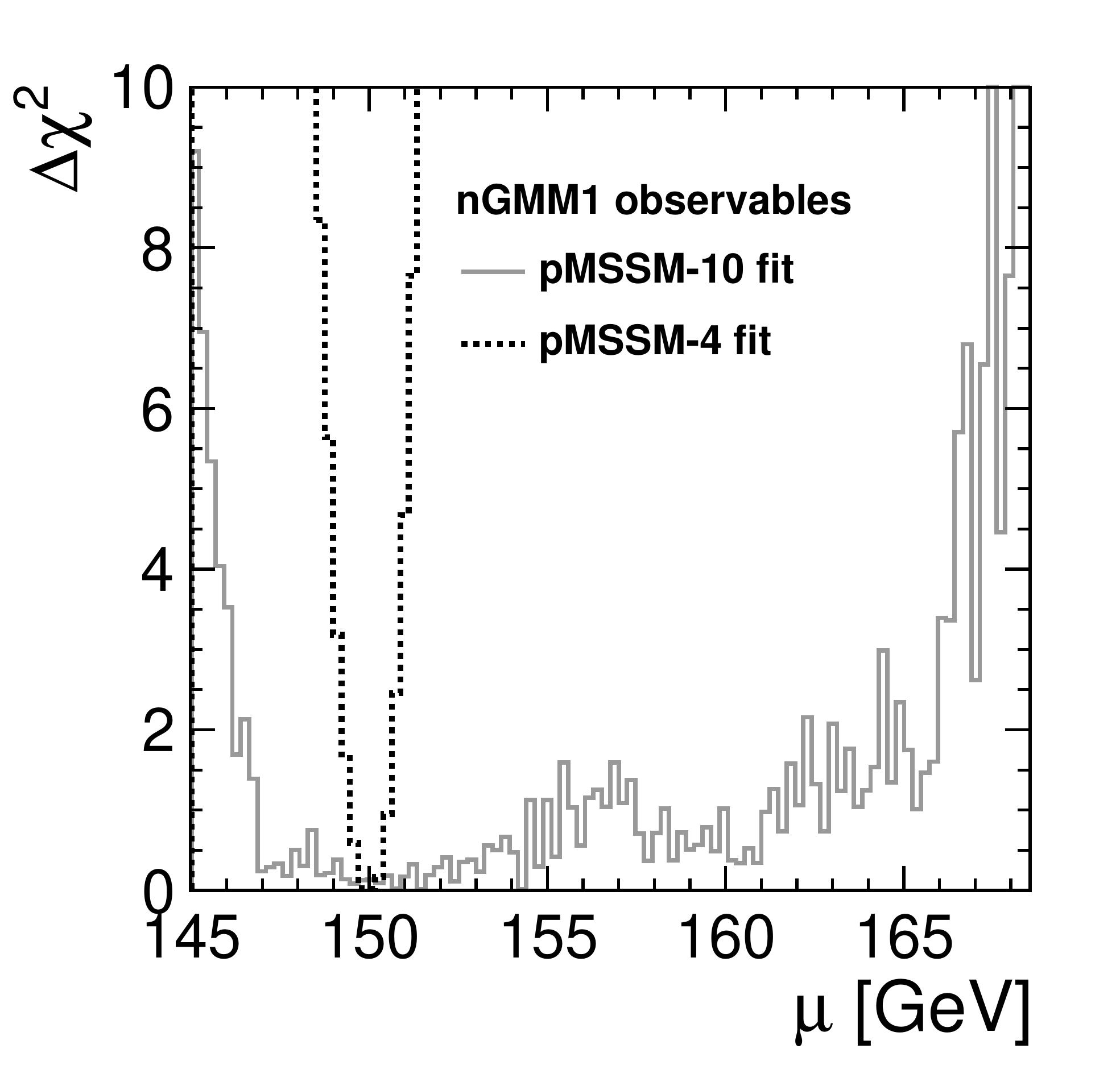}
\caption{}
\label{fig:nGMM1I20parabolae:mu}
\end{subfigure}
\begin{subfigure}{0.4\linewidth}
\includegraphics[width=\textwidth]{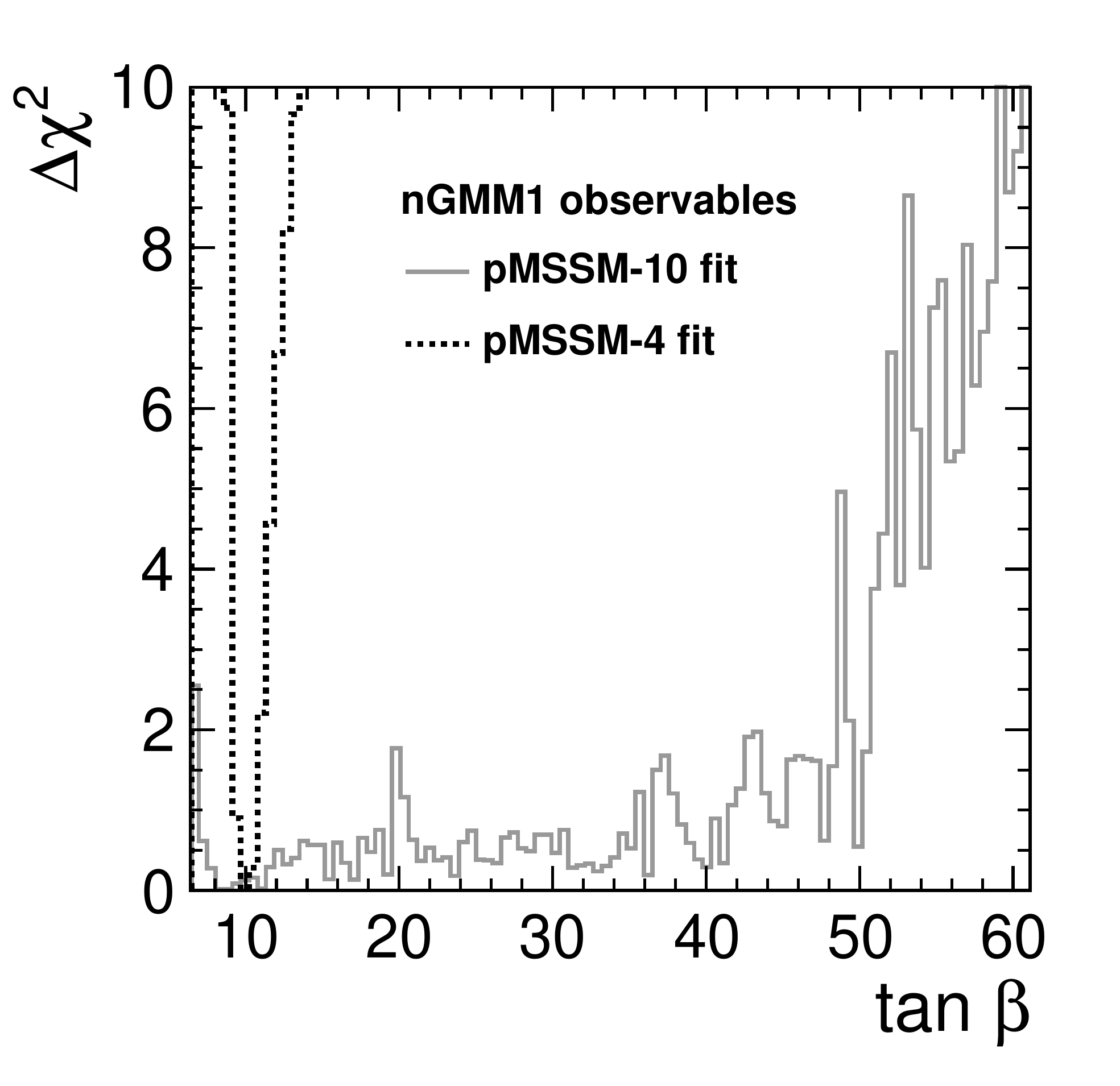}
\caption{}
\label{fig:nGMM1I20parabolae:tb}
\end{subfigure}
\end{center}
\caption{nGMM1: Minimum $\chi^2$ as a function of $M_1$, $M_2$ and $\mu$ and $\tan \beta$ in pMSSM-4 fit (dashed black line) and pMSSM-10 fit (solid grey line). For each bin, the minimum $\chi^2$ of all Markov chain points which have the $x$ axis quantity in that bin is plotted.}
\label{fig:nGMM1I20parabolae}
\end{figure}

Finally, Fig.~\ref{fig:nGMM1I20parabolae} shows the minimum $\chi^2$ as a function of $M_1$, $M_2$ and $\mu$ and 
$\tan \beta$ in the pMSSM-10 and pMSSM-4 fits to the nGMM1 observables. Also here, the much smaller parameter 
space to be sampled in case of the 4-dimensional fit leads to much smoother curves than in the 10-dimensional case. 
Again, the determinations of $\mu$ and $\tan{\beta}$ improve significantly. However $M_1$ and $M_2$ exchange 
their roles compared to the other benchmarks, so that now $M_1$ is somewhat better constrained in the pMSSM-4 fit, 
while the precision on $M_2$ is nearly identical in the two fits. However, it should be noted that
$M_1$ and $M_2$ are less well constrained than in the cases of the ILC1 and ILC2 benchmarks. This results from 
a combination of  the worse experimental resolutions and the larger absolute values of $M_1$ and $M_2$ in case 
of nGMM1. In this most challenging case, the mass splitting between  \charginoone\ and \LSP\ is only 2.5\,GeV, 
which corresponds to less than 2$\sigma$ of the experimental resolution. As discussed in Sec.~\ref{observables}, 
the mass {\em differences} $m_{\neutralinotwo}-m_{\LSP}$ 
and $m_{\charginoone}-m_{\LSP}$ are directly accessible experimentally as the endpoint of the di-lepton or di-jet 
invariant mass spectrum. 
Therefore, we consider in this case as alternative input these mass differences 
in addition to the \LSP\ mass, which also presents a set of observables with minimal correlations. 
The corresponding precisions are summarized in Tab.~\ref{tab:nGMM1massdifferenceinputs}.

\begin{table}[htpb]{}
\centering
$
\begin{array}{  c c c c  }
 \hline \hline 
 \text{observable} & \text{nGMM1 model value [GeV]}  & \text{precision} & \text{I20 precision} \\ \hline 
 m_{\LSP} & 154.9 & 1.7\% & 1.0\% \\ 
 m_{\neutralinotwo}-m_{\LSP} & 5.3 & 2.1\% & 1.4\% \\ 
   m_{\charginoone}-m_{\LSP} & 2.4 & 2.5\% & 1.2\%  \\  \hline \hline   
 \end{array}
$
\caption{Experimental precision on the higgsino mass differences in nGMM1 combined from 500\,GeV 500$^{-1}$ fb for both $\mathcal{P}(\pm 0.8, \mp 0.3)$, as well as scaled to 1600 fb$^{-1}$ for both polarisations at $\sqrt{s}=500 \GeV$, ignoring the data sets with other centre-of-mass energies in the I20 operating scenario. Again, it is assumed that the precisions obtained from the simulations based on the \texttt{Isajet} spectrum can be transfered to the \texttt{SPheno3.3.9beta} spectrum.}

\label{tab:nGMM1massdifferenceinputs}
\end{table}


\begin{figure}[htbp]
\begin{center}
\includegraphics[width=0.4\textwidth]{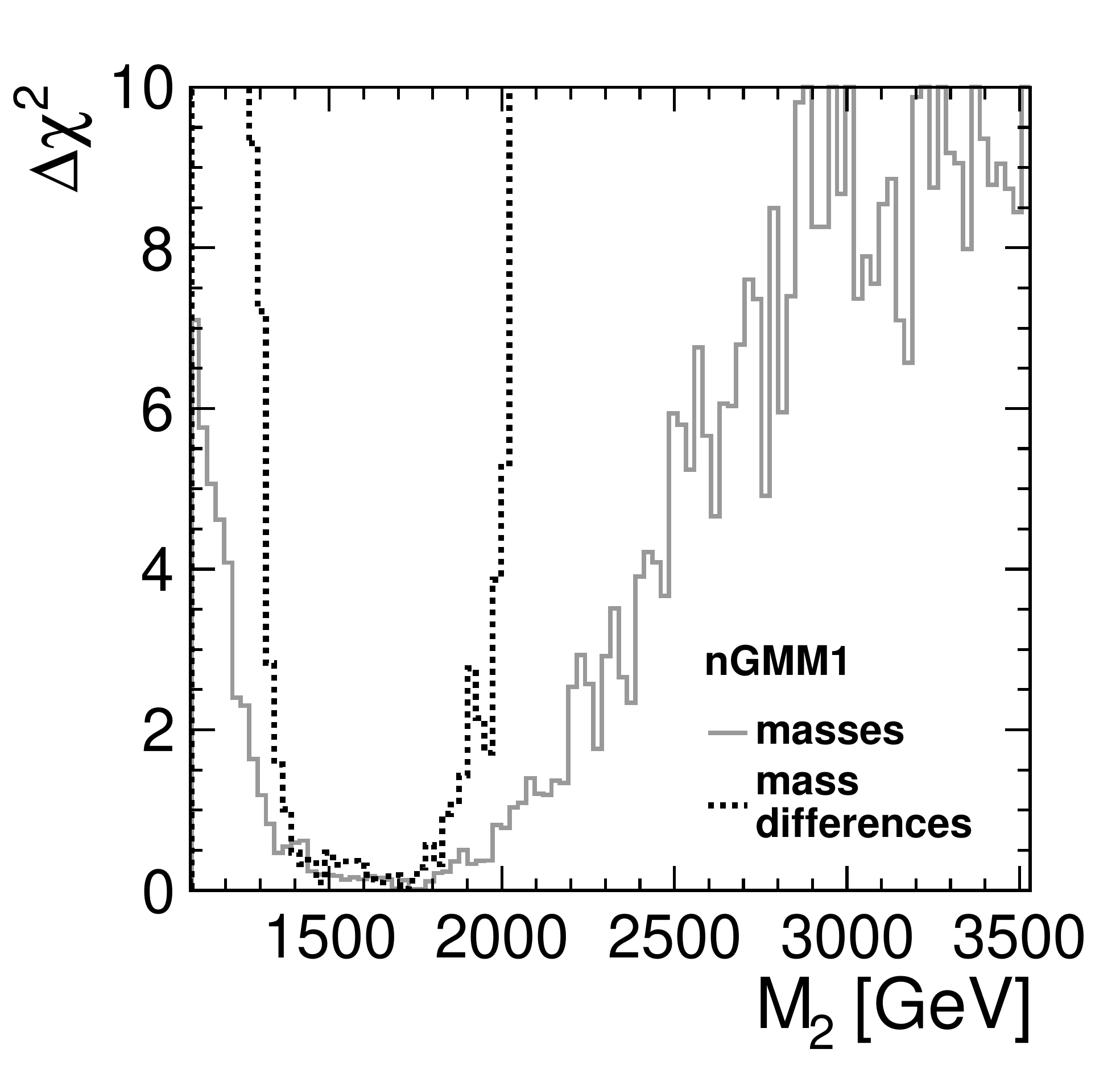}
\end{center}
\caption{nGMM1 $\chi^2$ parabola for $M_2$ in the 10-parameter fit with mass differences (black, dashed) and the same fit with masses as observables (grey, solid).}
\label{fig:nGMM1I2010paramparabolaemdif}
\end{figure}

As can be seen in Fig.~\ref{fig:nGMM1I2010paramparabolaemdif}, the determination of $M_2$ in the 10-parameter fit 
improves significantly when instead of the absolute masses the mass differences are used as fit input, especially the upper bound. There is
no significant effect on $M_1$, $\mu$  or $\tan \beta$.

\begin{figure}[htbp]
\begin{subfigure}{0.49\linewidth}
\includegraphics[width=\textwidth]{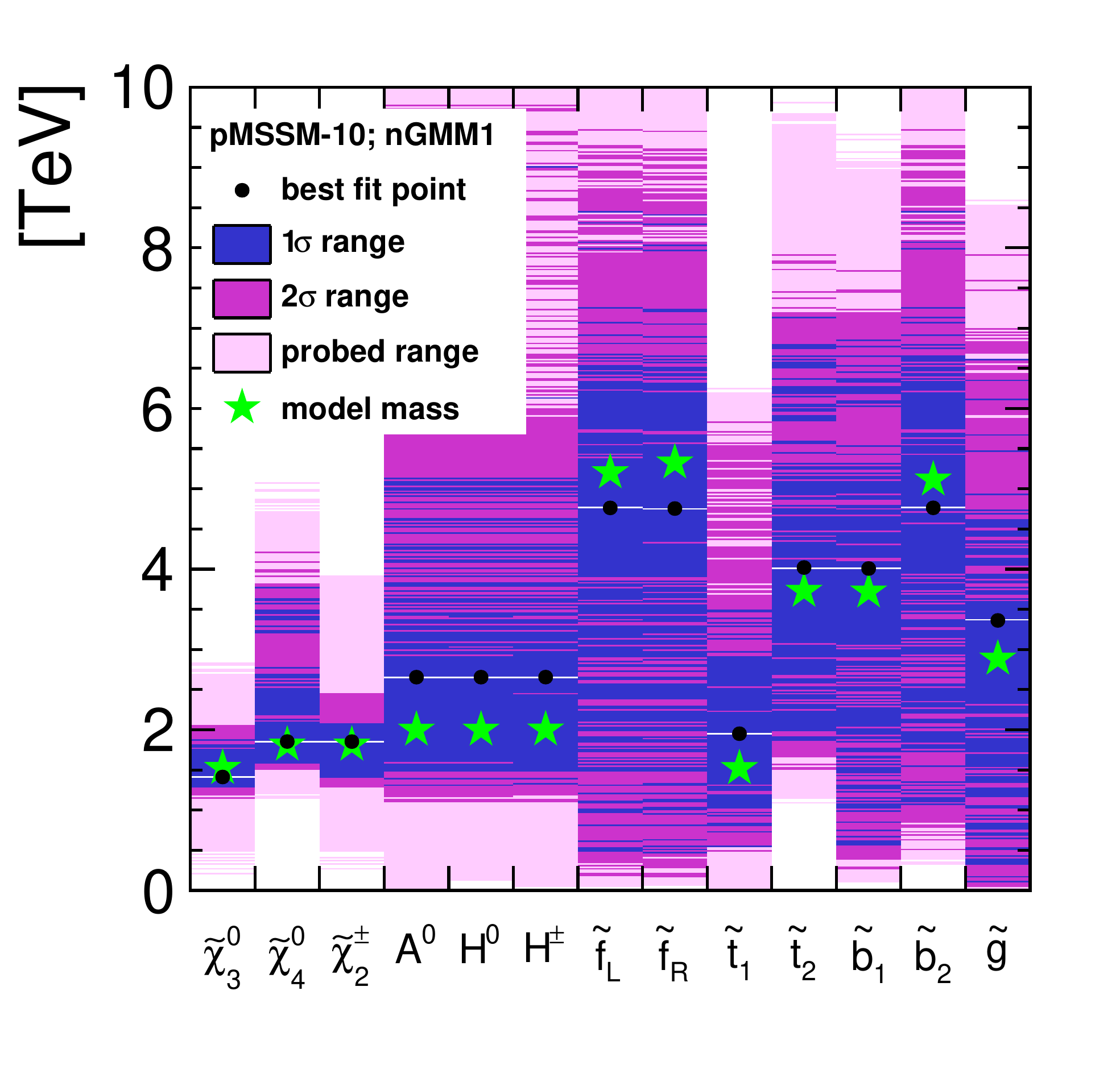}
\caption{Fitted masses, standard input}
\label{fig:nGMM1fittedmasses10p}
\end{subfigure}
\begin{subfigure}{0.49\linewidth}
\includegraphics[width=\textwidth]{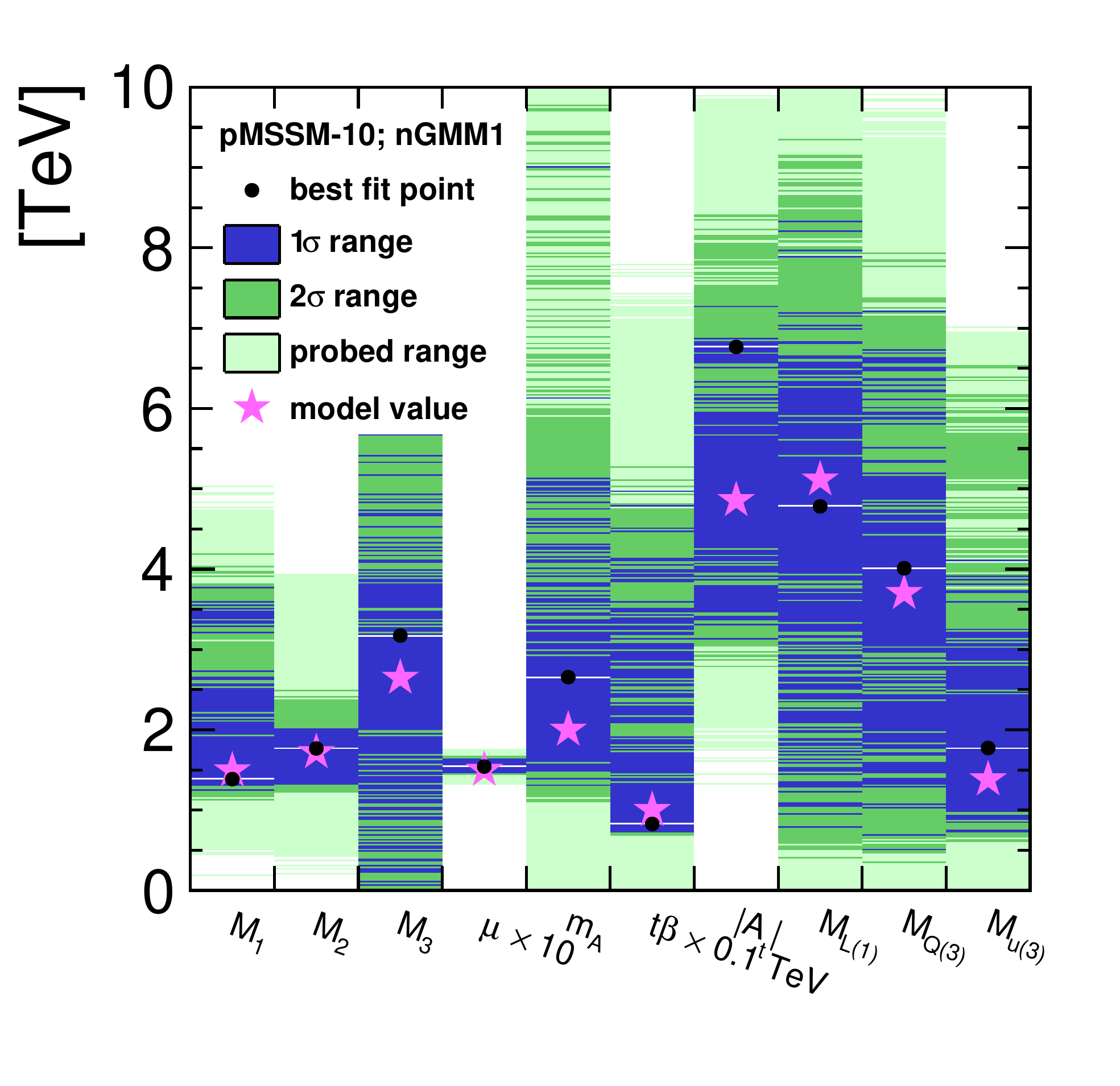}
\caption{Fitted parameters, standard input}
\label{fig:nGMM1fittedparam10p}
\end{subfigure}
\vspace{0.1cm}
\begin{subfigure}{0.49\linewidth}
\includegraphics[width=\textwidth]{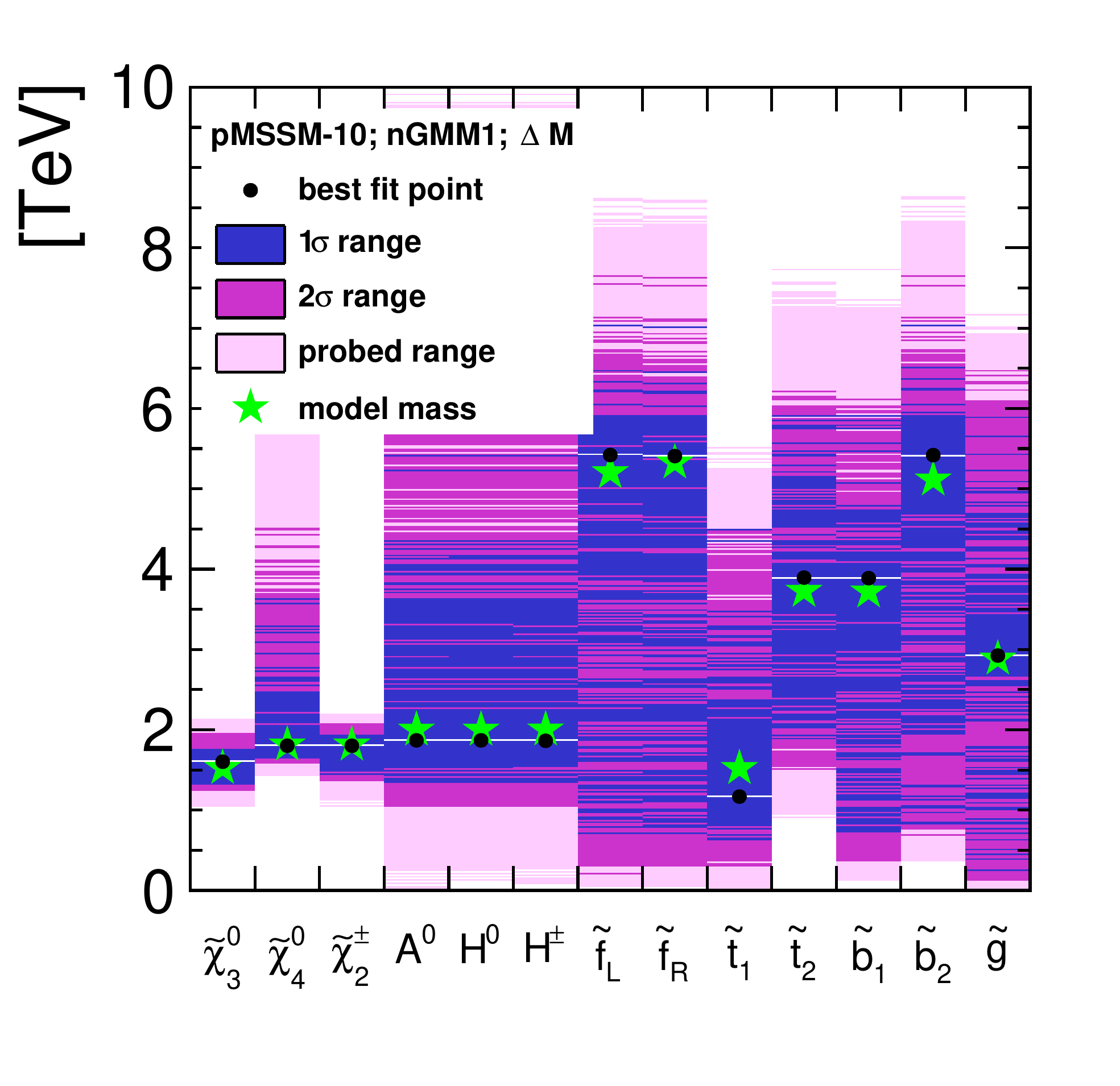}
\caption{Fitted masses, mass difference input}
\label{fig:nGMM1fittedmasses10pmdif}
\end{subfigure}
\begin{subfigure}{0.49\linewidth}
\includegraphics[width=\textwidth]{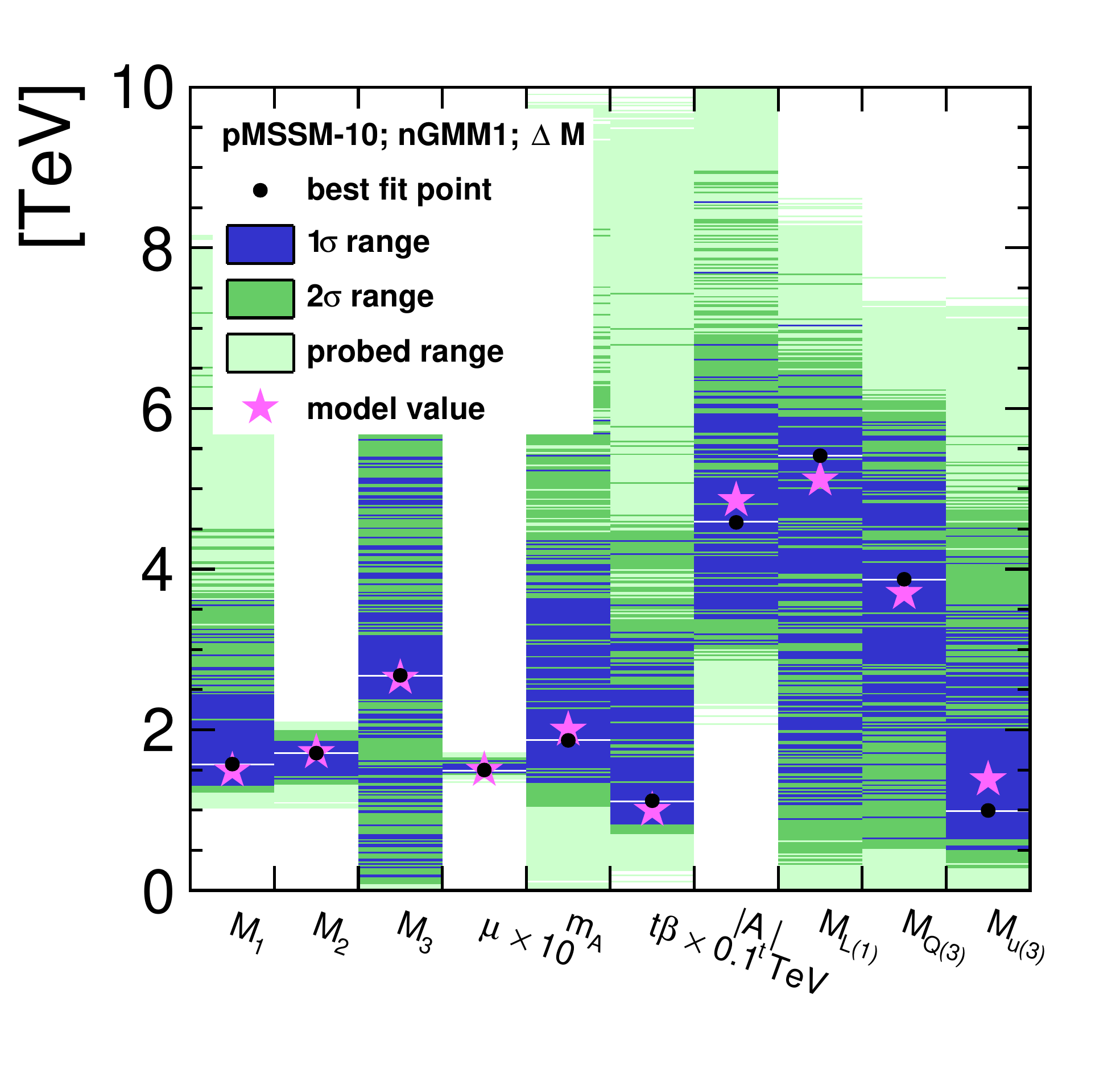}
\caption{Fitted parameters, mass difference input}
\label{fig:nGMM1fittedparam10pmdif}
\end{subfigure}
\caption{Predicted mass and SUSY parameter ranges from the pMSSM-10 fit to nGMM1 observables - 
including either $\neutralinotwo$ 
and $\charginoone$ masses or their mass differences with the LSP. The green/magenta star indicates the true model 
values, while the black dot shows the best fit point. 
}
\label{fig:nGMM1fitted10p}

\end{figure}

The resulting best fit values for the pMSSM parameters and their 1 and 2$\sigma$ intervals are compared 
to the input values in Tab.~\ref{tab:nGMM1fittedparam10p}, quantifying the effect which could already 
be seen qualitatively in Fig's.~\ref{fig:nGMM1I20parabolae} 
and~\ref{fig:nGMM1I2010paramparabolaemdif}. 
As before, constraints on the sfermion sector can be derived due to their loop contributions. 
In contrast to the perfect agreement of the best fit point 
with the input parameter values in ILC1, the best fit point for nGMM1 visibly overestimates the sfermion mass 
parameters. However the true values still remain within the 1$\sigma$ interval.
Using the mass differences as input instead of the absolute higgsino masses notably improves the precision on $M_2$,
as expected from the $\chi^2$ distribution, but it also significantly improves the agreement of the best fit point with 
the true model parameters. These improvements can also be seen in Fig.~\ref{fig:nGMM1fittedparam10p} and~\ref{fig:nGMM1fittedparam10pmdif}, in particular the better agreement in $m_A$ and $A_t$. 

\begin{table}[htpb]{}
\centering
\resizebox{0.99\textwidth}{!}{
$\begin{array}{  c c | c c c  | c c c  | c c c  }
\hline 
& & \multicolumn{3}{c|}{\text{pMSSM-4}} & \multicolumn{3}{c|}{\text{pMSSM-10}} & \multicolumn{3}{c}{\text{pMSSM-10 with mass differences}} \\
  \text{parameter}   & \text{true}  & \text{best fit point} & 1\sigma \text { CL} & 2\sigma \text { CL} & \text{best fit point} &  1\sigma \text { CL}&  2\sigma \text { CL} & \text{best fit point} &  1\sigma \text { CL}&  2\sigma \text { CL}  \\ \hline \hline
 M_1 & 1493 & 1501 & \ ^{+173}_{-149} & \ ^{+411}_{-280} &  1386  & \ ^{+2386}_{-145} & \ ^{+2830}_{-282} & 1573 & \ ^{+2091}_{-282} &  \ ^{+5650}_{-344} \\
  M_2 &  1720 & 1711 & \ ^{+220}_{-158} & \ ^{+530}_{-279}  & 1768 & \ ^{+254}_{-451} & \ ^{+717}_{-549}  & 1710 &  \ ^{+137}_{-313} &  \ ^{+277}_{-394} \\
   \mu &   150.0 & 150.0 & \ ^{+0.4}_{-0.4} & \ ^{+0.9}_{-0.9}  & 154.2 &\ ^{+7.4}_{-8.7} & \ ^{+12.9}_{-8.3}  & 149.9 &  \ ^{+11.5}_{-3.4} &  \ ^{+15.3}_{-4.2} \\
   \tan \beta &  10.0 & 10.0 & \ ^{+0.5}_{-0.3} & \ ^{+1.2}_{-0.6}  & 8.3 & \ ^{+41.9}_{-1.3} & \ ^{+44.6}_{-1.9} & 11.2 &  \ ^{+32.5}_{-3.4} &  \ ^{+63.9}_{-4.2} \\
  m_A &   2000 & & & &   2655 &\ ^{+6493}_{-1449} & \ ^{+11492}_{-1596} & 1868 &  \ ^{+4018}_{-567} &  \ ^{+6423}_{-867}  \\
  M_3 &  2646 & &  &   & 3173  &\ ^{+4229}_{-3168} & \ ^{+5347}_{-3168}  & 2677 &  \ ^{+3892}_{-2541} &  \ ^{+4550}_{-2614}\\
   M_{L} &  5115 & & & &  4781 & \ ^{+3589}_{-4077} & \ ^{+4630}_{-4456} & 5412 &  \ ^{+1629}_{-4581} &  \ ^{+2319}_{-5118}   \\
   M_{U(3)} &   1381 & & & & 1774 & \ ^{+2384}_{-1086} & \ ^{+4826}_{-1214} & 996 &  \ ^{+3540}_{-500} &  \ ^{+4686}_{-741} \\
   M_{Q(3)} &  3701 & & &   & 4011 & \ ^{+3254}_{-3535} & \ ^{+3982}_{-3697} & 3874 &  \ ^{+1983}_{-3245} &  \ ^{+2356}_{-3370}  \\
   A_t &   -4857 & & & &   -6766 & \ ^{+3698}_{-509} & \ ^{+4012}_{-1702} & -4582 &  \ ^{+1558}_{-4006} &  \ ^{+1750}_{-4390}\\  \hline
   \chi^2  & & 0.0138  & &  &  0.0927 & &  & 0.0668 & & \\ \hline \hline
 \end{array}$
}
\caption{Fitted parameters in nGMM1: pMSSM-4, pMSSM-10 and pMSSM-10 with mass differences as input. All units in GeV except for $\tan \beta$ and $\chi^2$.}
\label{tab:nGMM1fittedparam10p}
\end{table}

\begin{table}[htbp]{}
\centering
$\begin{array}{  c c | c c c | c c c }
  \hline \hline
    \text{prediction} & \multicolumn{3}{c|}{\text{nGMM1} M} & \multicolumn{3}{c}{\text{nGMM1} \Delta M} \\ 
    & \text{model mass} \text{best fit} & 1\sigma & 2 \sigma & \text{best fit} & 1\sigma & 2 \sigma 
 \\ \hline \hline
m_{\neutralinothree} & 1522 & 1412       &     \ _{-134 }^{+454 } &    \ _{-260 }^{+640}   & 1603       &     \ _{-283 }^{+149 } &    \ _{-349 }^{+347}     \\
m_{\neutralinofour}  & 1809 & 1854       &     \ _{-264 }^{+1920 } &   \ _{-336 }^{+2364}  & 1802       &     \ _{-146 }^{+1834 } &   \ _{-218 }^{+2710}    \\
m_{\charginotwo}     & 1808 &  1853  & \ _{-443 }^{+229 } &    \ _{-557 }^{+601}           & 1801       &     \ _{-349 }^{+137 } &    \ _{-433 }^{+275}     \\
m_{H_0}              & 2000 & 2655       &     \ _{-1355 }^{+6365 } &  \ _{-1555 }^{+7125}                 & 1868       &     \ _{-528 }^{+3992 } &   \ _{-828 }^{+6372}    \\
m_{A_0}              & 2000 &  2655  & \ _{-1355 }^{+6365 } &  \ _{-1555 }^{+7125}                 & 1868       &     \ _{-528 }^{+3992 } &   \ _{-828 }^{+6372}    \\
m_{H^{\pm}}          & 2002 & 2656       &     \ _{-1336 }^{+6364 } &  \ _{-1556 }^{+7124}         & 1863       &     \ _{-523 }^{+3997 } &   \ _{-823 }^{+6377}    \\
m_{\supL}            & 5121 & 4762       &     \ _{-4282 }^{+3698 } &  \ _{-4582 }^{+4718}         & 5421  &  \ _{-4721 }^{+1619 } &  \ _{-5221 }^{+2239}           \\
m_{\supR}            & 5110 & 4754       &     \ _{-4294 }^{+3666 } &  \ _{-4594 }^{+4706}         & 5408       &     \ _{-4708 }^{+1612 } &  \ _{-5108 }^{+2232}   \\
m_{\stopone}         & 1519 &  1951  & \ _{-1411 }^{+1549 } &  \ _{-1471 }^{+3889}                 & 1168       &     \ _{-548 }^{+3332 } &   \ _{-868 }^{+3332}    \\
m_{\stoptwo}         & 3782 & 4029       &     \ _{-2160 }^{+3120 } &  \ _{-2520 }^{+3900}         & 3894  &  \ _{-2014 }^{+2026 } &  \ _{-2394 }^{+2326}           \\
m_{\sbottomone}      & 3774 & 4008       &     \ _{-3448 }^{+2852 } &  \ _{-3748 }^{+3712}         & 3888       &     \ _{-3168 }^{+2032 } &  \ _{-3528 }^{+2232}   \\
m_{\sbottomtwo}      & 5154 &  4763  & \ _{-3703 }^{+3697 } &  \ _{-4263 }^{+4717}         & 5419  &  \ _{-4599 }^{+1621 } &  \ _{-4739 }^{+2241}           \\
m_{\gluino}          & 2846 & 3361       &     \ _{-3261 }^{+3259 } &  \ _{-3361 }^{+4559}         & 2924       &     \ _{-2684 }^{+2976 } &  \ _{-2804 }^{+3556}   \\
  \hline \hline
  \end{array}$
\caption{nGMM1: True and fitted masses and their uncertainties from pMSSM-10 fits with the standard set of 
observables as well as with the higgsino mass differences replacing the $\neutralinotwo$ and $\charginoone$ 
masses as observables. All values in GeV.}
\label{tab:nGMM1mssm10pfit-fittedmasses}
\end{table}

Again, the determined parameters can be used to predict the masses of the yet unobserved sparticles, as shown for the pMSSM-10 fit in Tab.~\ref{tab:nGMM1mssm10pfit-fittedmasses} and in Fig.~\ref{fig:nGMM1fittedmasses10p} and~\ref{fig:nGMM1fittedmasses10pmdif} with standard input and when using the mass differences instead. As expected, the improved precision on $M_2$ when using the mass differences as input leads to improved predictions of the $\neutralinothree$ and $\charginotwo$ masses. In addition, the agreement between the
best fit predictions for the heavy Higgs boson masses as well as for all the sfermion masses with their true value improves significantly due to the better agreement in $m_A$ and $A_t$.

\begin{figure}[htbp]
 \begin{subfigure}{0.49\linewidth}
\includegraphics[width=1.\textwidth]{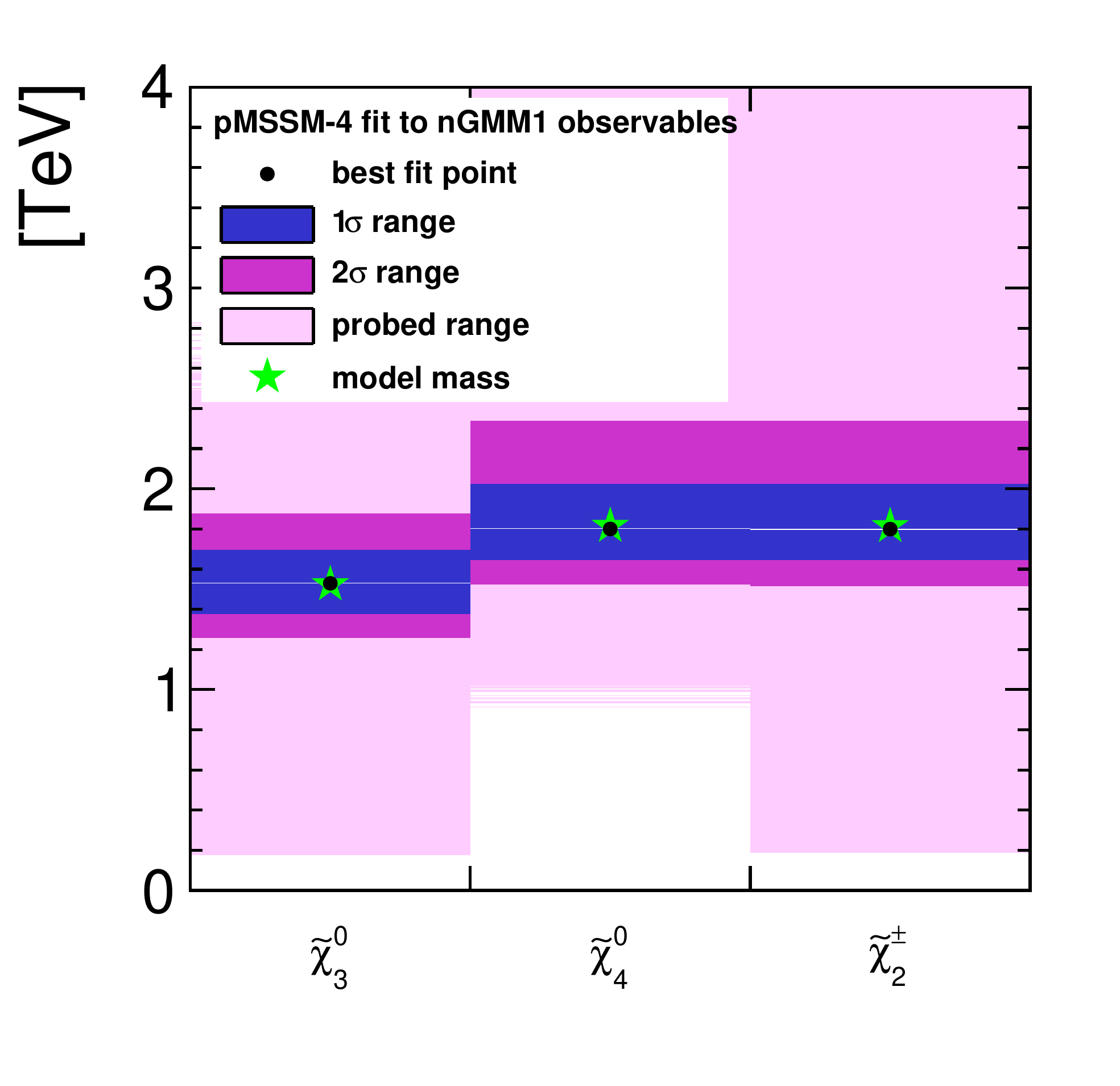}
\caption{Fitted masses}
\label{fig:nGMM1fittedmasses4p}
\end{subfigure}
\begin{subfigure}{0.49\linewidth}
\includegraphics[width=1.\textwidth]{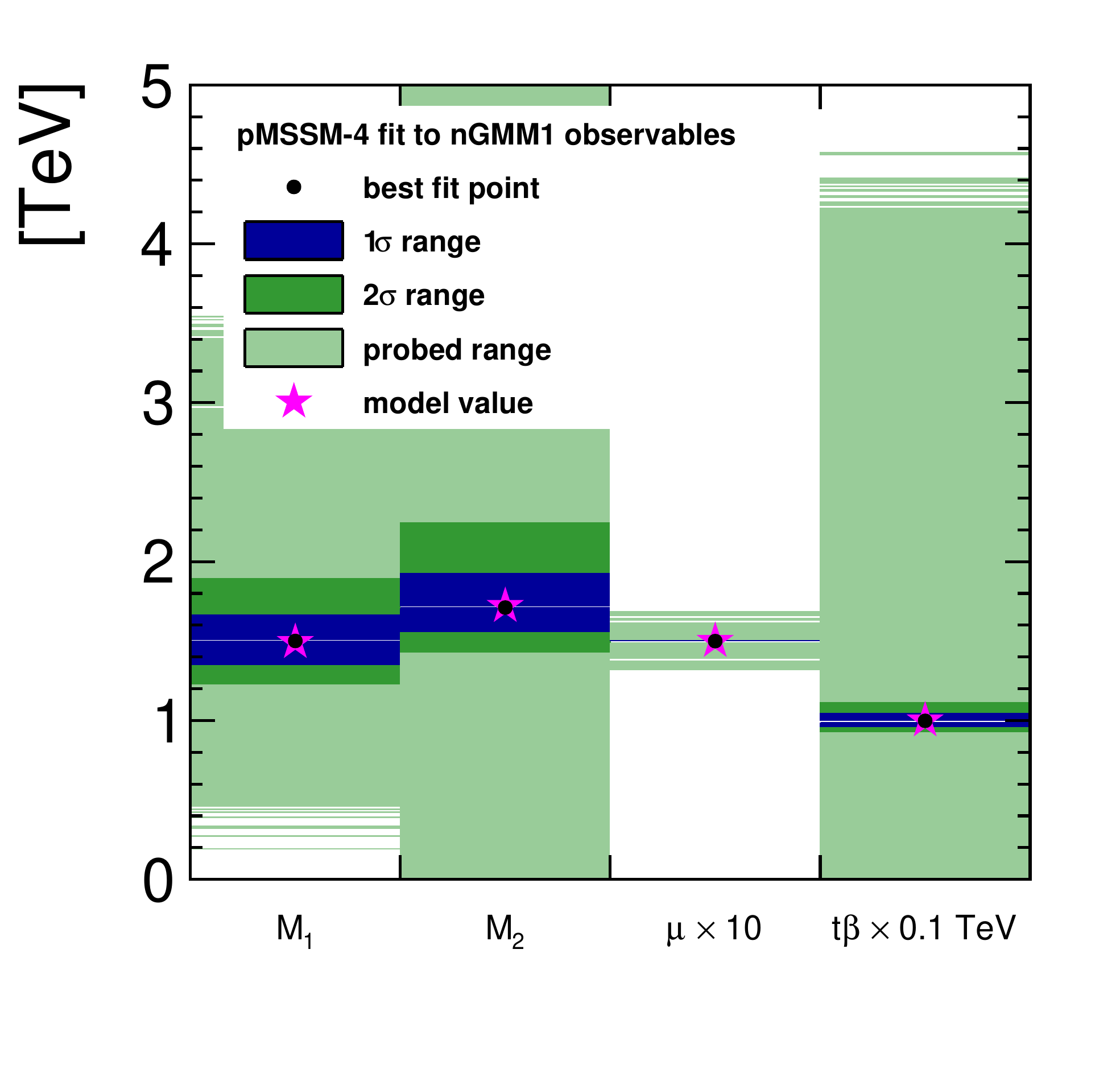}
\caption{Fitted parameters}
\label{fig:nGMM1fittedparam4p}
\end{subfigure}
\caption{Predicted mass and SUSY parameter ranges from the pMSSM-4 fit to nGMM1. The green/magenta 
star indicates the true model values, while the black dot shows the best fit point. }
\label{fig:nGMM1fitted4p}
\end{figure}

Finally, Fig.~\ref{fig:nGMM1fittedmasses4p} shows the result of the 4 parameter fit of
 $M_1$, $M_2$, $\mu$, $\tan \beta$. Again, the remaining parameters are fixed to their model values,
based on the assumption that the best fit point of the 10-parameter fit is sufficiently close to the
true point that the effect of fixing to the true values is negligible. The masses of the heavier electroweakinos
are predicted within an 1\,$\sigma$ uncertainty of about 150\,GeV. This fit has only been run with the standard input, further improvements could be expected when using the mass differences as input also in this fit.

%
%
%
%

\subsection{Dark Matter in Higgsino Fits}
\label{ssec:omegaDM}

An additional benefit from our fits to MSSM parameters is that it is possible to 
extract various WIMP dark matter related observables\cite{Baltz:2006fm}.
These include 1. the thermally-produced WIMP relic density 
$\Omega_{\tilde{\chi}}^{\rm TP}h^2$, 2. the spin-dependent (SD) and spin-independent (SI)
WIMP-nucleon scattering cross section (e.g.\ $\sigma^{\rm SI}(\chi p)$) 
which is constrained by WIMP direct detection search experiments, and 
3. the thermally-averaged WIMP-WIMP annihilation cross-section times relative velocity
(evaluated as $v\rightarrow 0$) $\langle\sigma v\rangle$ which is constrained
by indirect WIMP search results which look for cosmic WIMP-WIMP
annihilation to high energy photons and anti-matter. 
The theory predictions for these observables from \texttt{IsaReD}\cite{Baer:2002fv} and \texttt{IsaReS}\cite{Baer:2004qq} are listed in Tab.~\ref{tab:bm}. 
The higgsino-like WIMPs are thermally underproduced as dark matter and 
if their abundance is augmented via non-thermal WIMP production, 
then the higgsino-like WIMPs are excluded by direct and indirect 
WIMP search experiments~\cite{Baer:2018rhs}.
However, by requiring naturalness in the QCD sector ({\it i.e.} the axionic
solution to the strong $CP$ problem) as well as in 
the electroweak sector, then we are led to require the presence of axionic dark matter as well.
Thus, from naturalness, we expect two dark matter particles: axions as well
as higgsino-like WIMPs.
In fact, detailed calculations using eight coupled Boltzmann equations (which track
axion, WIMP, axino, saxion, gravitino and radiation abundances) suggest that
the axions usually dominate the dark matter abundance~\cite{Bae:2014rfa}. 
Then the diminished presence of higgsinos in the relic DM density leads to consistency with
WIMP search results since there are fewer higgsinos present in the relic abundance
(typically 10-20\%) than is usually assumed (100\%). 

\begin{figure}[htb]
\begin{subfigure}{0.33\linewidth}
\includegraphics[width=1.00\textwidth]{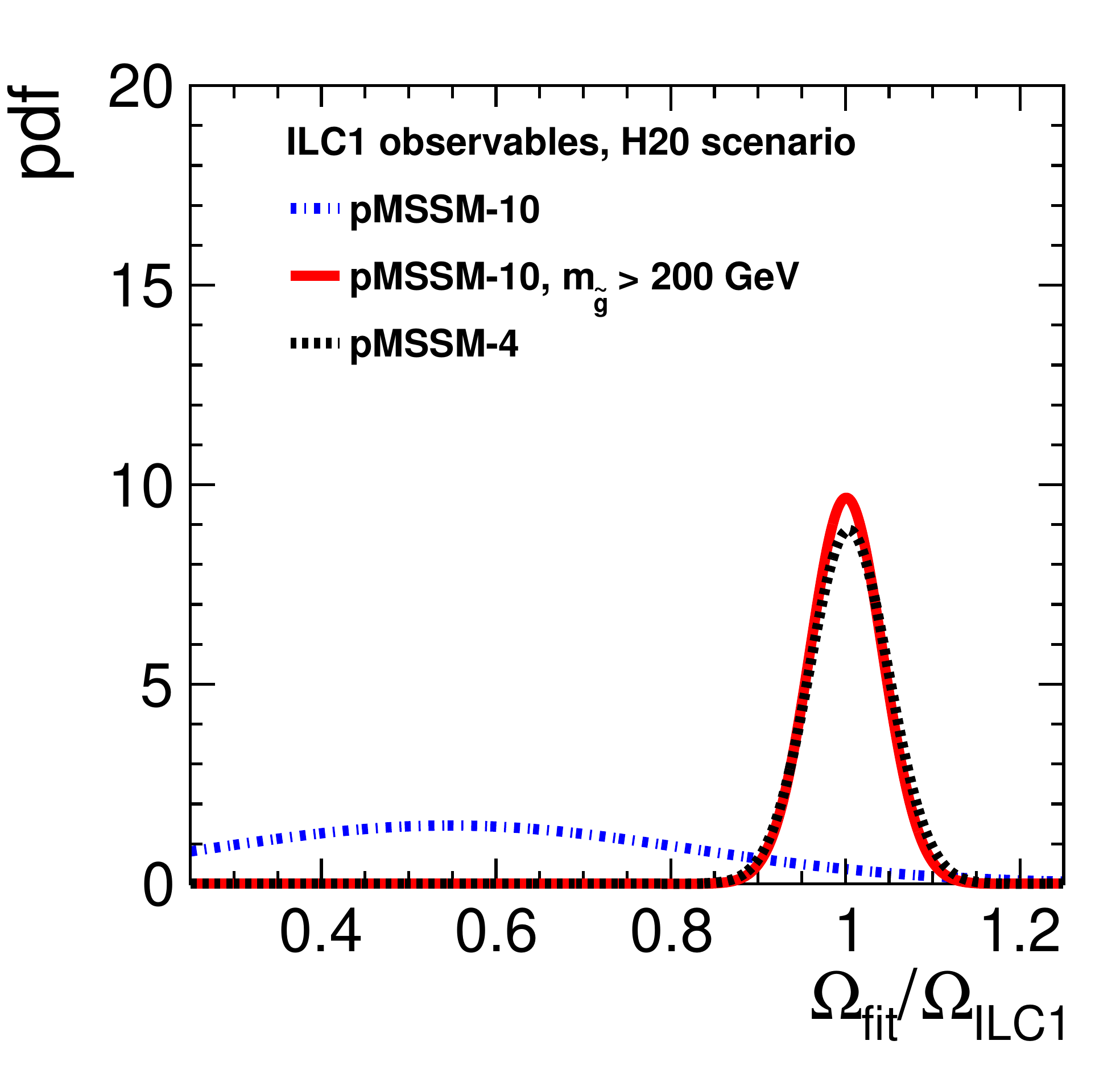}
\caption{ILC1\label{fig:Omega:ILC1}}
\end{subfigure}
\begin{subfigure}{0.33\linewidth}
\includegraphics[width=1.00\textwidth]{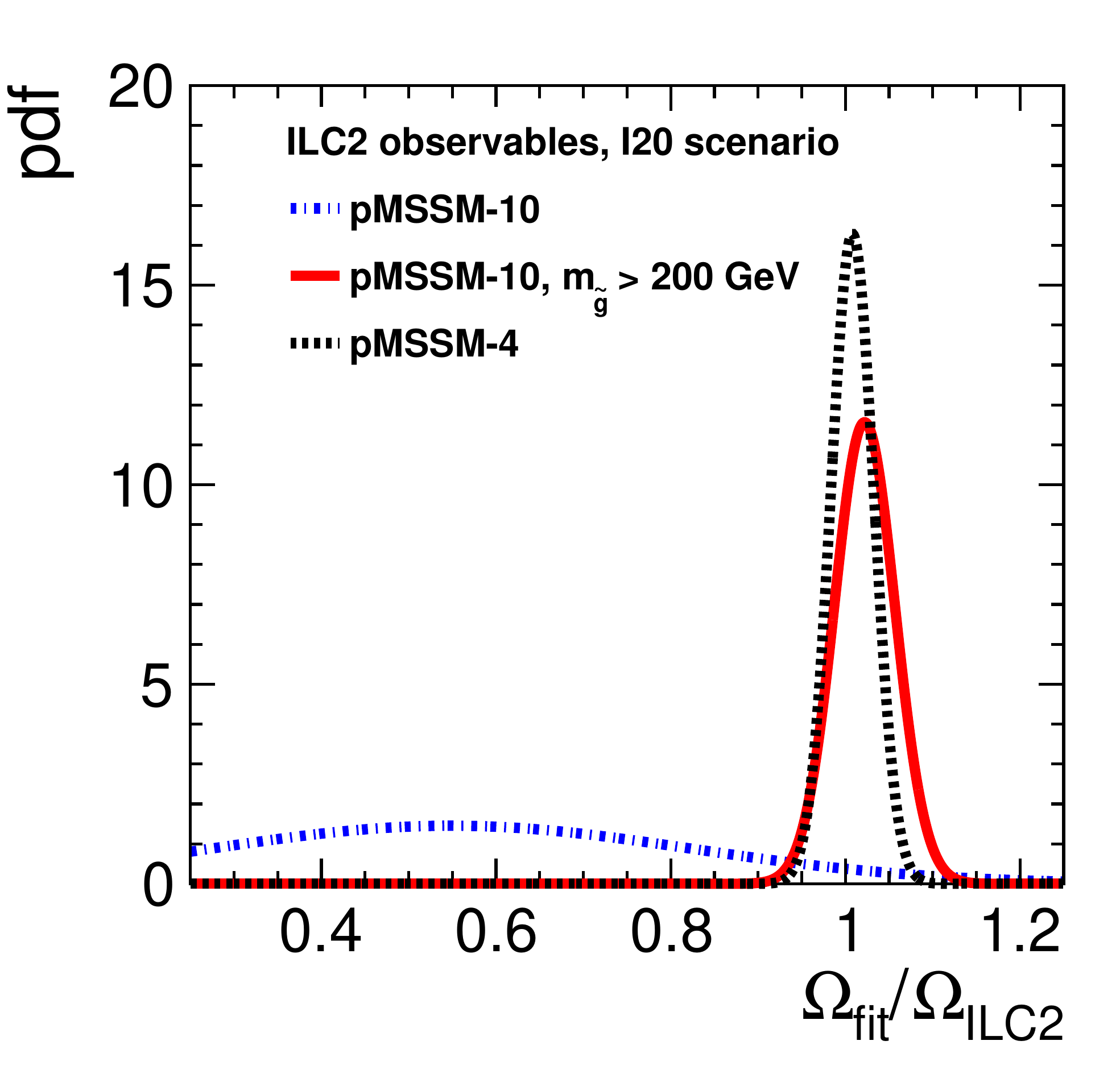}
\caption{ILC2\label{fig:Omega:ILC2}}
\end{subfigure}
\begin{subfigure}{0.33\linewidth}
\includegraphics[width=1.00\textwidth]{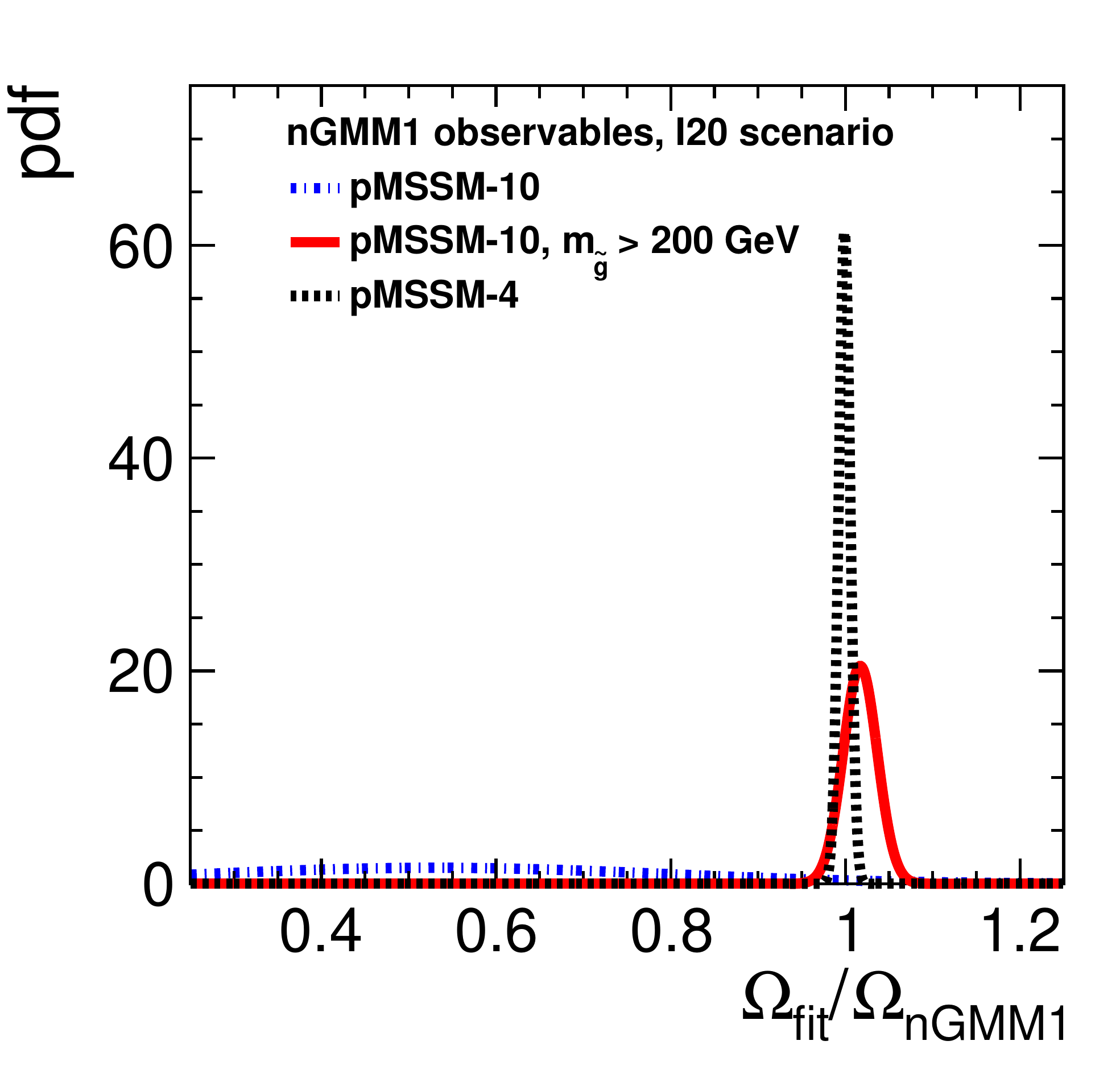}
\caption{nGMM1\label{fig:Omega:nGMM1}}
\end{subfigure}
\caption{Fitted relic densities in ILC1, ILC2 and nGMM1 fits.}
\label{fig:higgsinorelicdensities}
\end{figure}

To obtain these fitted values, we use \texttt{Fittino}~\cite{Bechtle:2004pc} together with
\texttt{MicrOmegas}~\cite{bib:micromegas} and \texttt{AstroFit}~\cite{Nguyen:2012rx}.
The fitted and scaled relic density is plotted and the 2$\sigma$ confidence interval 
has been extracted. 
The centre of the $2\sigma$ confidence level is calculated and used as the mean. 
The width of the $2\sigma$ range is divided by two to obtain the 1$\sigma$ 
width assuming the $\Delta \chi^2$ distribution is parabolic. 
The distributions are more flat than parabolic so this procedure gives a 
conservative estimate of the $1\sigma$ width. 
The relic density distribution from each fit is plotted, assuming a gaussian distribution, in 
Fig.~\ref{fig:higgsinorelicdensities}. In case of the pMSSM-10 fit without any further inputs,
the relic density is not sufficiently constrained. However this has been traced to be due to fit solutions
with extremely low gluino masses of less than 200\,GeV. Excluding these points, the blue dashed curves
are obtained, which show a very good determination of the relic density agreeing quite well with the theoretical value. The precision improves even further when the pMSSM4 fit is run after the pMSSM10 fit. Such a measurement of the relic density would clearly confirm a possible underabundance of higgsino-like WIMPs. 

We also fit the expected values of $\sigma^{\rm SI}(\chi p)$ and $\langle\sigma v\rangle$
which are listed in Tab.~\ref{tab:higgsinobenchmarks-astrofitresults} 
(these theory values are somewhat higher than those 
obtained in Tab.~\ref{tab:bm} using \texttt{IsaReS}~\cite{Baer:2004qq}
due to \texttt{Isajet}/\texttt{Spheno} spectrum differences and 
different coding algorithms for direct/indirect detection rates). 
The $\sigma^{\rm SI}$ values can be fit to an 
accuracy typically better than 1\% while the $\langle\sigma v\rangle$ values are
typically fit to $\sim 10\%$ or worse. 
By comparing the direct detection rates from WIMP detection experiments to
the ILC fitted values for a measured higgsino mass $m_{\chi}$,
a direct measurement of WIMP relic density can be made since the WIMP
direct detection rates are actually sensitive to $\xi\sigma^{\rm SI}$ where
$\xi$ is the ratio of actual WIMP abundance divided by the total measured
abundance $\Omega_{a+ \chi}h^2$. Such interplay between ILC results
and direct detection results offer direct confirmation that WIMPs
would comprise only a portion of dark matter. In addition, indirect WIMP
detection rates are proportional to $\xi^2$ since they search for WIMP-WIMP
annihilation. The interplay of ILC results with indirect WIMP detection
rates could offer further confirmation for multicomponent dark matter.


\begin{table}[htbp]{}
\centering
$
\renewcommand{\arraystretch}{1.4}
\begin{array}{  c c c c }
  \hline \hline
    \text{observable}  & \text{ILC1}
& \text{ILC2} & \text{nGMM1} \\ \hline
\sigma^{\rm SI} \text{ model } [ 10^{-9}  \text{ pb}] & 259.3  & 316.9 & 328.5 \\
\sigma^{\rm SI} \text{ best fit }[ 10^{-9}  \text{ pb}]   & 260.7^{+4.1}_{-6.9}  & 317.0^{+2.1}_{-2.1} &  328.5^{+1.5}_{-0.9} \\
  \langle \sigma v \rangle \text{ model } [10^{-27} \text{ cm}^3\text{s}^{-1}]  & 15.36 & 3.439 & 0.597  \\
   \langle \sigma v \rangle \text{ best fit } [10^{-27} \text{ cm}^3\text{s}^{-1}] & 15.01^{+1.52}_{-0.88} & 3.501^{+5.741}_{0.523}  & 0.621^{+0.994}_{-0.165} \\
  \hline \hline
  \end{array}
$
\caption{Relic density from \texttt{MicrOMEGAs} and \texttt{Astrofit}, and direct and indirect detection cross sections from \texttt{Astrofit} in the pMSSM-10 fits to ILC1, ILC2 and nGMM1 observables (without the gluino mass measurement). $\Omega_{\text{Planck}}$ is taken to be 0.1199 \cite{Ade:2015xua}.}
\label{tab:higgsinobenchmarks-astrofitresults}
\end{table}

\section{Testing gaugino mass unification}
\label{sec:test}

The pMSSM parameters which were extracted in Sec. \ref{ssec:weakscale_fits} 
were fitted at the energy scale $Q=1$\,TeV.
The scale dependence of the parameters is governed by their 
renormalization group equations or RGEs. 
Using the MSSM RGEs, the fitted parameters can be
evolved to higher energy scales in order to check hypotheses 
regarding unification. Specifically, we will test unification of the various
gaugino masses which are assumed to unify at 
$Q=m_{\rm GUT}\simeq 2\times 10^{16}$ GeV (the scale at which gauge couplings unify) 
in models like NUHM2 and NUHM3 but which would unify at
a lower scale in models such as nGMM1. Since in this work we do not subscribe to any particular GUT or string theory, GUT scale threshold corrections to gauge and Yukawa couplings and soft SUSY breaking terms are not imposed.

This section continues the program initiated by Blair {\it et al.} of
extracting tests of high scale unification from weak scale measurements of
SUSY particle properties at ILC\cite{Blair:2000gy,Blair:2002pg}. Since the estimates of the achievable precision for the experimental observables used in Sec.~\ref{sec:fit} are somewhat more pessimistic than the results obtained in Sec.~\ref{sec:event_selection}, we also discuss the expected impact of the experimental improvements taking the nGMM1 benchmark as an example.

\subsection{Method}

The running pMSSM-10 weak scale parameters and error bars are extracted using 
\texttt{Fittino} and \texttt{SPheno3.3.9beta} at $Q=1$\,TeV.
Then a random scan of $10^4$ samples of the 10 parameters is performed,
approximating the parameter PDFs as either Gaussian or flat within $\pm 1\sigma$, depending on
the shape of their $\chi^2$ distribution, see e.g.\ Fig.~\ref{fig:nGMM1I20parabolae}.
For each of the sampled points, \texttt{SPheno} was used to calculate the running parameters at each of 21 energy scales between $91$ and $10^{19}$\,GeV. The mean and standard distribution of these parameters' distributions at each energy scale were used to define confidence bands, as shown in e.g.\ Fig.~\ref{fig:nGMM1fittedparam4p}.
The unification scale $Q_{\mathrm{unif}}$ is determined by fitting linear functions 
the running parameters in a range close to the visible intersection and extracting the intersection point. With each value for $Q_{\mathrm{unif}}$, a corresponding estimate of $M_{1/2}$ is determined.
Gaussian functions can be fitted to the
distributions of the resulting values for $Q_{\mathrm{unif}}$ and $M_{1/2}$ in order to obtain central values and uncertainties.

For the gluino mass, several scenarios are considered: the determination from loop contributions to the higgsino observables only, a direct observation at the LHC resulting in a precision of $10\%$ on the physical gluino mass, or simply by assuming gaugino mass unification.
In the latter case, the extracted mean $M_{1/2}$ and $Q_{\mathrm{unif}}$ values 
can be used to determine the 
value of $M_3(Q=1\ {\rm TeV})$ and consequently the physical gluino mass. 
In this case, predictions for the expected value of $m_{\tg}$ may be made
which can serve as a target for future hadron collider searches or compared to the mass of an already-discovered gluino.

\subsection{Running gaugino masses for ILC1}
\begin{figure}[htbp]
 \begin{subfigure}{0.48\linewidth}
\includegraphics[width=\textwidth]{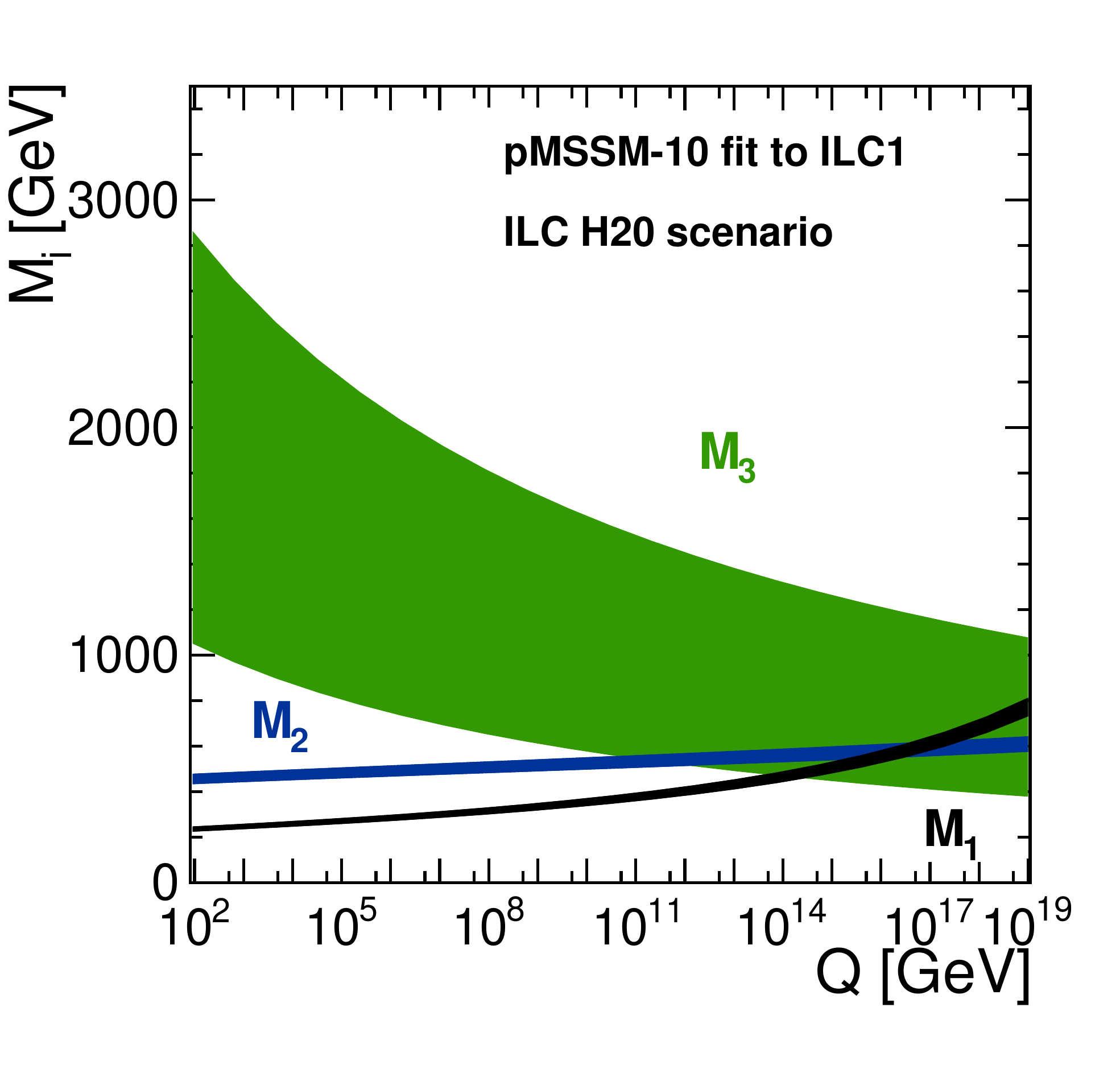}
\caption{}
\label{fig:ILC1-H20-running:M1M2M3}
\end{subfigure}
\begin{subfigure}{0.48\linewidth}
\includegraphics[width=\textwidth]{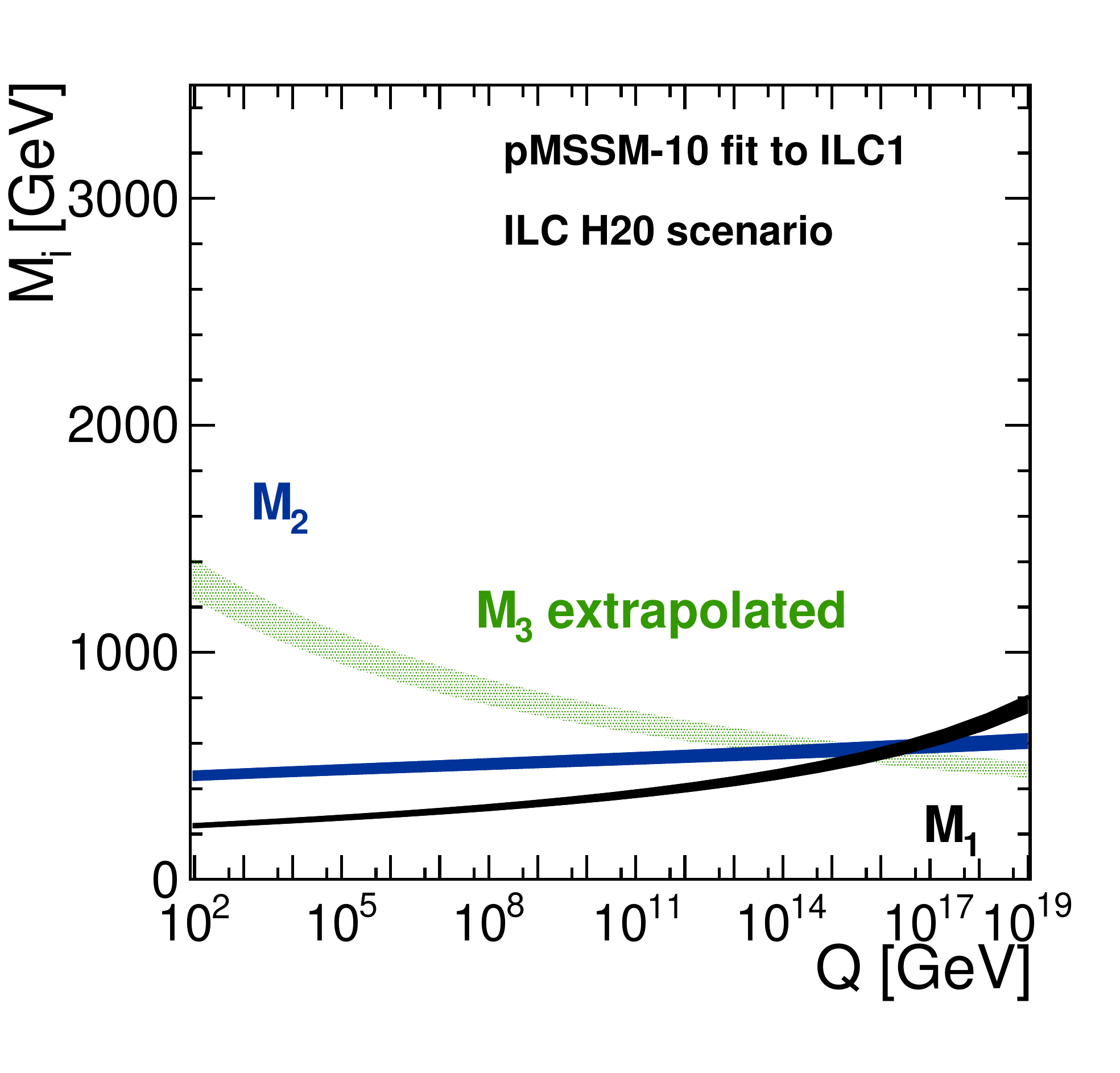}
\caption{}
\label{fig:ILC1-H20-running:M3extrapolated}
\end{subfigure}
\caption{The running gaugino masses $M_i$ based on the pMSSM-10 fit to ILC1 observables. The bands correspond to one standard deviation. (a) Using $M_3$ at the weak scale as constrained from ILC measurements (b) $M_3$ is assumed to unify with $M_1$ and $M_2$ and then run to the weak scale to gain a prediction for $M_3$($Q = 1$\,TeV).}
\label{fig:ILC1-H20-running}
\end{figure}

\begin{figure}[htbp]
 \begin{subfigure}{0.48\linewidth}
\includegraphics[width=\textwidth]{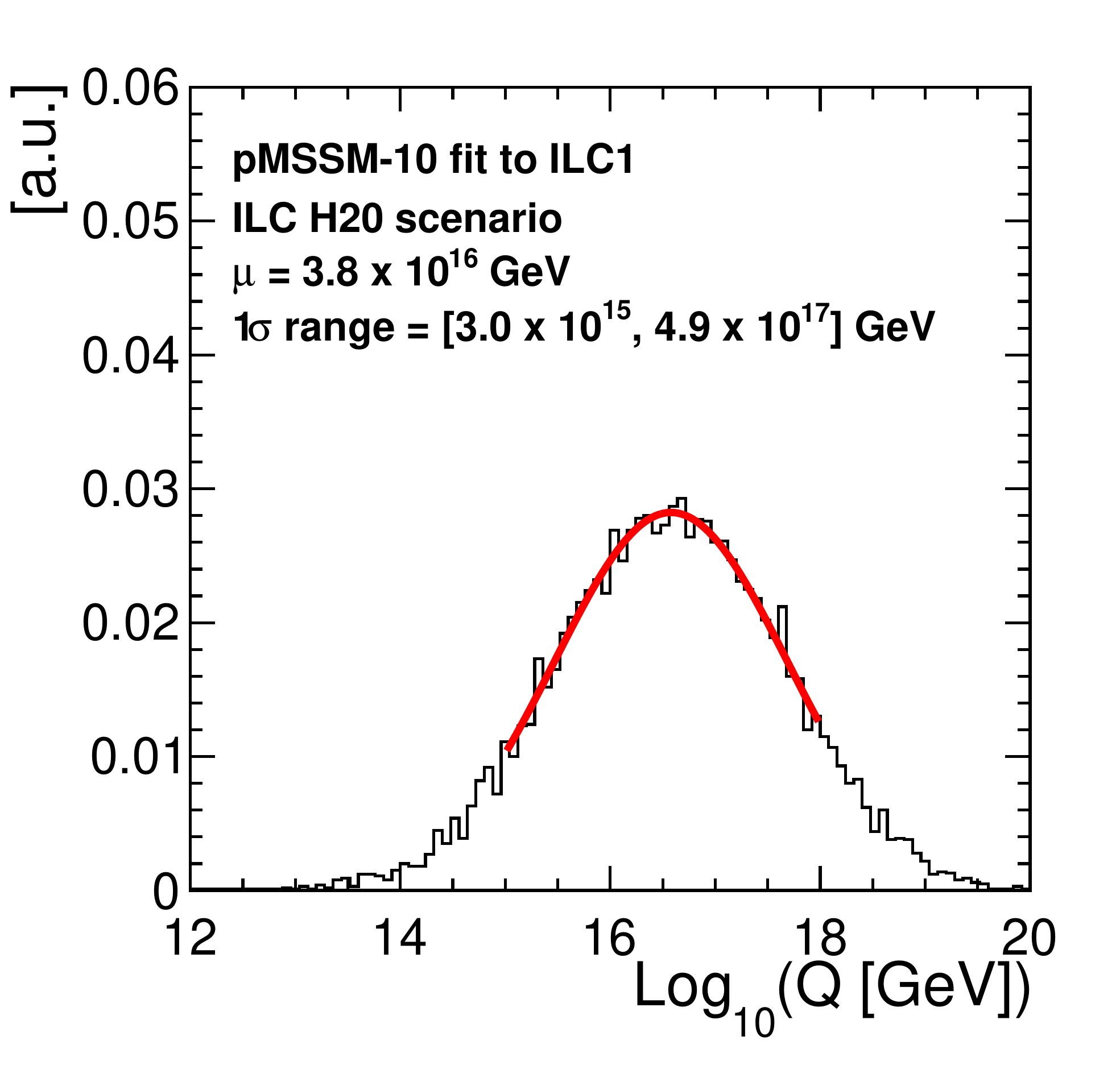}
\caption{Unification scale $Q_{\mathrm{unif}}$}
\label{fig:ILC1-H20-running:qscale}
\end{subfigure}
\begin{subfigure}{0.48\linewidth}
\includegraphics[width=\textwidth]{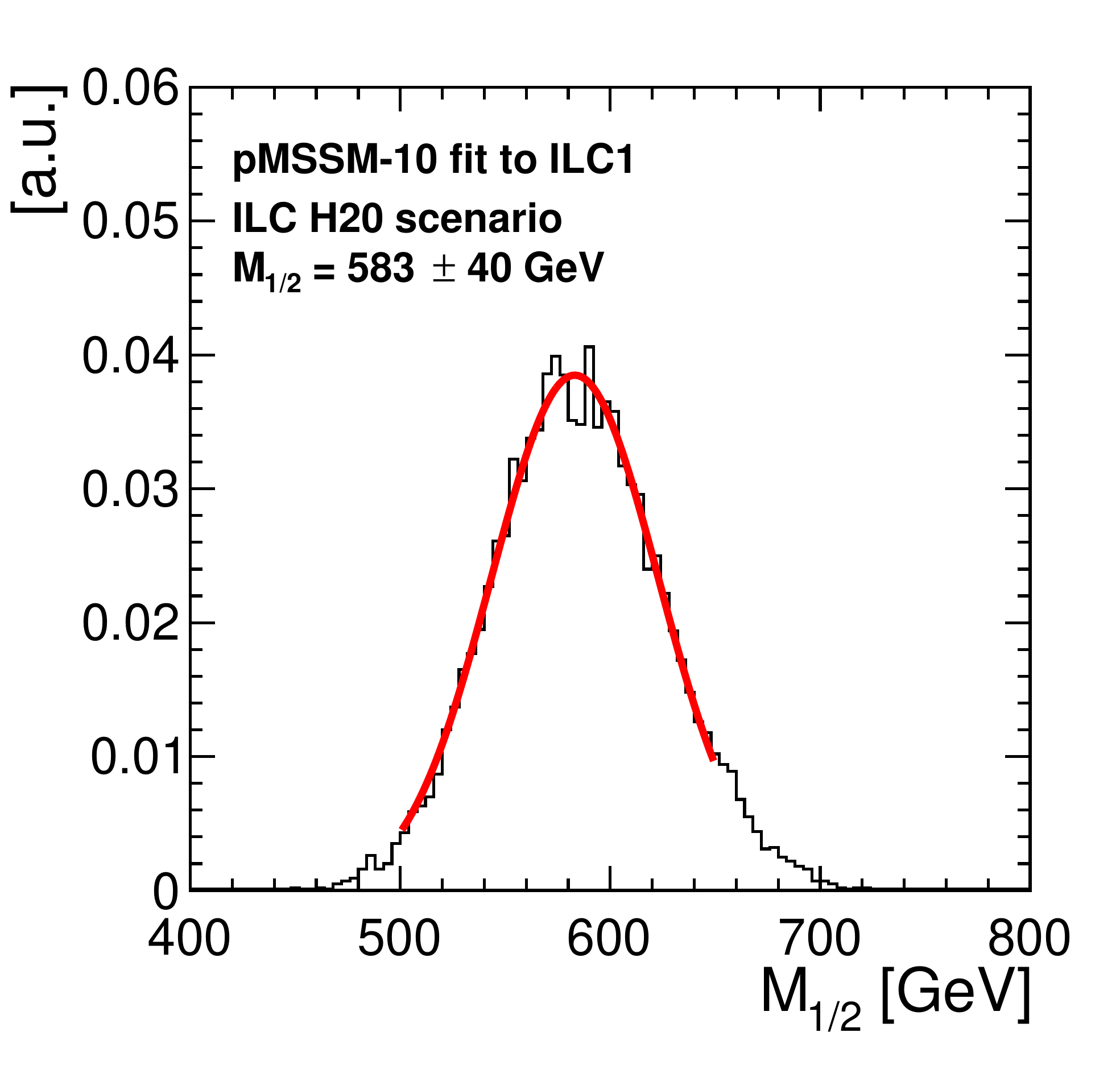}
\caption{unified gaugino mass 
$M_{1/2}$}
\label{fig:ILC1-H20-running:M12}
\end{subfigure}
\caption{
Distribution of the unification scale $Q_{\mathrm{unif}}$ and unified gaugino mass 
$M_{1/2}$ obtained from the running parameters $M_1$ and $M_2$ and their uncertainties from 
the pMSSM-10 fit to ILC1 observables.
}
\label{fig:fittedGutscalesILC1}
\end{figure}

The weak scale ILC1 parameters are sampled according to Gaussian distributions for $M_1$ and $M_2$ and uniformly within the $1\sigma$ range for $M_3$, motivated by the shape of the $\chi^2$ distributions obtained in the pMSSM-10 fits discussed in Sec.~\ref{ssec:weakscale_fits:ILC1}.
The resulting running of the gaugino mass parameters in the ILC1 pMSSM-10 fit is plotted in Fig.~\ref{fig:ILC1-H20-running:M1M2M3}.
From the plot, it can be seen that $M_1$ and $M_2$ cross near $10^{16}$ GeV
which would verify the prediction of a SUSY GUT model. 
The uncertainty band for $M_3$ is quite wide but is
consistent with the hypothesis of unification of all three gaugino masses 
at the same energy scale.
The extracted unification scale $Q_{\mathrm{unif}}$ for $M_1$ and $M_2$ is 
plotted in Fig.~\ref{fig:ILC1-H20-running:qscale} from which it can
be seen that the distribution follows a Gaussian. 
The gaugino mass unification scale is
found to be $Q_{\mathrm{unif}}=3.8\times 10^{16}$\,GeV with a 68\% confidence range 
of $[3.0\times 10^{15} , 4.9\times 10^{17}]$\,GeV.
From Fig. \ref{fig:ILC1-H20-running:M12}, the unified gaugino 
mass parameter is found to be $M_{1/2}=583\pm 40$ GeV in agreement with
the GUT scale model model fit. 

If it is then assumed that the unification is due to an NUHM2 model, 
and true model parameter values are assumed for parameters other than 
$M_{1/2}$, then instead $M_3$ can be extracted by running down in energy 
to find the running value of $M_3(Q=1\ {\rm TeV})$.
From Fig. \ref{fig:ILC1-H20-running:M3extrapolated} we obtain 
$M_3(Q = 1\ {\rm TeV}) = 1216\pm 76$\,GeV (which agrees
with the the weak scale fitted value). 
Consequently, a {\em prediction} for the physical gluino mass can be obtained: 
$m_{\tg}= 1467\pm 80$\,GeV which could then be checked against 
results from hadron collider searches.

\subsection{Running gaugino masses for ILC2}

\begin{figure}[htbp]
\begin{subfigure}{0.48\linewidth}
  \includegraphics[width=\textwidth]{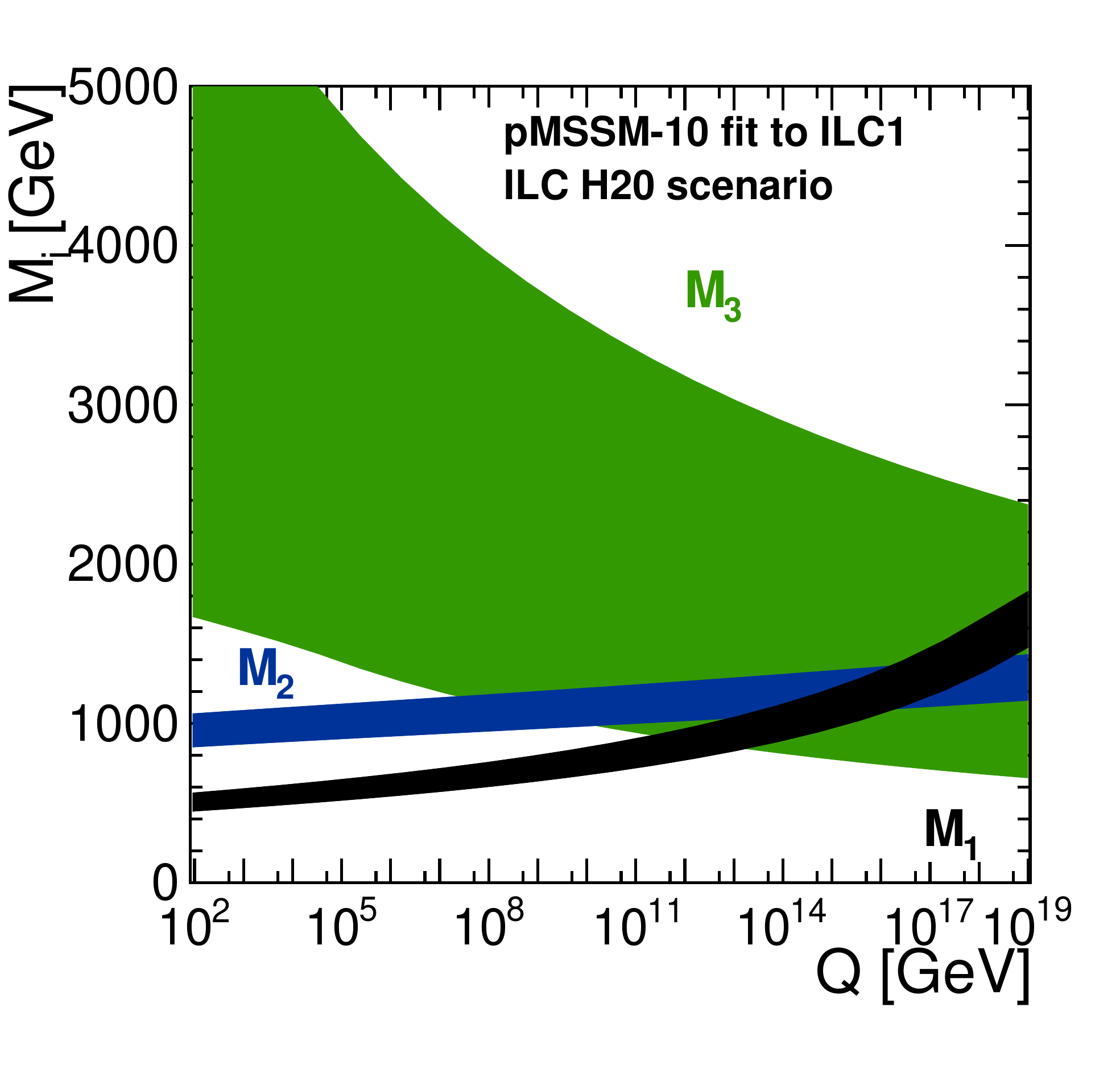}
  \caption{}
  \label{fig:ILC2-running:M1M2M3}
\end{subfigure}
 \begin{subfigure}{0.48\linewidth}
  \includegraphics[width=\textwidth]{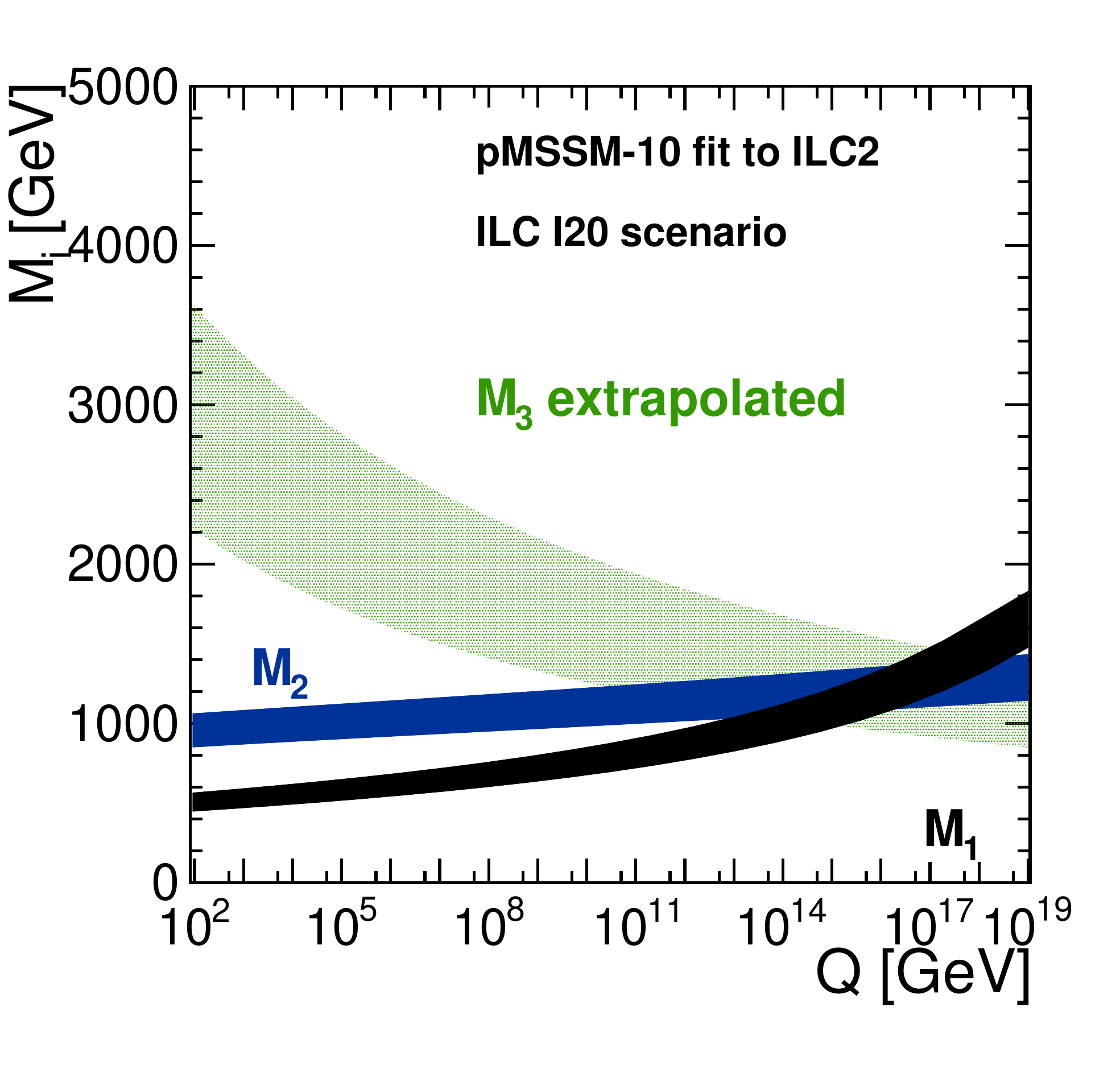}
  \caption{}
  \label{fig:ILC2-running:M3extrapolated}
\end{subfigure}
\label{fig:ILC2-running}
\caption{The running gaugino masses $M_i$ based on the pMSSM-10 fit to ILC2 observables. 
         The bands correspond to one standard deviation. 
         (a) Using $M_3$ at the weak scale as constrained from ILC measurements 
         (b) $M_3$ is assumed to unify with $M_1$ and $M_2$ and then run to the weak scale 
         to gain a prediction for $M_3$($Q = 1$\,TeV).}

\end{figure}

The uncertainties of the weak scale gaugino mass fit parameters are 
larger in the case of ILC2 as compared to ILC1. Still, the weak scale ILC2 parameters are sampled according to Gaussian distributions for $M_1$ and $M_2$ and uniformly within the $1\,\sigma$ range for $M_3$, motivated by the shape of the $\chi^2$ distributions obtained in the pMSSM-10 fits discussed in Sec.~\ref{ssec:weakscale_fits:ILC2}. The larger uncertainties are
 reflected in the running gaugino mass plots in 
Fig. \ref{fig:ILC2-running:M1M2M3} and \ref{fig:ILC2-running:M3extrapolated}. 
Nevertheless, it is still possible to verify that $M_1$ and $M_2$ unify 
near the GUT scale. 
For ILC2, the fitted weak scale error band for $M_3$ is so wide that 
it is consistent  with unification with $M_1$ or $M_2$ at almost any scale.

\begin{figure}[htbp]
 \begin{subfigure}{0.48\linewidth}
\includegraphics[width=\textwidth]{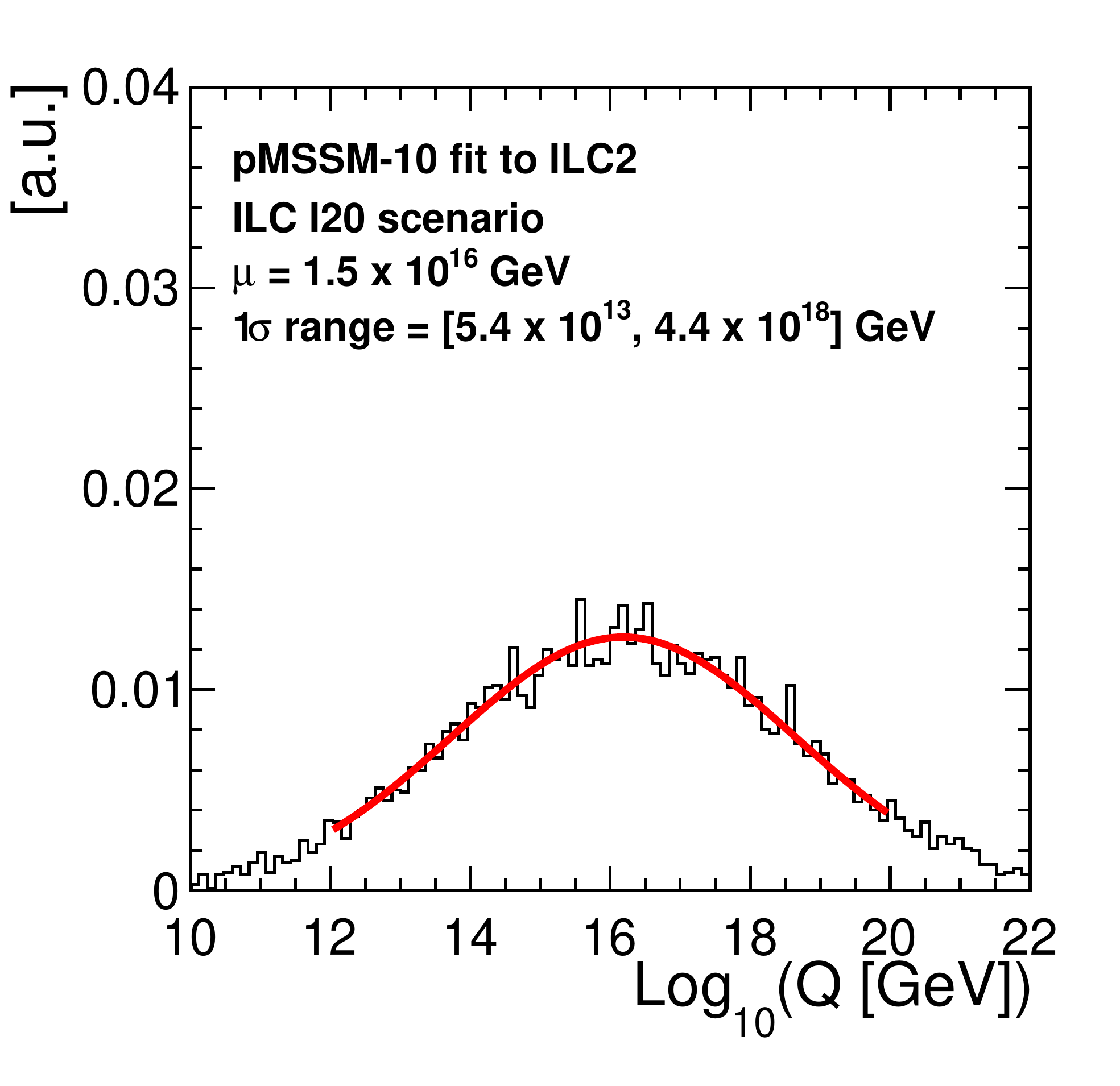}
\caption{Unification scale $Q_{\mathrm{unif}}$}
\label{fig:ILC2-I20-running:qscale}
\end{subfigure}
\begin{subfigure}{0.48\linewidth}
\includegraphics[width=\textwidth]{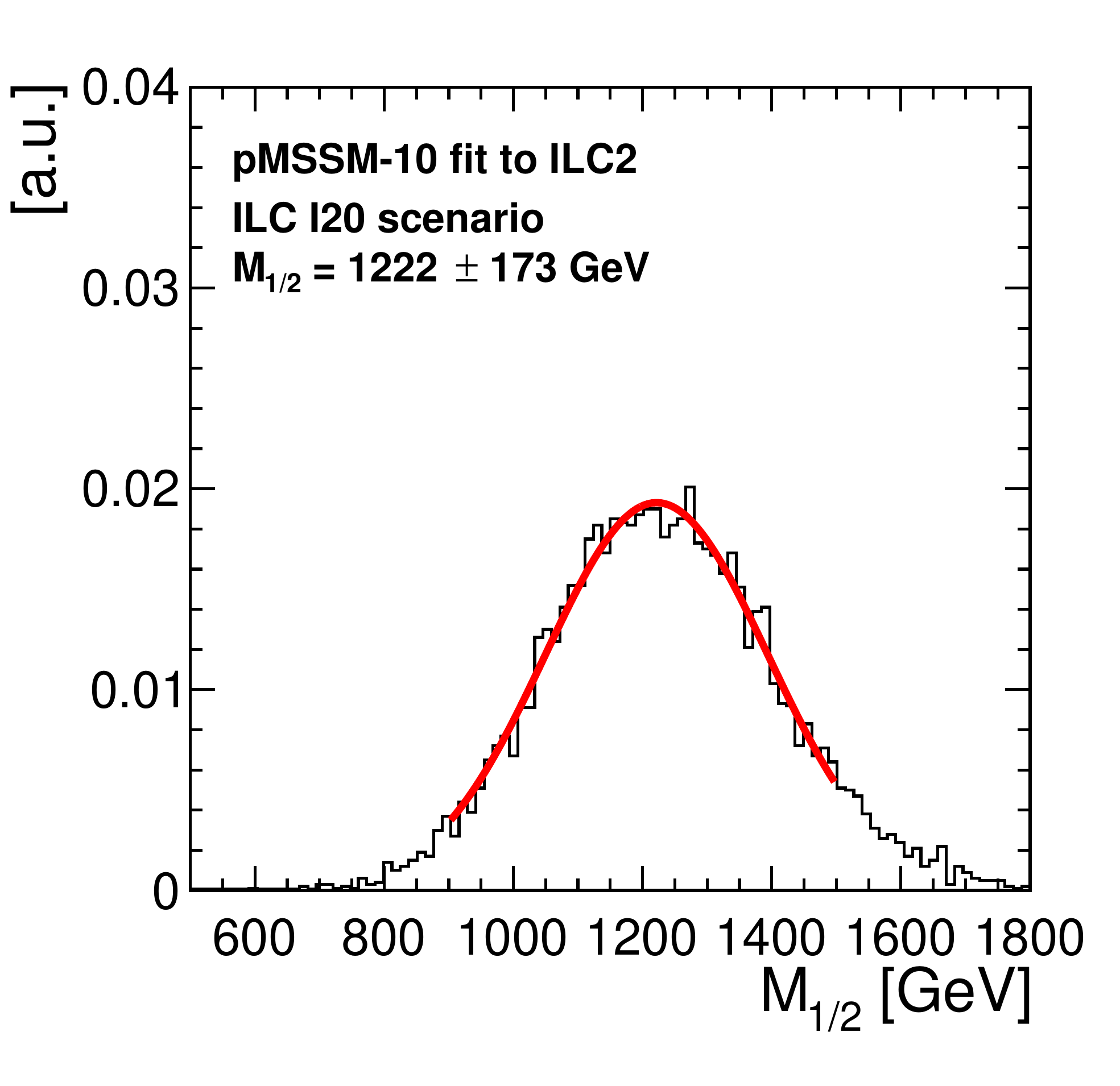}
\caption{unified gaugino mass 
$M_{1/2}$}
\label{fig:ILC2-I20-running:M12}
\end{subfigure}
\caption{
Distribution of the unification scale $Q_{\mathrm{unif}}$ and unified gaugino mass 
$M_{1/2}$ obtained from the running parameters $M_1$ and $M_2$ and their uncertainties from 
the pMSSM-10 fit to ILC2 observables.
}
\label{fig:fittedGutscalesILC2}
\end{figure}

Using the same methodology as for ILC1, the unification scale for ILC2 
where $M_1 = M_2$ is found to be Gaussian with a mean of 
$Q_{\mathrm{unif}}=1.5\times 10^{16}$\,GeV with a 68\% confidence interval 
of $[5.4\times 10^{13}, 4.4\times 10^{18}]$\,GeV, as shown in
Fig. \ref{fig:ILC2-I20-running:qscale}. 
The unified value of $M_{1/2}$ is found in 
Fig. \ref{fig:ILC2-I20-running:M12} to be Gaussian with 
$M_{1/2} = 1220\pm 170$\,GeV which corresponds to the GUT scale fit model value. 
If $M_3$ is instead {\em assumed} to unify with $M_1$ and $M_2$ 
at $Q_{\mathrm{unif}}$ and the  NUHM2 model is adopted, then
the extrapolated value of $M_3$ at $1$\,TeV is found to be 
$M_3 (Q = 1\ {\rm TeV}) = 2616\pm 582$\,GeV while the physical 
gluino mass is found to be $m_{\tg}=2872\pm 605$\,GeV. 
Such a large value may serve as a target for gluino pair searches at 
upgraded hadron colliders.

\subsection{Running gaugino masses for nGMM1}
\label{ssec:test:nGMM1}

The running of the gaugino mass parameters in the nGMM1 benchmark model 
differs from the running in the ILC1 and ILC2 models.
There are two reasons: 1.\ the underlying model is now a mirage unification 
model where the gaugino mass parameters unify at an intermediate energy scale
and 2.\ the determination of $M_1$ and $M_2$ from the weak scale fits 
is much less accurate in nGMM1 as compared to the ILC1 and ILC2 benchmark
models.

\begin{figure}[htbp]
 \begin{subfigure}{0.48\linewidth}
\includegraphics[width=\textwidth]{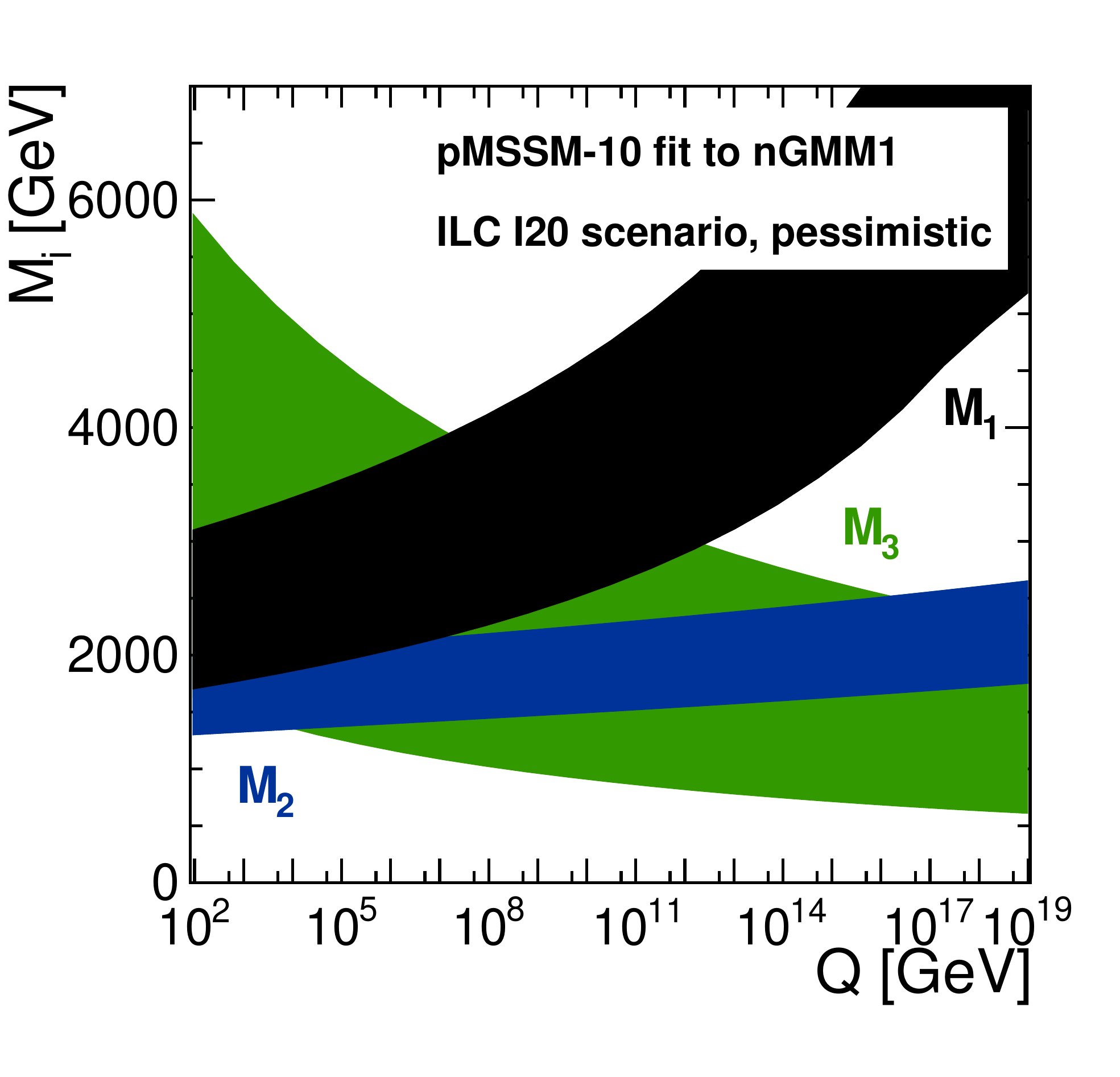}
\caption{}
\label{fig:nGMM1-running:M1M2M3}
\end{subfigure}
\begin{subfigure}{0.48\linewidth}
\includegraphics[width=\textwidth]{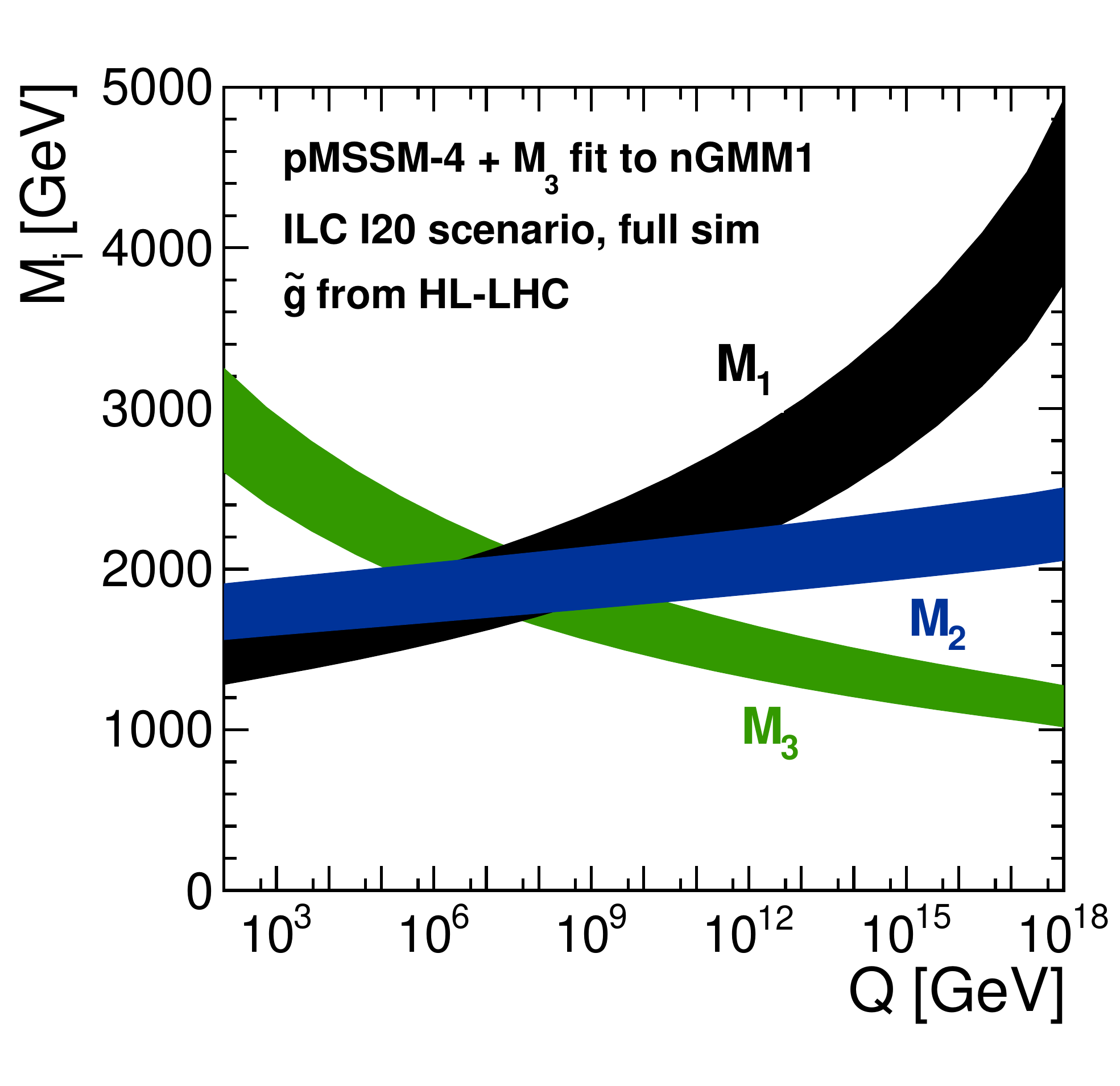}
\caption{}
\label{fig:nGMM1-running:improved}
\end{subfigure}
\caption{The running of the gaugino masses after extracting their weak scale 
values from a fit to nGMM1 observables. 
The bands correspond to one standard deviation. 
(a) pMSSM-10 fit result with absolute masses as input from Sec.~\ref{ssec:weakscale_fits:nGMM1} 
(b) estimated effect of improvement from using the full simulation results 
from Sec.~\ref{ssec:fitinputs}, and from including a 10\% measurement of the 
gluino mass from HL-LHC (or other future hadron collider). 
In addition a fit of the pMSSM-4 parameters, and $M_3$ as a fit parameter, is run, 
as discussed at the beginning of Sec.~\ref{ssec:weakscale_fits}. The Markov chain had a length of $10^5$ points.}
\label{fig:nGMM1-running}
\end{figure}

Figure~\ref{fig:nGMM1-running:M1M2M3} shows the running gaugino masses resultant from the pMSSM-10 fit with absolute masses as input as described in Sec.~\ref{ssec:weakscale_fits:nGMM1}. Even in this most conservative case, the plot is certainly inconsistent with any sort of GUT scale unification of gaugino masses. From a closer look we notice that the hierarchy between $M_1$ and $M_2$ at $Q = 1$\,TeV is not well defined, and that actually the $M_1$ band seems to start above the lower rim of the $M_2$ band. This effect occurs since, motivated by the shape of the $\chi^2$ landscape of the pMSSM-10 fit (c.f.\ Fig.~\ref{fig:nGMM1I20parabolae}), $M_1$ and $M_3$ are sampled from a uniform distribution and only $M_2$ is treated with a Gaussian. In addition the $1\,\sigma$ interval for $M_1$ is very asymmetric around the best fit point (c.f.\,Tab.~\ref{tab:nGMM1fittedparam10p}). In combination with the flat sampling, the $1\,\sigma$ band for $M_1$ seems to start much higher than the best fit value for $M_1$ would indicate.

A substantial improvement of the precision can be seen in 
Fig.~\ref{fig:nGMM1-running:improved}, which shows the analogous 
result obtained when using the improved experimental precisions presented 
in Sec.~\ref{ssec:fitinputs} plus a 10\% measurement of the gluino mass 
from the HL-LHC (or other future hadron collider). In addition to the improved inputs,
the parameter extraction has also been refined: the estimates of $M_1$, $M_2$ and $M_3$ 
at the weak scale are obtained from a fit of only the pMSSM-4 parameters and $M_3$, which could be run
subsequently to an initial pMSSM-10 fit as outlined at the beginning of Sec.~\ref{ssec:weakscale_fits}. 
In this case, all parameters can be sampled from Gaussian distributions, 
as can be seen from Fig.~\ref{fig:nGMM1I20parabolae}. 
The weak scale hierarchy between $M_1$ and $M_2$ is now well determined, 
and a clear crossing of all three bands is found at a scale much lower 
than the GUT scale: around $10^7- 10^8$\,GeV, consistent with the theory 
mass unification scale for the model point which occurs at $10^7$\,GeV.

This is not even the most optimistic case, since further improvements can be expected from using the
higgsino mass differences instead as input 
(c.f.\ Fig.~\ref{fig:nGMM1fittedparam10pmdif}) 
and from more precise \charginoone\ and \neutralinotwo\ masses extracted 
from scanning the thresholds of \charginoonep \charginoonem\ 
and \neutralinotwo \neutralino\ production, respectively. 
In addition, the consideration of further constraints from improved EWPOs, 
flavour physics, direct search limits etc.\ is expected to further 
improve the weak scale parameter determination.

\begin{figure}[htbp]
 \begin{subfigure}{0.48\linewidth}
\includegraphics[width=\textwidth]{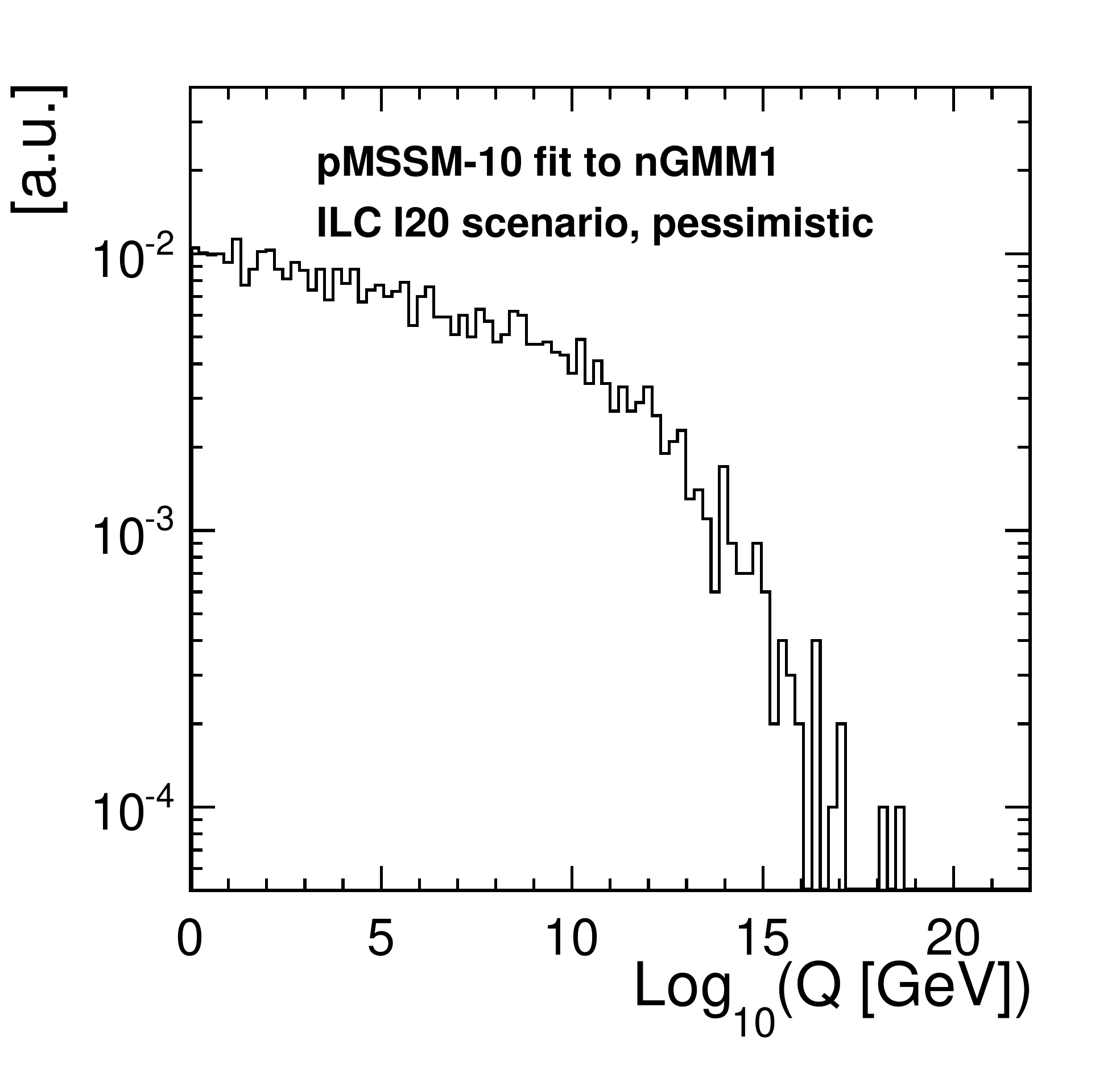}
\caption{Unification scale $Q_{\mathrm{unif}}$}
\label{fig:nGMM1-I20-running:qscale}
\end{subfigure}
\begin{subfigure}{0.48\linewidth}
\includegraphics[width=\textwidth]{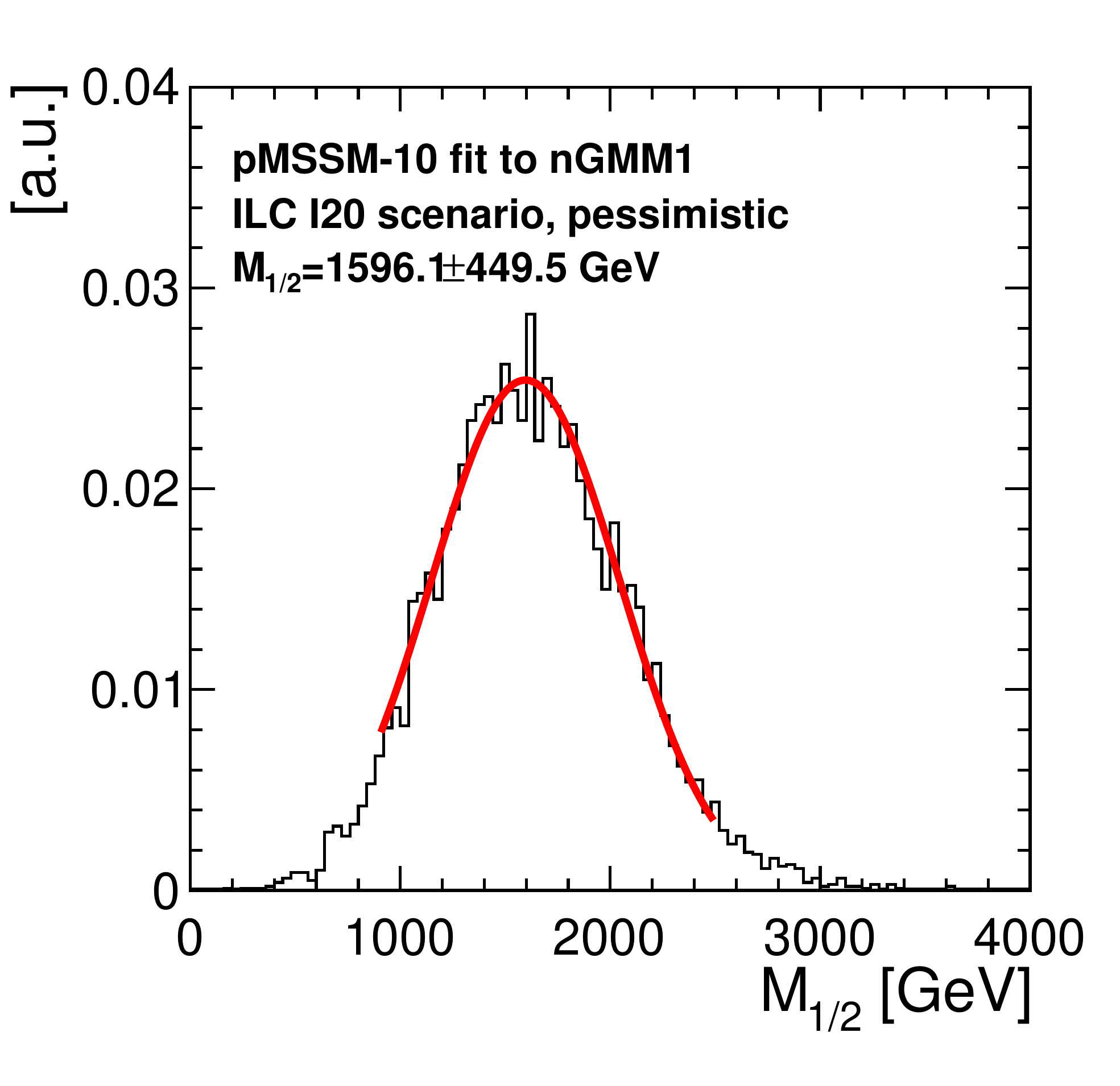}
\caption{unified gaugino mass 
$M_{1/2}$}
\label{fig:nGMM1-I20-running:M12}
\end{subfigure}
\caption{
Distribution of the unification scale $Q_{\mathrm{unif}}$ and unified gaugino mass 
$M_{1/2}$ obtained from the running parameters $M_1$ and $M_2$ and their uncertainties from 
the pMSSM-10 fit to nGMM1 observables.
}
\label{fig:fittedGutscalesnGMM1}
\end{figure}

For the conservative version of the running masses in Fig.~\ref{fig:nGMM1-running:M1M2M3}, we quantify the constraints on the unification
scale in Fig.~\ref{fig:fittedGutscalesnGMM1}a. While the distribution of obtained $Q_{\mathrm{unif}}$ values has no clear peak, it increases towards
lower unification scales, away from the GUT scale. A unification at $\sim 10^{16}$ GeV is
excluded with 99.9\% probability. 
The most probable unified value of $M_{1/2}$ was found to be 
$1600\pm 450$ GeV in Fig.~\ref{fig:fittedGutscalesnGMM1}b. 
The drastic improvement in Fig.~\ref{fig:nGMM1-running:improved} compared to Fig.~\ref{fig:nGMM1-running}, illustrates the substantial impact which
can be expected from further refinements of the underlying analysis.

Due to the extracted gaugino mass unification scale not matching with 
the GUT scale, there would be important implications for SUSY model building. 
It is noteworthy that the pMSSM fit and the fit parameter evolution indicate
that the underlying model does not have gaugino mass unification, 
even though the fit of NUHM2 parameters to the nGMM1 observables presented in Sec.~\ref{ssec:gutscale_fits}
does not entirely rule NUHM2 out as a possible model.
\section{Summary and conclusions}
\label{sec:conclude}

Supersymmetry with radiatively-driven naturalness is especially compelling in that
it reconciles electroweak naturalness with (multi-TeV) LHC 
sparticle mass limits and Higgs boson mass measurements.
The most fundamental consequence of radiatively-driven natural SUSY is
the prediction of four light higgsinos $\chi_1^\pm$, $\chi_{1,2}^0$ 
with mass $\sim 100-300$ GeV (the lower the better).
Such light higgsinos are difficult (but perhaps not impossible) to see at LHC,
but would be easily visible at ILC operating with $\sqrt{s}>2m(higgsino)$.
In this case, the ILC, initially constructed as a Higgs factory, would turn out
to be a higgsino factory! Thus, for this highly motivated scenario, ILC could serve
as both a SUSY discovery (or confirmation) machine, and a precision microscope.

In this paper we have examined the capability of experiments at the
ILC to both discover (or confirm) 
supersymmetry and to make precision measurements of superparticle 
properties that would probe the superpotential higgsino mass
parameter $\mu$ via direct sparticle mass measurements and in addition provide a
measurement of SUSY-breaking gaugino mass parameters via the higgsino
mass splittings. 

When these measurements are combined with precision Higgs
boson measurements, precision fits to both weak scale SUSY and high scale
SUSY model parameters can be made. 
We have investigated the capability of ILC to discover light higgsinos in three
natural SUSY benchmark models: two with unified gaugino masses and one 
with mirage unification of gaugino masses at an intermediate mass scale between
$m_{\rm GUT}$ and $m_{weak}$. Our calculations implement  a detailed ILD detector simulation along
with event generation from Whizard. 

By measuring 
$e^+e^-\rightarrow \tilde{\chi}_1^+\tilde{\chi}_1^-\rightarrow 
(\ell\nu_{\ell}\tilde{\chi}_1^0)+(q\bar{q}^\prime\tilde{\chi}_1^0)$ 
we are able to extract $m_{\tilde{\chi}_1^\pm}$
and $m_{\tilde{\chi}_1^0}$ via the $m(jj)$ and $E(jj)$ distributions, 
typically to percent level accuracy. 
By measuring the dilepton mass and energy distributions 
from $e^+e^-\rightarrow \tilde{\chi}_2^0\tilde{\chi}_1^0$ followed by
$\tilde{\chi}_2^0\rightarrow \ell^+\ell^-\tilde{\chi}_1^0$, we are able to
measure $m_{\tilde{\chi}_1^0}$ and $m_{\tilde{\chi}_2^0}$ to typically percent level accuracy.
We combine the higgsino mass measurements with precision
higgsino pair production cross section measurements using 
different beam polarizations.

When these precision higgsino measurements are combined with precision Higgs
boson measurements, precision fits to both weak scale SUSY and high scale
SUSY model parameters can be made.
In particular, an indirect measurement of wino and bino SUSY breaking masses can be
extracted from the higgsino mass splittings. 
When extrapolated to high energies, the hypothesis of gaugino mass unification can be
tested. If combined with LHC gluino mass measurements, the unification of all three
gaugino masses may be explored. Such measurements will shed light on different possibilities
for SUSY breaking as may be expected in SUSY GUT models or in models with mixed
moduli- and anomaly- (mirage) mediation. 
In addition, fits of SUSY dark matter observables may shed light on the nature of dark 
matter, such as confirming or ruling out multi-component dark matter as expected 
from natural SUSY where both higgsino-like WIMPs and axions are expected to be produced in
the early universe.

Thus, in assessing the ILC capabilities in this compelling SUSY extension of the SM,
we conclude that ILC can indeed serve as a SUSY discovery machine and precision
microscope, offering a window into the intricacies of SUSY breaking and fundamental 
particle physics and providing insights into the nature of dark matter and cosmology.

\section{Acknowledgments}

This work was supported in part by the Office of Science, US Department of Energy and by the Deutsche Forschungsgemeinschaft (DFG) through the Collaborative Research Centre SFB 676 ``Particles,
Strings and the Early Universe'', project B1.
We would like to thank the LCC generator working group and the ILD software working group for providing the simulation and reconstruction
tools and producing the Monte Carlo samples used in this study.
This work has benefited from computing services provided by the ILC Virtual Organization, supported by the national
resource providers of the EGI Federation and the Open Science GRID,
and of those of the German National Analysis Facility (NAF).

\appendix

\section{Additional Figures and Tables}
\label{sec:appendix_event_selection}.

\begin{table}[htbp]
\caption{Expected number of events for chargino signal and major backgrounds
for the electron final state and beam polarization $\mathcal{P}_{-+}$.
The integrated luminosity is assumed to be 500~fb$^{-1}$.
For each benchmark model, the background refers to the other SUSY backgrounds.
}
\label{tab:chargino_cutflow_electron_left}
\centering
\resizebox{0.99\textwidth}{!}{
\begin{tabular}{l|rr|rr|rr|rrrrr}
\hline\hline
$\widetilde\chi^+_1\widetilde\chi^-_1 \rightarrow
\widetilde\chi^0_1\widetilde\chi^0_1q\bar{q}^\prime e\nu_e$
& \multicolumn{2}{c|}{ILC1} & \multicolumn{2}{c|}{ILC2} & \multicolumn{2}{c|}{nGMM1} & \multicolumn{5}{c}{SM bkg.} \\
  500 GeV, 500 fb$^{-1}$, $\mathcal{P}_{-+}$
& Signal & Bkg. & Signal & Bkg. & Signal & Bkg. & $e^+e^-\rightarrow$2f & $e^+e^-\rightarrow$4f & $\gamma\gamma \rightarrow $ 2f & $e\gamma\rightarrow$3f & $\gamma\gamma\rightarrow$4f \\
\hline
Preselection                  	&   53963 &  423992	&   41962 &  322011	&   66118 &  476646	&11906936 &14941264 &307189572 &65344394 &   61765\\
Lepton selection              	&    4926 &   11922	&    2733 &    7676	&    4453 &   12325	&  543911 &  914027 &93465142 &21607557 &    1905\\
BeamCal veto                  	&    4869 &   11752	&    2707 &    7602	&    4414 &   12188	&  495890 &  748137 & 1284355 & 3964924 &    1772\\
$p_{T}>5$~GeV                 	&    3146 &    2323	&    1242 &    1110	&    1337 &    1109	&  226624 &  506571 &  967020 & 3804929 &    1328\\
$N_{\rm trk,jet}\geq 2$       	&    2285 &     324	&     667 &     108	&     515 &      98	&   42892 &  220378 &   65284 & 1745715 &     627\\
$|\cos\theta_{j}|<0.98$       	&    2225 &     314	&     652 &     106	&     504 &      97	&   15612 &  168407 &   50786 & 1323463 &     513\\
$\Delta\phi<1.0$              	&    1544 &     122	&     411 &      30	&     296 &      19	&    1507 &   34570 &   11157 &  533787 &      51\\
$|\cos\theta_{jj}|<0.2$       	&    1535 &      90	&     405 &      24	&     293 &      17	&    1360 &   32195 &    9471 &  483002 &      40\\
$E_{\rm vis}<80$              	&    1496 &      87	&     402 &      24	&     291 &      17	&      59 &     403 &    1810 &    7835 &     2.9\\
$E_{\rm miss}>400$            	&    1485 &      87	&     402 &      24	&     291 &      17	&      12 &      69 &     7.1 &      48 &     2.0\\
$|\cos\theta_{\rm miss}|<0.99$	&    1463 &      85	&     392 &      23	&     283 &      15	&     5.9 &      64 &     0.0 &      22 &     2.0\\

\hline\hline
\end{tabular}
}
\end{table}

\begin{table}[htbp]
\caption{Expected number of events for chargino signal and major backgrounds
for the muon final state and beam polarization $\mathcal{P}_{-+}$.
The integrated luminosity is assumed to be 500~fb$^{-1}$.
For each benchmark model, the background refers to the other SUSY backgrounds.
}
\label{tab:chargino_cutflow_muon_left}
\centering
\resizebox{0.99\textwidth}{!}{
\begin{tabular}{l|rr|rr|rr|rrrrr}
\hline\hline
$\widetilde\chi^+_1\widetilde\chi^-_1 \rightarrow
\widetilde\chi^0_1\widetilde\chi^0_1q\bar{q}^\prime\mu\nu_\mu$
& \multicolumn{2}{c|}{ILC1} & \multicolumn{2}{c|}{ILC2} & \multicolumn{2}{c|}{nGMM1} & \multicolumn{5}{c}{SM bkg.} \\
  500 GeV, 500 fb$^{-1}$, $\mathcal{P}_{-+}$
& Signal & Bkg. & Signal & Bkg. & Signal & Bkg. &  $e^+e^-\rightarrow$2f & $e^+e^-\rightarrow$4f & $\gamma\gamma \rightarrow $ 2f & $e\gamma\rightarrow$3f & $\gamma\gamma\rightarrow$4f \\
\hline
Preselection                  	&   53459 &  424497	&   41714 &  322259	&   65104 &  477660	&11906936 &14941264 &307189572 &65344394 &   61765\\
Lepton selection              	&    5748 &   32945	&    3497 &   21394	&    6194 &   34867	& 1125893 & 1297965 &42676970 & 2497567 &    2716\\
BeamCal veto                  	&    5683 &   32500	&    3462 &   21165	&    6134 &   34476	& 1025945 & 1049378 &  420779 &  325406 &    2475\\
$p_{T}>5$~GeV                 	&    3677 &    3141	&    1566 &    1720	&    1832 &    1794	&   99197 &  345356 &  101920 &  146861 &    1430\\
$N_{\rm trk,jet}\geq 2$       	&    2612 &     710	&     805 &     225	&     690 &     228	&   19319 &  183151 &     197 &   10945 &     509\\
$|\cos\theta_{j}|<0.98$       	&    2544 &     688	&     784 &     221	&     672 &     223	&   11089 &  150507 &      28 &    7906 &     331\\
$\Delta\phi<1.0$              	&    1972 &     259	&     532 &      53	&     412 &      40	&     755 &   37957 &      28 &     874 &      55\\
$|\cos\theta_{jj}|<0.2$       	&    1954 &     118	&     526 &      29	&     406 &      32	&     471 &   37320 &     0.0 &     174 &      51\\
$E_{\rm vis}<80$              	&    1905 &     110	&     523 &      28	&     404 &      31	&      59 &     379 &     0.0 &     0.0 &      22\\
$E_{\rm miss}>400$            	&    1889 &     110	&     523 &      28	&     404 &      31	&      33 &      39 &     0.0 &     0.0 &     8.0\\
$|\cos\theta_{\rm miss}|<0.99$	&    1862 &     108	&     509 &      28	&     389 &      29	&      33 &      37 &     0.0 &     0.0 &     7.0\\

\hline\hline
\end{tabular}
}
\end{table}

\begin{table}[htbp]
\caption{Expected number of events for chargino signal and major backgrounds
for the electron final state and beam polarization $\mathcal{P}_{+-}$.
The integrated luminosity is assumed to be 500~fb$^{-1}$.
For each benchmark model, the background refers to the other SUSY backgrounds.
}
\label{tab:chargino_cutflow_electron_right}
\centering
\resizebox{0.99\textwidth}{!}{
\begin{tabular}{l|rr|rr|rr|rrrrr}
\hline\hline
$\widetilde\chi^+_1\widetilde\chi^-_1 \rightarrow
\widetilde\chi^0_1\widetilde\chi^0_1q\bar{q}^\prime e\nu_e$
& \multicolumn{2}{c|}{ILC1} & \multicolumn{2}{c|}{ILC2} & \multicolumn{2}{c|}{nGMM1} & \multicolumn{5}{c}{SM bkg.} \\
  500 GeV, 500 fb$^{-1}$, $\mathcal{P}_{+-}$
& Signal & Bkg. & Signal & Bkg. & Signal & Bkg.  & $e^+e^-\rightarrow$2f & $e^+e^-\rightarrow$4f & $\gamma\gamma \rightarrow $ 2f & $e\gamma\rightarrow$3f & $\gamma\gamma\rightarrow$4f \\
\hline
Preselection                  	&   13276 &  163541	&   10382 &   97217	&   17159 &  180558	& 7839612 & 4800015 &307189572 &64002532 &   61765\\
Lepton selection              	&    1251 &    5394	&     660 &    2546	&    1153 &    5528	&  434539 &  449786 &93465142 &20689292 &    1905\\
BeamCal veto                  	&    1238 &    5320	&     653 &    2518	&    1143 &    5467	&  395552 &  293541 & 1284355 & 3493039 &    1772\\
$p_{T}>5$~GeV                 	&     834 &     787	&     298 &     310	&     350 &     323	&  210050 &  199506 &  967020 & 3341264 &    1328\\
$N_{\rm trk,jet}\geq 2$       	&     615 &     116	&     161 &      30	&     140 &      27	&   39447 &   50256 &   65284 & 1297669 &     627\\
$|\cos\theta_{j}|<0.98$       	&     600 &     113	&     157 &      29	&     137 &      26	&   13665 &   24867 &   50786 &  900538 &     513\\
$\Delta\phi<1.0$              	&     423 &      47	&     100 &     6.9	&      77 &     6.1	&    1376 &    4950 &   11157 &  386411 &      51\\
$|\cos\theta_{jj}|<0.2$       	&     421 &      25	&      99 &     5.1	&      76 &     5.2	&    1275 &    4411 &    9471 &  358781 &      40\\
$E_{\rm vis}<80$              	&     409 &      24	&      98 &     5.1	&      76 &     5.2	&      29 &      59 &    1810 &    7315 &     2.9\\
$E_{\rm miss}>400$            	&     408 &      24	&      98 &     5.1	&      76 &     5.2	&     7.8 &      19 &     7.1 &      22 &     2.0\\
$|\cos\theta_{\rm miss}|<0.99$	&     404 &      23	&      96 &     4.6	&      73 &     5.1	&     7.4 &      16 &     0.0 &     8.0 &     2.0\\

\hline\hline
\end{tabular}
}
\end{table}

\begin{table}[htbp]
\caption{Expected number of events for chargino signal and major backgrounds
for the muon final state and beam polarization $\mathcal{P}_{+-}$.
The integrated luminosity is assumed to be 500~fb$^{-1}$.
For each benchmark model, the background refers to the other SUSY backgrounds.
}
\label{tab:chargino_cutflow_muon_right}
\centering
\resizebox{0.99\textwidth}{!}{
\begin{tabular}{l|rr|rr|rr|rrrrr}
\hline\hline
$\widetilde\chi^+_1\widetilde\chi^-_1 \rightarrow
\widetilde\chi^0_1\widetilde\chi^0_1q\bar{q}^\prime\mu\nu_\mu$
& \multicolumn{2}{c|}{ILC1} & \multicolumn{2}{c|}{ILC2} & \multicolumn{2}{c|}{nGMM1} & \multicolumn{5}{c}{SM bkg.} \\
  500 GeV, 500 fb$^{-1}$, $\mathcal{P}_{+-}$
& Signal & Bkg. & Signal & Bkg. & Signal & Bkg.  & $e^+e^-\rightarrow$2f & $e^+e^-\rightarrow$4f & $\gamma\gamma \rightarrow $ 2f & $e\gamma\rightarrow$3f & $\gamma\gamma\rightarrow$4f \\
\hline
Preselection                  	&   13222 &  163594	&   10352 &   97248	&   16876 &  180842	& 7839612 & 4800015 &307189572 &64002532 &   61765\\
Lepton selection              	&    1502 &   16551	&     869 &    7667	&    1619 &   18398	&  783612 &  536319 &42676970 & 2358203 &    2716\\
BeamCal veto                  	&    1487 &   16318	&     860 &    7574	&    1604 &   18183	&  710583 &  302815 &  420779 &  307086 &    2475\\
$p_{T}>5$~GeV                 	&     993 &    1108	&     388 &     521	&     480 &     613	&   76202 &   72875 &  101920 &  140719 &    1430\\
$N_{\rm trk,jet}\geq 2$       	&     733 &     276	&     199 &      69	&     182 &      65	&   14401 &   18374 &     197 &    6910 &     509\\
$|\cos\theta_{j}|<0.98$       	&     712 &     270	&     194 &      67	&     175 &      64	&    8258 &   11680 &      28 &    4533 &     331\\
$\Delta\phi<1.0$              	&     555 &     134	&     134 &      20	&     107 &      10	&     622 &    3238 &      28 &     343 &      55\\
$|\cos\theta_{jj}|<0.2$       	&     550 &      37	&     132 &     9.1	&     106 &     8.6	&     373 &    3093 &     0.0 &      66 &      51\\
$E_{\rm vis}<80$              	&     536 &      34	&     131 &     8.7	&     105 &     8.3	&      29 &      38 &     0.0 &     0.0 &      22\\
$E_{\rm miss}>400$            	&     532 &      34	&     131 &     8.7	&     105 &     8.3	&     8.2 &     9.5 &     0.0 &     0.0 &     8.0\\
$|\cos\theta_{\rm miss}|<0.99$	&     524 &      34	&     127 &     8.5	&     101 &     8.2	&     8.2 &     7.2 &     0.0 &     0.0 &     7.0\\

\hline\hline
\end{tabular}
}
\end{table}

\begin{figure}[htbp]
\begin{subfigure}{0.24\linewidth}
{\includegraphics[width=1.\textwidth]{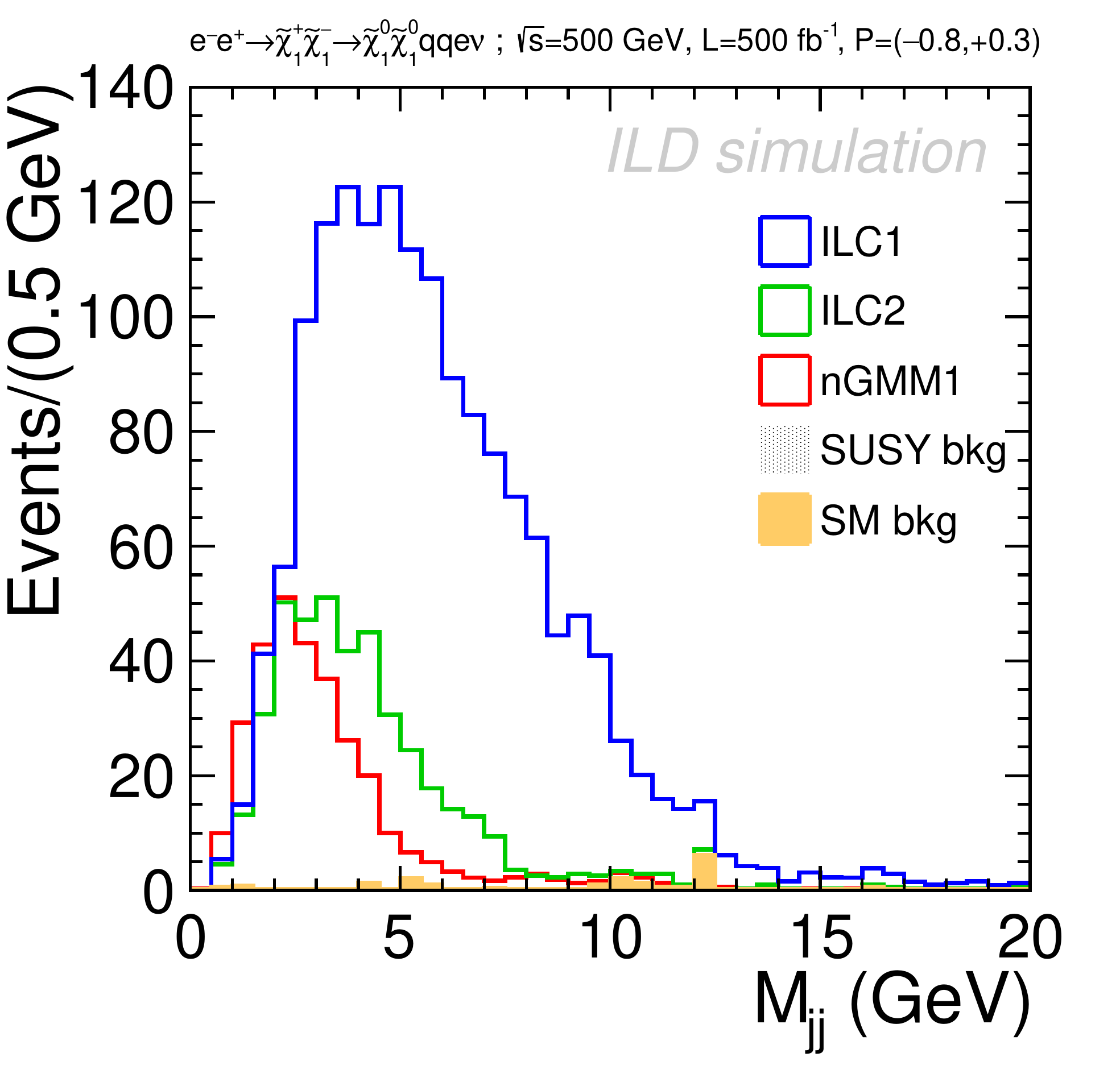}}
\caption{$\tilde{\chi}_1^0\tilde{\chi}_1^0q\bar{q}^\prime e \nu_{e}$; $\mathcal{P}_{-+}$ \label{fig:appendix_chargino_measurement:mass:a}}
\end{subfigure}
\begin{subfigure}{0.24\linewidth}
{\includegraphics[width=1.\textwidth]{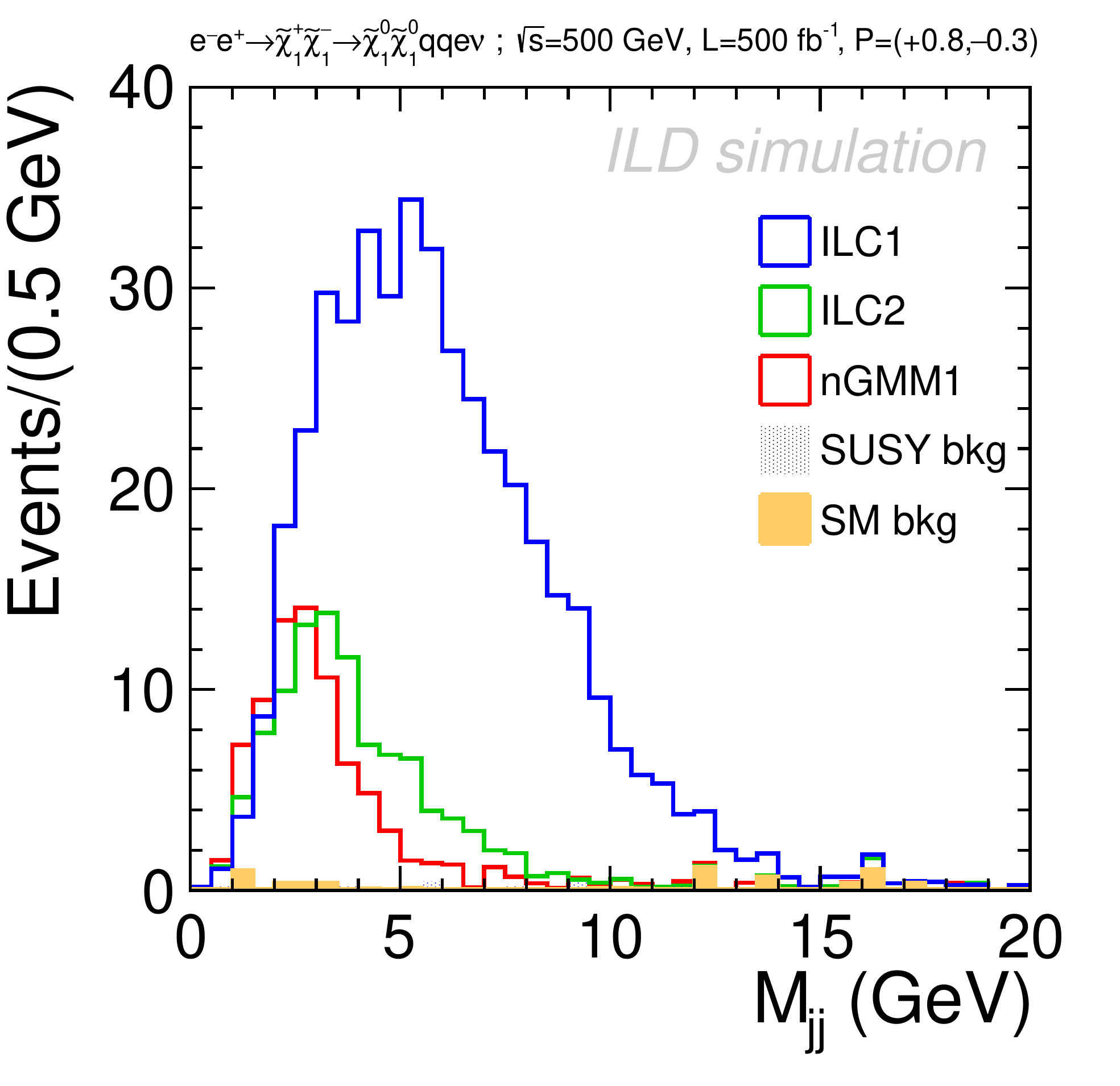}}
\caption{$\tilde{\chi}_1^0\tilde{\chi}_1^0q\bar{q}^\prime e \nu_{e}$; $\mathcal{P}_{+-}$ \label{fig:appendix_chargino_measurement:mass:b}}
\end{subfigure}
\begin{subfigure}{0.24\linewidth}
{\includegraphics[width=1.\textwidth]{combinedC_MmL.pdf}}
\caption{$\tilde{\chi}_1^0\tilde{\chi}_1^0q\bar{q}^\prime \mu \nu_{\mu}$; $\mathcal{P}_{-+}$ \label{fig:appendix_chargino_measurement:mass:c}}
\end{subfigure}
\begin{subfigure}{0.24\linewidth}
{\includegraphics[width=1.\textwidth]{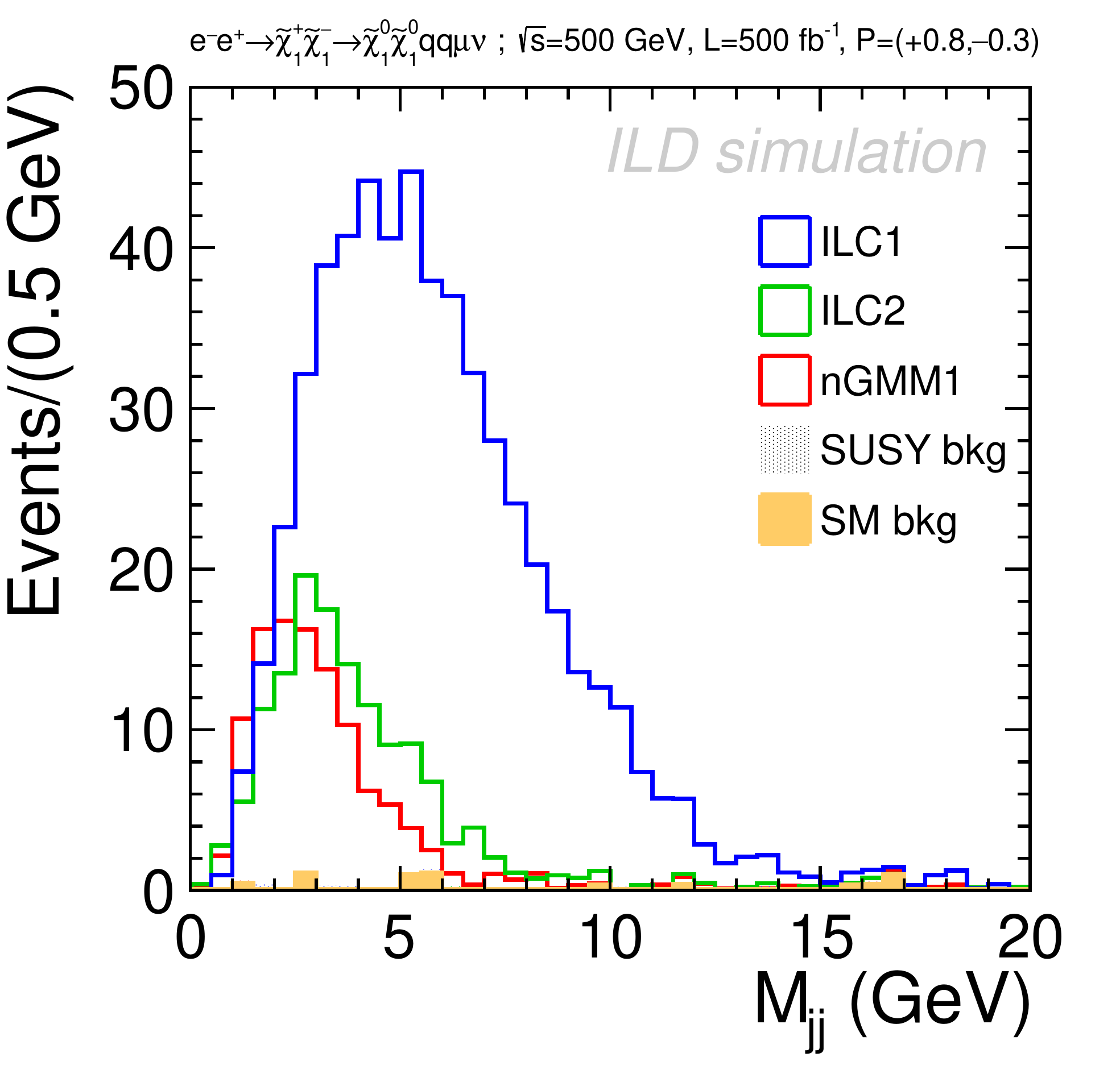}}
\caption{$\tilde{\chi}_1^0\tilde{\chi}_1^0q\bar{q}^\prime \mu \nu_{\mu}$; $\mathcal{P}_{+-}$ \label{fig:appendix_chargino_measurement:mass:d}}
\end{subfigure}

\caption{Reconstructed di-jet mass in the chargino channel
$e^+e^-\rightarrow \tilde{\chi}_1^+\tilde{\chi}_1^-\rightarrow 
\tilde{\chi}_1^0\tilde{\chi}_1^0q\bar{q}^\prime\ell\nu_{\ell}$ for $500$\,fb$^{-1}$ at $\sqrt{s}=500$\,GeV. In all cases, the background contributions are very small. The signal histograms are stacked on top of the backgrounds.
}
\label{fig:appendix_chargino_measurement:mass}
\end{figure}

\begin{figure}[htbp]
\begin{subfigure}{0.24\linewidth}
{\includegraphics[width=1.\textwidth]{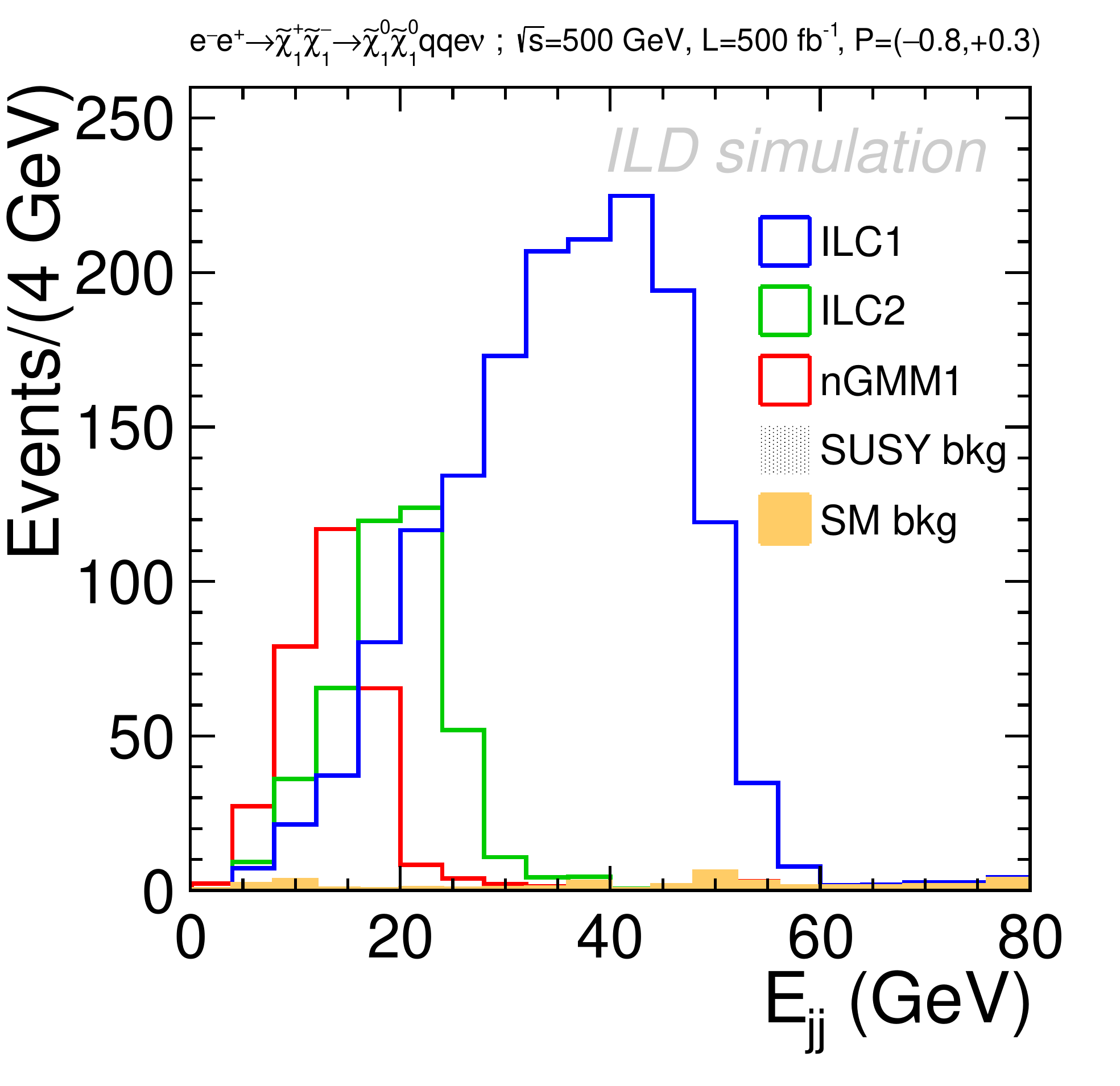}}
\caption{$\tilde{\chi}_1^0\tilde{\chi}_1^0q\bar{q}^\prime e \nu_{e}$; $\mathcal{P}_{-+}$ \label{fig:appendix_chargino_measurement:energy:a}}
\end{subfigure}
\begin{subfigure}{0.24\linewidth}
{\includegraphics[width=1.\textwidth]{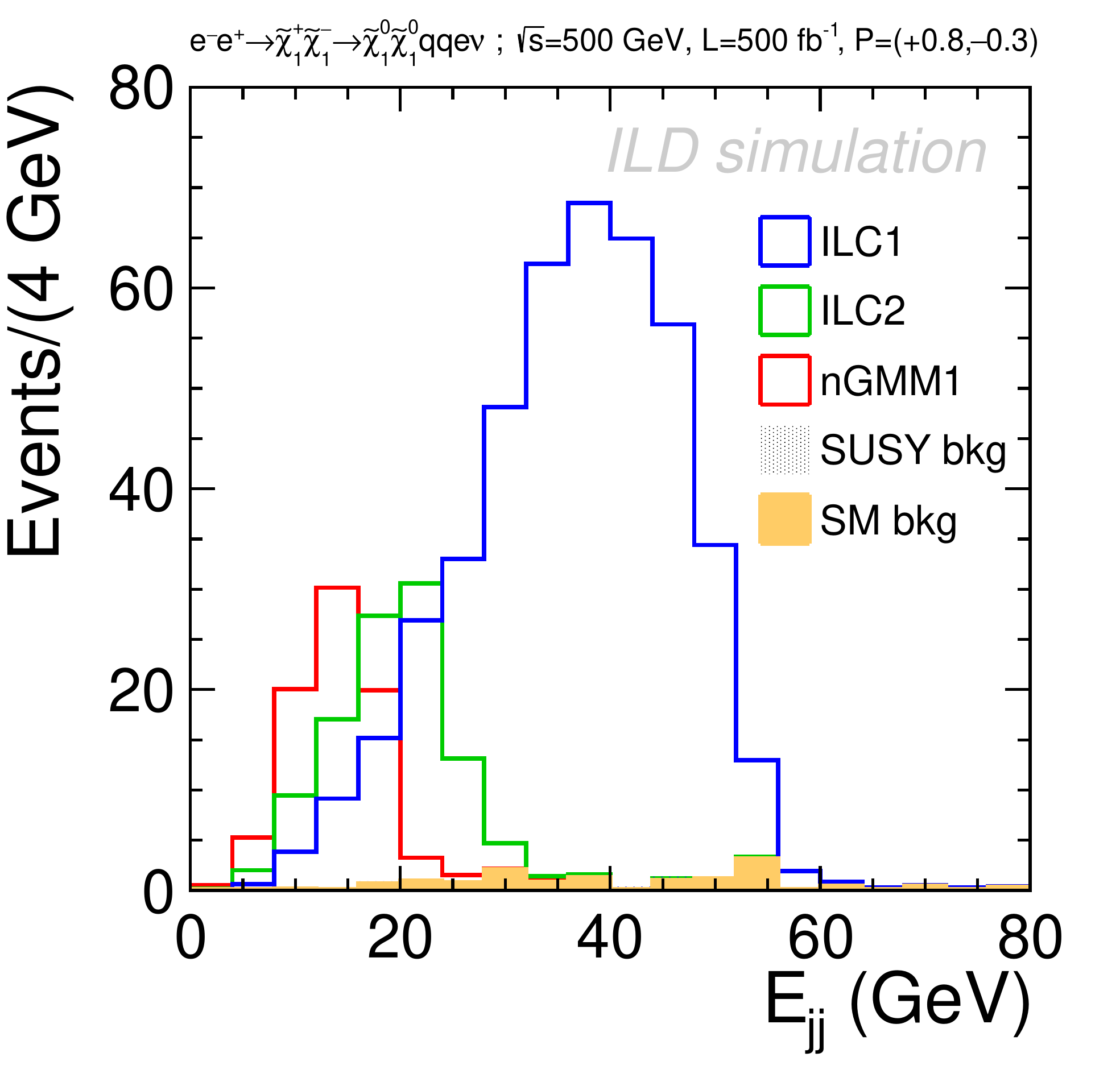}}
\caption{$\tilde{\chi}_1^0\tilde{\chi}_1^0q\bar{q}^\prime e \nu_{e}$; $\mathcal{P}_{+-}$ \label{fig:appendix_chargino_measurement:energy:b}}
\end{subfigure}
\begin{subfigure}{0.24\linewidth}
{\includegraphics[width=1.\textwidth]{combinedC_EmL.pdf}}
\caption{$\tilde{\chi}_1^0\tilde{\chi}_1^0q\bar{q}^\prime \mu \nu_{\mu}$; $\mathcal{P}_{-+}$ \label{fig:appendix_chargino_measurement:energy:c}}
\end{subfigure}
\begin{subfigure}{0.24\linewidth}
{\includegraphics[width=1.\textwidth]{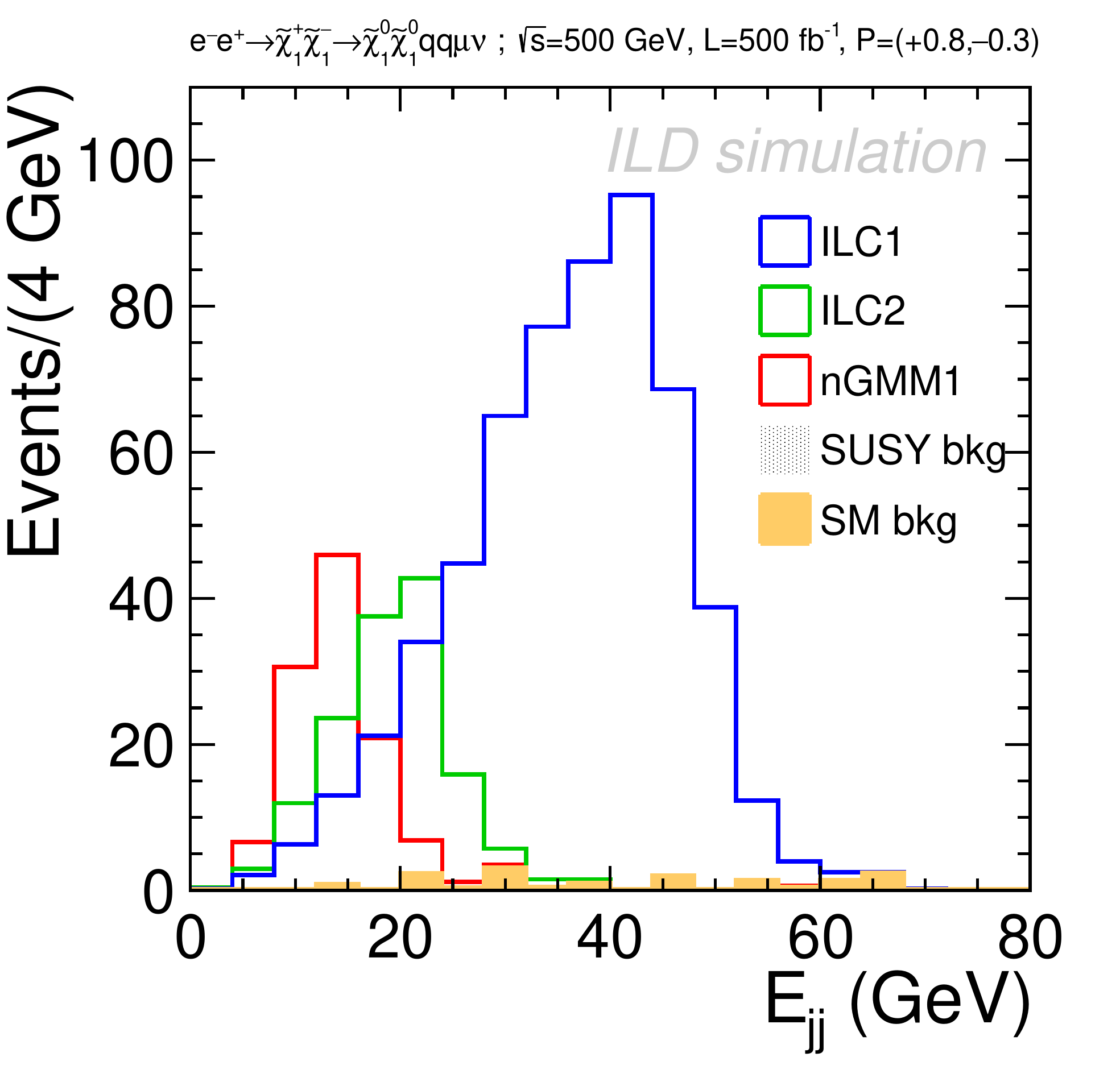}}
\caption{$\tilde{\chi}_1^0\tilde{\chi}_1^0q\bar{q}^\prime \mu \nu_{\mu}$; $\mathcal{P}_{+-}$ \label{fig:appendix_chargino_measurement:energy:d}}
\end{subfigure}

\caption{Reconstructed di-jet energy in the chargino channel
$e^+e^-\rightarrow \tilde{\chi}_1^+\tilde{\chi}_1^-\rightarrow 
\tilde{\chi}_1^0\tilde{\chi}_1^0q\bar{q}^\prime\ell\nu_{\ell}$ for $500$\,fb$^{-1}$ at $\sqrt{s}=500$\,GeV. In all cases, the background contributions are very small. The signal histograms are stacked on top of the backgrounds.
}
\label{fig:appendix_chargino_measurement:energy}
\end{figure}

\begin{figure}[htbp]
\begin{subfigure}{0.24\linewidth}
{\includegraphics[width=1.\textwidth]{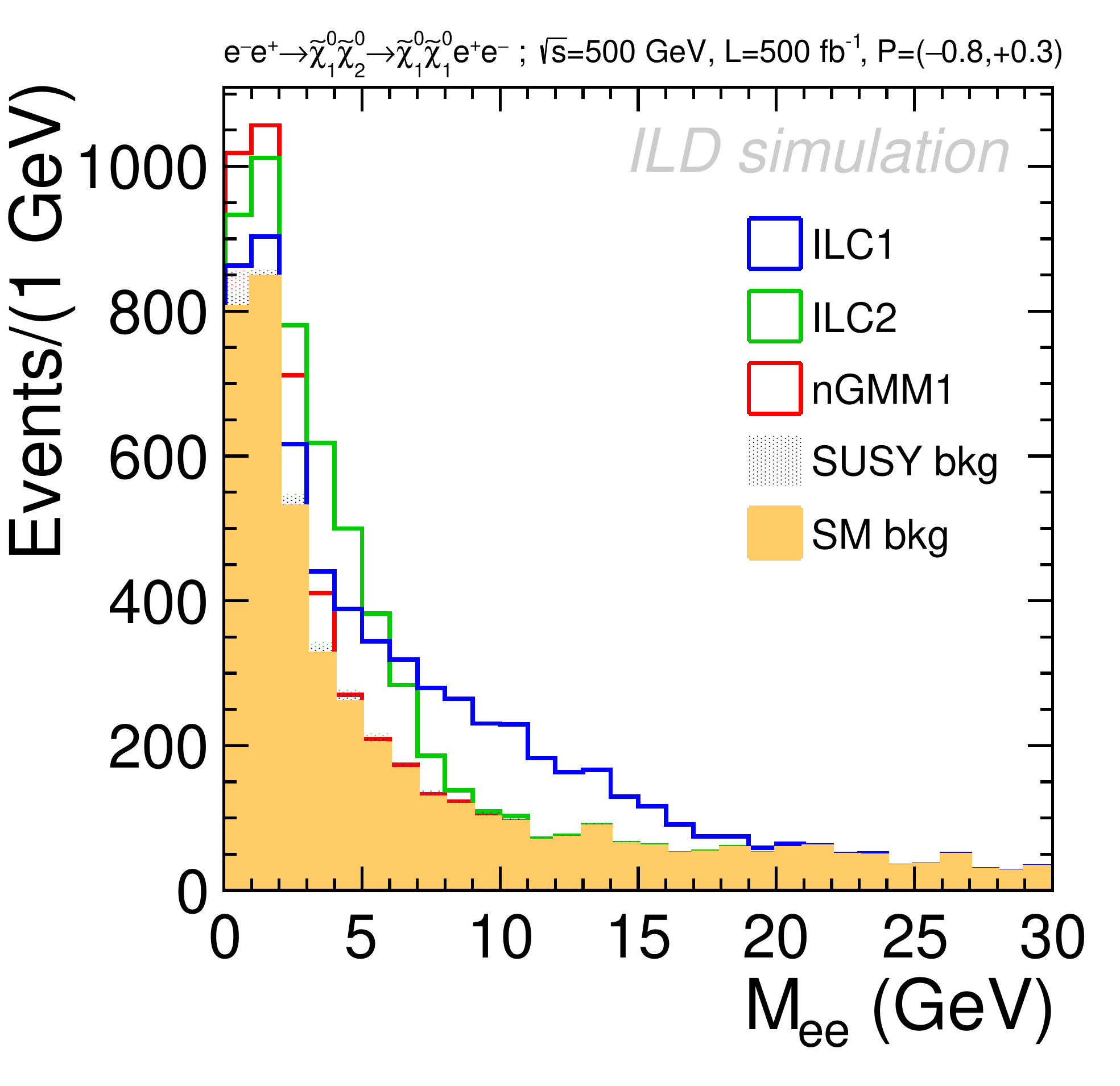}}
\caption{$\tilde{\chi}_1^0\tilde{\chi}_1^0 e^+e^-$; $\mathcal{P}_{-+}$ \label{fig:appendix_neutralino_measurement:mass:a}}
\end{subfigure}
\begin{subfigure}{0.24\linewidth}
{\includegraphics[width=1.\textwidth]{combined_MeR.pdf}}
\caption{$\tilde{\chi}_1^0\tilde{\chi}_1^0e^+e^-$; $\mathcal{P}_{+-}$ \label{fig:appendix_neutralino_measurement:mass:b}}
\end{subfigure}
\begin{subfigure}{0.24\linewidth}
{\includegraphics[width=1.\textwidth]{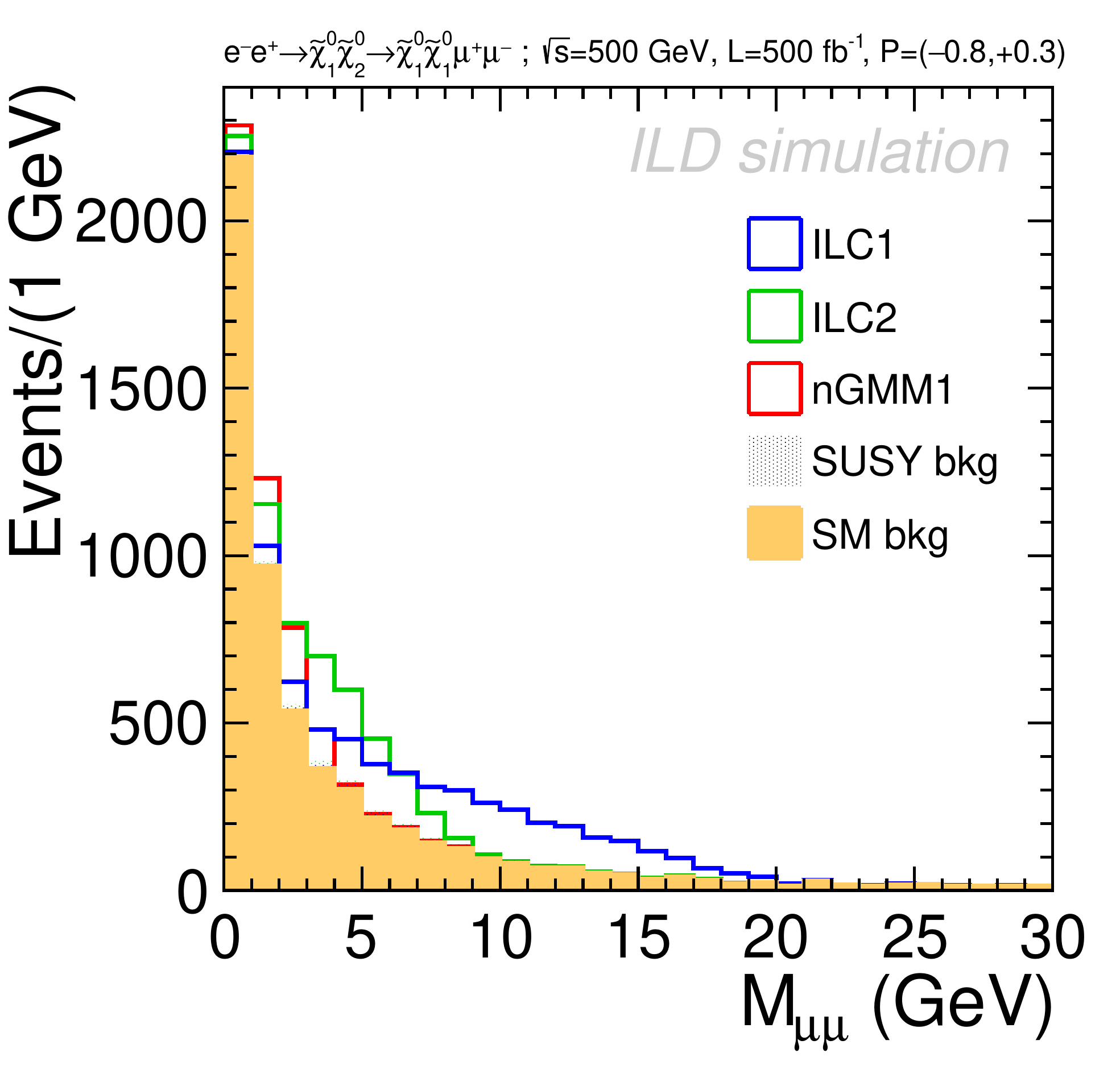}}
\caption{$\tilde{\chi}_1^0\tilde{\chi}_1^0\mu^+\mu^-$; $\mathcal{P}_{-+}$ \label{fig:appendix_neutralino_measurement:mass:c}}
\end{subfigure}
\begin{subfigure}{0.24\linewidth}
{\includegraphics[width=1.\textwidth]{combined_MeR.pdf}}
\caption{$\tilde{\chi}_1^0\tilde{\chi}_1^0\mu^+\mu^-$; $\mathcal{P}_{+-}$ \label{fig:appendix_neutralino_measurement:mass:d}}
\end{subfigure}

\caption{Reconstructed di-lepton mass in the neutralino channel
$e^+e^-\rightarrow \tilde{\chi}_2^0\tilde{\chi}_1^0\rightarrow 
\tilde{\chi}_1^0\tilde{\chi}_1^00\ell^+\ell^-$ for $500$\,fb$^{-1}$ at $\sqrt{s}=500$\,GeV. In all cases, the SUSY background contributions are very small. The signal histograms are stacked on top of the backgrounds.
}
\label{fig:appendix_neutralino_measurement:mass}
\end{figure}

\begin{figure}[htbp]
\begin{subfigure}{0.24\linewidth}
{\includegraphics[width=1.\textwidth]{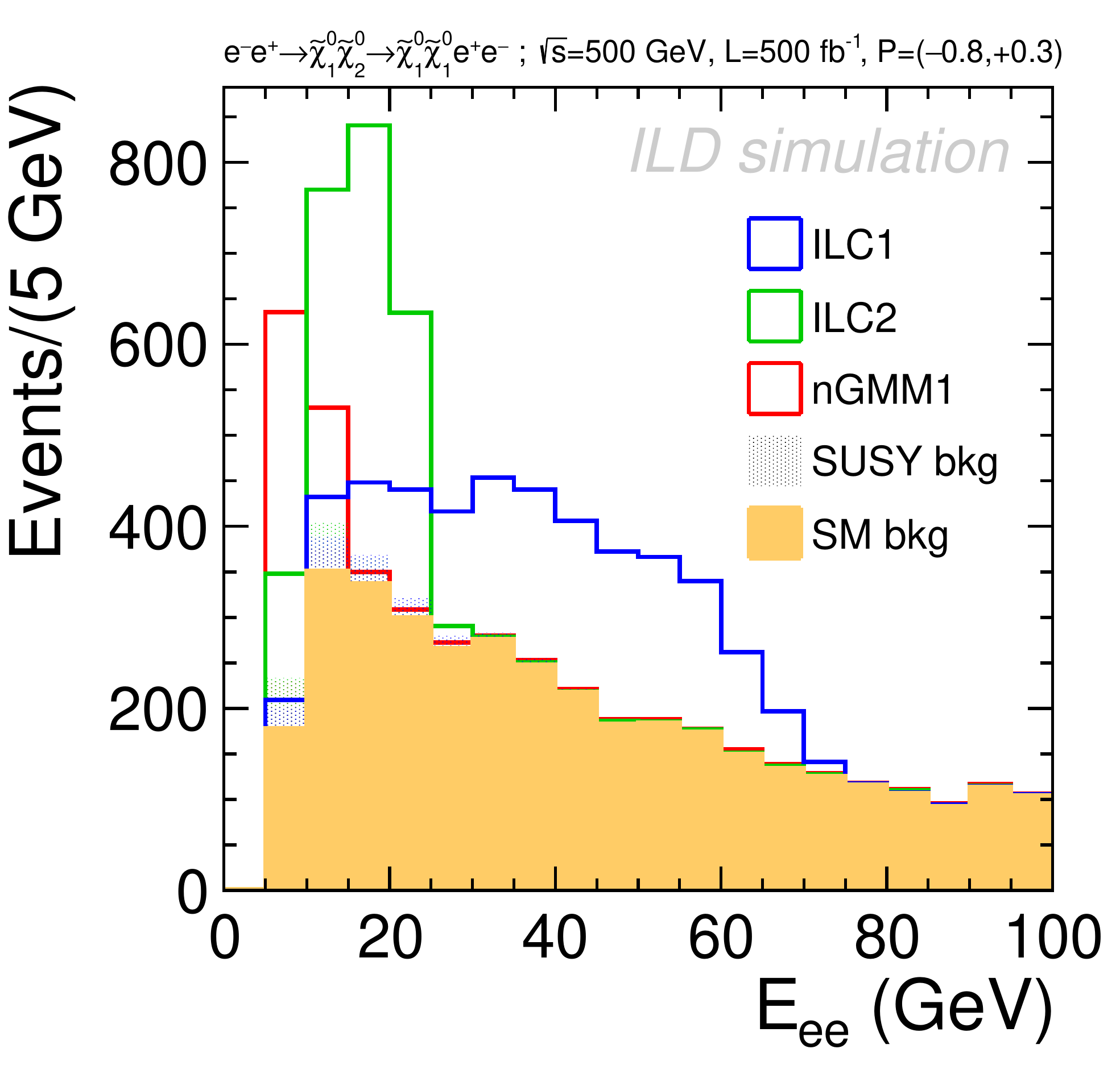}}
\caption{$\tilde{\chi}_1^0\tilde{\chi}_1^0 e^+e^-$; $\mathcal{P}_{-+}$ \label{fig:appendix_neutralino_measurement:energy:a}}
\end{subfigure}
\begin{subfigure}{0.24\linewidth}
{\includegraphics[width=1.\textwidth]{combined_EeR.pdf}}
\caption{$\tilde{\chi}_1^0\tilde{\chi}_1^0e^+e^-$; $\mathcal{P}_{+-}$ \label{fig:appendix_neutralino_measurement:energy:b}}
\end{subfigure}
\begin{subfigure}{0.24\linewidth}
{\includegraphics[width=1.\textwidth]{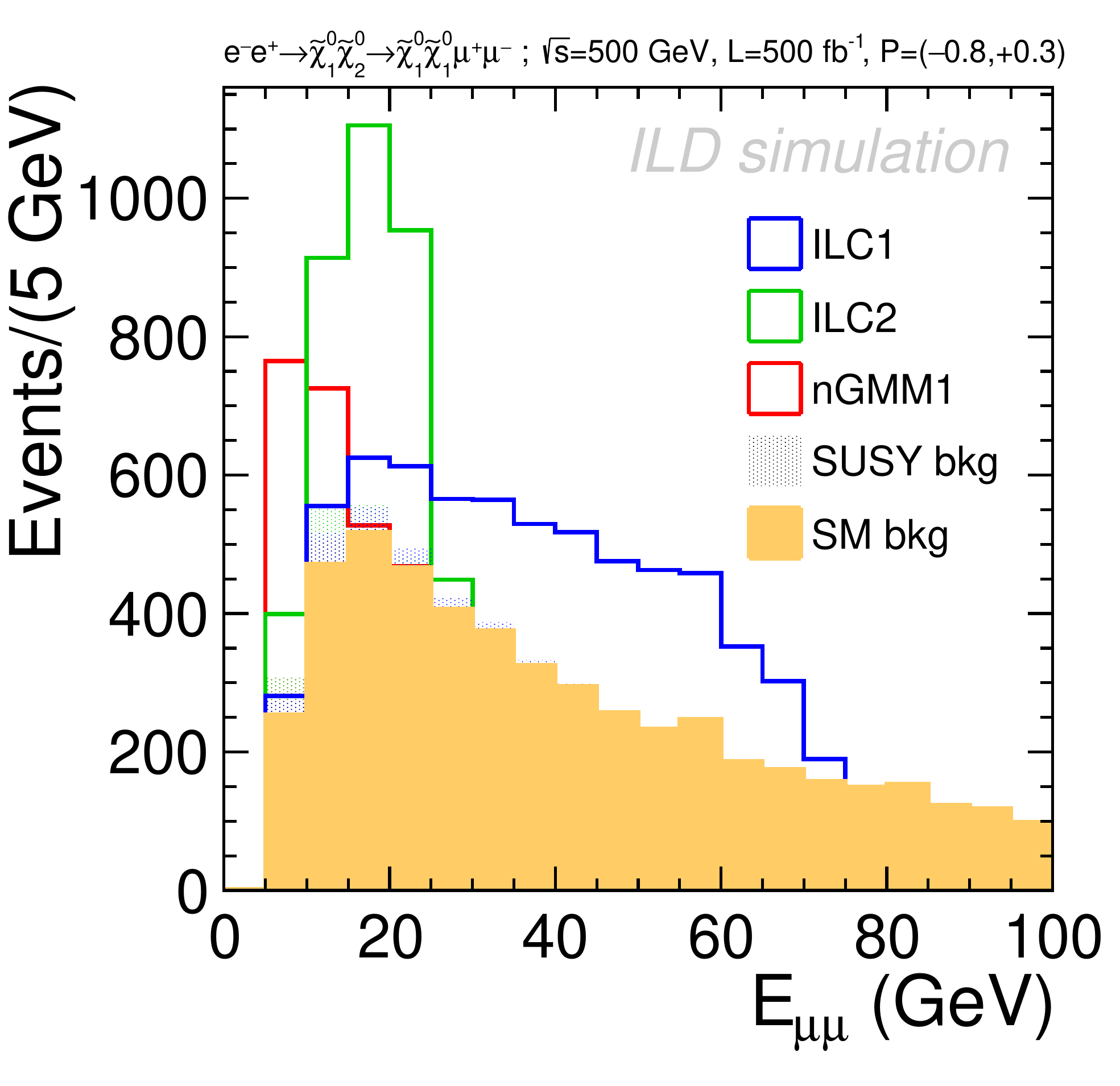}}
\caption{$\tilde{\chi}_1^0\tilde{\chi}_1^0\mu^+\mu^-$; $\mathcal{P}_{-+}$ \label{fig:appendix_neutralino_measurement:energy:c}}
\end{subfigure}
\begin{subfigure}{0.24\linewidth}
{\includegraphics[width=1.\textwidth]{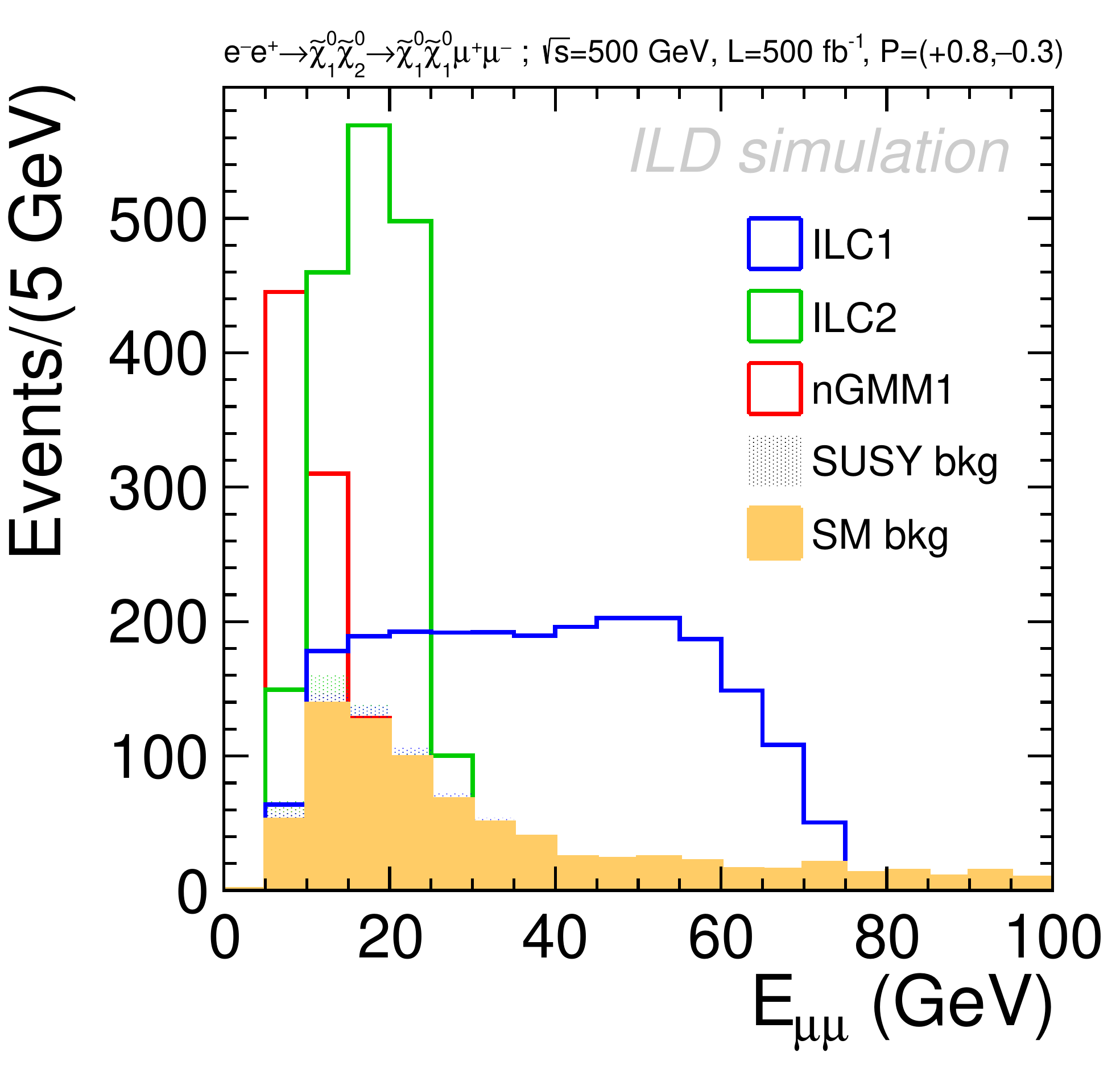}}
\caption{$\tilde{\chi}_1^0\tilde{\chi}_1^0\mu^+\mu^-$; $\mathcal{P}_{+-}$ \label{fig:appendix_neutralino_measurement:energy:d}}
\end{subfigure}

\caption{Reconstructed di-lepton energy in the neutralino channel
$e^+e^-\rightarrow \tilde{\chi}_2^0\tilde{\chi}_1^0\rightarrow 
\tilde{\chi}_1^0\tilde{\chi}_1^00\ell^+\ell^-$ for $500$\,fb$^{-1}$ at $\sqrt{s}=500$\,GeV. In all cases, the SUSY background contributions are very small. The signal histograms are stacked on top of the backgrounds.
}
\label{fig:appendix_neutralino_measurement:energy}
\end{figure}

\begin{table}[htbp]
\caption{Expected number of events for neutralino signal and major backgrounds
for the electron final state and beam polarizations $P(e^-,e^+)=(-0.8,+0.3)$.
The integrated luminosity is assumed to be 500~fb$^{-1}$.
For each benchmark model, the background refers to the other SUSY backgrounds.
}
\label{tab:neutralino_cutflow_electron_left}
\centering
\resizebox{0.99\textwidth}{!}{
\begin{tabular}{l|rr|rr|rr|rrrrr}
\hline\hline
$\LSP\neutralinotwo(e^-e^+)$
& \multicolumn{2}{c|}{ILC1} & \multicolumn{2}{c|}{ILC2} & \multicolumn{2}{c|}{nGMM1} & \multicolumn{4}{c}{SM bkg.} \\
$P(e^-,e^+)=(-0.8,+0.3)$
& sig. & bkg. & sig. & bkg. & sig. & bkg.  &  $e^+e^-\rightarrow$2f & $e^+e^-\rightarrow$4f & $\gamma\gamma \rightarrow $ 2f & $e\gamma\rightarrow$3f & $\gamma\gamma\rightarrow$4f \\
\hline
Preselection                     	&    4370 &   15977	&    3098 &   12393	&    1076 &    8301	&  261999 & 1115296 &   87581 &  313496 &   14260\\
Lepton selection, $N_{\rm trk}=2$	&    4028 &    4039	&    2866 &    3576	&     994 &    2837	&   23958 &  365653 &   22592 &   41791 &     192\\
BeamCal veto                     	&    3965 &    3977	&    2831 &    3531	&     986 &    2808	&   18100 &  152375 &   22592 &   40935 &     176\\
$p_{T}>2.3$~GeV                  	&    3822 &    3638	&    2504 &    3106	&     728 &    2329	&   16543 &  141410 &   21961 &   38709 &     103\\
$|\cos\theta_{\rm \ell}|<0.95$   	&    3601 &    3443	&    2455 &    2997	&     727 &    2257	&    9108 &   75460 &   21885 &   37151 &      82\\
$\Delta\phi<0.8$                 	&    2384 &     327	&    1696 &     371	&     653 &     365	&    3809 &   32251 &   14662 &   28218 &      34\\
$E_{\rm vis}<25$~GeV             	&    1621 &     189	&    1253 &     235	&     492 &     239	&      14 &    4768 &      15 &     669 &      25\\
$E_{\rm miss}>300$~GeV           	&    1621 &     189	&    1253 &     235	&     492 &     239	&      14 &    4034 &      15 &     489 &      25\\
$|\cos\theta_{\rm miss}|<0.98$   	&    1621 &     189	&    1251 &     232	&     491 &     238	&      14 &    4021 &      14 &     488 &      25\\
$M_{\ell\ell}$ selection         	&    1621 &     185	&    1250 &     226	&     490 &     207	&      14 &    3875 &      14 &     371 &      19\\

\hline\hline
\end{tabular}
}
\end{table}

\begin{table}[htbp]
\caption{Expected number of events for neutralino signal and major backgrounds
for the muon final state and beam polarizations $P(e^-,e^+)=(-0.8,+0.3)$.
The integrated luminosity is assumed to be 500~fb$^{-1}$.
For each benchmark model, the background refers to the other SUSY backgrounds.
}
\label{tab:neutralino_cutflow_muon_left}
\centering
\resizebox{0.99\textwidth}{!}{
\begin{tabular}{l|rr|rr|rr|rrrrr}
\hline\hline
$\LSP\neutralinotwo(\mu^-\mu^+)$
& \multicolumn{2}{c|}{ILC1} & \multicolumn{2}{c|}{ILC2} & \multicolumn{2}{c|}{nGMM1} & \multicolumn{4}{c}{SM bkg.} \\
$P(e^-,e^+)=(-0.8,+0.3)$
& sig. & bkg. & sig. & bkg. & sig. & bkg. & 2f & 4f & aa\_2f & ae\_3f & aa\_4f \\
\hline
Preselection                     	&    4895 &   15452	&    3705 &   11786	&    1427 &    7950	&  261999 & 1115296 &   87581 &  313496 &   14260\\
Lepton selection, $N_{\rm trk}=2$	&    4532 &    3887	&    3436 &    3731	&    1325 &    3013	&   21615 &  294934 &   64989 &  112098 &     251\\
BeamCal veto                     	&    4461 &    3814	&    3395 &    3681	&    1312 &    2978	&   20370 &  121784 &   64989 &  111493 &     233\\
$p_{T}>2.3$~GeV                  	&    4348 &    3556	&    3060 &    3283	&     977 &    2508	&   19939 &  115883 &   62899 &  106535 &     133\\
$|\cos\theta_{\rm \ell}|<0.95$   	&    4067 &    3373	&    2997 &    3185	&     977 &    2477	&   13041 &   69986 &   62893 &  106527 &     102\\
$\Delta\phi<0.8$                 	&    2676 &     271	&    2024 &     292	&     868 &     193	&   11796 &   26636 &   42316 &   80441 &      11\\
$E_{\rm vis}<25$~GeV             	&    1939 &     180	&    1498 &     210	&     645 &     136	&     0.0 &    6569 &      84 &     105 &      11\\
$E_{\rm miss}>300$~GeV           	&    1939 &     180	&    1498 &     210	&     645 &     136	&     0.0 &    5595 &      84 &     105 &      11\\
$|\cos\theta_{\rm miss}|<0.98$   	&    1939 &     180	&    1496 &     208	&     640 &     135	&     0.0 &    5574 &      77 &     105 &      11\\
$M_{\ell\ell}$ selection         	&    1939 &     176	&    1496 &     197	&     640 &      91	&     0.0 &    5506 &      77 &     100 &     9.6\\

\hline\hline
\end{tabular}
}
\end{table}

\begin{table}[htbp]
\caption{Expected number of events for neutralino signal and major backgrounds
for the electron final state and beam polarizations $P(e^-,e^+)=(+0.8,-0.3)$.
The integrated luminosity is assumed to be 500~fb$^{-1}$.
For each benchmark model, the background refers to the other SUSY backgrounds.
}
\label{tab:neutralino_cutflow_electron_right}
\centering
\resizebox{0.99\textwidth}{!}{
\begin{tabular}{l|rr|rr|rr|rrrrr}
\hline\hline
$\LSP\neutralinotwo(e^-e^+)$
& \multicolumn{2}{c|}{ILC1} & \multicolumn{2}{c|}{ILC2} & \multicolumn{2}{c|}{nGMM1} & \multicolumn{4}{c}{SM bkg.} \\
$P(e^-,e^+)=(+0.8,-0.3)$
& sig. & bkg. & sig. & bkg. & sig. & bkg. & 2f & 4f & aa\_2f & ae\_3f & aa\_4f \\
\hline
Preselection                     	&    3486 &    6769	&    2495 &    5364	&     865 &    3144	&  166524 &  879484 &   87581 &  300006 &   14260\\
Lepton selection, $N_{\rm trk}=2$	&    3214 &    1042	&    2308 &    1024	&     800 &     892	&   23031 &  348421 &   22592 &   38695 &     192\\
BeamCal veto                     	&    3160 &    1024	&    2275 &    1014	&     794 &     880	&   17315 &  136363 &   22592 &   38347 &     176\\
$p_{T}>2.3$~GeV                  	&    3047 &     938	&    2021 &     889	&     602 &     727	&   15819 &  126608 &   21961 &   36737 &     103\\
$|\cos\theta_{\rm \ell}|<0.95$   	&    2872 &     882	&    1985 &     852	&     602 &     694	&    8867 &   65285 &   21885 &   36359 &      82\\
$\Delta\phi<0.8$                 	&    1904 &     121	&    1382 &     169	&     541 &     184	&    3743 &   25036 &   14662 &   27658 &      34\\
$E_{\rm vis}<25$~GeV             	&    1284 &      69	&    1020 &     113	&     412 &     128	&      13 &     858 &      15 &     117 &      25\\
$E_{\rm miss}>300$~GeV           	&    1284 &      69	&    1020 &     113	&     412 &     128	&      13 &     530 &      15 &      97 &      25\\
$|\cos\theta_{\rm miss}|<0.98$   	&    1284 &      69	&    1017 &     113	&     409 &     126	&      13 &     529 &      14 &      96 &      25\\
$M_{\ell\ell}$ selection         	&    1284 &      69	&    1017 &     111	&     409 &     119	&      13 &     508 &      14 &      83 &      19\\

\hline\hline
\end{tabular}
}
\end{table}

\begin{table}[htbp]
\caption{Expected number of events for neutralino signal and major backgrounds
for the muon final state and beam polarizations $P(e^-,e^+)=(+0.8,-0.3)$.
The integrated luminosity is assumed to be 500~fb$^{-1}$.
For each benchmark model, the background refers to the other SUSY backgrounds.
}
\label{tab:neutralino_cutflow_muon_right}
\centering
\resizebox{0.99\textwidth}{!}{
\begin{tabular}{l|rr|rr|rr|rrrrr}
\hline\hline
$\LSP\neutralinotwo(\mu^-\mu^+)$
& \multicolumn{2}{c|}{ILC1} & \multicolumn{2}{c|}{ILC2} & \multicolumn{2}{c|}{nGMM1} & \multicolumn{4}{c}{SM bkg.} \\
$P(e^-,e^+)=(+0.8,-0.3)$
& sig. & bkg. & sig. & bkg. & sig. & bkg. & 2f & 4f & aa\_2f & ae\_3f & aa\_4f \\
\hline
Preselection                     	&    3856 &    6399	&    2972 &    4887	&    1122 &    2887	&  166524 &  879484 &   87581 &  300006 &   14260\\
Lepton selection, $N_{\rm trk}=2$	&    3573 &     950	&    2762 &     973	&    1039 &     844	&   20711 &  281254 &   64989 &  112161 &     251\\
BeamCal veto                     	&    3513 &     932	&    2727 &     960	&    1028 &     835	&   19521 &  108836 &   64989 &  111489 &     233\\
$p_{T}>2.3$~GeV                  	&    3421 &     869	&    2451 &     862	&     777 &     691	&   19069 &  103802 &   62899 &  106529 &     133\\
$|\cos\theta_{\rm \ell}|<0.95$   	&    3209 &     824	&    2402 &     836	&     777 &     681	&   11940 &   60407 &   62893 &  106521 &     102\\
$\Delta\phi<0.8$                 	&    2099 &      73	&    1633 &      93	&     692 &      72	&   10844 &   19083 &   42316 &   80441 &      11\\
$E_{\rm vis}<25$~GeV             	&    1522 &      50	&    1223 &      70	&     518 &      51	&     0.0 &    1213 &      84 &     105 &      11\\
$E_{\rm miss}>300$~GeV           	&    1522 &      50	&    1223 &      70	&     518 &      51	&     0.0 &     686 &      84 &     105 &      11\\
$|\cos\theta_{\rm miss}|<0.98$   	&    1521 &      50	&    1222 &      69	&     516 &      51	&     0.0 &     678 &      77 &     105 &      11\\
$M_{\ell\ell}$ selection         	&    1521 &      49	&    1222 &      67	&     516 &      40	&     0.0 &     672 &      77 &     100 &     9.6\\

\hline\hline
\end{tabular}
}
\end{table}



\begin{footnotesize}


\end{footnotesize}



\begin{thebibliography}{99}

\bibitem{lhc_h}  G.~Aad {\it et al.} [ATLAS Collaboration],
  Phys.\ Lett.\ B {\bf 716} (2012) 1;
S.~Chatrchyan {\it et al.} [CMS Collaboration],
  Phys.\ Lett.\ B {\bf 716} (2012) 30.

\bibitem{Susskind:1978ms}   L.~Susskind,
  Phys.\ Rev.\ D {\bf 20} (1979) 2619.
  doi:10.1103/PhysRevD.20.2619


\bibitem{Ellis:1986yg}
  J.~R.~Ellis, K.~Enqvist, D.~V.~Nanopoulos and F.~Zwirner,
  Mod.\ Phys.\ Lett.\ A {\bf 1} (1986) 57.
  doi:10.1142/S0217732386000105


\bibitem{Barbieri:1987fn}
  R.~Barbieri and G.~F.~Giudice,
  Nucl.\ Phys.\ B {\bf 306} (1988) 63.
  doi:10.1016/0550-3213(88)90171-X

\bibitem{Dimopoulos:1995mi}
  S.~Dimopoulos and G.~F.~Giudice,
  Phys.\ Lett.\ B {\bf 357} (1995) 573
  doi:10.1016/0370-2693(95)00961-J
  [hep-ph/9507282].

\bibitem{Carena:2002es}
  M.~Carena and H.~E.~Haber,
  Prog.\ Part.\ Nucl.\ Phys.\  {\bf 50} (2003) 63
  doi:10.1016/S0146-6410(02)00177-1
  [hep-ph/0208209].

\bibitem{Kitano:2006gv}
  R.~Kitano and Y.~Nomura,
  Phys.\ Rev.\ D {\bf 73} (2006) 095004
  doi:10.1103/PhysRevD.73.095004
  [hep-ph/0602096].

\bibitem{Papucci:2011wy}
  M.~Papucci, J.~T.~Ruderman and A.~Weiler,
  JHEP {\bf 1209} (2012) 035
  doi:10.1007/JHEP09(2012)035
  [arXiv:1110.6926 [hep-ph]].


\bibitem{Arkani-Hamed:2015vfh}
  N.~Arkani-Hamed, T.~Han, M.~Mangano and L.~T.~Wang,
  Phys.\ Rept.\  {\bf 652} (2016) 1
  doi:10.1016/j.physrep.2016.07.004
  [arXiv:1511.06495 [hep-ph]].



\bibitem{Baer:2012cf}
  H.~Baer, V.~Barger, P.~Huang, D.~Mickelson, A.~Mustafayev and X.~Tata,
  Phys.\ Rev.\ D {\bf 87} (2013) no.11,  115028
  doi:10.1103/PhysRevD.87.115028
  [arXiv:1212.2655 [hep-ph]].

\bibitem{Baer:2015rja}
  H.~Baer, V.~Barger and M.~Savoy,
  Phys.\ Rev.\ D {\bf 93} (2016) no.3,  035016
  doi:10.1103/PhysRevD.93.035016
  [arXiv:1509.02929 [hep-ph]].


\bibitem{Baer:2013gva}
  H.~Baer, V.~Barger and D.~Mickelson,
  Phys.\ Rev.\ D {\bf 88} (2013) no.9,  095013
  doi:10.1103/PhysRevD.88.095013
  [arXiv:1309.2984 [hep-ph]].
  
\bibitem{Ross:2016pml}
  G.~G.~Ross, K.~Schmidt-Hoberg and F.~Staub,
  Phys.\ Lett.\ B {\bf 759} (2016) 110
  doi:10.1016/j.physletb.2016.05.053
  [arXiv:1603.09347 [hep-ph]].

\bibitem{Martin:1999hc}
  S.~P.~Martin,
  Phys.\ Rev.\ D {\bf 61} (2000) 035004
  doi:10.1103/PhysRevD.61.035004
  [hep-ph/9907550].

 
\bibitem{Choi:2004sx}
  K.~Choi, A.~Falkowski, H.~P.~Nilles, M.~Olechowski and S.~Pokorski,
  JHEP {\bf 0411} (2004) 076
  doi:10.1088/1126-6708/2004/11/076
  [hep-th/0411066].

\bibitem{Choi:2005ge}
  K.~Choi, A.~Falkowski, H.~P.~Nilles and M.~Olechowski,
  Nucl.\ Phys.\ B {\bf 718} (2005) 113
  doi:10.1016/j.nuclphysb.2005.04.032
  [hep-th/0503216].

\bibitem{Choi:2007ka}
  K.~Choi and H.~P.~Nilles,
  JHEP {\bf 0704} (2007) 006
  doi:10.1088/1126-6708/2007/04/006
  [hep-ph/0702146 [HEP-PH]].

\bibitem{Baer:2014ica}
  H.~Baer, V.~Barger, D.~Mickelson and M.~Padeffke-Kirkland,
  Phys.\ Rev.\ D {\bf 89} (2014) no.11,  115019
  doi:10.1103/PhysRevD.89.115019
  [arXiv:1404.2277 [hep-ph]];
A.~Mustafayev and X.~Tata,
  Indian J.\ Phys.\  {\bf 88} (2014) 991
  doi:10.1007/s12648-014-0504-8
  [arXiv:1404.1386 [hep-ph]].


\bibitem{Baer:2006rs} 
  For a review, see {\it e.g.} 
  H.~Baer and X.~Tata,
  Cambridge, UK: Univ. Pr. (2006) 537 p;
S.~P.~Martin,
  [hep-ph/9709356].

\bibitem{Aaboud:2017leg}
  M.~Aaboud {\it et al.} [ATLAS Collaboration],
  Phys.\ Rev.\ D {\bf 97} (2018) no.5,  052010
  doi:10.1103/PhysRevD.97.052010
  [arXiv:1712.08119 [hep-ex]].

\bibitem{Sirunyan:2018iwl}
  A.~M.~Sirunyan {\it et al.} [CMS Collaboration],
  Phys.\ Lett.\ B {\bf 782} (2018) 440
  doi:10.1016/j.physletb.2018.05.062
  [arXiv:1801.01846 [hep-ex]].

\bibitem{Baer:2012vr}
  H.~Baer, V.~Barger, A.~Lessa and X.~Tata,
  Phys.\ Rev.\ D {\bf 86} (2012) 117701
  doi:10.1103/PhysRevD.86.117701
  [arXiv:1207.4846 [hep-ph]].

\bibitem{ATLAS:2013hta}
  [ATLAS Collaboration],
  arXiv:1307.7292 [hep-ex].

\bibitem{nuhm2} D.~Matalliotakis and H.~P.~Nilles,
  Nucl.\ Phys.\ B {\bf 435} (1995) 115;
P.~Nath and R.~L.~Arnowitt,
  Phys.\ Rev.\ D {\bf 56} (1997) 2820;
J.~R.~Ellis, T.~Falk, K.~A.~Olive and Y.~Santoso,
  Nucl.\ Phys.\ B {\bf 652} (2003) 259;
H.~Baer, A.~Mustafayev, S.~Profumo, A.~Belyaev and X.~Tata,
  JHEP {\bf 0507} (2005) 065.

\bibitem{Bae:2013bva}
  K.~J.~Bae, H.~Baer and E.~J.~Chun,
  Phys.\ Rev.\ D {\bf 89} (2014) no.3,  031701
  doi:10.1103/PhysRevD.89.031701
  [arXiv:1309.0519 [hep-ph]].

\bibitem{Baer:2016usl}
  H.~Baer, V.~Barger, M.~Savoy and X.~Tata,
  Phys.\ Rev.\ D {\bf 94} (2016) no.3,  035025.

\bibitem{Baer:2012up}
  H.~Baer, V.~Barger, P.~Huang, A.~Mustafayev and X.~Tata,
  Phys.\ Rev.\ Lett.\  {\bf 109} (2012) 161802
  doi:10.1103/PhysRevLett.109.161802
  [arXiv:1207.3343 [hep-ph]].

\bibitem{Baer:2018hwa}
  H.~Baer, V.~Barger and D.~Sengupta,
  Phys.\ Rev.\ D {\bf 98} (2018) no.1,  015039
  doi:10.1103/PhysRevD.98.015039
  [arXiv:1801.09730 [hep-ph]].

\bibitem{Baer:2018hpb}
  H.~Baer, V.~Barger, J.~S.~Gainer, D.~Sengupta, H.~Serce and X.~Tata,
  Phys.\ Rev.\ D {\bf 98} (2018) no.7,  075010
  doi:10.1103/PhysRevD.98.075010
  [arXiv:1808.04844 [hep-ph]].

\bibitem{Baer:2014yta}
  H.~Baer, V.~Barger, D.~Mickelson, A.~Mustafayev and X.~Tata,
  JHEP {\bf 1406} (2014) 172
  doi:10.1007/JHEP06(2014)172
  [arXiv:1404.7510 [hep-ph]].

\bibitem{Baer:2013yha}
  H.~Baer, V.~Barger, P.~Huang, D.~Mickelson, A.~Mustafayev, W.~Sreethawong and X.~Tata,
  Phys.\ Rev.\ Lett.\  {\bf 110} (2013) no.15,  151801
  doi:10.1103/PhysRevLett.110.151801
  [arXiv:1302.5816 [hep-ph]].

\bibitem{Baer:2014kya}
  H.~Baer, A.~Mustafayev and X.~Tata,
  Phys.\ Rev.\ D {\bf 90} (2014) no.11,  115007
  doi:10.1103/PhysRevD.90.115007
  [arXiv:1409.7058 [hep-ph]].

\bibitem{Baer:2016wkz}
  H.~Baer, V.~Barger, J.~S.~Gainer, P.~Huang, M.~Savoy, D.~Sengupta and X.~Tata,
  Eur.\ Phys.\ J.\ C {\bf 77} (2017) no.7,  499
  doi:10.1140/epjc/s10052-017-5067-3
  [arXiv:1612.00795 [hep-ph]].

\bibitem{Baer:2016hfa}
  H.~Baer, V.~Barger, H.~Serce and X.~Tata,
  Phys.\ Rev.\ D {\bf 94} (2016) no.11,  115017
  doi:10.1103/PhysRevD.94.115017
  [arXiv:1610.06205 [hep-ph]].


\bibitem{ILCSOFT} “Ilcsoft home page,” (2016).

\bibitem{Schulte:1999tx}
  D.~Schulte,
  CERN-PS-99-014-LP, CERN-PS-99-14-LP, CLIC-NOTE-387, CERN-CLIC-NOTE-387.

\bibitem{TDR} T.~Behnke {\it et al.}    
The International Linear Collider Technical Design Report - Volume 1: Executive Summary" (2013),
arXiv:1306.6327 [physics.acc-ph].

\bibitem{Whizard}  W.~Kilian, T.~Ohl, and J.~Reuter,
Eur. \ Phys.\  J {C71} (2011) 1742
arXiv:0708.4233 [hep-ph].

 
\bibitem{ILD} H.~Abramowicz {\it et al.} 
  ``The International Linear Collider Technical Design Report - Volume 4: Detectors,''(2013), arXiv:1306.6329 [physics.ins-det].


\bibitem{PYTHIA} T.~Sjostrand, L.~Lonnblad, and S.~Mrenna, 
“PYTHIA 6.2: Physics and manual,” (2001), arXiv:hep-ph/0108264 [hep-ph].

\bibitem{ISAJET} F. Paige, S. Protopopescu, H.~Baer and X. Tata, 
"ISAJET 7.85 A Monte Carlo Event Generator for $pp$,$\overline{p}p$, and $e^{+}e^{−}$ Reactions"

\bibitem{Mokka} P. Mora de Freitas and H. Videau, 
in Linear colliders. Proceedings, International Workshop on physics and experiments with future electron-positron linear
colliders, LCWS 2002, Seogwipo, Jeju Island, Korea, August 26-30, 2002 (2002) pp. 623–627.

\bibitem{Marlin} F. Gaede,
Nucl.\  Instrum. \ Meth. {A559} (2006)  177–180 

\bibitem{Pandora} M. A. Thomson, 
 Nucl.\  Instrum. \ {A611} (2009) 25-40,  arXiv:0907.3577 [physics.ins-det].

\bibitem{Tanabashi:2018oca}
  M.~Tanabashi {\it et al.} [Particle Data Group],
  Phys.\ Rev.\ D {\bf 98} (2018) no.3,  030001.
  doi:10.1103/PhysRevD.98.030001

\bibitem{LiteHiggsinos} M.~Berggren  {\it et al.} 
Eur. \ Phys.\  J {C73} (2013) 2660

\bibitem{Catani:1993hr} 
  S.~Catani, Y.~L.~Dokshitzer, M.~H.~Seymour and B.~R.~Webber,
  Nucl.\ Phys.\ B {\bf 406}, 187 (1993).
  doi:10.1016/0550-3213(93)90166-M

\bibitem{Ellis:1993tq} 
  S.~D.~Ellis and D.~E.~Soper,
  Phys.\ Rev.\ D {\bf 48}, 3160 (1993)
  doi:10.1103/PhysRevD.48.3160
  [hep-ph/9305266].

\bibitem{Catani:1991hj} 
  S.~Catani, Y.~L.~Dokshitzer, M.~Olsson, G.~Turnock and B.~R.~Webber,
  Phys.\ Lett.\ B {\bf 269}, 432 (1991).
  doi:10.1016/0370-2693(91)90196-W

\bibitem{Abramowicz:2010bg} 
  H.~Abramowicz {\it et al.},
  JINST {\bf 5}, P12002 (2010)
  doi:10.1088/1748-0221/5/12/P12002
  [arXiv:1009.2433 [physics.ins-det]].






\bibitem{Baer:2016ucr}
  H.~Baer, V.~Barger and H.~Serce,
  Phys.\ Rev.\ D {\bf 94} (2016) no.11,  115019
  doi:10.1103/PhysRevD.94.115019
  [arXiv:1609.06735 [hep-ph]].

\bibitem{bib:micromegas}
G.~Belanger, F.~Boudjema, A.~Pukhov and A.~Semenov,
  Comput.\ Phys.\ Commun.\  {\bf 149}, 103 (2002)
  [hep-ph/0112278].
  
  \bibitem{bib:feynhiggs}
  M.~Frank, T.~Hahn, S.~Heinemeyer, W.~Hollik, H.~Rzehak and G.~Weiglein,
  JHEP {\bf 0702} (2007) 047
  [hep-ph/0611326]. \\
  G.~Degrassi, S.~Heinemeyer, W.~Hollik, P.~Slavich and G.~Weiglein,
  Eur.\ Phys.\ J.\ C {\bf 28} (2003) 133
  [hep-ph/0212020].\\
  S.~Heinemeyer, W.~Hollik and G.~Weiglein,
  Eur.\ Phys.\ J.\ C {\bf 9} (1999) 343
  [hep-ph/9812472].\\
  S.~Heinemeyer, W.~Hollik and G.~Weiglein,
  Comput.\ Phys.\ Commun.\  {\bf 124} (2000) 76
  [hep-ph/9812320].
  
\bibitem{Caiazza:2018suv}
  S.~Caiazza,
  ``The $GridGEM module$ for the ILD TPC and A new algorithm for kinematic edge determination,'' PhD Thesis, Universit\"at Hamburg, 2018, 
  doi:10.3204/PUBDB-2018-05289

\bibitem{Porod:2003um}
  W.~Porod,
  Comput.\ Phys.\ Commun.\  {\bf 153}, 275 (2003)
  [arXiv:hep-ph/0301101].
  

\bibitem{Nguyen:2012rx}
  N.~Nguyen, D.~Horns and T.~Bringmann,
  ``AstroFit: An Interface Program for Exploring Complementarity in Dark Matter Research,''
  doi:10.1142/9789814405072\_0154
  arXiv:1202.1385 [astro-ph.HE].
  
\bibitem{CMS:2016rqf}
  CMS Collaboration [CMS Collaboration],
  CMS-PAS-HIG-15-004.

\bibitem{Aaboud:2016dig}
  M.~Aaboud {\it et al.} [ATLAS Collaboration],
  Phys.\ Lett.\ B {\bf 759} (2016) 555
  doi:10.1016/j.physletb.2016.06.017
  [arXiv:1603.09203 [hep-ex]].
  
\bibitem{Aaboud:2016cre}
  M.~Aaboud {\it et al.} [ATLAS Collaboration],
  Eur.\ Phys.\ J.\ C {\bf 76} (2016) no.11,  585
  doi:10.1140/epjc/s10052-016-4400-6
  [arXiv:1608.00890 [hep-ex]].

\bibitem{ATLAS:2017vjw}
  The ATLAS collaboration [ATLAS Collaboration],
  ATLAS-CONF-2017-021.

\bibitem{ATLAS:2017msx}
  The ATLAS collaboration [ATLAS Collaboration],
  ATLAS-CONF-2017-037.

  
  
\bibitem{Seidel:2013sqa}
  K.~Seidel, F.~Simon, M.~Tesar and S.~Poss,
  Eur.\ Phys.\ J.\ C {\bf 73} (2013) no.8,  2530
  doi:10.1140/epjc/s10052-013-2530-7
  [arXiv:1303.3758 [hep-ex]].

\bibitem{Berggren:2013vfa}
  M.~Berggren, F.~Br\"ummer, J.~List, G.~Moortgat-Pick, T.~Robens, K.~Rolbiecki and H.~Sert,
  Eur.\ Phys.\ J.\ C {\bf 73} (2013) no.12,  2660
  doi:10.1140/epjc/s10052-013-2660-y
  [arXiv:1307.3566 [hep-ph]].
  
\bibitem{Bechtle:2004pc}
  P.~Bechtle, K.~Desch and P.~Wienemann,
  Comput.\ Phys.\ Commun.\  {\bf 174} (2006) 47
  doi:10.1016/j.cpc.2005.09.002
  [hep-ph/0412012].
  
\bibitem{Baer:2004fu}
  H.~Baer, A.~Mustafayev, S.~Profumo, A.~Belyaev and X.~Tata,
  Phys.\ Rev.\ D {\bf 71} (2005) 095008
  doi:10.1103/PhysRevD.71.095008
  [hep-ph/0412059].

\bibitem{Lehtinen:PhDThesis}  
S.-L.~Lehtinen, ``Supersymmetry parameter determination at the International Linear Collider,'' PhD Thesis, Hamburg University, 2018. 
  
  \bibitem{Fujii:2015jha}
  K.~Fujii {\it et al.},
  ``Physics Case for the International Linear Collider,''
  arXiv:1506.05992 [hep-ex].
  
  \bibitem{Barklow:2015tja}
  T.~Barklow, J.~Brau, K.~Fujii, J.~Gao, J.~List, N.~Walker and K.~Yokoya,
  ``ILC Operating Scenarios,''
  arXiv:1506.07830 [hep-ex].

\bibitem{Yan:2016xyx}
  J.~Yan, S.~Watanuki, K.~Fujii, A.~Ishikawa, D.~Jeans, J.~Strube, J.~Tian and H.~Yamamoto,
  Phys.\ Rev.\ D {\bf 94} (2016) no.11,  113002
  doi:10.1103/PhysRevD.94.113002
  [arXiv:1604.07524 [hep-ex]].
  
  
\bibitem{Baltz:2006fm}
  E.~A.~Baltz, M.~Battaglia, M.~E.~Peskin and T.~Wizansky,
  Phys.\ Rev.\ D {\bf 74} (2006) 103521
  doi:10.1103/PhysRevD.74.103521
  [hep-ph/0602187].

\bibitem{Baer:2018rhs}
  H.~Baer, V.~Barger, D.~Sengupta and X.~Tata,
  Eur.\ Phys.\ J.\ C {\bf 78} (2018) no.10,  838
  doi:10.1140/epjc/s10052-018-6306-y
  [arXiv:1803.11210 [hep-ph]].

\bibitem{Bae:2014rfa}
  K.~J.~Bae, H.~Baer, A.~Lessa and H.~Serce,
  JCAP {\bf 1410} (2014) no.10,  082
  doi:10.1088/1475-7516/2014/10/082
  [arXiv:1406.4138 [hep-ph]].

\bibitem{Baer:2002fv}
  H.~Baer, C.~Balazs and A.~Belyaev,
  JHEP {\bf 0203} (2002) 042.

\bibitem{Baer:2004qq}
  H.~Baer, A.~Belyaev, T.~Krupovnickas and J.~O'Farrill,
  JCAP {\bf 0408} (2004) 005
  doi:10.1088/1475-7516/2004/08/005
  [hep-ph/0405210].

\bibitem{Ade:2015xua}
  P.~A.~R.~Ade {\it et al.} [Planck Collaboration],
  Astron.\ Astrophys.\  {\bf 594} (2016) A13
  doi:10.1051/0004-6361/201525830
  [arXiv:1502.01589 [astro-ph.CO]].

\bibitem{Blair:2000gy}
  G.~A.~Blair, W.~Porod and P.~M.~Zerwas,
  Phys.\ Rev.\ D {\bf 63} (2001) 017703
  doi:10.1103/PhysRevD.63.017703
  [hep-ph/0007107].

\bibitem{Blair:2002pg}
  G.~A.~Blair, W.~Porod and P.~M.~Zerwas,
  Eur.\ Phys.\ J.\ C {\bf 27} (2003) 263
  doi:10.1140/epjc/s2002-01117-y
  [hep-ph/0210058].



\bibitem{Bambade:2019fyw}
  P.~Bambade {\it et al.},
  ``The International Linear Collider: A Global Project,''
  arXiv:1903.01629 [hep-ex].



\end{thebibliography}
\end{document}